\newcommand{\ismainfile}{}

\documentclass[aps,prx,onecolumn,nofootinbib,11pt,superscriptaddress]{revtex4-2}
\usepackage{cmap}
\RequirePackage{array}[=2016-10-06]
\RequirePackage{longtable}[=2020-01-07]
\usepackage{setspace}
\input{preamble}

\begin{document}

\title{Magic Quantum Code Surgery}

\author{Kathleen~(Katie)~Chang}
\thanks{These authors contributed equally to this work.}
\affil{Yale University, Department of Physics, New Haven, Connecticut 06520, USA}
\affil{Yale Quantum Institute, Yale University, New Haven, Connecticut 06511, USA}
\email{katie.chang@yale.edu}

\author{Anasuya~Lyons}
\thanks{These authors contributed equally to this work.}
\affil{Department of Physics, Harvard University, Cambridge, Massachusetts 02138, USA}
\email{anasuya\_lyons@g.harvard.edu}

\author{Yuanjie~Ren}
\thanks{These authors contributed equally to this work.}
\affil{Department of Physics, Massachusetts Institute of Technology, Cambridge, Massachusetts 02139, USA}
\email{yuanjie@mit.edu}

\author{Harald~Putterman}
\affil{Department of Physics, Harvard University, Cambridge, Massachusetts 02138, USA}

\author{Nathanan~Tantivasadakarn}
\affil{C. N. Yang Institute for Theoretical Physics, Stony Brook University, Stony Brook, NY 11794, USA}

\author{Victor~V.~Albert}
\affil{Joint Center for Quantum Information and Computer Science, NIST/University of Maryland, College Park, MD 20742}

\author{Benjamin~J.~Brown}
\affil{IBM Quantum, T. J. Watson Research Center, Yorktown Heights, New York 10598, USA}
\affil{IBM Denmark, Sundkrogsgade 11, 2100 Copenhagen, Denmark}

\author{Dominic~J.~Williamson}
\affil{School of Physics, University of Sydney, Sydney, NSW 2006, Australia}

\begin{abstract}
\noindent

Non-Clifford operations required for universal quantum computation can be implemented by measuring logical Clifford operators. We present a generalized framework to fault-tolerantly measure transversal Clifford logical operators in quantum low-density parity check (QLDPC) codes. Our protocol takes any QLDPC code and deforms the code across additional ancillary qubits such that a product of stabilizers measures the desired logical Clifford operator. 
We prove that the deformed, Clifford-stabilized code remains LDPC and that its distance is preserved up to a constant factor independent of the number of physical qubits.
We further prove that the Clifford measurement has a fault distance against data qubit and measurement errors that grows linearly with the distance of the original code. We apply our methods to implement logical non-Clifford gates in algorithmically relevant subroutines and to prepare magic states in quantum stabilizer codes.
\end{abstract}

\maketitle

\tableofcontents

\section{Introduction}\label{sec:intro}\ifdefined\ismainfile
\else
    \documentclass[11pt]{article}

    \input{preamble.tex}
    
    \begin{document}
    \title{Introduction}
    \date{\today}

    \maketitle

\fi

Quantum low-density parity-check (QLDPC) codes provide a promising avenue for scaling up fault-tolerant quantum computers (FTQC). 
High-performance practical instances~\cite{Bravyi2024faulttolerant} and general families with good asymptotic scaling~\cite{panteleev2022asymptotically, dinur2022goodquantumldpccodes, leverrier2022quantum} demonstrate the potential for QLDPC codes with high thresholds and low overheads. 
To achieve universal FTQC, a QLDPC code must be equipped with protocols to perform a set of universal quantum logic gates. 
These protocols should be designed to preserve the threshold and overhead scaling of the QLDPC code. 
The conventional approach to performing universal quantum computation is via a generating set of Clifford gates, Pauli-$Z$ state preparation and measurement, and the preparation of magic states such as the $\ket{T}=\frac{1}{\sqrt{2}}(\ket{0}+\sqrt{i}\ket{1})$ state~\cite{bravyi2005universal}. 

There has been a recent surge of progress in the design of general purpose, practical and efficient Clifford logic gates for QLDPC codes~\cite{cohen2022low,cross2025improvedqldpcsurgerylogical,Ide_2025,williamson2024lowoverheadfaulttolerantquantumcomputation,baspin2025fast,cowtan2025fast,zheng2025high,swaroop2025universaladaptersquantumldpc,Zhang_2025,chang2026constant,tan2025single,li2026transversal}. 
It is equally crucial to develop methods for logical magic state preparation for QLDPC codes. These methods aim to produce encoded magic states such as $\ket{T}$ while preserving the threshold and overhead of the QLDPC code as it is scaled up to suppress errors on the magic states.
Furthermore, it is desirable to have a magic state preparation procedure that is practical in terms of resource requirements and adaptable to fit any desired QLDPC memory. 
The standard approach to magic state preparation is magic state distillation (MSD), which requires logical Clifford operations and the preparation of noisy logical magic states on a chosen QLDPC code~\cite{knill2004faulttolerantpostselectedquantumcomputation, bravyi2005universal, meier2012magicstatedistillationfourqubitcode, Campbell_2017}.
After distilling a high-fidelity magic state on an ancillary code block, this state is teleported into a QLDPC memory via a joint Clifford operation using standard techniques. 
However, implementing MSD in practice comes with a costly overhead, which has motivated the study of alternative methods for high-fidelity magic state preparation.

Another approach is to perform transversal non-Clifford logic gates, such as $CCZ$, directly on a QLDPC memory. 
This is an elegant solution, however, this method requires codes that obey strict conditions which can be hard to satisfy while maintaining encoding efficiency~\cite{Zhu_2025, Breuckmann_2026, jacob2026singleshotdecodingfaulttolerantgates,Menon_2026,bauer2026findingdiagonallogicalgates,sahay2026computingqldpccodesclimbing}.
These transversal gates can be implemented in low-dimensional settings with a just-in-time decoder~\cite{bombin2018transversal,brown2020fault, Scruby_2022}.

Finally, another paradigm is to fault-tolerantly prepare logical magic states in stabilizer codes via measuring transversal Clifford gates~\cite{Chamberland_2019, chamberland2020very,brown2020fault,gidney2024magic,gupta2024encoding,hetenyi2026constant,davydova2025universal,sajith2025noncliffordgatesstabilizercodes,williamson2026fast,huang2025generatinglogicalmagicstates,huang2026hybridlatticesurgerynonclifford,zhu2026nonabelianQLDPCtqftformalism,christos2026nonabelianquantumlowdensityparity}.
Here, fault tolerance means that the minimum number of data qubit and measurement errors required to cause a nontrivial logical error grows with the distance of the code on which the magic state is prepared.
This method is appealing because finding efficient codes with transversal Clifford gates is much easier than finding instances with transversal non-Clifford gates.
Recent work~\cite{davydova2025universal, huang2026hybridlatticesurgerynonclifford, huang2025generatinglogicalmagicstates, Kobayashi_2026, warman2026constantdepthcliffordhierarchygatesnonabelian, manjunath2026universalquantumcomputationgroup} uses this method to prepare magic on topological codes, however the BPT bound~\cite{bravyi2010tradeoffs} guarantees that topological codes scale sub-optimally.
This raises the challenge of constructing highly-efficient QLDPC codes with transversal Clifford gates that are well suited to magic state preparation via measurement. 

\begin{figure}
    \centering
    \includegraphics[width=0.75\linewidth]{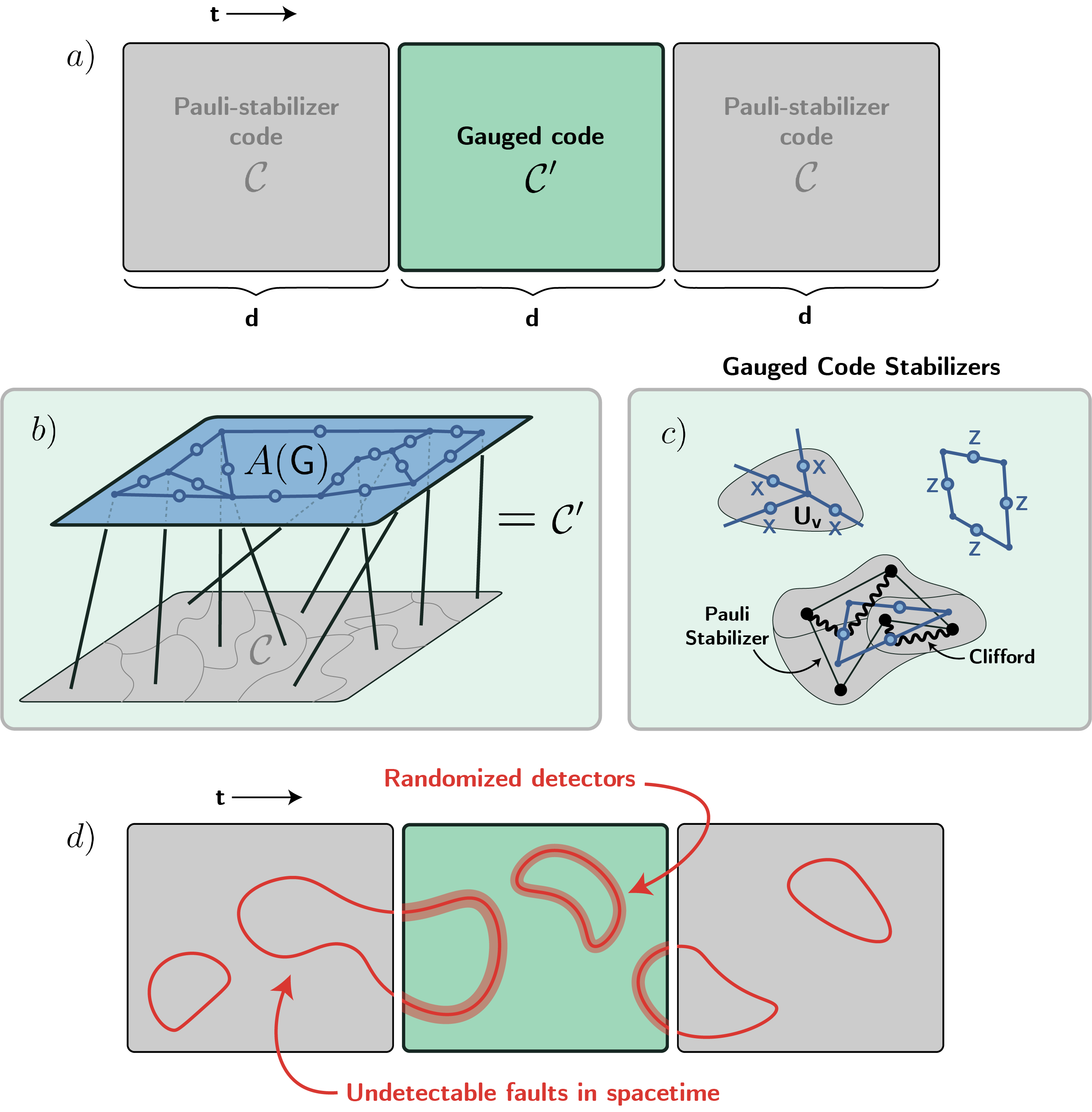}
    \caption{
    (a) An abstract depiction of the spacetime volume for the protocol to measure a Clifford operator. Time flows from left to right. We begin with $d$ rounds of syndrome extraction in a Pauli-stabilized code $\cC$ with distance $d$, where $\cC$ has a transversal Clifford operator $\T$ that we wish to measure. We switch into a gauged code $\cC'$ and measure the stabilizers of the gauged code for $d$ rounds. Then, we ungauge, returning to the original code, now projected onto an eigenspace of $\T$. We finish the gadget by measuring $d$ rounds of the $\cC$. (b) The gauged code $\cC'$. The gauged code is constructed by taking the base code $\cC$ and attaching a graph $A(\graphG)$ (blue) with an ancilla at every edge.
    The connection between the base code $\cC$ and the ancilla graph is abstractly depicted with thick black lines that couple to a patch (``vertex set'') of the base code. Each vertex set is associated to a vertex of the ancilla graph, where an ancilla will be placed to measure the gauged checks. These ``vertex'' ancilla can also be viewed as degrees of freedom in an intermediate, ``symmetry-enriched'' version of the base stabilizer code: measuring these ancilla out yields the final deformed Clifford model on the base code and edge ancillas (Sec.~\ref{sec:gauging-procedure}). (c) The attachment of the ancilla graph yields new deformed Clifford stabilizers. 
    Top left: A ``vertex'' stabilizer, the product of all such stabilizers gives the Clifford operator $\T$ being measured. Top right: A ``cycle'' check arising from a cycle of the ancilla graph. Bottom: a deformed stabilizer of the base code, which picks up Clifford decoration spanning the original code degrees of freedom and the ancilla system (Section~\ref{sec:gauging-procedure}).
    (d) We prove that the spacetime protocol in (a) for qubit codes and Clifford operators that satisfy $\T^2=\id$ has a fault distance against data qubit and measurement errors of at least a constant times $d$ (Section~\ref{sec:spacetime-distance}). In the gauged code, certain faults may randomize some detectors in addition to deterministically flipping others. The key observation is that spacetime faults which flip no detectors (represented by red loops in spacetime) only cause detectors in a constant radius around their support to return random outcomes (thickened red outline). }
    \label{fig:summary}
\end{figure}
\subsection{Summary of main results}

In this work, we introduce a method to fault-tolerantly prepare logical magic states in QLDPC codes via measuring transversal Clifford gates. 
The particular QLDPC surgery method we develop is a variant of the gauging logical measurement from Ref.~\cite{williamson2024lowoverheadfaulttolerantquantumcomputation}. 
First, we present a general construction to gauge or measure a transversal Clifford gate of order $p$, $\T^p=\id$, where $p\ge 2$.
This gauging is very adaptable in that the input code can be any \textit{modular} qudit QLDPC code with dimension $q$.
Our construction takes an approach that mirrors generalized code surgery; we introduce auxiliary \textit{gauge qudits} in the $\ket{0}$ state and perform local measurements that deform the initial code into a \textit{gauged code} that contains deformed stabilizers of the original input code, and new stabilizers that include $\T$ in the group they generate. As a result, the measurement of the stabilizers in this gauged code also measures $\T$. 
The code after gauging is a gauged code with a non-Abelian stabilizer group, generalizing the nilpotent and solvable quantum double models~\cite{Kitaev_2003,Tantivasadakarn_2024} that appear when a Clifford logical measurement is performed on a 2D topological code via gauging~\cite{davydova2025universal, sajith2025noncliffordgatesstabilizercodes}.
Our method is efficient in that the number of qudits of this gauged code can be made within a multiplicative logarithmic factor of the block-length of the original code.

\begin{theorem}
    \emph{Space overhead of the deformation (informal; see Section~\ref{sec:defining-gauging-graph}).} Take any QLDPC modular $q$-ary qudit code ($q\ge 2$) encoded in $n$ physical qudits with a transversal Clifford gate $\T$ such that $\T^p=\id$ where $p\ge 2$.  The gauged, QLDPC code that includes $\T$ in its stabilizer group constructed via the procedure defined in Section~\ref{sec:gauging-procedure} 
    has $O(n\log(n))$ qudits for a family of such codes with growing $n$.
\end{theorem}

\noindent In addition to this general gauging recipe, our main theoretical contributions are two-fold.

First, we prove that if the distance of the original QLDPC qudit code is $d$, then the distance of the gauged code is at least $d$ up to a constant factor.
This result applies to any QLDPC code on qudits of any dimension $q$, and transversal gates of any order $p$.

\begin{theorem}
    \emph{Space distance preservation (informal version of Thm.~\ref{thm:space-distance}).}
    Let $\cC$ be a QLDPC modular $q$-ary qudit code, with $q\ge 2$ and distance $d$, with a transversal Clifford gate $\T$. 
    Let $\cC'$ be the gauged code obtained from the procedure in Section~\ref{subsec:gauging-protocol}, using a gauging graph that satisfies the criteria in Definition~\ref{def:graph-desiderata}.
    Then, $\cC'$ is QLDPC and its distance is at least a constant fraction of the distance of $\cC$:
    \[
    d_{\mathrm{g}}\geq \frac{d}{\nu}
    \]
    for some constant $\nu\ge 1$ depending only on the local structure of the gauging construction.
\end{theorem}

\noindent
Along with the distance bound, Section~\ref{sec:space-distance} shows that the gauged code is isomorphic to the symmetric $\T = +1$ subspace of the original code.

While the gauging logical measurement applies to general operators and QLDPC codes~\cite{williamson2024lowoverheadfaulttolerantquantumcomputation}, we focus on the fault tolerance of gauging Clifford operators on qubit stabilizer codes for fault-tolerant magic state preparation.
In practice, the measurement of the Clifford operator $\T$ must occur in the presence of measurement errors. 
To achieve fault-tolerance to measurement and data qubit errors, we must repeat rounds of syndrome extraction in the gauged code at least $d$ times, just as in regular code surgery. 
We define the full gauging measurement procedure as follows: measure stabilizers of the original LDPC code for $d$ rounds, then measure the stabilizers of the gauged code for $d$ noisy rounds, and finally, measure stabilizers of the original code for $d$ rounds.   
Crucially, our results establish that gauging measurement can be made fault-tolerant for the measurement of order-two symmetries, $\T^2 = \id$. This condition is easily satisfied by most Clifford symmetries we are interested in gauging for magic state preparation, such as Hadamard, $\CNOT$, and $\SWAP$.
By fault-tolerant, we mean that a logical fault must contain at least $d/\gamma$ such errors, where $\gamma>0$ is a constant independent of the gauged code parameters.
We prove the following result.
\begin{theorem}
    \emph{Spacetime distance of gauging (informal version of Thm~\ref{thm:spacetime-distance}). Let $\cC$ be a QLDPC qubit stabilizer code with parameters $[[n,k,d]]$ and a transversal Clifford gate $\T$ satisfying $\T^2=\id$.
    The fault-tolerant protocol for measuring $\T$ defined in Section~\ref{sec:spacetime-distance} has adversarial spacetime fault distance 
    \[ d_{\mathrm{f}}\ge \frac{d}{\gamma}
    \]
    where $\gamma\ge 1$ is the a factor independent of code parameters defined in Thm~\ref{thm:spacetime-distance}.
    Equivalently, any undetectable logical failure requires at least $d/\gamma$ elementary faults, including Pauli data qubit errors and measurement faults.}
\end{theorem}

\noindent Establishing fault-tolerance of the spacetime gauging procedure is more complicated for a logical Clifford operator than for a logical Pauli operator for several reasons.

First, by construction, the gauged code must contain non-Pauli stabilizer generators.
As a result, Pauli errors on data qubits neither commute nor anticommute with some neighboring stabilizers.
Therefore, the measurement of these stabilizers is inherently non-deterministic.
Furthermore, the stabilizer generators in the gauged code are only guaranteed to commute within the code space. 
This means that as data qubit errors accumulate in spacetime, detectors of the spacetime volume become non-deterministic and are unusable for detecting errors. 
In this setting, we show that the errors in the spacetime volume of the gauged code can be cleaned in a well-behaved manner due to the structure of the QLDPC Clifford stabilizer group. Specifically, the gauged code is \emph{nilpotent}: all non-Abelian syndromes fuse to the vacuum or to an Abelian syndrome. 
We leave fault-tolerance of gauging $\mathbb{Z}_p$ where $p>2$ Clifford operators in the presence of measurement errors to future work.

In Section~\ref{sec:examples} we construct concrete examples of the gauging procedure.
We gauge a transversal symmetry $W^{\otimes n}$, where $W=XS$, on a \textit{self-dual bivariate bicycle} code constructed by concatenating a bivariate bicycle code with a $[[4,2,2]]$ code as the inner code; we gauge the transversal $\CNOT$ between two bivariate bicycle codes; and we gauge the transversal $XS$ and $HS$ gates on the color code, along with charge conjugation on the qutrit version of the $[[144,12,12]]$ code.
As an example of gauging two symmetries that do not act on all qubits of the code, we gauge a transversal $\CZ$ between two stacks of $3D$ surface codes; finally, we present an algorithm for constructing the gauging graph and the gauged stabilizer group for a fold-transversal $\mathbb{Z}_2$ symmetry on a $2$-dimensional hypergraph product (HGP) code.

An important aspect of this work is constructing concrete examples where measuring Clifford operators on high-rate codes is computationally useful. This is highly nontrivial, due to the difficulty in addressing logical qubits within a high-rate code block. 
In Section~\ref{sec:applicationsA} we introduce an algorithm that gauges the transversal $\CNOT$ between two identical, CSS stabilizer codeblocks to obtain a product state of $k$ copies of the $+1$ eigenstate of $\CNOT$, $\ket{+\CX}^{\otimes k}$.
These eigenstates can subsequently be converted into $O(k)$ copies of Toffoli states, which are the resource state for performing Toffoli gates.
This method is interesting in that it allows the generation of high-fidelity magic states in any CSS LDPC code without distillation or criteria of the LDPC code, such as having a triple intersection point or local multiplication property~\cite{Zhu_2025}.
In Section~\ref{sec:ctrl-t-gadget} we introduce a gadget that uses a sequence of measurements, including a single Clifford measurement, such that we obtain the unitary gate controlled-$U$ where $U$ is a transversal Clifford gate of some QLDPC code with $U^2=\pm 1 $. This gadget is particularly useful, for instance, if we choose $U=\SWAP$, which is readily available between any two identical code blocks, as this enables us to implement a $\SWAP$ network on $N$ registers from requiring $O(N)$ $T$ states to $O(\log_2(N))$ Clifford measurements. 
Using the same gadget, we can also reduce the logical circuit depth of an algorithm that solves the \textit{hidden cut problem}~\cite{bouland2024statehiddensubgroupproblem} by a factor of $\mathcal{O}(n/\epsilon^2)$, where $n$ is the number of qubits of state $\ket{\psi}$, which is $\epsilon$-far from a product state across an unknown bipartition. 
The aim of the hidden cut problem is to find this bipartition given some number of copies of $\ket{\psi}$.
These applications of $\CSWAP$ are an exciting starting point for exploring various structured circuits that are easily compiled into Clifford measurements.
Our results lay down an alternative road to practical and efficient universal FTQC with QLDPC codes. 
For example, our logical magic state preparation protocol via gauging $\CNOT$ can be directly integrated into the \textit{Extractor} architecture~\cite{he2025extractorsqldpcarchitecturesefficient} to produce a fixed-connectivity QLDPC architecture for universal FTQC completely comprised of augmented QLDPC code blocks.

\subsection{Comparison to existing methods for magic state preparation}

Our results lay down an alternative route to scalable logical magic state preparation in QLDPC codes that offers several advantageous features. First, our scheme is flexible as it only requires the existence of a transversal Clifford gate and hence is applicable to any CSS code. 
Our procedure can also be applied in parallel to Clifford gates with sparse and locally commuting overlap to perform more than one Clifford measurement at a time. 
Second, our scheme is scalable with the size of the code family to which it is applied. The spacetime protocol thus generated relies exclusively on sparse checks and detectors. The fault-distance of our procedure grows linearly with the distance of the target code family. In principle, the distance can be grown a single step at a time. 
The growing distance, combined with the sparse checks, means that our scheme is expected to exhibit a quantum error correcting threshold. That is, a physical error rate below which the logical error rate is suppressed exponentially as a linear function of the distance. Third, our scheme maintains a low overhead as it requires $O(n \log n)$ ancilla qubits~\cite{williamson2024lowoverheadfaulttolerantquantumcomputation,williamson2026fast, yuan2026parsimonious}.

Importantly, in contrast to methods such as magic state cultivation and distillation,  our procedure does not require post-selection. 
This translates directly into a lower qubit footprint for the production of a fixed number of magic states per round when compared to a similar procedure with a lower yield. The above features point to the potential utility of our scheme in the early to mid term fault tolerant era, when physical error rates are relatively close to threshold but the target logical error rate is lower than can be easily achieved through post-selection due to constraints on the number of physical qubits available. 

Magic state distillation is the most prominent approach for the preparation of logical magic states. It relies on the ability to inject noisy physical magic states into a code block upon which logical Pauli stabilizer measurements can be performed. Both the injection and the logical stabilizer measurement stages require post-selection and consequently the yield may be low, although it approaches a constant at the cost of considerable overhead. While the qubit overhead of recently discovered schemes is constant~\cite{wills2024constantoverheadmagicstatedistillation}, for the most widely considered schemes it is polylogarithmic. Distillation also requires multiple logical Pauli measurements, which is expected to incur an additional overhead of $O(n \log n)$ ancilla qubits for $O(d)$ rounds for a generic QLDPC code. There is a further overhead required for logical measurements that cannot be made simultaneously due to high overlap. In comparison to magic state distillation, our scheme provides an alternative path to logical magic states that provides higher yields, a potentially lower qubit footprint, and more fine-grained control over instance sizes. 

Magic state cultivation~\cite{Chamberland_2019,chamberland2020very,gidney2024magic,vaknin2026efficientmagicstatecultivation} is a widely considered approach to preparing logical magic states in the early fault tolerant era.
Cultivation is based on measuring logical Clifford operators to prepare magic states, and relies on post-selection to provide fault tolerance. Consequently, the yield of cultivation exponentially approaches zero making it insufficient for a stand alone scalable solution. In addition, while cultivation can still use a fault-tolerantly prepared cat state to perform measurements on codes with higher distances, it is unclear if the post-selection rates will make this a scalable strategy. In comparison to cultivation, our gauging Clifford measurement scheme has a constant yield that can approach 1 exponentially fast in some cases. Our scheme is scalable and achieves a growing distance while maintaining QLDPC checks and is expected to achieve exponential suppression of the logical error rate as a linear function of the distance for physical error rates below a threshold value~\cite{delafuente2026highthresholddecodingnonpaulicodes}. 

Transversal non-Clifford gates are another approach to preparing logical magic states, or performing non-Clifford operations. They have the advantage of being fault-tolerant with a constant time overhead, no qubit overhead, and yield 1. 
This comes at the cost of requiring complicated code structure that has proven hard to achieve simultaneously with good code properties~\cite{Zhu_2025, Breuckmann_2026, jacob2026singleshotdecodingfaulttolerantgates,Menon_2026,bauer2026findingdiagonallogicalgates,sahay2026computingqldpccodesclimbing}. 
In particular, to take advantage of the constant time overhead of a transversal non-Clifford gate for magic state preparation requires the ability to do single-shot state preparation. Transversal gates on high-rate QLDPC codes also suffer from a targetability problem on the logical qubits. 
In contrast to transversal gates, our approach does not require special code structure for CSS codes, as we can always find a transversal Clifford operator to measure on a CSS stabilizer code. We achieve many of the attractive features of transversal gates, including yield 1, and a low qubit overhead. 

The gauging Clifford measurement was previously considered for 2D topological codes, as part of a general family of fault-tolerant non-Clifford operations~\cite{Bauer_2026}. The advantage of restricting to 2D topological codes is the relative ease of implementation and the addressability of logical qubits~\cite{Fowler_2012}. This comes at the cost of low rates and relative distances due to the constraint of local checks. In comparison to the topological protocols, our work applies to more general QLDPC codes and can be integrated directly into a QLDPC architecture without requiring additional topological code blocks. 
Gauging in high-rate codes means that one can leverage the efficient memory of the QLDPC codes while preparing magic states.

For example, the gauged QLDPC code can be used to simultaneously store computational qubits during the Clifford gauging measurement (Section~\ref{sec:applications}). In particular, even if one $\ket{+\CX}$ magic state is being generated per Clifford measurement, the extra space-overhead used to encode more logical qubits may be utilized in logical compilation. 
Furthermore, we can consider a QLDPC code architecture, in which our magic state factory consists of gauging $\CZ$ on copies of the surface code as in~\cite{brown2020fault,delafuente2026highthresholddecodingnonpaulicodes}. 
Suppose, for example, that a logical compilation requires $K$ magic states of distance $d$ per logical cycle ($d$ syndrome extraction rounds), which would require $O(K)$ surface codes each of size $d\times d$.
Then, in comparison, using the $\CNOT$ gauging scheme on good QLDPC codes~\cite{panteleev2022asymptotically,leverrier2022quantum,dinur2022goodquantumldpccodes} to produce $K$ $\ket{+\CX}$ states per logical cycle has a space overhead of $O(K\times d\log(d))$, which is a multiplicative factor of $d/\log d$ more efficient in spacetime overhead.

\subsection{Related work}

At the heart of this work is an in-depth development of the gauging logical measurement introduced in Ref.~\cite{williamson2024lowoverheadfaulttolerantquantumcomputation} for Clifford operators on stabilizer codes. This follows earlier work on implementing gauging via an adaptive quantum circuit~\cite{Williamson_2021,Tantivasadakarn_2024,Tantivasadakarn_2023}. The procedure introduced in Ref.~\cite{williamson2024lowoverheadfaulttolerantquantumcomputation} already applied to general non-Pauli operators and codes. However, fault-tolerance was only shown for the measurement of Pauli logical operators on stabilizer codes. The simultaneous work in Ref.~\cite{Ide_2025} found a similar result restricted to CSS codes. 
The proof of spacetime fault tolerance and the construction of explicit applications for Clifford measurements on QLDPC codes set this work apart from the existing literature.

Prior to this work, Ref.~\cite{davydova2025universal} introduced a fault-tolerant procedure for measuring Clifford operators on topological stabilizer codes in two dimensions via gauging. The fault-tolerance of this procedure relied heavily on the topological properties of two-dimensional topological codes and the anyonic syndromes they support. The results shown here were not derived via a simple extension of the techniques in Ref.~\cite{davydova2025universal}. On the other hand, our constructions and examples provide QLDPC analogs of non-Abelian anyon models such as solvable twisted quantum double models. 
A series of follow-up works developed an array of generalized logic gates based on non-Abelian anyons without proving fault-tolerance~\cite{huang2025generatinglogicalmagicstates,sajith2025noncliffordgatesstabilizercodes}. Further follow-up works have performed numerical simulations of decoding non-Abelian anyons~\cite{delafuente2026highthresholddecodingnonpaulicodes,jing-intrinsic-heralding, jing2026integerlinearprogrammingdecoder}. Extending these ideas to design fault-tolerant gates for QLDPC codes is an interesting future direction. 

Independent of this work, Ref.~\cite{zhu2026nonabelianQLDPCtqftformalism} introduced a procedure to measure Clifford operators via gauging on a large class of QLDPC codes derived from a generalization of the code-to-manifold construction introduced by Freedman and Hastings~\cite{freedman2021buildingmanifoldsquantumcodes}. These codes retain a topological structure that allows arguments similar to those used for two dimensional topological codes to be applied. Ref.~\cite{zhu2026nonabelianQLDPCtqftformalism} argued for a static code distance, but did not establish spacetime fault-tolerance of the logical measurement procedure. They also considered Clifford gates derived from higher-form topological symmetries, and used this to find representative Clifford operators that could be gauged in parallel. Here we have focused on 0-form Clifford gates, however our gauging measurement can as well be applied to any representative of a higher-form Clifford gate and to hypergraphs with no local elements in their kernel. 

In a separate concurrent work, Ref.~\cite{williamson2026fast} introduced a notion of higher-form transversal gates for QLDPC codes together with a fast and fault-tolerant logical measurement procedure via gauging such gates. This procedure is distinct from ours in that it focuses on higher-form transversal gates that provide sufficient meta checks to ensure the gauging procedure is fault tolerant when performed for a single time step. It is a much stricter requirement for a QLDPC code to support a higher-form Clifford gate as compared to a standard 0-form Clifford gate, and so our results here are more generic and apply to all CSS codes. This leads to structurally different fault-tolerance properties, as the deformed checks of the intermediate non-Pauli code do not need to be measured in that case. It is an interesting future direction to combine our techniques with those of Ref.~\cite{williamson2026fast} to develop fault-tolerant higher-form gauging measurements that require more than a constant number of rounds of syndrome extraction in the deformed code, generalizing the partial block reading results derived for Pauli codes in Ref.~\cite{cowtan2025fast}. 

In another concurrent work, Ref.~\cite{christos2026nonabelianquantumlowdensityparity} considered non-Pauli QLDPC codes that arise from gauging Clifford logical operators. They present two distinct gauging methods: a homological method and a graph cluster-state based method. We use the same graph based method, previously introduced in Refs.~\cite{Williamson_2021,Tantivasadakarn_2024,verresen2022efficientlypreparingschrodingerscat,Tantivasadakarn_2023} in a topological setting. Reference~\cite{christos2026nonabelianquantumlowdensityparity} do not prove fault-tolerance of the graph gauging method and neither do they provide specific details of the Clifford measurement protocol implemented via gauging or give particular practical applications of this method. As discussed above, the original gauging measurement procedure in Ref.~\cite{williamson2024lowoverheadfaulttolerantquantumcomputation} was already applicable to non-Pauli gates and codes. Deriving the form of the deformed codes when a non-Pauli gate is gauged follows from a standard recipe which we have used here, see for example the explanation in Ref.~\cite{Bulmash_2019,Prem_2019,verresen2022efficientlypreparingschrodingerscat}.

\subsection{Organization of the paper}

The remainder of this paper is organized as follows.
Section~\ref{sec:gauging-procedure} presents the procedure for measuring logical Cliffords, and derives the stabilizers of the intermediate gauged code.

Section~\ref{sec:space-distance} establishes the properties of the gauged code, namely its isomorphism with the symmetric $\T= +1$ subspace of the original code, and the fact that code distance is preserved under our gauging protocol.
Section~\ref{sec:spacetime-distance} addresses fault tolerance of the gauging procedure, proving a lower bound on the spacetime fault distance of the gauging measurement of a $\mathbb{Z}_2$ symmetry $\T$ of a qubit CSS code.
Section~\ref{sec:examples} presents examples of gauging on bivariate bicycle codes, the color code, and hypergraph product codes.
Section~\ref{sec:applications} presents applications: (1) a scheme to prepare multiple $\tof$ states on a single CSS QLDPC codeblock using our fault-tolerant gauging protocol, and (2) a gadget that uses a single Clifford measurement of transversal $\SWAP$ to perform many $\CSWAP$ operations in parallel.
We conclude with a discussion, and some more detailed derivations are collected in Appendix~\ref{app:gauging-procedure-proofs}.

\subsubsection*{Reader's Guide }

Section \ref{sec:gauging-procedure} is a pre-requisite for all following sections. However, a reader not interested in following the proof of the fault distance for the protocol can skip directly from Section \ref{sec:gauging-procedure} to Section~\ref{sec:examples} to understand the examples of the construction, or Section~\ref{sec:applications} to see applications of Clifford measurement in magic state preparation and algorithmic subroutines.
Sections \ref{sec:space-distance} and Section \ref{sec:spacetime-distance} may be read independently.

\subsection*{Acknowledgments}

Part of this work was performed while AL and KC were summer interns at IBM Quantum.
KC and AL thank Mackenzie Shaw, Louis Golowich, Adam Wills, Zhiyang He, Christopher Pattison and Alfred for encouraging discussions and support. 
KC and BJB thank Sergey Bravyi for bringing to our attention an application of our gadgets to the hidden cut problem.
KC thanks Shifan Xu for enlightening discussions on the SWAP network circuit.
Part of this work was done during the ``Noise-robust Phases of Quantum Matter'' program at the Kavli Institute for Theoretical Physics (KITP).
BJB is grateful for the hospitality of the Niels Bohr Institute at the University of Copenhagen.
DJW is supported by the Australian Research Council Discovery Early Career Research Award (DE220100625). 
AL acknowledges the NSF GRFP (Grant No. DGE 2140743) and the Kavli Institute for Theoretical Physics Graduate Fellowship Program, supported in part by Grant No. NSF PHY-1748958 to the Kavli Institute for Theoretical Physics (KITP), the Heising-Simons Foundation, and the Simons Foundation (216179, LB).
YR was supported by the U.S.
Department of Energy, Office of Science, National Quantum Information Science Research Centers, Quantum Systems Accelerator, under Grant
number DOE DE-SC0012704.
HP was supported by the Department of Defense through the National Defense Science and Engineering Graduate (NDSEG) Fellowship Program.

\subsection*{AI statement}

The scientific content of this work was conceived by the authors without the use of generative AI. 
Generative AI tools (ChatGPT 5.6, GPT-6 Sol, GPT-6 Astra, Claude Fable 5) were used during the final stages of the manuscript preparation to verify proofs, and for language and presentation edits. AI assistance for basic spellchecking and grammar was provided by Grammarly. 
All
content, claims, and conclusions have been reviewed and
verified by the authors to ensure accuracy and originality. 
Certain products, commercial and otherwise, are
mentioned in this publication. These mentions are for
informational purposes only, and do not imply recommendation or endorsement by NIST.

\ifdefined\ismainfile
\else
    \end{document}
\fi

\subsection*{Recurring Notation}

We list notation that is commonly used throughout the paper below. More specialized notation is specified in the corresponding section.
Throughout, a \emph{constant} is a number independent of $n$, $k$, and $d$. It may depend on fixed local dimensions $q$ and $p$, uniform upper bounds on stabilizer weights, qudit degrees, vertex-set sizes, graph degrees, cycle weights and edge incidences, and decoration-path lengths, as well as a uniform positive lower bound on the relevant expansion. These bounds are held fixed as the code family grows.

\begin{longtable}{p{0.25\linewidth}p{0.71\linewidth}}
    \toprule
    Notation & Denotes \\
        \midrule
    \endhead
        $\cC $ & Original QLDPC Pauli stabilizer code over qudits with dimension $q$, with $n$ physical qudits, $k$ logical qudits, and distance $d$. The $n$ qudits are referred to as matter qudits.\\
        $\cC'$ & Gauged code obtained by deforming $\cC$.
        \\
        $\T = \bigotimes_{v\in \vertexV}\T_v, \quad \T^p = \id$ & Transversal logical Clifford operator $\T$. $\T_v$ is its constant-sized, local operators and $p$ is the order.
        \\
        $p$ & Order of $\T$ and the dimension of the gauge qudits.
        \\
        $\mathcal{M}$ & Set of matter qudits in the original code.
        \\
        $\graphG = (\vertexV, \edgeE, \faceF)$ & Gauging $2$-complex or gauging graph. Vertices correspond to local operators $\T_v$. Edges correspond to ancilla or gauge qudits. Faces correspond with cycle stabilizers.
        \\
        $\Sigma : \mathcal{M} \to \vertexV$ & Assignment of matter qudits in the original code to vertex sets. $\Sigma^{-1}(v)$ is the constant-sized set of qudits associated with $v\in \vertexV$, acted on by $\T_v$.
        \\
        $\Xp^a, \Zp^\chi$ & Generalized Pauli shift and clock operators for the Abelian symmetry $\Agrp$, with $a\in\Agrp$, $\chi\in\widehat{\Agrp}$, and $\Zp^\chi\Xp^a=\omega^\chi(a)\Xp^a\Zp^\chi$. The gauge qudits have dimension $|\Agrp|$.
        \\
        $X,Z$ & Generalized Pauli shift and clock operators on $q$-dimensional matter qudits where $ZX = XZ e^{2\pi i /q}$
        \\
        
        $\ComGrp{A}{B}$ & Group commutator $ABA^\dagger B^\dagger$. 
        \\
        
        $\av$ & Vertex stabilizer. The product of all vertex stabilizers is the transversal Clifford operator $\prod_{v\in \vertexV}\av = \T$.
        \\
        $\bp$ & Cycle stabilizers which are $\mathsf{Z}$-type Pauli operators with support on cycles $ f\in \faceF$.
        \\
        $S_i, \bar{S}_i$  & Stabilizer of $\cC$ and symmetrized stabilizer $\bar{S} = \frac{1}{p}\sum_{a = 0}^{p-1} \T^a S (\T^a)^\dagger$.
        \\

        $\mathbb{S}_i = \mathfrak{C}_2(S_i) \mathrm{\ or\ }\mathfrak{C}(S_i)$ & Decorated counterpart in the gauged code $\cC'$ of a stabilizer $S_i$ of the code $\cC$. $\mathfrak{C}_2$ denotes the decoration map when $p=2$. If $p>2$, the decoration map is $\mathfrak{C}$.\\

        $d_\mathrm{g},d_\mathrm{f}$ & $d_\mathrm{g}$ is the distance of $\cC'$ to data qudit Pauli errors. $d_\mathrm{f}$ denotes the spacetime fault distance of the gauging protocol (Sec.~\ref{sec:spacetime-distance}).\\
        \bottomrule
    \\
    \caption{Essential notation used throughout Sections~\ref{sec:gauging-procedure},~\ref{sec:space-distance},~\ref{sec:spacetime-distance}.}
    \label{tab:notation-lite}\\
\end{longtable}

\section{Procedure for measuring logical Clifford operators}\label{sec:gauging-procedure}\ifdefined\ismainfile
    \def\GaugingProcedureEndDocument{}
\else
    \documentclass[11pt]{article}

    \input{preamble.tex}
    
    \begin{document}
    \title{Procedure for measuring logical Clifford operators}
    \date{\today}

    \maketitle
    \def\GaugingProcedureEndDocument{\end{document}}
\fi

In this section, we describe how to implement a gauging measurement of a Clifford operator~$\T$. Starting with an LDPC code possessing a transversal logical Clifford operator, we can systematically produce a deformed---``gauged''--- LDPC code that contains $\T$ in its stabilizer group. The value of $\T$ can be inferred from the syndrome outcomes of the gauged code, and the system is left in an eigenstate of $\T$ after returning to the original code. We present some explicit examples of the application of our method in Sec.~\ref{sec:examples}. The reader that is not interested in the technical proofs of either the gauged code distance or the fault tolerance of the gauging measurement may proceed directly to Sec.~\ref{sec:examples} after this section.

We begin by enumerating the requirements on the initial code and its transversal gate, followed by the necessary properties of the ancilla system we  use to perform the gauging procedure. Then we write down the explicit gauging circuit (Definition \ref{def:gauging-circuit}). We demonstrate the action of this circuit through an explicit example: gauging $\CZbar$ on the $[[4, 2, 2]]$ code. Then, we derive closed-form expressions for the gauged code stabilizers for two cases: (i) an initial qubit stabilizer code and transversal gate $\T$ such that $\T^2 = \id$, and (ii) a generic qudit code with $\T^p = \id$ for some integer $p$.

\subsection{Gauging measurement protocol}\label{subsec:gauging-protocol}
We start with an LDPC code $\cC = \langle S_i \rangle$ and unitary transversal gate. 

\begin{definition}[Transversal Gate] \label{def:transversal-gate}
    A transversal gate $\T$ acting on a code $\cC$ with a set of qubits $\physq$ is a finite-depth circuit taking the following form:
\begin{equation}
    \T = \otimes_{v \subset \physq} \T_{v}, \quad \bigcup_v v = \physq, \quad v \cap v' = \emptyset
    \label{eq:block-form}
\end{equation}
where each $v$ is an $O(1)$ subset of the code qubits, and $\T_{v}$ is a unitary acting solely on this subset. 
\end{definition}
We note that if we set $k = \max_{v} |v|$, this definition is equivalent to a $k$-fold transversal gate as defined in \cite{chakraborty2026}.

In general, we may have a qudit code with qudits of dimension $q$ and a transversal gate with order $p$: $\T^p = \id$. We also assume that $\T_{v}^p = \id$ for all $\T_{v}$.\footnote{This is always possible via the introduction of local phases multiplying the $\T_{v}$ that cancel on the full transversal gate $\T$.} We act on qudits with clock and shift operators, $X$, $Z$:

\begin{align}
    ZX = \omega XZ, &\quad \omega = e^{2\pi i/q} \\
    X^q = &Z^q = \id
\end{align} When $q=2$, we recover the usual Pauli algebra.  

Since $p \neq q$ in general, and we need to introduce ancilla qudits of dimension $p$ to perform the gauging measurement, we distinguish clock and shift operators acting on these local degrees of freedom of dimension $p$ with a different font; $\cX,\cZ$ with $\cX^p = \cZ^p = \id$ and $\cZ \cX = \Omega \cX \cZ$ for phase $\Omega = e^{2\pi i/p} $. In the condensed matter literature, it is common to refer to the degrees of freedom of the initial LDPC code as \textit{matter} qudits, and the ancilla qudits as \textit{gauge} qudits. The states $\ket{+}$ and $\ket{0}$ denote the $+1$ eigenstates of the qudit shift and clock operators respectively, with the dimension of the qudit understood from context.

We write the support of an operator $O$ restricted to some subset $A$ of the physical degrees of freedom as $O_A$. We frequently make use of the group commutator: $\ComGrp{A}{B} = A B A^\dagger B^\dagger$. Finally, we denote a controlled-$U$ gate with control $a$ and target $b$ as follows:
\begin{equation}
    \control{a}{U}{b} = \control{a}{(U)}{b} \equiv \sum_{n = 0}^{d_a - 1} \dyad{n}_a \otimes U^{n}_b
\end{equation} where $d_a$ is the dimension of the control qudit $a$.\footnote{Note that this gate is a unitary only when the order of $U$ divides $d_a$. However, in the case considered here, we have $d_a = p$ and $U = \T_v$, and as such this gate is always unitary.}

We now provide a conceptual overview of the gauging protocol. We aim to measure a logical Clifford operator $\T$--- perhaps to prepare a magic state, or implement a logical gate teleportation scheme--- and we wish to do it in a fault-tolerant manner. By gauging the logical Clifford, we map our initial code $\cC$ onto a deformed code $\cC'$ with $\T$ in its stabilizer group. If this deformed code $\cC'$ is itself a QLDPC code (not, however, a Pauli stabilizer code), we can convert the measurement of $\T$ into the measurement of constant weight stabilizer generators instead.

\begin{definition}[Gauging the logical operator of a quantum code]
    Take a code $\cC$ and a transversal logical gate $\T$ satisfying $\T^p = \id$. Let the codespace of the \emph{symmetric code} be the intersection of the codespace of $ \cC $ and the $+1$ eigenspace of $\T$. This is, in general, a non-LDPC code, because $\T$ may act on extensively many qu(d)its. \emph{Gauging} maps the codespace of a symmetric code to an isomorphic codespace belonging to a QLDPC code, $\cC'$ called the \emph{gauged} code.
\end{definition}

The following section describes how to arrive at the generators of the gauged QLDPC code $\cC'$.
We first introduce an auxiliary connected graph $\graphG=(\vertexV,\edgeE)$ whose vertices are identified with the local gates $\T_{v}$. The properties of the graph influence the fault tolerance of the procedure. 
Since $\T$ may have extensive weight, and our goal is to measure it in a fault-tolerant way, we introduce new constant-weight operators $\av$ such that
\begin{equation}
    \prod_v \av = \T,
    \label{eq:gauged-sym}
\end{equation} 
where $\av=\T_v \prod_{e\in \delta v} \Xp_e$, and $\delta: \vertexV \to \edgeE$ is the coboundary map of $\graphG$. The introduction of these new operators allows us to break up the measurement of a large operator into small pieces, since measuring the $\av$ also measures $\T$. This measurement could be used to prepare a magic state or drive a measurement-based logic gate.

\begin{definition}[Gauging Measurement Protocol]
\label{def:gauging-measurement-protocol}
    The gauging measurement for a transversal logical operator $\T$ with $\T^p = \id$ is described by a simple procedure~\cite{williamson2024lowoverheadfaulttolerantquantumcomputation}:

\begin{enumerate}
    \item Initialize an auxiliary qudit of dimension $p$ in the state $\ket{0}_e$ for each edge $e\in \edgeE$.
    \item Measure all $\av$ operators.
    \item Measure all $\Zp_e$ operators. 
    \item Apply a feedforward operation of the form $\prod_{v} \T^{a_v}_v$ for some subset of the vertices $\{v\} \subseteq \vertexV$ and some set of $a_v \in \{0, \dots, p-1\}$.
\end{enumerate} 
\end{definition}
The above procedure can be applied to any operator $\T$ that satisfies $[\T_v,\T_{v'}]=0$ for all $v,v'\in \vertexV$. For simplicity, we focus on transversal gates $\T$ where each $\T_v$ is supported on a distinct, tensor factor (with $O(1)$ support).

The gauging measurement protocol implements the projective measurement of the target gate~$\T$. 
\begin{lemma}[Gauging measurement operation~\cite{williamson2024lowoverheadfaulttolerantquantumcomputation}]  
Let $a_v$ denote the measurement outcome of $\av$ in the gauging measurement protocol of Def.~\ref{def:gauging-measurement-protocol}.
This protocol performs a projective measurement of the symmetry $\T$ with outcome 
\[
\sigma = \prod_{v\in \vertexV} a_v .
\]
Equivalently, up to the feedforward operation in Def.~\ref{def:gauging-measurement-protocol}, the final state on the original code is given by: 
\[
\Pi_\sigma \ket{\psi_\mathrm{initial}}, \quad \Pi_\sigma = \frac{1}{p} \sum_{a=0}^{p-1} \sigma^{-a} \T^a.
\]
\end{lemma}
\begin{proof}
The proof is given in Ref.~\cite{williamson2024lowoverheadfaulttolerantquantumcomputation}, and the argument is discussed in Sec. \ref{sec:ungauging}.
\end{proof}

In general, the stabilizers of the original, ungauged code $\cC$ do not commute with each $\av$. As a result, after step~2, the code is deformed into a \textit{gauged code} $\cC'$. The gauged code has additional checks $\av$ and $\bp:=\prod_{e\in \partial p} \cZ_e$ where $\partial p$ form a cycle basis of $\graphG$. The original checks $S_i$ of code $\cC$ are updated into \emph{decorated stabilizers} $\gaugedS_i$ which are modified to ensure that they commute with the added $\av$ and $\bp$ operators in the codespace. Our goal is to construct a graph $\graphG$ such that the deformed code $\cC' = \langle \gaugedS_i, \av, \bp \rangle$ retains a high distance, and remains an LDPC code. We discuss the necessary properties of $\graphG$ in Sec. \ref{sec:defining-gauging-graph}.

We now consider a circuit-level implementation of steps 1 and 2 of Definition~\ref{def:gauging-measurement-protocol}. 
Given a suitable choice of gauging graph $\graphG$, we can write down the stabilizers of the gauged code in a systematic way:

\begin{enumerate}
    \item \textbf{Define vertex sets:} Break the support of the transversal gate $\T$ into \emph{vertex sets}. These are constant-sized subsets of the original code degrees of freedom which form a partition of the support of $\T$. This choice of subsets defines a map $\Sigma: \physq \to \vertexV$ from matter qudits to their corresponding vertex set $v \in \vertexV$. The minimal choice for a vertex set is the ``unit cell'' of the $\T$ action. For example, consider a transversal gate made up of single-qudit gates that acts identically on all qudits. A straightforward choice of partition is to make each matter qudit a distinct vertex set. We discuss how to choose vertex sets in more detail later in Sec. \ref{sec:defining-gauging-graph}.
    
    \item \textbf{Add vertex ancilla qudits (symmetry enrichment):} We add a \emph{vertex ancilla} for each vertex set $v \in \vertexV$, initialized in the state $\ket{+}_v$ (add stabilizers $\Xp_v$). The pre-image $\vertexset{v} \subseteq \physq$ gives the set of all matter qudits associated to a given vertex set/ancilla. Let $\T_v:=\T_{\vertexset{v}}$ denote the local factor of the transversal gate supported on $\vertexset{v}$; in general $\T_v$ may be a multi-qudit operator. Apply controlled gates to each vertex set that perform the following basis transformation: $$\Xp_v \xrightarrow {C_v \T_v}\Xp_v \T_v.$$ At this point, our wavefunction is a $+1$ eigenstate of $\T \otimes \prod_v \Xp_v$ (this is the product of all the newly added stabilizers).
    
    \item \textbf{Add edge ancilla qudits (cluster state entangler):} We aim to generate a state that is an eigenstate of $\T$ by itself, not of some joint operator on the matter qudits and the ancilla degrees of freedom. We want a circuit that takes $\prod_v \Xp_v \to \id$. To do this, we introduce a second set of \emph{edge} ancillas, $\edgeE$, each initialized in the $\ket{0}$ state. As suggested by the notation, we define the oriented graph $\graphG = (\vertexV, \edgeE)$---called the \emph{gauging graph}--- on which these edge ancillas live shortly. We apply a graph cluster state entangler between the vertex and edge ancilla, which performs the following basis transformation: $$\Xp_v \T_v \xrightarrow{\text{cluster entangler}} \Xp_v \T_v \prod_{e \in \delta v} \Xp^{\eta_v(e)}_e.$$ Here, $\eta_v(e) \in \{-1, 1\}$ is a function giving the orientation of an edge $e$ with respect to vertex~$v$.
    
    \item \textbf{Measuring out vertex ancillas:} Finally, we complete the gauging process by measuring out the vertex ancillas in the $\Xp$ basis, producing the desired $\av$ stabilizers, up to a phase: $$\av = \omega^{m_v} \T_v \prod_{e \in \delta v} \Xp^{\eta_v(e)}_e, \quad \omega = e^{2\pi i /p}.$$ 
    Here, $m_v \in \{0, \dots, p-1\}$ denotes the measurement outcome for $\Xp_v$. We can always apply a finite-depth circuit to correct all $m_v \neq 0$ outcomes to $m_v = 0$ \cite{Tantivasadakarn_2023, Ren_2025}. 
    Once corrected, the product of all these $\av$ terms is simply $\T$, since each edge $e$ on the gauging graph is shared by two vertices and the $\eta$ function ensures one vertex contributes $\Xp$ while the other contributes $\Xp^\dagger$. At this point we have successfully implemented the measurement of $\T$.
\end{enumerate}

Collected into a single expression, the above protocol is as follows:

\begin{definition}[Gauging Map (Circuit-Level)] \label{def:gauging-circuit}
Consider a code $\cC$ and associated transversal gate $\T$. Given a choice of vertex set (the map $\Sigma$) and gauging graph $\graphG = (\vertexV, \edgeE)$ with edge orientations given by $\eta_v(e)$, the gauging operation taking $\cC \to \cC'$ is implemented by the following three-part map $\tilde{\Gamma}$:
\begin{equation}
\widetilde{\Gamma} \equiv  \underbrace{\left (\prod_{v \in \vertexV} \dyad{+}_v \right )}_\textrm{P} \underbrace{\left ( \prod_{v \in \vertexV} \prod_{e \in \delta v} C_v \Xp^{\eta_v(e)}_e \right )}_{\textrm{C.E.}} \underbrace{\left ( \prod_{v \in \vertexV} C_v \T_v \right )}_\textrm{S.E.} \underbrace{\left ( \prod_{v \in \vertexV} \ket{+}_v \prod_{e \in \edgeE} \ket{0}_e \right )}_\textrm{Ancilla Init.}
    \label{eq:gauging-circuit}
\end{equation}

The labels for each part of the map stand for (0) ``Ancilla Initialization'', which we will normally take as given, (1) ``Symmetry Enrichment'', (2) ``Cluster Entangler'', and (3) ``Projection''. In practice, the final projection is implemented by an adaptive circuit consisting of measurement in the $\Xp$ basis on the vertex ancilla and feedback to correct erroneous ($\neq + 1$) measurement outcomes.
\end{definition}

In the following sections we describe each of these steps in more detail, and derive the general form for the resultant gauged code stabilizers. We note that our procedure builds on Refs.~\cite{Williamson_2021,verresen2022efficientlypreparingschrodingerscat,Tantivasadakarn_2023,Tantivasadakarn_2024,Ren_2025}, where the gauging circuit was developed for topological codes. The protocol presented here applies to any qudit LDPC stabilizer code and any suitable transversal gate $\T$.

\subsection{Defining the gauging ancilla system} \label{sec:defining-gauging-graph}

We now consider the design of the gauging graph. The first step is to pick the vertices $\vertexV$ of the gauging graph. We start with a code $\cC$ and a transversal gate of interest, $\T$. We then aim to choose a map $\Sigma: \physq \to \vertexV$ that satisfies the conditions below. 

\begin{enumerate}
    \item \textbf{Constant size:} We want to ensure that the gauged code $\cC'$ remains LDPC; this implies that we must have $|\vertexset{v}| \sim O(1)$ for all $v \in \vertexV$, or else the $\av$ terms may not be LDPC.
    \item \textbf{No overlap:} We also require that $\Sigma$ is a valid function: i.e. each physical qudit belongs to exactly one vertex set. This partition should respect the structure of $\T$, meaning the support of an elementary gate of $\T$ should not be split across vertex sets. As mentioned above, a minimal choice of vertex set for a single-qubit strongly transversal gate would be one qudit per vertex set. By contrast, a transversal $\CNOT$ acting between two codeblocks naturally lends itself to vertex sets consisting of one qubit from each codeblock, since a single qudit vertex set would not contain the full support of a single $\CNOT$. 
\end{enumerate}

Assuming we have satisfied these constraints, there are many valid choices for $\Sigma$. The particular structure of a given starting code $\cC$ informs the most convenient vertex set--- for instance, a good choice for code living on a tri-colorable lattice may be to make faces of a certain color vertex sets. We present such examples below. In general, however, we can always pick the ``minimal'' choice of vertex set, dependent on the structure of the transversal gate $\T$ itself.

\begin{definition}[Minimal Vertex Set]
Given a transversal gate $\T$ acting on a code $\cC$ with the form given in Def. \ref{def:transversal-gate}, the \emph{minimal vertex set} is chosen such that each vertex ancilla $v$ corresponds to a minimal local action $\T_v \subset \T$. In other words, no $\T_v$ can be written as a tensor product supported on a smaller subset of qudits.
\end{definition}

\begin{definition}[Valid Vertex Set]
    A \emph{valid vertex set} is any coarse-graining of the minimal vertex set (where coarse-graining merges some vertex sets into one larger vertex set).  
\end{definition}

We complete the specification of the graph $\graphG$ by adding edges connecting vertices $\vertexV$. These edges are required to obey certain restrictions to produce a valid gauging measurement protocol. Again, we want to ensure that the final code $\cC'$ is LDPC, which enforces that $\graphG$ should have $O(1)$ degree--- otherwise, the $\av$ operators may not be low-weight. We also require that $\graphG$ has a sparse cycle basis: we show below that the gauging protocol produces new stabilizers $\bp$ for every cycle of $\graphG$, and a sparse cycle basis ensures there is a generating set of $\bp$ that are LDPC. The last condition that ensures the gauged code is LDPC is a ``well-connected'' condition, which ensures any modifications to the original stabilizers are LDPC as well. To ensure we measure only $\T$ and not individual operators whose product is $\T$, we want $\graphG$ to consist of a single connected component. Finally, to ensure the code $\cC'$ has the same distance scaling as the code $\cC$, we want the homology of $\cC'$ to be completely determined by $\cC$, and thus $\graphG$ should be homologically trivial (this is actually already satisfied if we have specified a complete, sparse cycle basis).
We collect these constraints into a set of desiderata, below.

\begin{definition}[Graph Desiderata] \label{def:graph-desiderata}
When constructing the gauging graph $\graphG$, the following suffice to ensure a well-behaved gauged code.

\begin{enumerate}

\item \textbf{Connectedness:} $\graphG$ should be connected to ensure the desired logical action.

\item \textbf{Sparseness:}
\begin{enumerate}
    \item \emph{Bounded degree.} The degree of $\graphG$ should be bounded by a constant.
    \item \emph{Sparse cycle filling.} Each 2-cell $p\in\graphG_2$ has a boundary $\partial_2p$ of uniformly $O(1)$-bounded weight. Equivalently, $\supp{\gaugedB_p}$ is bounded by some constant for all $p$. This can be ensured by applying a decongestion procedure to $\graphG$ as described in Refs.~\cite{williamson2024lowoverheadfaulttolerantquantumcomputation,swaroop2025universaladaptersquantumldpc,yuan2026parsimonious}. 
    \item \emph{Well-connected.} Any pair of vertices whose corresponding vertex sets share support on at least one stabilizer are connected by a path that contains a constant number of edges. This is required to ensure the gauged code is LDPC.
\end{enumerate}

\item \textbf{Distance preserving:}
    \begin{enumerate}
        \item \emph{Trivial homology.} When the gauging graph is expressed as a chain complex (see Def. \ref{def:gauging-complex}), the homology $H_1$ should be trivial. This ensures that no additional, potentially low-weight, diagonal logical operators are introduced on the edge qudits. Enforcing a sparse cycle basis also ensures this condition.
        \item \emph{Expansion.} The generalized Cheeger constant of $\graphG$ (see Def.~\ref{def:generalized-cheeger})  should satisfy
        \[
        h_{\Agrp}(\graphG)\ge 1.
        \]
        This condition is only required before cycle decongestion. 
    \end{enumerate}
\end{enumerate}
    
\end{definition}

A straightforward choice of $\graphG$, given these constraints, is the following: place an edge between any two vertex ancillas whose associated vertex sets have support on the same stabilizer.

For each stabilizer $S \in C$ define an edge $e = (v, v') \in \edgeE$ for any pair $v, v' \in \invvertexset{\supp{S}}$, where $\supp{S}$ denotes the qudits in the support of the stabilizer $S$. The LDPC property of the input code ensures that the resulting graph is sparse.
We remark that this choice is somewhat redundant, as it suffices to include a spanning tree on the vertices in the support of any stabilizer to ensure the well-connected property. Furthermore, this choice does not suffice to guarantee connectedness or sufficient expansion. These properties can be guaranteed by adding further edges, such as those of an expander graph.

\begin{lemma}[Construction of a suitable graph]
We can always construct a graph satisfying the desiderata in Definition~\ref{def:graph-desiderata}. 
\end{lemma}
\begin{proof}
    This follows from the proof in Ref.~\cite{williamson2024lowoverheadfaulttolerantquantumcomputation} with a slight modification to the step ensuring well-connectedness. 
    A suitable graph $\graphG$ can be constructed by taking a sparse expander on the desired vertex set that satisfies $h_{\Agrp}(\graphG)\ge 1$ and adding edges to ensure the well-connected condition. The expander can be constructed randomly or deterministically via standard methods. Here, to ensure well-connectedness, instead of including the edges from a matching on the support of each check that intersects the logical nontrivially as was done in Ref.\cite{williamson2024lowoverheadfaulttolerantquantumcomputation}, we include a spanning tree on the support of each such check. In fact, any connected graph on the support suffices. The addition of these edges does not decrease the generalized Cheeger constant, and due to the LDPC property of the original checks, this step also preserves the bounded degree property of the initial sparse expander. 
    Next, we construct a sparse cycle basis by applying a decongestion procedure to $\graphG$ either following the sparsification/cellulation method described in Remark~4 of Ref.~\cite{williamson2024lowoverheadfaulttolerantquantumcomputation} and Fig.~4 of Ref.~\cite{swaroop2025universaladaptersquantumldpc}, or by using the parsimonious cone method from Ref.~\cite{yuan2026parsimonious}. The former leads to a qubit overhead of $|\vertexV| \log^3 |\vertexV|$, while the latter leads to a qubit overhead of $|\vertexV| \log |\vertexV|$. At this point all graph desiderata are satisfied. We remark that the expansion of the graph after decongestion may be reduced. However, the relative expansion required to ensure the distance of the gauged code is preserved, see Refs.~\cite{swaroop2025universaladaptersquantumldpc,he2025extractorsqldpcarchitecturesefficient}. 
\end{proof}

\begin{remark}[Spatial overhead] \label{rem:spatial-overhead}
    The number of additional qubits in the gauged code is determined by the number of edges in the graph $\graphG$ after decongestion. This is because the vertex ancilla qubits are measured out to reach the gauged code. In the case where the logical operator being measured is a Pauli operator, the final number of additional qubits in the gauged code scales as $O(|\vertexV| \log |\vertexV|)$ when the parsimonious cone method is used to decongest $\graphG$~\cite{williamson2024lowoverheadfaulttolerantquantumcomputation,yuan2026parsimonious}.
    
    The main difference in this work as compared to Ref.~\cite{williamson2024lowoverheadfaulttolerantquantumcomputation} is that we add more edges to the gauging graph to connect vertex sets that share a stabilizer via a spanning tree or some other connected graph. For a QLDPC code with constant sized vertex sets $|\vertexset{v}|$ there are at most $O(|\vertexV|)$ checks overlapping the target logical operator. This is due to the constant degree of the checks and the constant size of the vertex sets. Each of these checks may be split over a constant number of vertex sets due to the constant size of each check. As a result, the number of edge qubits in $\graphG$ is increased by at most a constant multiplicative factor when compared to the Pauli case considered in Ref.~\cite{williamson2024lowoverheadfaulttolerantquantumcomputation}. Thus, the Clifford gauging procedure retains the $O(|\vertexV| \log |\vertexV|)$ overhead scaling.
\end{remark}

\subsection{Warm-up: gauging the logical $CZ$ of the $[[4,2,2]]$ code}
\label{subsec:warmup-422}

Before deriving the gauged stabilizers in general, we work through a simple example: gauging the logical $CZ$ gate of the $[[4,2,2]]$ code.
The full system consists of three species of qubits
(Eq.~\ref{fig:warmup-labels}). 
The four \emph{matter qubits} $q_1, q_2, q_3, q_4$ carry
the code, with \emph{matter stabilizers} $$S_X=X_1X_2X_3X_4, \quad S_Z=Z_1Z_2Z_3Z_4$$

We consider vertex sets that consist of single qubits, and so we additionally define four
\emph{vertex ancillas} $v_1, v_2, v_3, v_4$, all prepared in $\ket{+}$ and contributing
initial \emph{ancilla stabilizers} $A_{v_i} = X_{v_i}$ where $i = 1, 2, 3, 4$. We then define four \emph{edge qubits}, which lie on the edges of the gauging graph (which we take to be a square). We label these ancilla $e_{ij}$, where the $ij$ subscript indicates the edge connects vertex ancilla $v_i$ and $v_j$. These edge ancilla are all prepared in $\ket{0}$ and contribute the \emph{edge
stabilizers} $Z_{e_{ij}}$ for all pairs $(i,j)\in \{(1,2), (2,3), (3, 4), (4, 1)\}$. All in all, we have twelve degrees of freedom in the following geometry:
\begin{equation}
    \raisebox{-36pt}{
    \begin{tikzpicture}
        \node[name=s, regular polygon, regular polygon sides = 4, inner sep = 0.6cm, draw, red, thick] at (0, 0) {};

        \node[draw, circle, fill=red!50!white, inner sep = 1.5pt, text=black] at (s.corner 2) {\tiny{$v_1$}};
        \node[draw, circle, fill=black, inner sep = 1pt, text=white] at ($(s.corner 2) + (315:0.5)$) {\tiny{$1$}};
        \foreach \anchor/\placement/\n/\label in
        {corner 1/below/1/2, corner 3/right/3/4, corner 4/above/4/3}
        {
        \node[draw, circle, fill=red!50!white, inner sep = 1.5pt, text=black] at (s.\anchor) {\tiny{$v_{\label}$}};
        \node[draw, circle, fill=black, inner sep = 1pt, text=white] at ($(s.\anchor) + (135 + \n*90:0.5)$) {\tiny{$\label$}};
        }
        \foreach \anchor/\placement/\n/\label in
        {side 1/below/1/12, side 2/below/2/41, side 3/right/3/34, side 4/above/4/23}
        {
        \node[draw, circle, fill=red!30!white!70!blue, inner sep = 1.5pt, text=black] at (s.\anchor) {\tiny{$e_{\label}$}};
        }
    \end{tikzpicture}
    }
    \label{fig:warmup-labels}
\end{equation}
The ten independent stabilizer generators define a codespace with two logical qubits:

\begin{equation}
    \cC = \text{span} \{ \ket{\bar{\psi}} \otimes \ket{+}^{\otimes 4}_\vertexV \otimes \ket{0}^{\otimes 4}_\edgeE \} \quad \ket{\bar{\psi}} = \ket{\bar{00}}_{422}, \ket{\bar{01}}_{422},  \ket{\bar{10}}_{422},  \ket{\bar{11}}_{422}
\end{equation}
The code's transversal gate is the alternating phase-gate pattern: 
\begin{equation}
    \T = \Sg_1 \Sg^\dagger_2 \Sg_3 \Sg^\dagger_4
\end{equation}
We consider the logical basis $$\bar{X}_1 = X_1 X_2, \quad \bar{Z}_1 = Z_1 Z_4, $$ $$\bar{X}_2 = X_1 X_4, \quad \bar{Z}_2 = Z_1 Z_2$$
The action of the transversal gate is:
\begin{equation}
\begin{aligned}
    \T \bar{X}_1 \T^\dagger &= (\Sg_1 X_1 \Sg_1^\dagger)( \Sg_2^\dagger X_2 \Sg_2) = -Y_1 Y_2 = X_1 X_2 (Z_1 Z_2) = \bar{X}_1 \bar{Z}_2 \\
    \T \bar{X}_2 \T^\dagger &= (\Sg_1 X_1 \Sg_1^\dagger) (\Sg_4^\dagger X_4 \Sg_4) = -Y_1 Y_4 = X_1 X_4 (Z_1 Z_4) = \bar{X}_2 \bar{Z}_1 \\
\end{aligned}
\end{equation} Thus, this transversal gate yields a logical $CZ$.

While the full transversal circuit $\T$ squares to the identity (up to a stabilizer), in our current presentation, the local action on each vertex set is order four. We can multiply the transversal circuit by the $S_X$ stabilizer to obtain: 
\begin{equation}
    \T=\prod_{v = 1}^4W_v=\CZbar,\qquad
    W_1=W_3=e^{-i\pi/4}XS,\qquad W_2=W_4=e^{+i\pi/4}XS^\dagger ,
    \label{eq:warmup-T}
\end{equation}
Now the action of the transversal gate on each vertex set, given by $W_i$, squares to the identity. Note we have added some additional phases to ensure that the final gauged code stabilizers admit $+1$ eigenspaces.

The gauging circuit consists of three layers (Definition~\ref{def:gauging-circuit}): (1) the symmetry
enrichment (S.E.), which entangles each ancilla with its vertex set by
$C_{v_i}(W_i)$, (2) the cluster entangler (C.E.), and (3) the measurement of the
vertex ancillas in the $X$ basis, which implements the projective measurement of the transversal gate.

The ancilla stabilizer $X_{v_1}$ is pushed through the circuit as follows:

\begin{equation}
\raisebox{-55pt}{
    \begin{tikzpicture}
        \node[name=s, regular polygon, regular polygon sides = 4, inner sep = 0.6cm, draw, red, thick] at (0, 0) {};

        \node[draw, circle, fill=red!50!white, inner sep = 1.5pt, text=black] at (s.corner 2) {\tiny{$X$}};
        \node[draw, circle, fill=black, inner sep = 1pt, text=white] at ($(s.corner 2) + (315:0.5)$) {\tiny{$\id$}};
        \foreach \anchor/\placement/\n in
        {corner 1/below/1, corner 3/right/3, corner 4/above/4}
        {
        \node[draw, circle, fill=red!50!white, inner sep = 1.5pt, text=black] at (s.\anchor) {\tiny{$\id$}};
        \node[draw, circle, fill=black, inner sep = 1pt, text=white] at ($(s.\anchor) + (135 + \n*90:0.5)$) {\tiny{$\id$}};
        }
        \foreach \anchor/\placement/\n in
        {side 1/below/1, side 2/below/2, side 3/right/3, side 4/above/4}
        {
        \node[draw, circle, fill=red!30!white!70!blue, inner sep = 1.5pt, text=black] at (s.\anchor) {\tiny{$\id$}};
        }

        \draw[->, thick] (1.5, 0) -- (2.5, 0) node [pos = 0.5, above] {S.E.};

        \node[name=s, regular polygon, regular polygon sides = 4, inner sep = 0.6cm, draw, red, thick] at (4, 0) {};

        \node[draw, circle, fill=red!50!white, inner sep = 1.5pt, text=black] at (s.corner 2) {\tiny{$X$}};
        \node[draw, circle, fill=black, inner sep = 1pt, text=white] at ($(s.corner 2) + (315:0.5)$) {\tiny{$W$}};
        \foreach \anchor/\placement/\n in
        {corner 1/below/1, corner 3/right/3, corner 4/above/4}
        {
        \node[draw, circle, fill=red!50!white, inner sep = 1.5pt, text=black] at (s.\anchor) {\tiny{$\id$}};
        \node[draw, circle, fill=black, inner sep = 1pt, text=white] at ($(s.\anchor) + (135 + \n*90:0.5)$) {\tiny{$\id$}};
        }
        \foreach \anchor/\placement/\n in
        {side 1/below/1, side 2/below/2, side 3/right/3, side 4/above/4}
        {
        \node[draw, circle, fill=red!30!white!70!blue, inner sep = 1.5pt, text=black] at (s.\anchor) {\tiny{$\id$}};
        }
        \draw[->, thick] (5.5, 0) -- (6.5, 0) node [pos = 0.5, above] {C.E.};

        \node[name=s, regular polygon, regular polygon sides = 4, inner sep = 0.6cm, draw, red, thick] at (8, 0) {};

        \node[draw, circle, fill=red!50!white, inner sep = 1.5pt, text=black] at (s.corner 2) {\tiny{$X$}};
        \node[draw, circle, fill=black, inner sep = 1pt, text=white] at ($(s.corner 2) + (315:0.5)$) {\tiny{$W$}};
        \foreach \anchor/\placement/\n in
        {corner 1/below/1, corner 3/right/3, corner 4/above/4}
        {
        \node[draw, circle, fill=red!50!white, inner sep = 1.5pt, text=black] at (s.\anchor) {\tiny{$\id$}};
        \node[draw, circle, fill=black, inner sep = 1pt, text=white] at ($(s.\anchor) + (135 + \n*90:0.5)$) {\tiny{$\id$}};
        }
        \foreach \anchor/\placement/\n in
        {side 3/right/3, side 4/above/4}
        {
        \node[draw, circle, fill=red!30!white!70!blue, inner sep = 1.5pt, text=black] at (s.\anchor) {\tiny{$\id$}};
        }
        \foreach \anchor/\placement/\n in
        {side 1/below/1, side 2/below/2}
        {
        \node[draw, circle, fill=red!30!white!70!blue, inner sep = 1.5pt, text=black] at (s.\anchor) {\tiny{$X$}};
        }

        \draw[->, thick] (9.5, 0) -- (10.5, 0) node [pos = 0.5, above] {meas. $X_v$};

        \node[name=s, regular polygon, regular polygon sides = 4, inner sep = 0.6cm, draw, red, thick] at (12, 0) {};
        
        \node[draw, circle, fill=black, inner sep = 1pt, text=white] at ($(s.corner 2) + (315:0.5)$) {\tiny{$W$}};
        \foreach \anchor/\placement/\n in
        {corner 1/below/1, corner 3/right/3, corner 4/above/4}
        {
        
        \node[draw, circle, fill=black, inner sep = 1pt, text=white] at ($(s.\anchor) + (135 + \n*90:0.5)$) {\tiny{$\id$}};
        }
        \foreach \anchor/\placement/\n in
        {side 3/right/3, side 4/above/4}
        {
        \node[draw, circle, fill=red!30!white!70!blue, inner sep = 1.5pt, text=black] at (s.\anchor) {\tiny{$\id$}};
        }
        \foreach \anchor/\placement/\n in
        {side 1/below/1, side 2/below/2}
        {
        \node[draw, circle, fill=red!30!white!70!blue, inner sep = 1.5pt, text=black] at (s.\anchor) {\tiny{$X$}};
        }

        \draw [decorate,decoration={brace,amplitude=5pt,mirror,raise=4ex}] (11,-0.5) -- (13,-0.5) node[midway,yshift=-3em]{$\mathbb{A}_{v_1}$};
    \end{tikzpicture}
}
    \label{eq:warmup-flow}
\end{equation}
The gauging graph and vertex ancilla are drawn in red, the edge ancilla in purple, and the original $[[4, 2, 2]]$ degrees of freedom in black. The final step of Eq. \ref{eq:warmup-flow} yields the vertex checks $\mathbb{A}_{v}=W_vX_eX_{e'}$ ($e,e'$ the two loop
edges at $v$), with $\prod_v\mathbb{A}_v=\CZbar$.
Measuring these ancillas thus measures the logical.

Likewise, the initial edge stabilizers $Z_e$
become the cluster checks $L_e=Z_vZ_eZ_{v'}$.
The vertex measurement destroys
them individually, but their product around the loop formed by the gauging graph survives as the flux check
$\bp=\prod_{e \in \edgeE}Z_e=\prod_e L_e$.
\begin{equation}
\raisebox{-32pt}{
    \begin{tikzpicture}
        \node at (0, 0) {$\bp = $};
        \node[name=s, regular polygon, regular polygon sides = 4, inner sep = 0.6cm, draw, red, thick] at (1.75, 0) {};

        \node[draw, circle, fill=black, inner sep = 1pt, text=white] at ($(s.corner 2) + (315:0.5)$) {\tiny{$\id$}};
        \foreach \anchor/\placement/\n in
        {corner 1/below/1, corner 3/right/3, corner 4/above/4}
        {
        \node[draw, circle, fill=black, inner sep = 1pt, text=white] at ($(s.\anchor) + (135 + \n*90:0.5)$) {\tiny{$\id$}};
        }
        \foreach \anchor/\placement/\n in
        {side 1/below/1, side 2/below/2, side 3/right/3, side 4/above/4}
        {
        \node[draw, circle, fill=red!30!white!70!blue, inner sep = 1.5pt, text=black] at (s.\anchor) {\tiny{$Z$}};
        }
        
    \end{tikzpicture}
}
\end{equation}

The matter stabilizers $S_Z,S_X$ do not commute with the $C_{v}(W_v)$ gates involved in the symmetry enrichment step, so their support is expanded onto the vertex qubits. The conjugation of a Pauli operator $P$ with each controlled-$W$ or controlled-$W^\dagger$ gate is as follows:
\begin{equation}
    C_v(W_v)\,P\,C_v(W_v)^\dagger = P\cdot C_v(W_vPW_vP)
\end{equation}
In this example, we use the following commutators:
\begin{equation}
\begin{aligned}
    W_{v_i}Z_iW_{v_i}Z_i=-\id, &\quad i = 1, 2, 3, 4 \\
    W_{v_{i}}X_{i}W_{v_{i}}X_i=-iZ_i, &\quad i = 1, 3 \\
    W_{v_{i}}X_{i}W_{v_{i}}X_i=+iZ_i, &\quad i = 2, 4
\end{aligned}
\label{eq:warmup-remnants}
\end{equation}
The first $Z$-type commutator is simple, and its control evaluates to $C_v(-\id)=Z_v$. The $X$ check instead picks up controlled-$Z$ remnants
dressed by phase gates, $C_v(\mp iZ)=\Sg_v^{\mp1}\,C_v(Z_{q_v})$. These phase gates arise from the local phases in the group commutators, and 
tracking them locally is a subtlety of the gauging procedure. Nevertheless, globally they cancel, $(-i)(+i)(-i)(+i)\,ZZZZ=S_Z$, yielding one of the code's stabilizers.

The symmetry-enrichment (S.E.) layer sends $S_Z\mapsto S_Z\prod_vZ_v$, since each
decoration is the control $C_v(-\id)=Z_v$. Enforcing the stabilizer constraint $L_e=+1$ pairwise
yields the additional constraints $Z_{v_1}Z_{v_2}=Z_{e_{12}}$ and $Z_{v_3}Z_{v_4}=Z_{e_{34}}$.
These allow us to move the vertex support onto the edges to yield $\gaugedS_Z=ZZZZ\cdot Z_{e_{12}}Z_{e_{34}}$.
In other words, our derivation is only concerned with ensuring commutation relations within the $+1$ eigenspace, 
and the form and algebraic relations of the gauged stabilizers can change outside of that space. Pictorially, the transformation of the $S_Z$ stabilizer under the gauging circuit is as follows:
\begin{equation}
\raisebox{-55pt}{
    \begin{tikzpicture}
        \node[name=s, regular polygon, regular polygon sides = 4, inner sep = 0.6cm, draw, red, thick] at (0, 0) {};

        \foreach \anchor/\placement/\n in
        {corner 1/below/1, corner 2/right/2, corner 3/right/3, corner 4/above/4}
        {
        \node[draw, circle, fill=red!50!white, inner sep = 1.5pt, text=black] at (s.\anchor) {\tiny{$\id$}};
        \node[draw, circle, fill=black, inner sep = 1pt, text=white] at ($(s.\anchor) + (135 + \n*90:0.5)$) {\tiny{$Z$}};
        }
        
        \foreach \anchor/\placement/\n in
        {side 1/below/1, side 2/below/2, side 3/right/3, side 4/above/4}
        {
        \node[draw, circle, fill=red!30!white!70!blue, inner sep = 1.5pt, text=black] at (s.\anchor) {\tiny{$\id$}};
        }

        \draw[->, thick] (1.5, 0) -- (2.5, 0) node [pos = 0.5, above] {S.E.};

        \node[name=s, regular polygon, regular polygon sides = 4, inner sep = 0.6cm, draw, red, thick] at (4, 0) {};

        \foreach \anchor/\placement/\n in
        {corner 1/below/1, corner 2/right/2, corner 3/right/3, corner 4/above/4}
        {
        \node[draw, circle, fill=red!50!white, inner sep = 1.5pt, text=black] at (s.\anchor) {\tiny{$Z$}};
        \node[draw, circle, fill=black, inner sep = 1pt, text=white] at ($(s.\anchor) + (135 + \n*90:0.5)$) {\tiny{$Z$}};
        }
        \foreach \anchor/\placement/\n in
        {side 1/below/1, side 2/below/2, side 3/right/3, side 4/above/4}
        {
        \node[draw, circle, fill=red!30!white!70!blue, inner sep = 1.5pt, text=black] at (s.\anchor) {\tiny{$\id$}};
        }
        \draw[->, thick] (5.5, 0) -- (6.5, 0) node [pos = 0.5, above] {$L_e = +1$};

        \node[name=s, regular polygon, regular polygon sides = 4, inner sep = 0.6cm, draw, red, thick] at (8, 0) {};

        \foreach \anchor/\placement/\n in
        {corner 1/below/1, corner 2/right/2, corner 3/right/3, corner 4/above/4}
        {
        \node[draw, circle, fill=red!50!white, inner sep = 1.5pt, text=black] at (s.\anchor) {\tiny{$\id$}};
        \node[draw, circle, fill=black, inner sep = 1pt, text=white] at ($(s.\anchor) + (135 + \n*90:0.5)$) {\tiny{$Z$}};
        }
        \foreach \anchor/\placement/\n in
        {side 1/below/1, side 3/right/3}
        {
        \node[draw, circle, fill=red!30!white!70!blue, inner sep = 1.5pt, text=black] at (s.\anchor) {\tiny{$Z$}};
        }
        \foreach \anchor/\placement/\n in
        {side 2/below/2, side 4/right/4}
        {
        \node[draw, circle, fill=red!30!white!70!blue, inner sep = 1.5pt, text=black] at (s.\anchor) {\tiny{$\id$}};
        }

        \draw[->, thick] (9.5, 0) -- (10.5, 0) node [pos = 0.5, above] {meas. $X_v$};

        \node[name=s, regular polygon, regular polygon sides = 4, inner sep = 0.6cm, draw, red, thick] at (12, 0) {};

        \foreach \anchor/\placement/\n in
        {corner 1/below/1, corner 2/right/2, corner 3/right/3, corner 4/above/4}
        {
        \node[draw, circle, fill=black, inner sep = 1pt, text=white] at ($(s.\anchor) + (135 + \n*90:0.5)$) {\tiny{$Z$}};
        }
        \foreach \anchor/\placement/\n in
        {side 1/below/1, side 3/right/3}
        {
        \node[draw, circle, fill=red!30!white!70!blue, inner sep = 1.5pt, text=black] at (s.\anchor) {\tiny{$Z$}};
        }
        \foreach \anchor/\placement/\n in
        {side 2/below/2, side 4/right/4}
        {
        \node[draw, circle, fill=red!30!white!70!blue, inner sep = 1.5pt, text=black] at (s.\anchor) {\tiny{$\id$}};
        }
        \draw [decorate,decoration={brace,amplitude=5pt,mirror,raise=4ex}] (11,-0.5) -- (13,-0.5) node[midway,yshift=-3em]{$\gaugedS_{Z}$};
    \end{tikzpicture}
}
    \label{eq:warmup-decorated-Z}
\end{equation}

The transformation of the $S_X$ stabilizers under the gauging circuit is a little more complicated. After the symmetry enrichment step, we find:
\begin{equation}
\raisebox{-32pt}{
    \begin{tikzpicture}
        \node[name=s, regular polygon, regular polygon sides = 4, inner sep = 0.6cm, draw, red, thick] at (0, 0) {};

        \foreach \anchor/\placement/\n in
        {corner 1/below/1, corner 2/right/2, corner 3/right/3, corner 4/above/4}
        {
        \node[draw, circle, fill=red!50!white, inner sep = 1.5pt, text=black] at (s.\anchor) {\tiny{$\id$}};
        \node[draw, circle, fill=black, inner sep = 1pt, text=white] at ($(s.\anchor) + (135 + \n*90:0.75)$) {\tiny{$X$}};
        }
        
        \foreach \anchor/\placement/\n in
        {side 1/below/1, side 2/below/2, side 3/right/3, side 4/above/4}
        {
        \node[draw, circle, fill=red!30!white!70!blue, inner sep = 1.5pt, text=black] at (s.\anchor) {\tiny{$\id$}};
        }

        \draw[->, thick] (1.5, 0) -- (2.5, 0) node [pos = 0.5, above] {S.E.};

        \node[name=s, regular polygon, regular polygon sides = 4, inner sep = 0.6cm, draw, red, thick] at (4, 0) {};

        \foreach \anchor/\text/\n in
        {corner 1/\Sg/1/, corner 2/\Sg^\dagger/2, corner 3/\Sg/3, corner 4/\Sg^\dagger/4}
        {
        \draw[thick, purple, dashed] (s.\anchor) to ($(s.\anchor) + (135 + \n*90:0.75)$); 
        \node[draw, circle, fill=red!50!white, inner sep = 1.5pt, text=black] at (s.\anchor) {\tiny{$\text$}};
        \node[draw, circle, fill=black, inner sep = 1pt, text=white] at ($(s.\anchor) + (135 + \n*90:0.75)$) {\tiny{$X$}};
        }
        \foreach \anchor/\placement/\n in
        {side 1/below/1, side 2/below/2, side 3/right/3, side 4/above/4}
        {
        \node[draw, circle, fill=red!30!white!70!blue, inner sep = 1.5pt, text=black] at (s.\anchor) {\tiny{$\id$}};
        }
    \end{tikzpicture}
    }
    \label{eq:warmup-decorated-X}
\end{equation}
Here the dashed purple line indicates a physical $CZ$. The phase gates and controls that now sit on the vertex ancilla must be moved off before those degrees of freedom are measured in the $X$ basis. After the cluster entangler step, we can use the fact our code subspace now has the constraint $L_e = +1$ to deform the controlled gates away from the vertex ancilla. Indeed, for $U^2=(-1)^m\id$, we have
\begin{equation}
    C_v(U)\;\xrightarrow{\;L_e=+1\;}\;C_e(U)\,C_{v'}(U)\,CZ^{\,m}_{e,v'} .
    \label{eq:warmup-move}
\end{equation}
where the edge $e$ connects vertices $v, v'$. In our case, we can write $\Sg CZ = C(iZ)$, so we have $m = 1$. Thus, for any vertex set, we can replace $C_{v_i}(\pm i Z)_i$ with 
\begin{equation}
    C_{v_i}(\pm i Z)_i \to C_{v_{j}} (\pm i Z)_i C_{e_{i, j}} (\pm i Z)_i  CZ_{e_{ij}, v_{j}}, \quad j = i \pm 1 ~(\text{mod } 4) 
\end{equation}
Starting from vertex $v_4$ and iterating the application of Eq. \eqref{eq:warmup-move} clockwise to the ``root'' vertex $v_1$ we find:
\begin{equation}
    \prod_{i = 1}^4 C_{v_i}(\pm i Z)_i \to C_{e_{34}}(iZ_4)\cdot C_{e_{23}}(Z_3Z_4Z_{e_{34}})\cdot
C_{e_{12}}(iZ_2Z_3Z_4Z_{e_{34}})\cdot C_{v_1}(\gaugedS_Z)
\end{equation}
The last term, $C_{v_1}(\gaugedS_Z)$, is a ``leftover'' piece we can discard because it acts trivially on the codespace where $\gaugedS_Z=+1$. Note that some of these $CZ$ gates can be further simplified; for instance, we have the combination 
$$C_4 (Z_{e_{12}} Z_{e_{23}} Z_{e_{34}} ) \xrightarrow{\bp = +1} C_4 (Z_{e_{41}})$$
Expanding (and performing the above simplification) the gauged $S_X$ stabilizer in terms of elementary gates yields
\begin{equation}
    \raisebox{-47pt}{
    \begin{tikzpicture}

        \node at (-2, 0) {$\gaugedS_X = $};
        \node[name=s, regular polygon, regular polygon sides = 4, inner sep = 0.6cm, draw, red, thick] at (0, 0) {};

        \foreach \anchor/\n in
        {corner 1/1/, corner 2/2, corner 3/3, corner 4/4}
        {
        \node[draw, circle, fill=red!50!white, inner sep = 1.5pt, text=black] at (s.\anchor) {\tiny{$\id$}};
        \node[draw, circle, fill=black, inner sep = 1pt, text=white] at ($(s.\anchor) + (135 + \n*90:0.75)$) {\tiny{$X$}};
        }
        \foreach \anchor/\placement/\n in
        {side 1/below/1, side 2/below/2, side 3/right/3, side 4/above/4}
        {
        \node[draw, circle, fill=red!30!white!70!blue, inner sep = 1.5pt, text=black] at (s.\anchor) {\tiny{$\id$}};
        }

        \draw [decorate,decoration={brace,amplitude=5pt,mirror,raise=4ex}] (-1,-0.5) -- (1,-0.5) node[midway,yshift=-3em]{$S_X$};

        \node at (1.5, 0) {$\times$};

        \node[name=s, regular polygon, regular polygon sides = 4, inner sep = 0.6cm, draw, red, thick] at (3, 0) {};
        
        \draw[thick, purple, dashed] (s.side 3) to (s.side 4);
        \draw[thick, purple, dashed] (s.side 3) to (s.side 1);
        
        \foreach \anchor/\n in
        {corner 1/1/, corner 2/2, corner 3/3, corner 4/4}
        {
        \node[draw, circle, fill=red!50!white, inner sep = 1.5pt, text=black] at (s.\anchor) {\tiny{$\id$}};
        \node[draw, circle, fill=black, inner sep = 1pt, text=white] at ($(s.\anchor) + (135 + \n*90:0.75)$) {\tiny{$\id$}};
        }
        \foreach \anchor/\placement/\n/\text in
        {side 1/below/1/\Sg, side 2/below/2/\id, side 3/right/3/\Sg, side 4/above/4/\id}
        {
        \node[draw, circle, fill=red!30!white!70!blue, inner sep = 1.5pt, text=black] at (s.\anchor) {\tiny{$\text$}};
        }

        \draw [decorate,decoration={brace,amplitude=5pt,mirror,raise=4ex}] (2,-0.5) -- (4,-0.5) node[midway,yshift=-3em]{Phase ``tail''};

        \node at (4.5, 0) {$\times$};

        \node[name=s, regular polygon, regular polygon sides = 4, inner sep = 0.6cm, draw, red, thick] at (6, 0) {};
        
        \draw[thick, purple, dashed] (s.side 1) to ($(s.corner 1) + (135 + 1*90:0.75)$);
        \draw[thick, purple, dashed] (s.side 1) to ($(s.corner 4) + (135 + 4*90:0.75)$);

        \draw[thick, purple, dashed] (s.side 4) to ($(s.corner 4) + (135 + 4*90:0.75)$);

        \draw[thick, purple, dashed] (s.side 2) to ($(s.corner 3) + (135 + 3*90:0.75)$);
        
        \foreach \anchor/\n in
        {corner 1/1/, corner 2/2, corner 3/3, corner 4/4}
        {
        
        \node[draw, circle, fill=red!50!white, inner sep = 1.5pt, text=black] at (s.\anchor) {\tiny{$\id$}};
        \node[draw, circle, fill=black, inner sep = 1pt, text=white] at ($(s.\anchor) + (135 + \n*90:0.75)$) {\tiny{$\id$}};
        }
        \foreach \anchor/\placement/\n/\text in
        {side 1/below/1/\id, side 2/below/2/\id, side 3/right/3/\id, side 4/above/4/\id}
        {
        \node[draw, circle, fill=red!30!white!70!blue, inner sep = 1.5pt, text=black] at (s.\anchor) {\tiny{$\text$}};
        }

        \draw [decorate,decoration={brace,amplitude=5pt,mirror,raise=4ex}] (5,-0.5) -- (7,-0.5) node[midway,yshift=-3em]{Deformed $CZ$s};

    \end{tikzpicture}
    }
\end{equation}

Every check above squares to one, and any two stabilizers commute, with a single exception: the decorated $X$ check and
the \emph{root} vertex check, whose group commutator,
$\mathbb{A}_{v_1}\,\gaugedS_X\,\mathbb{A}_{v_1}\gaugedS_X=\gaugedS_Z$, nevertheless remains in the stabilizer group.
These two checks therefore commute precisely on the codespace, where $\gaugedS_Z=+1$, and anticommute on the $\gaugedS_Z=-1$ sector. Note that the gauged codespace is \emph{three}-dimensional, the dimension of the $\CZbar=+1$ eigenspace rather than a power of two.

We return to (almost) the original code by measuring the gauge degrees of freedom in the computational basis; this will project out all the decorations we have added to the $[[4, 2, 2]]$ stabilizers, up to an application of some of the $W_{v_i}$. More specifically, since we had $\bp = +1$ prior to measuring out the edge ancillas, we should end up with an even number of $\ket{1}$ outcomes. For instance, suppose we measured $Z_{e_{12}} = Z_{e_{23}} = +1$ and $Z_{e_{34}} = Z_{e_{41}} = -1$. Then the decorated stabilizers become: 
\begin{equation}
    \raisebox{-35pt}{
    \begin{tikzpicture}

        \node at (-2, 0) {$\gaugedS_X \to $};
        \node[name=s, regular polygon, regular polygon sides = 4, inner sep = 0.6cm, draw, red, thick] at (0, 0) {};

        \foreach \anchor/\n in
        {corner 1/1/, corner 2/2, corner 3/3, corner 4/4}
        {
        \node[draw, circle, fill=red!50!white, inner sep = 1.5pt, text=black] at (s.\anchor) {\tiny{$\id$}};
        \node[draw, circle, fill=black, inner sep = 1pt, text=white] at ($(s.\anchor) + (135 + \n*90:0.75)$) {\tiny{$X$}};
        }
        \foreach \anchor/\placement/\n in
        {side 1/below/1, side 2/below/2, side 3/right/3, side 4/above/4}
        {
        \node[draw, circle, fill=red!30!white!70!blue, inner sep = 1.5pt, text=black] at (s.\anchor) {\tiny{$\id$}};
        }

        \node at (1.5, 0) {$\times ~(i) $};

        \node[name=s, regular polygon, regular polygon sides = 4, inner sep = 0.6cm, draw, red, thick] at (3.15, 0) {};
        
        \foreach \anchor/\n/\text in
        {corner 1/1/Z, corner 2/2/\id, corner 3/3/\id, corner 4/4/\id}
        {
        \node[draw, circle, fill=red!50!white, inner sep = 1.5pt, text=black] at (s.\anchor) {\tiny{$\id$}};
        \node[draw, circle, fill=black, inner sep = 1pt, text=white] at ($(s.\anchor) + (135 + \n*90:0.75)$) {\tiny{$\text$}};
        }
        \foreach \anchor/\placement/\n/\text in
        {side 1/below/1/\id, side 2/below/2/\id, side 3/right/3/\id, side 4/above/4/\id}
        {
        \node[draw, circle, fill=red!30!white!70!blue, inner sep = 1.5pt, text=black] at (s.\anchor) {\tiny{$\text$}};
        }

        \node at (5.5, 0) {$ = W_2 S_X W_2$};

    \end{tikzpicture}
    }
\end{equation}
\begin{equation}
    \gaugedS_Z \to - S_Z = W_2 S_Z W_2
\end{equation}
We obtain $S_X$ and $S_Z$ by applying $W_2$ again. It is straightforward to check for any other set of consistent edge measurement outcomes that the necessary correction is some combination of the $W_i$. 

Meanwhile, only the product $\CZbar = \prod_v \av$ is preserved when we measure out the edge ancilla, ensuring we end up in an eigenstate of $\CZbar$ in the \emph{original} codespace. 

Thus, measuring $\CZbar$ on it can prepare the $\ket{CZ}$ magic
states used in Section~\ref{sec:applications}, if we find the parity of the $\av$ stabilizers is $+1$. If we have an odd number of violated $\av$, we would infer $\CZbar = -1$, and thus prepare the (non-magical) logical state $\ket{11}$ when we return to the original code. In this case, we reset the logical state of the code to $\ket{++}$ and attempt the gauging process again. Only a constant number of tries should be necessary, scaling with the overlap of the initial logical state and the desired magic state.

\subsection{Qubit codes with $\mathbb{Z}_2$ Clifford transversal gates}

We now briefly consider the special case where $p = q = 2$, as this case demonstrates some additional structure that is not seen in the general case. Given that in this case the ancillary degrees of freedom have the same dimensionality as the local degrees of freedom of the code, we write all Pauli operators as $X$ and $Z$.
Take a qubit code $\cC$ with transversal gate $\T$ such that $\T^2 = \id$. It is straightforward to show that the stabilizers of $\cC$ fall into two groups, $\{S\}$ and $\{N\}$, such that 
\begin{equation}
\label{eq:global-z2-commutation}
    \ComGrp{S}{\T} = N_S \in \{N\}, \quad \ComGrp{N}{\T} = \id
\end{equation}
All stabilizers in $\{N\}$ commute with the operator $\T$, and any other stabilizer $S \in \{S\}$ commutes with $\T$ up to an element of $\{N\}$, which we will denote $N_S$.

We will show that the gauged code has stabilizers as given in Theorem \ref{thm:z2-gauged-stabilizers}.

\begin{theorem}[$\mathbb{Z}_2$ Gauged Stabilizers] \label{thm:z2-gauged-stabilizers}
Given a Pauli stabilizer code $\cC$, a transversal gate such that $\T^2 = \id$, a choice of vertex set $\Sigma$, and a gauging graph $\graphG = (\vertexV, \edgeE)$, the gauged code stabilizers have the following generic form:
    \begin{align}
    \gaugedS = \mathfrak{C}_{2}[S] = S \cdot \left ( \prod_{\substack{v \neq v_0 \\ v \in \invvertexset{S}}} \mathsf{S}^{\mu_v}_{p(v, v_0)} \prod_{e \in p(v, v_0)} C_e(N_S)_v \right ), \label{eq:z2-decorated-stabilizers} \\
    N_S = \ComGrp{S}{\T}, \quad i^{\mu_v} (N_S)_v = \ComGrp{S_v}{\T_v}
\end{align} 
\begin{equation}
    \av = \T_v \prod_{e \in \delta v} X_e, \quad \bp = \prod_{e \in \partial_2 f} Z_e, \quad f \text{ is a face of } \graphG
    \label{eq:av-bp-stabilizers}
\end{equation}
The set $\invvertexset{S} \subseteq \vertexV$ includes all vertex sets in the support of the stabilizer $S$, and the map $\partial_2$ takes a face of $\graphG$ to its boundary edges. We have chosen a \emph{root} vertex $v_0$ for every stabilizer of the initial code, and $p(v, v_0) \subseteq \edgeE$ is a path between vertices $v, v_0$ on the gauging graph. Additionally, $$\mathsf{S}_{p(v, v_0)} \equiv \frac{\id + \prod_{e \in p(v, v_0)} Z_e}{2} + i \frac{\id - \prod_{e \in p(v, v_0)}Z_e}{2}.$$
\end{theorem}
\begin{proof}
    See below.
\end{proof}

The ``invariant'' stabilizers $\{N\}$ are useful to define because they will partially control the non-Abelian (non-stabilizer) behavior of the gauged code: $\av$ and $\gaugedS$ stabilizers of the gauged code only commute up to the gauged version of some element of $\{N\}$ (see Corollary \ref{cor:z2-commutation-relations} below).

\begin{corollary}[Commutation relations of Clifford-decorated stabilizers]
\label{cor:z2-commutation-relations}
The nontrivial commutation relations between the stabilizers in Theorem \ref{thm:z2-gauged-stabilizers} are given by:
\begin{align}
    \ComGrp{\gaugedS}{\mathbb{A}_{v_0}}
    =\mathbb{N}_S,
    \qquad
    \mathbb{N}_S=\mathfrak{C}_2[N_S].
\end{align}
\begin{equation}
    \ComGrp{\gaugedS_i}{\gaugedS_j}
    =\mathbb{B}_{f(i,j)},
\end{equation}
for some cycle $f(i,j)$.  The remaining commutators are trivial on the full Hilbert space:
\begin{equation}
    \ComGrp{\bp}{\av}
    =
    \ComGrp{\bp}{\gaugedS}
    =
    \id.
\end{equation}
\end{corollary}
\begin{proof}
    See Appendix~\ref{app:gauging-procedure-proofs}, Proof~\ref{proof:clifford-decorated-commutation-relations}.
\end{proof}

We will now derive Theorem \ref{thm:z2-gauged-stabilizers}. We assume some choice of gauging graph $\graphG = (\vertexV, \edgeE)$ and some choice of vertex sets $\Sigma: \physq \to \vertexV$.

\textbf{Symmetry Enrichment:} The first step of the gauging map as defined in Eq. \ref{eq:gauging-circuit} is to apply $\T_v$ to each vertex set, controlled on the corresponding vertex ancilla: this is the ``symmetry enrichment'' step. All stabilizers of the original code $\cC$ will be deformed:

\begin{equation}
    \begin{aligned}
        S & \xrightarrow{S.E.} S \cdot \prod_{v \in \invvertexset{S}} C_v \ComGrp{S_v}{\T_v}\\
        N_S & \xrightarrow{S.E.} N_S \cdot \prod_{v \in \invvertexset{N_S}} C_v \ComGrp{(N_S)_v}{\T_v}
    \end{aligned}
\end{equation}
Although we defined the commutation relation between the stabilizers and the full transversal gate $\T$ in Eq. \ref{eq:global-z2-commutation}, the commutation relations when restricted to a given vertex set may differ by a phase. In particular, due to the fact we are dealing with Pauli stabilizers, a Clifford operator $\T$, and we know $\ComGrp{S}{\T} = N_S$, it is straightforward to show that
\begin{equation}
    \ComGrp{S_v}{\T_v} = i^{\mu_v} (N_S)_v, \quad \ComGrp{(N_S)_v}{\T_v} = i^{2\mu_v} = (-1)^{\mu_v} 
\end{equation}
for some $\mu_v = 0, 1, 2, 3$. Since for any stabilizer $S$, $\ComGrp{S}{\T}$ must be Hermitian, we must have $i^{\sum_v \mu_v} = 1 \implies \sum_v \mu_v = 0 ~(\text{mod } 4)$, summing over $v \in \invvertexset{S}$. 

The vertex ancilla are initialized in the $\ket{+}$ state, and so under the symmetry enrichment circuit their stabilizers are deformed as well:
\begin{equation}
    X_v \xrightarrow{S.E} X_v \T_v
\end{equation}

\textbf{Cluster Entangler:} The next step of the gauging map is to apply the graph cluster entangler between the vertex ancilla and the edge ancilla, which we initialize in the $\ket{0}$ state. The graph cluster entangler is defined as: 
\begin{equation}
    \prod_{v \in \vertexV} \prod_{e \in \delta v} C_v X_e
\end{equation}
Only the vertex stabilizers and the edge ancilla stabilizers are impacted: 
\begin{equation}
    X_v \T_v \xrightarrow{C.E.} X_v \T_v \prod_{e \in \delta v} X_e
\end{equation}
\begin{equation}
    Z_e \to L_e = Z_{v} Z_{e} Z_{v'}, \quad \partial e = \{v, v'\}
\end{equation}
Recall that $\delta: \vertexV \to \edgeE$ and $\partial: \edgeE \to \vertexV$ are the co-boundary and boundary maps of $\graphG$. 

\textbf{Measuring out vertices:} For the final step, we project the vertex ancilla back to the $\ket{+}$ state. The deformed vertex stabilizers become the $\av$ stabilizers:
\begin{equation}
    X_v \T_v \prod_{e \in \delta v} X_e \xrightarrow{\dyad{+}} \av = \T_v \prod_{e \in \delta v} X_e
\end{equation}
Let $p \subseteq \edgeE$ be a collection of edges forming a closed path. This defines the stabilizer $\bp$, which commutes with the projector $\dyad{+}$, although individual $L_e$ stabilizers do not.
\begin{equation}
    \prod_{e \in p, \partial p = 0} L_e = \prod_{e \in p} Z_e \xrightarrow{\dyad{+}} \bp 
\end{equation}
To find the surviving deformed stabilizers of the original code after the projection, we need to push the support of the control gates away from the vertex ancilla. For the invariant stabilizers $N_S$, we can use the $L_e$ stabilizers directly to push support away from the vertex ancilla:
\begin{equation}
\begin{aligned}
    N_S \cdot \prod_{v \in \invvertexset{N_S}} C_v \ComGrp{(N_S)_v}{\T_v}  &= N_S \cdot \prod_{v \in \invvertexset{N_S}} Z_v^{\mu_v} \\
    &\xrightarrow{L_e = +1} \mathbb{N}_S \equiv N_S \prod_{v \in \invvertexset{N_S} / \{v_0\}} \prod_{e \in p(v, v_0)} Z_e^{\mu_v}
\end{aligned}
\end{equation}
Here, $v_0$ is some ``root'' vertex we have chosen by convention, and $p(v, v_0) \subseteq \edgeE$ is a path connecting vertices $v, v_0$ on the gauging graph $\graphG$ ($\partial p(v, v_0) = \{v, v_0\}$). There is no longer any support on the vertex ancilla, and so under the projection, this stabilizer remains unchanged: $\mathbb{N}_S \xrightarrow{\dyad{+}} \mathbb{N}_S$.

For the other stabilizers $\{S\}$, we use the following identities, which are valid on the subspace where $L_e = +1$:

\begin{equation}
    C_v U \xrightarrow{L_e = +1} C_{u} U \prod_{e \in p(v, u)} C_e U \equiv C_{u} U \cdot C_{p(v, u)} U, \quad U = U^\dagger
    \label{eq:control-gate-pushing}
\end{equation}
\begin{equation}
    \mathsf{S}^{\mu_v}_{v} \mathsf{S}^{-{\mu_v}}_{u} \xrightarrow{L_e = +1} \mathsf{S}^{\mu_v}_{p(v, u)} \cdot \prod_{e \in p(v, u)} (C_uZ_{e})^{\mu_v}, \quad \mathsf{S}_{p(v, u)} \equiv \frac{1 + \prod_{e \in p(v, u)} Z_e}{2} + i \frac{1 - \prod_{e \in p(v, u)}Z_e}{2}
    \label{eq:S-vertex-edge-identity}
\end{equation}
Applying these to the deformed $S$ stabilizer after the cluster entangler step yields:
\begin{equation}
\begin{aligned}
    S \cdot \prod_{v \in \invvertexset{S}} C_v \ComGrp{S_v}{\T_v} &= S \cdot \prod_{v \in \invvertexset{S}} \Sg^{\mu_v}_{v} C_v (N_S)_v \\ 
    \xrightarrow{L_e = +1} S \cdot &C_{v_0} \left ( N_S \prod_{v \neq v_0} \prod_{e \in p(v, v_0)} Z_e^{\mu_v}  \right ) \cdot \left ( \prod_{v \neq v_0} \mathsf{S}^{\mu_v}_{p(v, v_0)} \prod_{e \in p(v, v_0)} C_e (N_S)_v \right ) \\
    &= S \cdot C_{v_0} \mathbb{N}_S \cdot \left ( \prod_{v \neq v_0} \mathsf{S}^{\mu_v}_{p(v, v_0)} \prod_{e \in p(v, v_0)} C_e(N_S)_v \right )
\end{aligned}
\end{equation}
Here, $v \neq v_0$ is understood to mean the set $\invvertexset{S} / \{v_0\}$. There is still support on the root vertex ancilla, $v_0$. However, it is in the form of a controlled $\mathbb{N}_S$ stabilizer. In the gauged code's codespace, $\mathbb{N}_S = +1$, and so this controlled-gate acts trivially. Thus, we can remove it to obtain a deformed stabilizer $\gaugedS$ that commutes with the projection $\dyad{+}$:
\begin{equation}
    S \cdot C_{v_0} \mathbb{N}_S \cdot \left ( \prod_{v \neq v_0} \mathsf{S}^{\mu_v}_{p(v, v_0)} \prod_{e \in p(v, v_0)} C_e(N_S)_v \right ) \xrightarrow{\mathbb{N}_S = +1} \gaugedS \equiv S \cdot \left ( \prod_{v \neq v_0} \mathsf{S}^{\mu_v}_{p(v, v_0)} \prod_{e \in p(v, v_0)} C_e(N_S)_v \right ) 
    \label{eq:z2-decorated-stabilizers-final}
\end{equation}
Note that this expression holds for $\mathbb{N}_S$ stabilizers as well. In this case, $N_{N_S} = \id$, killing the control gates, while $\mu_v = 0, 2$ for all vertex sets, turning the phase gates $\Sg$ into products of Pauli $Z$. This completes the proof of Theorem \ref{thm:z2-gauged-stabilizers}: we have derived the stated expressions for the gauged code stabilizers.

\begin{remark}[Measuring non-Hermitian stabilizers]\label{rem:measurement-circuit}
    From Eqs. \ref{eq:z2-decorated-stabilizers}~-~\ref{eq:av-bp-stabilizers} we see that the gauged stabilizers are always unitary, though not Hermitian in general. 
    It is always possible to ``measure'' a non-Hermitian but unitary operator by using an ancilla qudit to project onto one of its eigenspaces and measuring the ancilla to learn onto which eigenspace the state was projected.
    This is achievable through a generalized quantum phase estimation circuit~\cite{kitaev1995quantummeasurementsabelianstabilizer}.
\end{remark}

\subsection{Structure of generic gauged codes}

In the general case, where we have a qudit code and a $\mathbb{Z}_p$ symmetry, we cannot draw on the structure we took advantage of in the previous section. The resulting gauged stabilizers are not in general unitary, but we can still write down operators that define the gauged codespace.

\begin{theorem}[Generic Gauged Stabilizers]
    Consider a qudit code $\cC = \langle S \rangle$ with transversal gate $\T$ such that $\T^p = \id$. Assume some $\Sigma, \graphG = (\vertexV, \edgeE)$. We assign an ancillary qu(p)it to each vertex $v \in \vertexV$ and each edge $e \in \edgeE$. The vertices are initialized in the $\Xp = +1$ eigenstate while the edges are initialized in the $\Zp = +1$ computational basis state. 

    Define the symmetrized stabilizers 
    \begin{equation}
        \sym = \frac{1}{p} \sum_{a = 0}^{p-1} \T^a S (\T^a)^\dagger
    \end{equation}

    Then the gauged code $\cC'$ has stabilizers:
    \begin{equation}
        \av = \T_v \prod_{e \in \delta v} \Xp_e^{\eta_v(e)}, \quad \bp = \prod_{e \in p} \Zp^{\eta_p(e)}_e, \quad \bar{\mathbb{S}} = \Circo[\sym]
    \end{equation}
    \begin{equation}
        \Circo[\sym] = \sum_{\mathbf{a'}}  \prod_{v \neq v_0} \prod_{e \in p(v, v_0)} \dyad{\eta_p(e) a_v'}_e \otimes \T^{\mathbf{a'}}_{v \neq v_0} \sym (\T^{\mathbf{a'}}_{v \neq v_0})^\dagger 
    \end{equation}
    \label{thm:generic-gauged-stabilizers}
\end{theorem}
\begin{proof}
    The derivation of the $\av$ and $\bp$ stabilizers is much the same as the $\mathbb{Z}_2$ case.
\begin{equation}
    \av =  \T_v \prod_{e \in \delta v} \Xp_e^{\eta_v(e)}, \quad 
    \bp = \prod_{e \in \partial p} \Zp^{\eta_p(e)}_e 
\end{equation} where $\eta_v(e) = \pm 1$ captures whether an edge $e$ is going into or out of a given vertex $v$, while $\eta_p(e) = \pm 1$ denotes whether the edge $e$ is aligned or anti-aligned with the orientation of the path (here, a cycle) $p$. 

The bare stabilizers of the code $\cC$ can no longer be separated into invariant stabilizers $\{N\}$ and variant stabilizers $\{S\}$ under the action of $\T$. Instead, we consider ``symmetrized'' stabilizers:
\begin{align}
    \sym = \frac{1}{p} \sum_{a = 0}^{p-1} \T^a S (\T^a)^\dagger
\end{align} Immediately, we see that $\sym$ is invariant under the action of $\T$:
\begin{equation}
    \T^b \sym (\T^b)^\dagger = \sym, \quad b = 0, \dots, p-1
\end{equation} 
Under the symmetry enrichment step, we have: 
\begin{align}
    \sym &\xrightarrow{\textrm{S.E.}} \sum_{\mathbf{a}} \dyad{\mathbf{a}}_{\invvertexset{S}} \otimes \T^{\mathbf{a}} \sym (\T^{\mathbf{a}})^\dagger, \quad \mathbf{a} \in \mathbf{F}_{p}^{|\invvertexset{S}|}, \quad
    \T^{\mathbf{a}} \equiv \prod_{v \in \invvertexset{S}} \T^{a_v}_v
\end{align}
Note that this is not a trivial deformation, as $\sym$ is not necessarily invariant under the local action $\T_v$. The deformed $\sym$ is left unchanged by the cluster entangler, although we should remember that now the stabilizers $L_e =(\Zp_{v}^\dagger \Zp_e \Zp_{v'})^{\eta_v(e)}$ have been introduced.

Before we project out the vertex ancilla in the $\dyad{+}$ basis, we need to remove any support from the vertex ancilla that anti-commutes with this projection. Pick a root vertex $v_0 \in \invvertexset{S}$. We will re-index the sum over $\mathbf{a}$ in terms of its component $a_0$ on the vertex $v_0$:

\begin{align}
    \sum_{\mathbf{a}} \dyad{\mathbf{a}}_{\invvertexset{S}} &= \sum_{\mathbf{a'}} \sum_{a_0 = 0}^{p-1} \dyad{a_0}_{v_0} \otimes \dyad{\mathbf{a'}}_{v\neq v_0}, \quad \mathbf{a'} \in \mathbf{F}_p^{|\invvertexset{S}|-1}
\end{align} We use $v \neq v_0$ as shorthand for $\invvertexset{S}\backslash \{v_0\}$. The new vector $\mathbf{a'}$ captures the state of the other vertex ancilla, excluding the root. Plugging back into the expression for the decorated stabilizer yields:
\begin{align}
    \sym &\xrightarrow{\textrm{S.E.}} \sum_{\mathbf{a'}} \sum_{a_0} \dyad{a_0}_{v_0} \otimes \dyad{\mathbf{a'}}_{v\neq v_0} \otimes  \T^{\mathbf{a'}}_{v\neq v_0} \T^{a_0}_{v_0} \sym (\T^{a_0}_{v_0})^\dagger  (\T^{\mathbf{a'}}_{v\neq v_0} )^\dagger  \\
    &=\sum_{\mathbf{a'}} \sum_{a_0} \dyad{a_0}_{v_0} \otimes \dyad{\mathbf{a'}}_{v\neq v_0} \otimes \T^{\mathbf{a'} - a_0}_{v\neq v_0} \sym (\T^{\mathbf{a'} - a_0}_{v\neq v_0})^\dagger \\
    &=\sum_{\mathbf{a'}} \left ( \sum_{a_0} \dyad{a_0}_{v_0} \otimes \dyad{\mathbf{a'}+a_0}_{v\neq v_0}  \right )\otimes \T^{\mathbf{a'}}_{v\neq v_0} \sym (\T^{\mathbf{a'}}_{v\neq v_0})^\dagger
\end{align}
Here, $\mathbf{a}' \pm a_0$ indicates elementwise addition/subtraction of the constant $a_0$ (modulo $p$). We have used the invariance of $\sym$ to remove the dependence on $a_0$ from the action on the physical qudits. After the cluster state entangler, within the resultant codespace, we can further simplify the form of $\sym$. We take advantage of the following identity:
\begin{equation}
    \dyad{a} = \frac{1}{p} \sum_{b = 0}^{p-1} \omega^{ab} \Zp^b 
\end{equation}
After some algebra, using the fact that $L_e = (\Zp_{v}^\dagger \Zp_e \Zp_{v'})^{\eta_v(e)} = +1$ for all edges $e$ on the gauging graph, we find:
\begin{align}\label{eq:vertex-ancilla-dropout}
    \sum_{a_0 = 0}^{p-1} \dyad{a_0}_{v_0} \otimes \dyad{\mathbf{a'} + a_0}_{v \neq v_0} \xrightarrow{L_e = +1} \id_{v_0} \otimes \prod_{v \neq v_0} \prod_{e \in p(v, v_0)} \dyad{\eta_p(e) a'_v}_e
\end{align}

Recall $p(v, v_0)$ is a set of edges forming a path on the gauging graph between vertices $v, v_0$. The function $\eta_p(e) = \pm 1$ denotes whether a given edge is aligned with the path $p(v, v_0)$.
The support on the vertex ancilla has completely dropped out, and we are left with the following expression for the gauged, symmetrized stabilizers:
\begin{align}
\sym \xrightarrow{\textrm{Gauging Map}} \bar{\mathbb{S}} = \sum_{\mathbf{a'}}  \prod_{v \neq v_0} \prod_{e \in p(v, v_0)} \dyad{\eta_p(e) a_v'}_e \otimes \T^{\mathbf{a'}}_{v \neq v_0} \sym (\T^{\mathbf{a'}}_{v \neq v_0})^\dagger 
\end{align}
This completes the proof of Theorem \ref{thm:generic-gauged-stabilizers}.
\end{proof}

\begin{theorem}[$\mathbb{Z}_2$ Equivalence] \label{thm:z2-generic-equivalence}
For a qubit code $\cC$ and a $\mathbb{Z}_2$ symmetry $\T$, the stabilizers in Theorems \ref{thm:z2-gauged-stabilizers} and \ref{thm:generic-gauged-stabilizers} generate the same gauged codespace.
\end{theorem}
\begin{proof}
    Given the same root and path conventions, the $\av$ and $\bp$ stabilizers are clearly identical. Consider the generic form of the decorated stabilizers in the $\mathbb{Z}_2$ case. 
    
    Start with an invariant stabilizer $N$, and notice that by definition, it is already symmetrized.
    \begin{equation}
        \bar{N}_S = \frac{1}{2} \left ( N_S + \T N_S \T \right ) = N_S 
    \end{equation}
    Thus, we have:
    \begin{equation}
        \begin{aligned}
            \bar{\mathbb{N}}_S &= \sum_{\bm{a}} \prod_{v \neq v_0} \prod_{e \in p(v, v_0)} \dyad{\eta_p(e) a_v}_e \otimes \T^{\mathbf{a}}_{v \neq v_0}  N_S \T^{\mathbf{a}}_{v \neq v_0}\\
            &= \sum_{\bm{a}} \prod_{v \neq v_0} \prod_{e \in p(v, v_0)} \dyad{\eta_p(e) a_v}_e \otimes   (-1)^{\bm{a}\cdot \bm{\mu}} N_S \\
            &= \left ( \prod_{v \neq v_0} \sum_{a_v = 0, 1} (-1)^{a_v \mu_v} \prod_{e \in p(v, v_0)} \dyad{\eta_p(e) a_v}_e  \right ) \otimes N_S \\
            &= \left ( \prod_{v \neq v_0} \prod_{e \in p(v, v_0)} Z^{\mu_v}_e  \right ) \otimes N_S \\
        \end{aligned}
    \end{equation}
    which matches Theorem \ref{thm:z2-gauged-stabilizers}. Similarly, for the variant stabilizers: 
    \begin{equation}
        \sym = \frac{1}{2} \left ( S + \T S \T \right ) = \frac{S}{2} \left ( 1 + N_S \right )
    \end{equation}
    \begin{equation}
        S \T_v S \T_v = i^{\mu_v} (N_S)_v \implies \T_v S = i^{\mu_v} S (N_S)_v \T_v 
    \end{equation}
    Plugging into the general form: 
    \begin{equation}
        \begin{aligned}
            \bar{\gaugedS} &= \sum_{\bm{a}} \prod_{v \neq v_0} \prod_{e \in p(v, v_0)} \dyad{\eta_p(e) a_v}_e \otimes \frac{1}{2} \T^{\mathbf{a}}_{v \neq v_0} S (1 + N_S) \T^{\mathbf{a}}_{v \neq v_0}\\
            &= \sum_{\bm{a}} \prod_{v \neq v_0} \prod_{e \in p(v, v_0)} \dyad{\eta_p(e) a_v}_e \otimes \frac{i^{\bm{a} \cdot \bm{\mu}}}{2} S (N_S)^{\bm{a}}  (1 + (-1)^{\bm{a} \cdot \bm{\mu}} N_S) \\
            &= S \cdot \sum_{\bm{a}} \prod_{v \neq v_0} \prod_{e \in p(v, v_0)} \dyad{\eta_p(e) a_v}_e \otimes \frac{i^{\bm{a} \cdot \bm{\mu}}}{2} (N_S)^{\bm{a}}  (1 + (-1)^{\bm{a} \cdot \bm{\mu}} (-1)^{\bm{a} \cdot \bm{\mu}}) \\
            &=  S \cdot \sum_{\bm{a}} \prod_{v \neq v_0} \prod_{e \in p(v, v_0)} \dyad{\eta_p(e) a_v}_e \otimes i^{\bm{a} \cdot \bm{\mu}}(N_S)^{\bm{a}} \\
            &= S \cdot \prod_{v \neq v_0} \Sg_{p(v,v_0)}^{\mu_v} \prod_{e \in p(v, v_0)} C_e (N_S)_v = \Circo_2[S]
        \end{aligned}
    \end{equation}
This coincides with the decorated stabilizers derived in the $\mathbb{Z}_2$ section, concluding the proof.
\end{proof}

\subsection{Ungauging} \label{sec:ungauging}

We now consider steps 3 and 4 of Protocol. \ref{def:gauging-measurement-protocol}, which are the ``ungauging'' and correction steps which project us back onto our original code (but now in a logical eigenstate of $\T$). Ungauging consists of measuring out the edge ancillas in the $\Zp$ basis. This automatically reads out the $\bp$ stabilizers. Suppose just before step 3, we were in the $\bp = +1$ subspace. We discuss the scenario that takes errors leading to $\bp\neq 1$ outcomes into account in Sec. \ref{sec:spacetime-distance} when we prove the fault distance of the gauging measurement protocol.

First, consider the special case where all the edges are measured to be in the $\ket{0}$ state. In this case, we can see by inspection of the stabilizers in either Theorem \ref{thm:z2-gauged-stabilizers} or Theorem \ref{thm:generic-gauged-stabilizers} that projecting the edges to $\ket{0}$ projects the decorated stabilizers back onto the stabilizers of the original code. Even in the case where some edges are measured to not be in the $\ket{0}$ state, if all $\bp = +1$, the edge configuration we project onto should be related to the all $\ket{0}$ state by the application of some product of $\av$ stabilizers. This follows from the fact the gauging graph is homologically trivial; all operators with $\Xp$ support on the edge ancilla that also commute with the $\bp$ stabilizers must themselves be stabilizers. If they are stabilizers with $\Xp$ support, they must be some configuration of the $\av$ stabilizers.

Let $\ket{\mathbf{z}}_{\edgeE}$, $\mathbf{z} \in \mathbb{Z}_p^{|\edgeE|}$ denote the state of the edge ancilla after we measure them in the $\Zp$ basis, and let $\ket{\Psi}_{\physq}$ be the state on the matter qudits. Then, by the argument in the previous paragraph:
\begin{equation}
    \ket{\mathbf{z}}_{\edgeE} \ket{\Psi}_{\physq} = \left (\prod_{v \in \Omega \subseteq \vertexV} \av^{g_v} \right ) \ket{0}_{\edgeE}\ket{\Psi'}_{\physq}, \quad \ket{\Psi'}_{\physq} = \prod_{v \in \Omega} (\T^{g_v}_v)^\dagger \ket{\Psi}_{\physq}
\end{equation}
Here, $\ket{\Psi'}_{\physq}$ is a codestate of the \emph{original} code, and $g_v \in \{0, \dots, p-1\}$. The state $\ket{\Psi}_{\physq}$ belongs to a deformed code, which is related to our original code through the application of the local Clifford correction conjugation by this same set of $\av$ stabilizers: this is the local Clifford correction $\prod_{v \in \Omega} (\T^{g_v}_v)^\dagger$. Note that there are $p$ distinct products of the $\av$ with the same action on the edge ancilla, which differ on the matter qudits by some number of applications of the full symmetry operator $\T$.

No matter what the measured states of the edge ancilla, by measuring in the $\Zp$ basis, we also glue the $\av$ stabilizers together into $\T = \prod_v \av$. Thus, we also project the matter qudits into an eigenstate of $\T$ through this ungauging measurement. This ensures that the ambiguity in the Clifford correction we should apply doesn't matter, since the options differ by powers of $\T$, and thus only by a global phase corresponding to the relevant eigenvalue of $\T$.

\ifdefined\ismainfile
\else
\bibliographystyle{unsrt}
\bibliography{bib}
    \end{document}
\fi

\section{Algebraic formalism and distance of the gauged code}\label{sec:space-distance}\ifdefined\ismainfile
\else
    \documentclass[11pt]{article}
    
    \input{preamble.tex}
    
    \author{}
    \title{Algebraic formalism and distance of the gauged code}
    
    \begin{document}

    \maketitle

\fi

The gauging measurement protocol was defined in the section above, it remains to prove that this protocol is \emph{fault-tolerant}. Here, ``fault-tolerance'' means that the minimum un-detectable fault that causes a logical failure is of large weight. This includes the application of a physical logical operator to the code, and inferring an incorrect measurement outcome of the logical Clifford $\T$. As an intermediate step to the proof of large fault distance in both space and time, we now focus on characterizing the properties of the gauged code. Specifically, we demonstrate that the gauged code's distance satisfies $d_g\geq d/\nu$, where $\nu$ is the maximum size of a vertex set, so long as the graph desiderata in Def.~\ref{def:graph-desiderata} are satisfied. 

To enable the proof, we introduce an alternate, algebraic formulation of the gauging procedure from Section \ref{sec:gauging-procedure}, following Refs.~\cite{Bulmash_2019,Prem_2019}. After re-casting the gauging protocol in this language, we demonstrate that the gauged codespace and the $\T$-symmetric subspace of the original stabilizer code are isomorphic. This isomorphism facilitates the proof of the gauged code distance in Theorem \ref{thm:space-distance}.
 
\begin{table}[H]
    \centering
    \begin{tabular}{@{}p{0.30\linewidth}p{0.67\linewidth}@{}}
    \hline
        Notation & Denotes \\
    \hline
        $\Agrp = \langle \T_1, \dots \T_n \rangle$ & Finite Abelian group generated by symmetries being gauged. \\
        $\widehat{\Agrp}$ & Irreducible representations of $\Agrp$.\\
        $\graphG = (\vertexV, \edgeE, \faceF)$ & Gauging graph and associated vertices, edges, and faces. \\
        $\Agrp[\vertexV], \Agrp[\edgeE], \Agrp[\faceF]$ & Assignments of elements of $\Agrp$ to every vertex, edge, or face. \\
        $\widehat{\Agrp}[\vertexV], \widehat{\Agrp}[\edgeE], \widehat{\Agrp}[\faceF]$ & Assignments of elements of $\widehat{\Agrp}$ to every vertex, edge, or face. \\
        $\delta_i, \partial_i$ & Coboundary and boundary maps \\
        $\T^g$ & Clifford symmetry corresponding to group element $g \in \Agrp$.\\
        $\T^{\bm{a}}$ & Product of local factors $\T_v$ according to the assignment $\bm{a} \in \Agrp[\vertexV]$\\
        $\Xp, \Zp$ & Generalized Pauli operators for $\Agrp$\\
        $\Xp^{\bm{a}}$ & Product of $\Xp$ according to the group element assignment $\bm{a}$.\\
        $\Zp^{\bm{\chi}}$ & Product of $\Zp$ according to the irrep assignment $\bm{\chi}$. \\
        $\mathcal{H}_{\rm{sym.}}$ & Subspace determined by $\T = +1$. \\
        $\mathcal{H}_{\rm{gau.}}$ & Subspace determined by $\av = +1$, $\bp = +1$ for all vertices and faces. \\
        $\Gamma, \tilde{\Gamma}$ & The gauging map introduced in this section and the gauging circuit (Section \ref{sec:gauging-procedure}). \\
        $h_{\Agrp}(\graphG)$ & Generalized Cheeger constant of $\graphG$.
        
    \end{tabular}
    \caption{Notation used for this section.}
    \label{tab:sec-3-notation}
\end{table}

\subsection{Algebraic formalism for gauging}

We begin by reformulating the gauging procedure discussed in Sec.~\ref{sec:gauging-procedure} in terms of algebraic operations. Suppose we are gauging some set of commuting Clifford symmetries $\{\T_i\}$ with $\T_i^{p_i} = \id$. These Clifford operators generate a finite Abelian group $\Agrp = \times_i \mathbb{Z}_{p_i}$. For example, for symmetries satisfying $\T^2 = \id$, the corresponding Abelian group is $\Agrp = \mathbb{Z}_2$.

\begin{definition}[Gauging 2-complex]
\label{def:gauging-complex}
Let $\vertexV$, $\edgeE$, $\faceF$ denote the vertices, edges and faces of the gauging graph $\graphG=(\mathsf{V},\mathsf{E})$ from Section~\ref{sec:gauging-procedure}. For any finite Abelian group $A$, let $\Agrp[\vertexV]$, $\Agrp[\edgeE]$, $\Agrp[\faceF]$ denote assignments of an element of $A$ to every vertex, edge and face,  respectively.
The relevant coboundary maps are 
\begin{align}
    \graphG^\bullet = \Agrp[\vertexV] \mathop{\rightarrow}^{\delta_{1}} 
    \Agrp[\edgeE] \mathop{\rightarrow}^{\delta_2} 
    \Agrp[\faceF] .
\end{align}
Similarly, let $\widehat{\Agrp}[\vertexV]$, $\widehat{\Agrp}[\edgeE]$, $\widehat{\Agrp}[\faceF]$ denote assignments of an irreducible representation label of $\Agrp$ to every vertex, edge and face, respectively.
The relevant boundary maps are
\begin{align}
    \graphG_\bullet = \widehat{\Agrp}[\vertexV] \mathop{\leftarrow}^{\partial_{1}} 
    \widehat{\Agrp}[\edgeE] \mathop{\leftarrow}^{\partial_2} 
    \widehat{\Agrp}[\faceF] .
\end{align}
\end{definition}

We use $+$ to denote the group operation in $\Agrp$, and bold symbols to denote assignments of group element to vertices, edges or faces.
If $\bm{a}, \bm{b}\in \Agrp[\vertexV]$, then their sum, $\bm a+\bm b $ is defined vertex-wise $(\bm a + \bm b)_v =  a_v + b_v$.
For a group element $g \in \Agrp$, we abuse notation and write $ g \id \in \Agrp[\vertexV]$ for the constant vertex assignment where $(g\id)_v = g$ for every $v\in \vertexV$. We write $g \bm{e_v} \in \Agrp[\vertexV]$ for the assignment with group element $g$ on vertex $v \in \vertexV$, and the identity element $0$ elsewhere. Similarly $g\bm{e_f} \in \Agrp[\faceF]$ is the assignment with $g$ on face $f \in \faceF$ and identity elsewhere.

Furthermore, we define $\T^g$ where $g \in \Agrp$ as the logical Clifford operator representing the group element $g$. In the case $\Agrp = \mathbb{Z}_q = \langle \T ~|~ \T^p = \id \rangle$ for $p\geq 2$ is a cyclic group, we have $g = 0, \dots, p-1$. Then $\T^g$ is simply the sole generator $\T$ multiplied $g$ times. In general, if we have Abelian group $\Agrp = \mathbb{Z}_{p_1} \times \cdots \times \mathbb{Z}_{p_n}$ we define:
\begin{equation}
    \T^g = \prod_{i=1}^n \T_i^{g_i}, \quad g = (g_1, \dots, g_n) \in \Agrp.
\end{equation}
When we write $\T^{\bm{a}}$, where the superscript is bolded, this denotes the multiplication of different local symmetry operators for each vertex set, determined by the value of $\bm{a} \in \Agrp[\vertexV]$ at each vertex.
\begin{equation}
    \T^{\bm{a}} = \prod_{v \in \vertexV} (\T_v)^{a_v}, \quad a_v =((a_v)_1, \dots, (a_v)_n) \in \Agrp
\end{equation}

We also generalize the clock and shift operators defined for $\Agrp = \mathbb{Z}_p$ to the case $\Agrp = \mathbb{Z}_{q_1} \times \cdots \times \mathbb{Z}_{q_n}$. For each cyclic group component $\mathbb{Z}_{p_i}$, the shift operator is $\Xp^{(i)}$ and the clock operator is $\Zp^{(i)}$. Shift operators are associated with group elements (through the $+$ group operation) while clock operators evaluate an irreducible representation of $\mathbb{Z}_{p_i}$ on a given state.
\begin{equation}
    \Zp^{(i)} \Xp^{(i)} = \omega_i \Xp^{(i)} \Zp^{(i)}, \quad \omega_i = \exp{2\pi i/p_i}
\end{equation}
The clock and shift operators for the full group are constructed from the tensor product of these elementary generators, with the corresponding commutation relation:
\begin{equation}
    \Xp \Zp = \omega \Zp \Xp, \quad \omega = \prod_{i = 1}^n \omega_i = \exp{2 \pi i \sum_{i = 1}^n 1/p_i} .
\end{equation}

The generalized Pauli operators $\Xp$ and $\Zp$ can be labeled by bolded superscripts indicating their support on the gauging graph. These superscripts may be elements of $\Agrp[\vertexV]$ or $\Agrp[\edgeE]$, if we are discussing Pauli operator support on the vertex ancillas or the edge ancillas respectively. For example, a generic configuration of shift operators on the vertex ancilla may be written:
\begin{equation}
    \Xp^{\bm{a}} = \prod_{v \in \vertexV} \Xp_v^{a_v}, \quad \bm{a} \in \Agrp[\vertexV]
\end{equation}
while a set of clock operators supported on the edges is labeled by a choice of irreducible representation for each edge:
\begin{equation}
    \Zp^{\bm{b}} = \prod_{e \in \edgeE} \Zp_e^{b_e}, \quad \bm{b} \in \widehat{\Agrp}[\edgeE]
\end{equation}

We call the symmetric subspace of our initial stabilizer code $\cC$ under $\T$ the \textit{symmetric code}. 

\begin{definition}[Symmetric code]
\label{def:symmetric-code}
Let
\begin{align}
    \mathcal H_{\rm sym.} := \{\ket{\psi} : \T^a\ket{\psi}=\ket{\psi} \quad \forall a\in \Agrp\}.
\end{align}
be the symmetric subspace. The projector
\begin{align}
    \Pi_{\rm sym.} = \frac{1}{|\Agrp|} \sum_{a \in \Agrp} \T^a
\end{align}
projects onto $\mathcal H_{\rm sym.}$. The symmetric code is the intersection of this subspace with the codespace of the initial code, $\cC$. The symmetric code is defined by $\T$ and the symmetrized versions of the original code stabilizers: 
\begin{equation}
    \sym = \Pi_{\rm sym.} S \Pi_{\rm sym.} = \frac{1}{|\Agrp|} \sum_{g \in \Agrp} (\T^g)^\dagger S \T^g.
\end{equation}
\end{definition}

As we demonstrate below, the gauging protocol, when viewed as a formal mathematical object, is an isomorphism between the symmetric code and the gauged code $\cC'$. A convenient description of the gauging map for demonstrating this isomorphism is the following:

\begin{definition}[Gauging map]
\label{def:gauging-map}
Let $\mathcal{H}_\mathrm{m}$ be the Hilbert space of the data qudits of the initial code $\cC$, and let $\mathcal{H}_\mathrm{g}$ be the Hilbert space of gauge qudits on edges $e\in \edgeE$.
For $\ket{\psi}\in \mathcal H_{\rm sym.}$, we define the gauging map $\Gamma:\mathcal{H}_\mathrm{sym}\rightarrow \mathcal{H}_\mathrm{m}\otimes \mathcal{H}_\mathrm{g}$. 
\begin{align}
\label{eq:gauging-map}
     \Gamma \ket{\psi} &:= \frac{1}{|\Agrp|^{(|\vertexV|-1)/2}}\sum_{\bm{a}\in \Agrp[\vertexV]/\Agrp} \T^{\bm{a}} \ket{\psi}\ket{\delta_1\bm{a}}
\end{align}
\end{definition}

The derivation of equation~\eqref{eq:gauging-map} is the following.
Start with the state $\ket{\psi}\ket{0}^{\otimes|\edgeE|}$, where every edge qudit is initialized in the $\ket{0}$ state.
For each vertex $v\in \vertexV$, the projector onto the $\av=1$ subspace is 
\[
\Pi_{\av}^{+} = \frac{1}{|\Agrp|} \sum_{g\in \Agrp} \av^g, \quad \av^g = \T_v^g \otimes \Xp^{\delta_1 g \bm{e_v}}
\]
Multiplying these projectors over all vertices $v$, we have
\[
\prod_{v\in \vertexV} \Pi_{\av}^{+} = \prod_{v\in \vertexV} \frac{1}{|\Agrp|} \sum_{g\in \Agrp} \av^g= \frac{1}{|\Agrp|^{|\vertexV|}}\sum_{\bm{a}\in \Agrp[\vertexV]} A(\bm{a})
\]
where $\bm{a}=(a_v)_{v\in \vertexV}\in \Agrp[\vertexV]$ runs over every possible choice of assigning a group element to each vertex, and $A(\bm{a}):=\prod_{v\in \vertexV} \av^{a_v}=\T^{\bm{a}}\otimes \Xp^{\delta_1\bm{a}}$.
It follows that 
\begin{align}
    \prod_{v\in \vertexV} \Pi_{\av}^{+}\ket{\psi}\ket{0}^{\otimes |\edgeE|} = \frac{1}{|\Agrp|^{|\vertexV|}}\sum_{\bm{a}\in \Agrp[\vertexV]}
    \T^{\bm{a}}\ket{\psi}\ket{\delta_1\bm{a}}.
    \label{eq:unsimplified-gauging}
\end{align}
A further simplification can be performed when multiple $\bm{a}$ have the same coboundary ($\delta_1\bm{a}=\delta_{1}\bm{a}'$).
In particular, this is true for $\bm{a}'=\bm{a}+g\id \ \forall g\in \Agrp$ because $g\id\in \Agrp[\vertexV]$ has the same value from both ends of each edge in a fully connected graph, $\delta_1(g\id)=0$.
In other words, two vertex labelings have the same coboundary when they differ by a constant $g\id$.
Since the states are identical when two labelings have the same coboundary, the final, normalized state after applying the gauging map is Eq.~\eqref{eq:gauging-map}.

Let $\bm{\bm{\chi}}=(\bm{\chi}_v)_{v\in \vertexV}\in \widehat{\Agrp}[\vertexV]$ be an assignment of an irreducible representation of $\Agrp$ to each vertex.
For $\bm{a}=(a_v)_{v\in\vertexV}\in \Agrp[\vertexV]$, define 
\begin{align}
    \bm{\bm{\chi}}(\bm{a}) :=\prod_{v\in \vertexV} \bm{\chi}_v(a_v),
\end{align}
which is the product of vertexwise eigenvalues associated with group element configuration $\bm{a}$.

We now define the gauged form of a matter operator $M$, meaning an operator supported entirely on matter qudits. We first decompose any $M$ into terms $M_{\bm{\bm{\chi}}}$, each of which has eigenvalue $\bm{\bm{\chi}}(\bm{a})$ under conjugation by $\T^{\bm{a}}$ for every $\bm{a}\in \Agrp[\vertexV]$\footnote{For readers familiar with the condensed matter theory literature, $\bm{\bm{\chi}}(\bm{a})$ is the charge configuration of the operator $M$ under the action of $\T^{\bm{a}}$.}.
We refer to this decomposition as the eigenoperator decomposition of a matter operator $M$.

\begin{definition}[Eigenoperator decomposition] \label{def:eigen-decomp}

For any matter operator $M$, its \emph{eigenoperator components} $M_{\bm{\bm{\chi}}}$ are
\begin{align}
    M_{\bm{\bm{\chi}}} := \frac{1}{|\Agrp|^{|\vertexV|}}\sum_{\bm{a}\in\Agrp[\vertexV]}\bm{\bm{\chi}}(\bm a)^*\T^{\bm{a}}M (\T^{\bm{a}})^\dagger,
    \label{eq:eigencomponent}
\end{align}
where each eigenoperator component satisfies
\begin{align}
    \T^{\bm{a}}M_{\bm{\bm{\chi}}} (\T^{\bm{a}})^\dagger = \bm{\bm{\chi}}(\bm a ) M_{\bm{\bm{\chi}}}.
\end{align}
Then the \emph{eigenoperator decomposition} of $M$ is
\begin{align}
    M=\sum_{\bm{\bm{\chi}} }M_{\bm{\bm{\chi}}}.
    \label{eq:eigenoperator-decomposition}
\end{align}
\end{definition}
One can verify that the $M_{\bm{\bm{\chi}}}$ defined in Eq.~\eqref{eq:eigencomponent} are orthogonal components of $M$ and satisfy $\sum_{\bm{\bm{\chi}}} M_{\bm{\bm{\chi}}}=M$. Once an operator is decomposed into its eigencomponents, it is straightforward to apply the gauging map.

\begin{definition}[Gauged Operator]
\label{def:gauged-operator}
Let $M=\sum_{\bm{\bm{\chi}}}M_{\bm{\bm{\chi}}}$ be a matter operator such that 
$M_{\bm{\bm{\chi}}} = 0$ when $\bm{\bm{\chi}}\notin\mathop{Im} (\partial_1)$. In other words,
\begin{align}
M=\sum_{\bm{\bm{\chi}} \in \mathop{Im} \partial_1} M_{\bm{\bm{\chi}}}  .
\end{align}
For each $\bm{\bm{\chi}} \in \mathop{Im} (\partial_1)$, choose a $\bm{p}_{\bm{\bm{\chi}}}\in \widehat{\Agrp}[\edgeE]$, or a path of edges labeled by irreps of $A$ such that
\begin{align}
    \partial_1\bm{p}_{\bm{\bm{\chi}}} = \bm{\bm{\chi}}.
\end{align}

We define the \emph{gauged operator} of $M$ as $\Gamma[M]$:
\begin{align}
    \Gamma[M]&= \sum_{\bm{\bm{\chi}} \in \mathop{Im} \partial_1} M_{\bm{\bm{\chi}}} \otimes \Zp^{\bm{p}_{\bm{\bm{\chi}}}}
\end{align}

\end{definition}

We have fully introduced the algebraic gauging formalism; however, we have not yet demonstrated the equivalence with the gauging circuit we introduced in Section \ref{sec:gauging-procedure}. 

\begin{lemma}[Equivalence of the gauging circuit and gauging map] \label{lem:gauging-circuit-map-equivalence}
    Consider the gauging circuit $\widetilde{\Gamma}$ introduced in the previous section (Definition \ref{def:gauging-circuit}). This circuit acts identically to the gauging map defined in Def. \ref{def:gauged-operator} within the $\bp = +1$ subspace. 
\end{lemma}
\begin{proof}
    Consider the action of the gauging circuit on a single eigen-operator component $M_{\bm{\bm{\chi}}}$.
    \begin{equation}
        \begin{aligned}
        \widetilde{\Gamma} = \underbrace{\left (\prod_{v \in \vertexV} \dyad{+}_v \right )}_\textrm{P} \underbrace{\left ( \prod_{v \in \vertexV} \prod_{e \in \delta v} C_v \Xp^{\eta_v(e)}_e\right )}_{\textrm{C.E.}} \underbrace{\left ( \prod_{v \in \vertexV} C_v \T_v \right )}_\textrm{S.E.} \underbrace{\left ( \prod_{v \in \vertexV} \ket{+}_v \prod_{e \in \edgeE} \ket{0}_e \right )}_\textrm{Ancilla Init.} 
        \end{aligned}
    \end{equation}
    \begin{equation}
        \begin{aligned}
        \widetilde{\Gamma} M_{\bm{\bm{\chi}}} \widetilde{\Gamma}^\dagger &= \mathrm{P} (\mathrm{C.E.}) (\mathrm{S.E.}) (\mathrm{Init.}) M_{\bm{\bm{\chi}}} (\mathrm{Init.})^\dagger(\mathrm{S.E.})^\dagger   (\mathrm{C.E.})^\dagger \mathrm{P}^\dagger \\
        &= \mathrm{P} (\mathrm{C.E.}) (\mathrm{S.E.}) M_{\bm{\bm{\chi}}} (\mathrm{S.E.})^\dagger   (\mathrm{C.E.})^\dagger \mathrm{P}^\dagger \\
        &= \mathrm{P} (\mathrm{C.E.}) [ M_{\bm{\bm{\chi}}} \otimes \Zp(\bm{\bm{\chi}})] (\mathrm{C.E.})^\dagger \mathrm{P}^\dagger \\
        &= \mathrm{P}  [ M_{\bm{\bm{\chi}}} \otimes \Zp^{\bm{\bm{\chi}}} ] \mathrm{P}^\dagger \\
        &= M_{\bm{\bm{\chi}}} \otimes \Zp^{\bm{p}_{\bm{\bm{\chi}}}}, \quad \partial_1 \bm{p}_{\bm{\bm{\chi}}} = \bm{\bm{\chi}} 
        \end{aligned}
    \end{equation}
    In the last step, we have used the edge stabilizers generated during the cluster entangling step to push $\Zp^{\bm{\bm{\chi}}}$ off the vertices onto the edges. This leads to the constraint $\partial_1 \bm{p}_{\bm{\bm{\chi}}} = \bm{\bm{\chi}}$. By linearity, we must have 
    \begin{equation}
        \widetilde{\Gamma}[M] = \sum_{\bm{\bm{\chi}} \in \mathrm{Im}\partial_1} \widetilde{\Gamma}[M_{\bm{\bm{\chi}}}] = \sum_{\bm{\bm{\chi}} \in \mathrm{Im}\partial_1} M_{\bm{\bm{\chi}}} \otimes  \Zp^{\bm{p}_{\bm{\bm{\chi}}}} \stackrel{\bp = +1}{=} \Gamma[M].
    \end{equation}
    In the last equality, emphasize that the action is equivalent inside the $\bp = +1$ space, as we may choose different path conventions.
\end{proof}

Recall the form of the gauged code obtained via the gauging circuit; Lemma \ref{lem:gauging-circuit-map-equivalence} assures that one obtains an equivalent code using the gauging map.
\begin{definition}[Gauged code]
\label{def:gauged-code-stabs}
    For ease of the reader, we copy the gauged code stabilizers derived in Section \ref{sec:gauging-procedure} here. We have included a translation of the stabilizers into the notation introduced for generic Abelian groups $\Agrp$.

    \begin{equation}
        \av^g = \T_v^g \Xp^{\delta_1 g \bm{e_v}}, \quad \bp^g = \Zp^{\partial_2 g \bm{e_f}}, \quad \bar{\mathbb{S}}  = \Gamma[\sym] 
    \end{equation}
    
    \begin{equation}
        \Gamma[\sym] = \sum_{\bm{\bm{\chi}} \in \mathop{Im} \partial_1} \sym_{\bm{\bm{\chi}}} \otimes \Zp^{\bm{p}_{\bm{\bm{\chi}}}}
    \end{equation}
    where $\partial_1\bm{p}_{\bm{\bm{\chi}}} = \bm{\bm{\chi}}.$
\end{definition}

\begin{remark}[Path independence of the gauging map]

We note that while the explicit form of the gauged operators presented in Definition \ref{def:gauged-operator} is dependent on a choice of edges $\bm{p}_{\bm{\bm{\chi}}}$, any other valid choice $\bm{q}_{\bm{\bm{\chi}}}$ differs by a cycle: 
\begin{equation}
    \partial_1 (\bm{p}_{\bm{\bm{\chi}}} - \bm{q}_{\bm{\bm{\chi}}}) = \bm{\bm{\chi}} - \bm{\bm{\chi}} = 0 
\end{equation}
Recall this cycle $\partial_2 \bm{f} = \bm{p}_{\bm{\bm{\chi}}} - \bm{q}_{\bm{\bm{\chi}}} \in \widehat{\Agrp}[\edgeE]$ is associated to a cycle stabilizer of the gauged code $\mathbb{B}_f = \Zp^{\partial_2 \bm{f}}$. Thus, in the subspace where all cycle stabilizers $\mathbb{B}_f = +1$, all paths with the same boundary are equivalent choices. In practice, there are other considerations: for example, when writing down the decorated stabilizers, we want our path choices to be constant weight for a given stabilizer generator, to ensure LDPC-ness of the gauged code.
\end{remark}

\subsection{The gauging map is an isomorphism}

We now prove an isomorphism between the $+1$ eigenspace of $\T$, $\mathcal{H}_{\rm{sym.}}$, and the subspace determined by $\av = \bp = +1$ constraints, $\mathcal{H}_{\rm{gau.}}$. The proof follows from basic algebraic manipulations, using the form of the gauging map in Def. \ref{def:gauging-map}.

\begin{lemma}[The gauging map is an isomorphism]
\label{lem:gauging-map-isomorphism}
    Let $\graphG$ be connected so $H^1(\graphG^\bullet)=0$. 
    Then, $\Gamma$ is an isomorphism,
    \begin{align}
        \Gamma:\mathcal H_{\rm sym.}\xrightarrow[]{\cong} \ \mathcal{H}_{\rm gau.} \equiv \{\ket{\phi}: \av\ket{\phi}=\ket{\phi}\ \forall v,\; \bp\ket{\phi}=\ket{\phi}\ \forall p\}, 
    \end{align}
    and 
    \begin{align}
        \Gamma[M]\Gamma\ket{\psi} = \Gamma(M\ket{\psi})
        \label{eq:gauging-map-associativity}
    \end{align}
    for every $M$ for which $\Gamma[M]$ in Def.~\ref{def:gauged-operator} is defined and every $\ket{\psi}\in \mathcal H_{\rm sym.}$ (Def.~\ref{def:symmetric-code}).
\end{lemma}
\begin{proof}
    First we have that 
\begin{align}
     \Gamma^\dagger \Gamma &= \frac{1}{|\Agrp|^{|\vertexV|+1}} \sum_{\bm{a}, \bm{b} \in \Agrp[\vertexV]} (\T^{\bm{b}})^\dagger \T^{\bm{a}} \otimes \langle \delta_1 \bm{b} \ket{\delta_1 \bm{a}} \\
     &= \frac{1}{|\Agrp|^{|\vertexV|+1}} \sum_{\bm{a} \in \Agrp[\vertexV]} \sum_{\substack{\bm{b} \text{ s.t. } \\ \delta_1 (\bm{b} - \bm{a}) = 0}} \T^{\bm{a} - \bm{b}} \label{eq:isomorphism-1}\\
     &= \frac{1}{|\Agrp|^{|\vertexV|+1}} \sum_{\bm{a} \in \Agrp[\vertexV]} \sum_{g \in \Agrp} \T^g\label{eq:isomorphism-2}\\
     &= \frac{1}{|\Agrp|} \sum_{g \in \Agrp} \T^g = \Pi_{\rm sym.}\label{eq:isomorphism-3}
\end{align}
    From Eq.~\eqref{eq:isomorphism-1} to Eq.~\eqref{eq:isomorphism-2}, we have used that $\graphG$ is connected and therefore $\delta_1\bm a =\delta_1\bm b$ iff $\bm a =\bm b + c\id$ for some $c\in \Agrp$. Eq.~\eqref{eq:isomorphism-3} means that acting the gauging map and then undoing it implements a projection onto the symmetric subspace $\mathcal H_{\rm sym.}$. Furthermore, it follows that the gauging map preserves the inner product between any two $\ket{\psi},\ket{\phi}\in \mathcal H_{\rm sym.}$, since $\bra{\phi}\Gamma^\dagger\Gamma\ket{\psi} = \bra{\phi}\Pi_{\rm sym.}\ket{\psi}=\langle\phi\ket{\psi}$.

    The reverse procedure: starting from the gauged degrees of freedom, undoing the gauging, and then re-doing it, yields the projector onto the subspace with $\av = +1$ and $\bp = +1$: 
\begin{equation}
\begin{aligned}
    \Gamma \Gamma^\dagger
    &= \frac{1}{|\Agrp|^{|\vertexV|+1}} \sum_{\bm{a}, \bm{b} \in \Agrp[\vertexV]} \T^{\bm{a}-\bm{b}}  \otimes \ket{\delta_1 \bm{a}} \bra{\delta_1 \bm{b}} \\
    &=  \frac{1}{|\Agrp|^{|\vertexV|+1}} \sum_{\bm{a}, \bm{b} \in \Agrp[\vertexV]} \T^{\bm{a}-\bm{b}}  \otimes \Xp^{\delta_1 (\bm{b} - \bm{a})} \dyad{\delta_1 \bm{b}} \\
    &= \frac{1}{|\Agrp|^{|\vertexV|}} \left ( \sum_{\bm{a'}} (\T^{\bm{a'}})^\dagger \otimes \Xp^{\delta_1\bm{a'}}\right ) \frac{1}{|\Agrp|} \left ( \sum_{\bm{b}} \dyad{\delta_1 \bm{b}} \right ) \\
    &= \prod_{v\in \vertexV} \Pi_{\av}^{+} \prod_{p } \Pi_{\bp}^+
\end{aligned}
\end{equation}
Here we have defined the projector onto all $\bp = +1$ as follows:
    \begin{equation}
    \begin{aligned}
        \prod_{f \in \faceF} \Pi_{\bp}^+ &= \prod_{f \in \faceF} \frac{1}{|\Agrp|} \sum_{\chi \in \widehat{A}} \Zp^{\partial_2 \bm{e_f}} \\
        &= \frac{1}{|\Agrp|^{|\faceF|}} \sum_{\bm{\chi} \in \widehat{A}[\faceF]} \Zp^{\partial_2 \bm{\chi}}\\
        &= \sum_{\bm{\epsilon} \in \Agrp[\edgeE]}\left( \frac{1}{|\Agrp|^{|\faceF|}}  \sum_{\bm{\chi} \in \widehat{A}[\faceF]} \omega^{\bm{\chi}(\bm{\epsilon})} \right )\dyad{\bm{\epsilon}}\\
        &=\sum_{\bm{\epsilon} \in \Agrp[\edgeE]} [0 \text{ unless } \bm{\epsilon} \in \text{Im} \delta_1]\dyad{\bm{\epsilon}}\\
        &=\frac{1}{|\Agrp|} \sum_{\bm{a} \in \Agrp[\vertexV]} \dyad{\delta_1 \bm{a}}
    \end{aligned}
\end{equation}
In the last line, we have divided by $|\Agrp|$ to remove over-counting of configurations $\bm{a}$ that differ by an element of $\ker \delta_1$. Thus $\Gamma$ is an isomorphism between the symmetric subspace and the subspace generated by the $\av$ and $\bp$ stabilizers. 

Now we show Eq.~\eqref{eq:gauging-map-associativity}.
We compute
\begin{align}
    \Gamma[M]\Gamma\ket{\psi}&= \frac{1}{|\Agrp|^{(|\vertexV|-1)/2}}\sum_{\bm a\in \Agrp[\vertexV]/\Agrp}(\sum_{\bm{\bm{\chi}}}M_{\bm{\bm{\chi}}})\T^{\bm{a}}\ket{\psi}\otimes \Zp^{\bm{p}_{\bm{\bm{\chi}}}}\ket{\delta_1\bm a}\label{eq:isomorphism-4}\\
    &= \frac{1}{|\Agrp|^{(|\vertexV|-1)/2}}\sum_{\bm a\in \Agrp[\vertexV]/\Agrp}\sum_{\bm{\chi}} \bm{\chi}(\bm a)M_{\bm{\chi}} \T^{\bm a}\ket{\psi}\ket{\delta_1\bm a}\label{eq:isomorphism-5}\\
    &=\frac{1}{|\Agrp|^{(|\vertexV|-1)/2}}\sum_{\bm a\in \Agrp[\vertexV]/\Agrp}\sum_{\bm{\chi}} \bm{\chi}(\bm a)\bm{\chi}(\bm a)^{-1} \T^{\bm a}M_{\bm{\chi}}\ket{\psi}\ket{\delta_1\bm a}\label{eq:isomorphism-6}\\
    &=\frac{1}{|\Agrp|^{(|\vertexV|-1)/2}}\sum_{\bm a\in \Agrp[\vertexV]/\Agrp} \T^{\bm a}(\sum_{\bm{\chi}}M_{\bm{\chi}})\ket{\psi}\ket{\delta_1\bm a}\notag\\
    &=\frac{1}{|\Agrp|^{(|\vertexV|-1)/2}}\sum_{\bm a\in \Agrp[\vertexV]/\Agrp} \T^{\bm a}M \ket{\psi}\ket{\delta_1\bm a} = \Gamma(M\ket{\psi}).\notag
\end{align}
From Eq.~\eqref{eq:isomorphism-4} to Eq.~\eqref{eq:isomorphism-5} we used  $\partial_1 \bm{p}_{\bm{\chi}} = \bm{\chi}$, which implies that
$\Zp^{\bm{p}_{\bm{\bm{\chi}}}}\ket{\delta_1\bm a} = \bm{\chi} (\bm a) \ket{\delta_1 \bm a}$, and from Eq.~\eqref{eq:isomorphism-5} to Eq.~\eqref{eq:isomorphism-6} we used $M_{\bm{\chi}} \T^{\bm{a}} = \bm{\chi}(\bm a)^{-1} \T^{\bm{a}}M_{\bm{\chi}}$.

\end{proof}

From Lemma~\ref{lem:gauging-map-isomorphism} we directly have the following isomorphism between the symmetric code (Def.~\ref{def:symmetric-code}) and the gauged code from Section~\ref{sec:gauging-procedure}.
\begin{corollary}[Isomorphism of gauged code and symmetric code]
\label{cor:sym-gau-isomorphism}
    Let $\T$ be the transversal Clifford symmetry in Def.~\ref{def:transversal-gate}, and assume that the gauging graph $\graphG$ satisfies the graph desiderata in Def.~\ref{def:graph-desiderata}. 
    Let the corresponding gauged code $\cC'$ have stabilizers defined in Thm.~\ref{thm:generic-gauged-stabilizers}.
    Then, the symmetric codespace $\cC_{\rm sym.}$ and the gauged codespace $\cC'$ are isomorphic. The gauging map $\Gamma$ implements this isomorphism.
\end{corollary}
\begin{proof}
Lemma \ref{lem:gauging-map-isomorphism} tells us that $\mathcal H_{\rm sym.}$ is isomorphic to the $+1$ eigenspace of the $\av$ and $\bp$ stabilizers. The extra constraint to arrive at the symmetric codespace and the gauged codespace is to project onto the $+1$ eigenspace of the symmetrized stabilizers and gauged symmetrized stabilizers, respectively. But these are naturally isomorphic through the action of the gauging map through the second statement of Lemma \ref{lem:gauging-map-isomorphism}. 
\end{proof}

We can summarize Lemma \ref{lem:gauging-map-isomorphism} and Corollary \ref{cor:sym-gau-isomorphism} with the following commutative diagram:
\begin{equation}\label{eq:gauging_cube}
\begin{tikzcd}[row sep=3em, column sep=3em, ampersand replacement=\&]
\mathcal H_{\rm sym.}
    \arrow[r, "\Pi_{\cC_{\rm sym.}}"]
    \arrow[d, "\Gamma"', "\cong"]
\& \cC_{\rm sym.}
    \arrow[d, "\Gamma"', "\cong"]
\\
\mathcal H_{\rm gau.}
    \arrow[r, "\Pi_{\cC'}"']
\& \cC'
\end{tikzcd}
\end{equation}

\begin{remark}

We note that the above isomorphisms also apply to the gauged code defined by the unitary $\mathbb{Z}_2$ stabilizers in Theorem \ref{thm:z2-gauged-stabilizers}, since by Theorem \ref{thm:z2-generic-equivalence} this code is equivalent to that obtained in the general case. 
\end{remark}

\subsection{Distance of the gauged code}

Armed with Lemma \ref{lem:gauging-map-isomorphism} we now establish distance preservation in the gauged code.

We define the support of an operator $O$ as $\supp{O}$, the set of physical qudits on which
$O$ is not the identity, and
\[
    \Wt(O):=|\supp{O}|
\]
is its physical weight.

A key property appearing in recent distance proofs for quantum code surgery is the expansion of $\graphG$.
Since we are working with a chain complex on the graph $\graphG$ with coefficients in the finite group $\Agrp$ rather than in $\mathbb{F}_2$, we require a generalized notion of expansion. Intuitively, the Cheeger constant bounds the relationship between the size of a set of vertices (a $0$-chain) and the size of its coboundary (a $1$-chain). 

For an equivalence class $[\bm{g}] = \{\bm{g} + a \bm{1}\}$, let $\tilde{\bm{g}} \in [\bm{g}]$ be the minimal weight representative.

\begin{definition}[Generalized Cheeger constant]
\label{def:generalized-cheeger}
Consider a graph $\graphG$ and coefficient group $\Agrp$. Then the \emph{generalized} Cheeger constant is defined:
\begin{equation}
    \begin{aligned}
    h_{\Agrp}(\graphG):=
    &\min_{\bm g\notin\ker\delta_1}
    \frac{|\supp{\delta_1\bm g}|}{|\supp{\tilde{\bm g}}|}
\end{aligned}.
\end{equation}
\end{definition}

\begin{theorem}[Gauged Code Distance]
\label{thm:space-distance}
Let $d$ be the distance of $\cC$, and let $\nu:=\max_{v\in\vertexV}|\vertexset{v}|$. Let $\Pi_{\cC'}$ denote a projector onto $\cC'$, the gauged code whose stabilizers are defined in Thm.~\ref{thm:generic-gauged-stabilizers} and Def. \ref{def:gauged-code-stabs}. Let $\mathrm{Wt}()$ count the number of matter and gauge qudits on which an operator acts nontrivially. 
Define 
\[
d_g := \min\{\mathrm{Wt}(\mathbb{L}):\Pi_{\cC'} \mathbb{L}\Pi_{\cC'}\not\propto \Pi_{\cC'}\}.
\] 
If the gauging graph satisfies Def.~\ref{def:graph-desiderata}, with generalized Cheeger constant 
\[
h_{\Agrp}(\graphG) \geq 1\]
then
\begin{align}
    d_g\ge d/\nu.
\end{align}
\end{theorem}
\begin{proof}

At a high level, our proof strategy is to break a generic logical $\mathbb{L}$ on the gauged code into its eigen-decomposition on the matter degrees of freedom, and its Pauli support on the gauge degrees of freedom. The weight of the total operator is bounded by the weight of any nontrivial constituent operator, and so we focus on a single component. At this point, we can apply similar arguments to previous work \cite{williamson2024lowoverheadfaulttolerantquantumcomputation, williamson2026fast} to show this component is bounded in weight, namely that it can always be cleaned into the form of a logical on the ungauged code, up to some decorations on the gauging graph. 

Pick an arbitrary $\mathbb{L}$ such that $\Pi_{\cC'}\mathbb{L}\Pi_{\cC'} \not\propto \Pi_{\cC'}$. 
Expand the support of $\mathbb{L}$ on the gauge qudits in term of generalized Paulis, then decompose each matter operator into its eigenoperator decomposition,
    \[
    \mathbb{L} = \sum_{\bm a,\bm b} M_{\bm a,\bm b}\otimes \Xp^{\bm a}\Zp^{\bm b} = \sum_{\bm{\chi}}\sum_{\bm a ,\bm b} (M_{\bm a, \bm b})_{\bm{\chi}} \otimes \Xp^{\bm a} \Zp^{\bm b}= \sum_{i}L_i,\qquad L_i = M_i \otimes \Xp^{\bm a_i}\Zp^{\bm b_i}
    \]
where $\bm a\in \Agrp[\edgeE]$, $\bm b\in \widehat{\Agrp}[\edgeE]$ and $\bm{\chi} \in \widehat{\Agrp}[\vertexV]$, and we relabel $i=(\bm a, \bm b, \bm{\chi})$.
    $M_i$ is a matter eigenoperator with eigenvalue $\bm{\chi}$,    $ \bm a_i\in \Agrp[\edgeE]$ and $ \bm b_i\in \widehat{\Agrp}[\edgeE]$.

    We use the fact that $\mathbb{L}$ is expanded in terms of a tensor product operator basis. On the matter qudits, we have expanded in terms of the eigenoperators of the symmetry action on each vertex set, and thus these basis operators act on at most $\nu$ matter qudits.
    On the edge ancilla, we have expanded in the onsite Pauli basis. 
    The basis for operators over the full system is the tensor product of these local basis operators.
    The support of any non-zero $L_i$ is the identity away from the edge ancilla and matter vertex sets that $\mathbb{L}$ is supported on. Thus, \begin{align}
        \mathrm{Wt}(\mathbb{L})\ge \frac{1}{\nu} \mathrm{Wt}(M_i) + \mathrm{Wt}(\Xp^{\bm{a}_i} \Zp^{\bm{b}_i}) \label{eq:eigendecomp-enlarged-support}
    \end{align}

    for any $L_i$. 
    The factor of $1/\nu$ comes from the fact the eigenoperator expansion may increase the support of $L_i$ in relation to $\mathbb{L}$, but only by the maximum number of sites in a vertex set.

    By linearity of the projector $\Pi_{\cC'}$, if $\Pi_{\cC'}\mathbb{L}\Pi_{\cC'}\not \propto \Pi_{\cC'}$,
    then at least one $L_i$ must satisfy $\Pi_{\cC'}L_i\Pi_{\cC'}\not \propto \Pi_{\cC'}$.

    Separately, we know that every $L_i$ is a simultaneous eigenoperator under conjugation of $\av, \bp$ for $v,p$:
    \[
    \av L_i \av^\dagger = \lambda_{i,v}L_i\qquad \bp L_i \bp^\dagger = \sigma_{i,p}L_i,\qquad\qquad \lambda_{i,v}, \sigma_{i,p}\in U(1).
    \]
    This follows from our choice of operator basis. On matter qudits, $L_i$ consists of a basis operator $M_i$ that satisfies $\T^{\bm c} M_i (\T^{\bm c})^\dagger = \bm{\chi}_i(\bm c) M_i$ for any $\bm c\in \Agrp[\vertexV]$. 
    On gauge qudits, $L_i$ is a generalized Pauli operator. Generalized Pauli operators are eigenoperators under Pauli conjugation, so conjugation by $\bp$ multiplies $L_i$ by a phase. 
    Conjugation by
    $\av$ acts as $\T_v$ on the matter qudits and as a generalized Pauli $\Xp$ on the gauge qubits, each of which multiply $L_i$ by a phase.
        
    Now that we have established that $L_i\ \forall i$ has well-defined eigenvalues under conjugation by $\av,\bp$, we show that any $L_i$ with $\Pi_{\cC'}L_i\Pi_{\cC'}\neq 0$ must satisfy $\lambda_{i,v}=\sigma_{i,p}=1$ for all $v,p$. Since every state in $\cC'$ is stabilized by $\av$ and $\bp$, 
    \[
    \Pi_{\cC'}L_i\Pi_{\cC'} = \Pi_{\cC'}\av L_i\av \Pi_{\cC'}  = \lambda_{i,v}\Pi_{\cC'}L_i\Pi_{\cC'} .
    \]
    If $\lambda_{i,v}\neq 1$, then $\Pi_{\cC'}L_i\Pi_{\cC'} =0$. A similar argument holds for $\bp$: $\sigma_{i,p}\neq 1$ implies that $\Pi_{\cC'}L_i\Pi_{\cC'} =0$.
    
    In summary, $\mathbb{L}$ has at least one $L_i$ such that $\Pi_{\cC'}L_i\Pi_{\cC'}\not \propto \Pi_{\cC'}$, and that $L_i$ must have eigenvalues $\lambda_{i,v}=\sigma_{i,p}=1$ under conjugation by $\av,\bp$ for all $v,p$.
    If this $L_i=M_i\otimes\Xp^{\bm a_i}\Zp^{\bm b_i}$ commutes with all $\bp$, then that means that $\bm a_i\in \ker\delta_2$. Using $H^1(\graphG^\bullet)=0$, this implies that $\ker\delta_2 = \mathop{Im}\delta_1$. Thus, there must exist some $\bm d_i\in \Agrp[\vertexV]$ such that $\bm a_i=\delta_1 \bm d_i$.
    Recall the stabilizer $A(\bm a) = \T^{\bm{a}}\otimes \Xp^{\delta_1\bm a}$ and let its inverse be denoted by $A(\bm a)^{-1} := A(-\bm a) = \T^{-\bm a}\otimes \Xp^{-\delta_1\bm a}$.
    If we multiply $L_i$ by the stabilizer $A(\bm d_i)^{-1}$, then the $\Xp$ support on all gauge qudits cancels out:
    \[
    L_i':=L_iA(\bm d_i)^{-1} = M_i\T^{-\bm d_i}\otimes \Xp^{\bm a_i-\delta_1 \bm d_i}\Zp^{\bm b_i} = M_i\T^{-\bm d_i}\otimes \Zp^{\bm b_i}.
    \]
    Since we only multiplied $L_i$ by a stabilizer $A(\bm d_i)^{-1}$, the logical action of $L_i$ on the codespace is the same as the logical action of $L_i'$ ($\Pi_{\cC'}L_i \Pi_{\cC'} =  \Pi_{\cC'}L_i' \Pi_{\cC'}$), so $L_i'$ has nontrivial action on the gauged code.
    We refer to an $L_i'$ as surviving when $\Pi_{\cC'}L_i' \Pi_{\cC'}\neq 0$.
    
    Next, we show that every surviving $L_i'$ is equivalent, up to stabilizers of $\cC'$, to $\Gamma[N_i]$ for some matter operator $N_i$; in other words, $L_i'$ and $\Gamma[N_i]$ have the same action on $\cC'$.
    Write $L_i' = N_i \otimes \Zp^{\bm b_i}$ where $N_i:= M_i\T^{-\bm d_i}$, and let $\bm{\chi}_i\in \widehat{\Agrp}[\vertexV]$ describe how $N_i$ transforms under the local symmetry, 
    \begin{align}
        \T^{\bm c}N_i (\T^{\bm c})^\dagger     = \bm{\chi}_i(\bm c) N_i,\qquad \forall \bm c\in \Agrp[\vertexV]\notag.
    \end{align}
    By Def.~\ref{def:gauged-operator}, $\Zp(\bm b_i)$ is a valid decoration of $N_i$ if  
    \begin{align}
        \partial_1 \bm b_i = \bm{\chi}_i.
        \label{eq:decorated-requirement}
    \end{align}
    This condition follows from commutation with the $\av$ stabilizers. For $g\in \Agrp$, $\av^g = \T^g_v\otimes \Xp^{\delta_1 \bm e_g}$ where $\bm e_{g,v}$ is $g$ at $v$, and zero elsewhere.
    Conjugating $L_i'$ by this local symmetry gives
    \begin{align}
        \av^gL_i'({\av^g})^\dagger = (\bm{\chi}_i)_v(g)\ (\partial_1\bm b_i)_v(g)^{-1}L_i'.
    \end{align}
    Since a surviving $L_i'$ must commute with every $\av^g$, the phases must cancel for every $v$ and $g$ and therefore $\partial_1\bm b_i = \bm{\chi}_i$.
    Thus, $\Zp^{\bm b_i}$ is a valid decoration of $N_i$.

    The particular decoration $\Zp^{\bm b_i}$ need not be equal to the one chosen in Def.~\ref{def:gauged-operator}.
    But since the two decorations have the same boundary, their difference is a closed $1$-chain.
    Since $H_1(\graphG_\bullet) = 0$, this difference corresponds to a product of $\bp$ stabilizers.
    Therefore, $L_i'$ and $\Gamma[N_i]$ have the same action on $\cC'$, or $\Pi_{\cC'}L_i'\Pi_{\cC'} = \Pi_{\cC'}\Gamma[N_i]\Pi_{\cC'}$.

    Since $L_i'$ has nontrivial action on $\cC$, so does $\Gamma[N_i]$. By the gauging isomorphism (Lemma~\ref{lem:gauging-map-isomorphism}), the action of $\Gamma[N_i]$ on $\cC'$ is the same as the action of $N_i$ on the symmetric code. Since every global symmetry $\T^a$ acts as identity on the symmetric code, $N_i\T^a$ has the same nontrivial action on the symmetric code for every $a\in \Agrp$. 
    Since the symmetric code is a subspace of $\cC$, $N_i\T^a$ cannot act trivially on the original codespace $\cC$. By the definition of the distance $d$ of $\cC$, 
    \begin{align}
        \Wt (N_i\T^a)\ge d\qquad \forall a\in \Agrp.
    \end{align}
    Recall that $L_i=L_i'A(\bm d_i) = N_i\otimes \Zp^{\bm b_i}A(\bm d_i)$.
    Choose $a_i\in \Agrp$ minimizing $|\supp{\bm d_i - a_i \id}|$.
    This is the reduced support of $\bm d_i$ that appears in the definition of the generalized Cheeger constant.
    The matter support $M_i = N_i\T^{\bm d_i}$ can differ from that of $N_i\T^{a_i}$ by at most $\nu |\supp{\bm d_i - a_i \id}|$, while the gauge support has at least $|\supp{\delta_1\bm d_i}|$. 
    Thus,
    \begin{align}
        \Wt(M_i) \ge \Wt(N_i \T^{a_i}) - \nu |\supp{\bm d_i - a_i \id}|
    \end{align}
    and on the gauge qudits, 
    \begin{align}
        \Wt\!\left(\Xp^{\bm a_i}\Zp^{\bm b_i}\right) \ge |\supp{\delta_1\bm d_i}| \ge |\supp{\bm d_i-a_i\id}|,
    \end{align}
    The second inequality uses the expansion condition that $h_{\Agrp}(\graphG) \ge 1$.

    Finally, the eigenoperator decomposition can enlarge matter support by at most a factor of $\nu$ (see Eq.~\eqref{eq:eigendecomp-enlarged-support}).
    Thus,
    \begin{align}
        \Wt(\mathbb{L}) &\ge \frac{1}{\nu}
        \Wt(M_i)
        + \Wt \!\left(\Xp^{\bm a_i}\Zp^{\bm b_i}\right)\\
        &\ge \frac{1}{\nu}
        \left(\Wt(N_i \T^{a_i})
        - \nu |\supp{\bm d_i -a_i \id }|\right) + 
        |\supp{\bm d_i -a_i \id}|\\
        &\ge \frac{d}{\nu}.
    \end{align}

\end{proof}

\begin{remark}[The $1/\nu$ factor in the distance lower-bound]
    The factor of $1/\nu$ in the bound above $d_\mathrm{g}\ge d/\nu$ arises from the eigenoperator decomposition and is specific to our argument for gauging a Clifford logical operator.
    Let $L$ be an operator that acts nontrivially on the gauged code. 
    After taking the eigenoperator decomposition of $L$, the rest of the proof shows that at least one term must have nontrivial action and has weight at least the original code distance $d$.
    However, for a general Clifford symmetry, using the eigenoperator decomposition on $L$ can enlarge the matter support within each vertex set.
    Since a vertex set contains at most $\nu$ matter qudits, the weight of the selected term can be as large as $\nu\Wt(L)$. Consequently, $\nu\Wt(L)\ge d$, leading to the distance bound shown.

    \vspace{6pt}
    The $1/\nu$ factor does not arise for Pauli surgery.
    This is because Pauli operators already have definite eigenvalues under conjugation by a Pauli symmetry, and hence the operator expansion does not enlarge physical qudit support. In this case, if we require $h_{\Agrp}(\graphG) \geq \nu $, the argument above leads to a gauged code distance satisfying $d_\mathrm{g}\ge d$, even if a vertex set contains multiple matter qudits. This is similarly true for surgery on any operator that is written in a definite charge basis on each vertex.
\end{remark}

\begin{remark}
    For some symmetries such as the transversal CNOT, a minimal vertex set contains one qubit from each codeblock, and thus $\nu=2$. 
    Nevertheless, we expect the distance of the gauged code to be preserved ($d_\mathrm{g}\ge d$). This is because even though the eigenoperator expansion extends support between codeblocks, the support on any single codeblock does not grow. We do not establish a proof of this claim here, we expect such a proof would require careful tracking of support between the two separate codeblocks.
\end{remark}

The following corollary is useful in the proof of the fault distance for the full spacetime gauging protocol in the next section. It demonstrates that we are free to add relative phases when we are breaking up the symmetry $\T$. When we perform the gauging circuit in practice, we start in a random configuration of $\av$ measurement outcomes. These outcomes may not correspond to $\T = +1$; the following corollary ensures that the resulting projected state still has distance $d_g\geq d/\nu$.

\begin{corollary}[Distance in the presence of random $\av$] 
    The codespace defined by $\T = \omega^n$ for $n = 0, \dots, p-1$ and a random set of $\av$ eigenvalues $a_v$ such that $\prod_v a_v = \omega^n$ has distance $d_g \geq d/\nu$, if the graph desiderata (Def. \ref{def:graph-desiderata}) are met.
\end{corollary}
\begin{proof}
    This follows from a re-definition of the symmetry $\T$ and the $\av$ stabilizers: 
    \[
    \T \to \omega^{-n} \T, \quad \av \to a_v^* \av
    \]
    Now we have $\T = \prod_v \av = +1$. Define the gauged code in terms of the redefined symmetry operator and $\av$ terms. Apply Theorem \ref{thm:space-distance}.
\end{proof}

\ifdefined\ismainfile
\else

\bibliography{bib}
    \end{document}
\fi

\section{Fault tolerance of the gauging procedure}\label{sec:spacetime-distance}
\ifdefined\ismainfile
\else
    \documentclass[11pt]{article}
    
    \input{preamble.tex}
    
    \author{}
    \title{Fault tolerance of the gauging procedure}
    
    \begin{document}

    \maketitle

\fi

In this section, we establish our main result: a lower-bound on the spacetime fault distance of the gauging procedure (Def.~\ref{def:detailed-gauging-procedure}).
Two features distinguish this setting from the fault-distance analysis in Pauli codes, requiring new machinery. 
First, the measured checks of the gauged code $\cC'$ are unitary but in general neither Hermitian nor Pauli, so a data fault need not commute or anti-commute with a check. Such faults are labeled ``non-Abelian'', and they act to randomize certain detectors. 
Second, a pair of decorated checks commute only up to a Pauli element of the check group (Theorem~\ref{cor:z2-commutation-relations}) and so outside of the codespace, the measurement of one check
can randomize another. 
Both of these mechanisms create would-be detectors whose values are 
random rather than deterministic, which we refer to as \emph{dropped detectors} 
(Definition~\ref{def:detectors}).

Here, we focus on lower bounding the weight of an undetectable fault configuration, where undetectable means that no deterministic detectors are violated. Such fault configurations are called \emph{loop-like} (Def. ~\ref{def:loop-like-fault}).
We prove that dropped detectors are localized to a constant-sized bounded region around any loop-like fault (Lemma ~\ref{lem:dropped-detector-localization}).
Further, we introduce a new variant of the standard cleaning argument (Section~\ref{sec:cleaning}) that deals with multiple measurement trajectories consistent with the same fault pattern. Using this cleaning argument, we map any loop-like fault on the gauging protocol, that is below a certain weight, onto a logically trivial operator on the ungauged code (Thm. \ref{thm:spacetime-distance}). We provide an informal statement of the main theorem and a proof sketch below.
\vspace{3pt}

\begingroup
\setlength{\LTleft}{0pt plus 1fill}
\setlength{\LTright}{0pt plus 1fill}
\begin{longtable}{@{}p{0.30\linewidth}p{0.67\linewidth}@{}}
    \hline
        Notation & Denotes \\
    \hline
    \endfirsthead
    \hline
        Notation & Denotes \\
    \hline
    \endhead
    \endfoot
    \caption{Notation used for this section.}
    \label{tab:sec-4-notation}
    \endlastfoot
        d & distance of $\cC$\\
        $O(\cdot)$ & Landau notation with respect to a family of codes with $d\to\infty$.\\
        $d_f$ & Fault distance of the gauging protocol. \\
        $\mathbb{C} \in \cA,\mathbb{N} \in \cN$ & Measured Abelian and non-Abelian checks of $\cC'$. $\cA$-checks detect Abelian faults. $\cN$-checks detect non-Abelian faults.\\
        $f, f_N, f_A$ & Elementary faults, non-Abelian elementary faults, Abelian elementary fault.\\
        $F, F_N, F_A$ & Fault configuration, non-Abelian component, Abelian component. \\
        $\lambda(Q)$ & Spectrum of operator $Q$. \\
        $(Q,t)$ & Measurement event of operator $Q$ at time $t$.\\
        $r(Q, t)$ & Measurement outcomes for a particular quantum trajectory in the absence of noise.\\
        $r_F(Q, t)$ & Noisy measurement outcomes in the presence of a fault configuration $F$. \\
        $E_F(t)$ & Support of fault $F$ on timestep $t$. \\
        $M_t(r, r_F)$ & Projective measurements performed at timestep $t$ to recreate measurement trajectory $r(Q, t)$. \\
        $K(F, r)$, $K(\emptyset, r)$ & Kraus operator corresponding to trajectory $(F, r)$. When $F = \emptyset$, it indicates a fault-free trajectory. \\
        $m_\T(r_F)$ & Inferred value of the operator $\T$ from faulty measurement trajectory $r_F$. \\
        $D_Q^t(r_F)$ & Detector value for operator $Q$ at time $t$, given faulty trajectory $r_F$. \\
        $\cA$-detector, $\cN$-detector & detectors made of measurement outcomes of $\cA$-checks and $\cN$-checks, respectively.\\
        $\text{Drop}(F)$ & Set of detectors randomized by the presence of some fault. Always $\cA$-detectors. \\
        $\mathcal{H} = (V_\mathcal{H}, E_\mathcal{H})$ & Fault location hypergraph. Vertices are fault locations, hyperedges are detectors whose outcome is impacted by its boundary faults. \\
        $\mathsf{N}_\xi (D)$ & Neighborhood around detector $D$ on the fault location hypergraph $\mathcal{H}$ of radius $\xi$. \\
        $\mathrm{Err}(R)$ & All linear combinations of all products of faults generated by some set of elementary spatial faults $R$. \\
        $\cB_t$ & The set of all Abelian byproducts that arise when we move non-Abelian faults supported at time $t$ forward or backward in time. \\
        
\end{longtable}
\endgroup

\subsection{Proof sketch}

We restrict to an order-two transversal symmetry $\T$, so that $\T^2 = \id$. As a result, this restricts to the case of gauge \textit{qubits} instead of qudits of dimension $p$, but for consistency, we still use $\cX,\cZ$ to denote Paulis on gauge qubits.
Let $\cC$ be the base code with distance $d$ and let $\cC'$ be its gauged code with gauging graph $\graphG =(\vertexV,\edgeE)$ that satisfies the desiderata in Def.~\ref{def:graph-desiderata}.

From the stabilizers defined in Thm.~\ref{thm:z2-gauged-stabilizers} and Cor.~\ref{cor:z2-commutation-relations}, take $\{\aS{i}\}_i = \{\av\} \sqcup (\{\gaugedS\}/\{\mathbb{N}_S\})$.
Throughout this section, we consider measuring the following stabilizer generators of $\cC'$,
\begin{align}
\label{eq:measured-checks}
    \cA := \{\aS{i}\}_{i},\qquad \cN:=\{\nS{j}\}_j = \{\bp\} \sqcup \{\mathbb{N}_S\} \cup \{\grpcom{\aS{i}}{\aS{k}}:i<k,\  \grpcom{\aS{i}}{\aS{k}} \neq \id
    )\} .
\end{align}
We refer to stabilizers in $\cA$ and $\cN$ as Abelian and non-Abelian stabilizers, respectively (see Table~\ref{tab:checks}).
Using Cor.~\ref{cor:z2-commutation-relations}, $\grpcom{\aS{i}}{\aS{k}}\in \langle \{\nS{j}\}_j \rangle$, so stabilizers in $\{\grpcom{\aS{i}}{\aS{k}}:i<k,\  \grpcom{\aS{i}}{\aS{k}} \neq \id)\} $ are redundant and do not change $\cC'$. 
Furthermore, they are LDPC since $\supp{\grpcom{\aS{i}}{\aS{k}}}\subseteq \supp{\aS{i}}\cup \supp{\aS{k}}$. 
It can be verified using Cor.~\ref{cor:z2-commutation-relations} that, on the full Hilbert space, the measured checks in Eq.~\eqref{eq:measured-checks} satisfy
\begin{align}
    \grpcom{\nS{j}}{\nS{l}}=\grpcom{\nS{j}}{\aS{i}}=\id,\qquad \grpcom{\aS{i}}{\aS{k}}\in \langle \cN \rangle.
\end{align}
\noindent

\begin{theorem}[Spacetime fault distance, informal] 
Consider any possible execution of the gauging procedure containing fewer than $d/\gamma$ elementary faults.

If every detector that is not dropped has value $1$, then the execution agrees with some fault-free execution both on the protected logical information and measured outcome of $\T$.

Therefore, the gauging procedure in Def.~\ref{def:detailed-gauging-procedure} admits a constant $\gamma = O(1)$ such that the adversarial spacetime fault distance satisfies 
\[
d_\mathrm{f} \geq d/\gamma.
\]

\end{theorem}

\textbf{Proof sketch.} 
It suffices to consider fault configurations $F$ that violate no deterministic detectors. 
$\cN$-detectors, comprising the product of $\cN$ stabilizer outcomes, remain deterministic regardless of $F$.
We show every non-deterministic or dropped $\cA$-detector lies near a non-Abelian fault, and therefore the number of dropped detectors is $O(\Wt(F))$.
For any fault configuration $F$ with $\Wt(F)\le d/\gamma$ that violates no deterministic detector, we show there must exist a clean `cut' or connected set of qubits and measurement outcomes that are perfect in each of the pre-gauging, gauged, and post-gauging spacetime regions. This ensures that no logical readout error of the Clifford measurement has occurred. 

Cleaning faults away from these cuts rewrites the effect of $F$ as a linear combination of
errors at four rounds: the first round of $\cC$, the last round of $\cC$ before gauging, the first round of $\cC$ after gauging, and the last round of $\cC$. 
All terms in this linear combination are supported on common sets $(R_0,R_1,R_2,R_f)$ satisfying
\begin{equation}\label{eq:cleaning-common-support}
|R_0|+|R_1|+|R_2|+|R_f|=O(\Wt(F)).
\end{equation}
For a suitable constant $\gamma$, $O(\Wt(F))$ is smaller than the code distance of $\cC$. As these now purely spatial errors are sandwiched between clean codespace projectors, they must be correctable. 

The error spaces at these four rounds are therefore correctable, so no non-trivial logical action can occur on the codespace. Combined with the fact that no logical measurement error occurred, we conclude the fault we started with was trivial. Thus, any nontrivial fault must be larger than $d/\gamma$.

\subsection{Spacetime gauging procedure, faults, and trajectories} \label{sec:spacetime-setup}

\begin{table}
    \centering
    \begin{tabular}{|c|c|c|}
        \hline
         & \textbf{Detects} & \textbf{Commutation} \\
        \hline
        Abelian check $\in \cA$ & Abelian faults & Grp commutator $\in \cN$ (commutes in codespace)  \\
        Non-abelian check $\in \cN$ & Non-Abelian faults & Commuting \\
        \hline
    \end{tabular}
    \caption{Distinctions between Abelian checks $\cA$ and non-Abelian checks $\cN$. Abelian checks are called Abelian as they detect Abelian faults, although they do not commute with one another. Non-abelian checks commute, but they detect non-Abelian faults.}
    \label{tab:checks}
\end{table}

We now introduce the notation to define a measurement schedule for Abelian stabilizers that do not commute over the full Hilbert space. Define the non-commutation graph $\Gamma_\cA = (V_\cA, E_\cA)$ as 
\[
    V_\cA = \cA,\qquad E_\cA :=\{(\aS{i},\aS{k}): \grpcom{\aS{i}}{\aS{k}} \neq \id
    )\}.
\]
Fix a vertex-coloring of this graph, 
\[
\cA = \bigsqcup_{r=1}^\kappa \cA_r.
\]
A valid coloring is guaranteed to exist with $\kappa=O(1)$ colors because $\Gamma_\cA$ has $O(1)$-bounded degree; each generator $\aS{i}$ can only fail to commute over the full Hilbert space with a constant number of other stabilizers, because $\cC'$ is LDPC.

\begin{definition}[Fault-tolerant gauging procedure. Extended from Def.~\ref{def:gauging-measurement-protocol}]\label{def:detailed-gauging-procedure}
Let $\{S_i\}$ be the measured stabilizer generators of $\cC$.
Let $t$ be an integral time indexing the measurement round. 
The fault-tolerant gauging protocol is:
\begin{enumerate}
    \item For each $t=1,\dots,d$, measure all $S_i$ simultaneously.
    \item\label{step:gauging} After $t=d$ initialize $\ket{0}_e$ for each $e\in \edgeE$, and measure all $\av$ stabilizers simultaneously, following steps $1-2$ of Def.~\ref{def:gauging-measurement-protocol}.
    \item For each $t=d+1,\dots, 2d$, measure the $\cC'$ checks in subrounds $r=0,\dots ,\kappa$. In subround $r=0$, measure all checks in $\cN$ simultaneously. For subrounds $r=1,\dots,\kappa$, measure checks in $\cA_r$. (At round $t=d+1$, omit the $\av$ measurements that were already made in Step~2.)
    \item \label{step:ungauging} After $t=2d$, measure every $\cZ_e,\ e\in \edgeE$. 
    Then perform the feedforward local Clifford operation $U_{c(z)}$ (defined in Eq.~\eqref{eq:feedforward-clifford}).
    \item For each $t=2d+1,\dots,3d$,  measure each $S_i$.
\end{enumerate}
The protocol reports as its $\T$ outcome the product of all $\av$ measurement outcomes at $t=2d$.
\end{definition}

For completeness, we explicitly describe the Clifford feedforward operation in the presence of errors--- see Sec. \ref{sec:ungauging} for a discussion of the ideal, fault-free case. 

In step \ref{step:ungauging} of the gauging protocol, we measure out all of the edge ancilla in the $\Zp$ basis. Let $z\in \mathbb{F}_2^{|\edgeE|}$ be the recorded edge measurement string.

In the absence of faults, we have $\bp = +1$, which constrains the post-measurement state of the edge ancilla to look like $$\ket{z} = \prod_{v \in \Omega} \T_v \av \ket{0}$$ 
for some subset of vertices $\Omega \subseteq \vertexV$. 
In other words, the binary string $z$ must be a coboundary of the gauging graph: $\delta_2 z = 0$, where $z = \delta_1 \Omega$.
If some readout errors flipped the outcomes of the measurement, or there were some $\bp$ violations present prior to the measurement, we find $\delta_2 z \neq 0$. 

In the case of measurement errors, the true measurement outcome string $z^*$ is a coboundary, and we must simply determine the minimum weight correction that converts our faulty string $z$ into a coboundary.

Fix an ordering $e_1,\dots,e_{|\edgeE|}$ of the edges, and let $\prec$ denote this ordering on $\mathbb{F}_2^{|E|}$. 
Define
    \[
    \hat{e}(z) = \min_{\prec}\bigl(\rm{argmin}_{e\in \mathbb{F}_2^{|\edgeE|}}\bigl\{|e|:\delta_2(z+e)=0\bigr\}\bigr).
    \]
The purpose of the edge ordering is to ensure that $\hat{e}()$ is a deterministic function of $z$ when there are multiple minimum weight corrections.
Now fix a reference vertex $v_0$. Since $H^1(\graphG)=0$ and $\graphG$ is connected, there is a unique vector $c(z) \in \mathbb{F}_2^{|\vertexV|}$ satisfying
\[
    \delta_1 c(z)  = z+ \hat{e}(z),\qquad c(z)_{v_0} = 0.
\]
The Clifford feedforward correction applies the following operator on the matter qubits,
\begin{equation}
\label{eq:feedforward-clifford}
    U_{c(z)} :=\prod_{v:c(z)_v=1}\T_v.
\end{equation} which is simply the matter support of the product of $\av$ corresponding to the edge qubit coboundary.

In the second case where we also have some $\bp = -1$ violations, these violations are converted into deformations of the ungauged code~\cite{lyons2026}. Essentially, the state of the system after the ungauging measurement may have some amplitude outside of the $+1$ eigenspace for certain Pauli stabilizers of the base code. Thus, when we perform the first round of stabilizer measurements after measuring out the edge ancillas, we may observe some stabilizer violations. However, since we spend $d$ rounds in the ungauged code after step \ref{step:gauging}, there is ample time to identify such errors and correct them. We still want to perform the minimum-weight correction and Clifford feedback described above, however, to ensure that the weight of errors produced on the ungauged code remains bounded.

\begin{definition}[Elementary faults]
\label{def:elementary-faults}
An \emph{elementary spatial fault} at time $t\in \{1,\dots, 3d\}$ is a nonidentity Pauli $f$, inserted immediately before round $t$, whose support lies entirely within one vertex set $\vertexset{v}$ or on one gauge qubit. We denote such spatial faults by $(t,f)$.

An \emph{elementary temporal fault} is a tuple $(t, Q, \pi)$, where an operator 
\[
Q\in \{S_i\}_i\cup \cN \cup \cA \cup \{\cZ_e\}_{e\in \edgeE}
\]
is measured at round $t$ in Def.~\ref{def:detailed-gauging-procedure}. Let $\lambda(Q)$ be the spectrum of eigenvalues of $Q$.
A temporal fault causes a permutation on measurement outcomes:
\[
\pi:\lambda(Q)\longrightarrow\lambda(Q).
\] 
In particular, if the projected eigenspace corresponds with the eigenvalue $r\in \lambda(Q)$, the recorded outcome is $\pi(r)$.
\end{definition}

\begin{definition}[Fault configuration]
    Let $\Lambda$ be the set of elementary faults allowed by Def.~\ref{def:detailed-gauging-procedure}.
    A \emph{fault configuration} is a finite set $F\subseteq \Lambda$ containing at most one temporal fault for each measurement. Its weight is $\Wt(F) := |F|$.
\end{definition}

\begin{definition}[Abelian and non-Abelian faults]
\label{def:abelian-non-abelian-faults}
    Consider a spatial fault $(t,f)$ where $t=[d+1,2d]$ (after Step~\ref{step:gauging} and before Step~\ref{step:ungauging} in Def.~\ref{def:detailed-gauging-procedure}). 
    This fault $f = f_A$ is \emph{Abelian} if 
    \begin{align}
        f_A\nS{j}=\nS{j}f_A,\qquad \nS{j}\in \cN.\label{eq:Fa-commutes-nS}
    \end{align}
    and 
    \begin{align}
        f^\dagger_A \aS{i} f_A \aS{i}^\dagger \in \{\pm \id \},\qquad\ \aS{i}\in \cA.
        \label{eq:abelian-condition-2}
    \end{align}
    A spatial fault is \emph{non-Abelian} otherwise, and denoted $f_N$.
    
    A temporal fault $(t, Q, \pi)$ is \emph{non-Abelian} when $Q\in \cN$ and \emph{Abelian} if $Q\in \cA$.
\end{definition}

\begin{remark}[Elementary faults on the ungauged code]
    Although there is no distinction between Abelian and non-Abelian faults on the base stabilizer code, it is convenient to consider the same elementary fault basis (in particular, the basis elements acting on the matter qubits) as we have defined above. Hence, we also label the faults as ``Abelian'' and ``non-Abelian'' on the base code. 
\end{remark}

Therefore, an Abelian spatial fault commutes with every $\nS{j}\in\cN$ and commutes or anticommutes with every $\aS{i}\in \cA$.
Consequently, it maps each $\aS{i}$-eigenspace to another eigenspace deterministically as opposed to a linear combination of eigenspaces, as with a regular Pauli fault.

Since every spatial fault $f$ and every $\nS{j}\in \cN$ are Pauli (recall the decorations to $\cN$ are simply $\Zp$ operators on the gauge edges), a non-Abelian spatial fault also maps each $\nS{j}$-eigenspace to another eigenspace deterministically. In contrast, it has unspecified action on the eigenspaces of checks in $\cA$.

During rounds in the base code $\cC$ ($t\notin [d+1,2d]$), we do not classify spatial or temporal faults as Abelian or non-Abelian. 
Let $\fn\subseteq F$ and $\fa\subseteq F$ denote non-Abelian and Abelian faults in $F$, respectively.

The next lemma shows that commuting an elementary non-Abelian spatial fault through a check in $\cA$ produces only a constant number of elementary Abelian spatial faults.
This is used below to bound the fault growth during the cleaning procedure.
\begin{lemma}[Non-Abelian faults leave behind a constant number of Abelian faults]
\label{lem:byproducts}
    There is a constant $\beta = O(1)$ such that, for every elementary non-Abelian spatial fault $(t, f_N)$ where $t\in \{d+1,\dots, 2d\}$ and every $\aS{i}\in \cA$, 
    \[
    f^\dagger_N \aS{i} f_N \aS{i}^\dagger = f_{A,1} f_{A,2} \cdots f_{A,m},\qquad m\le \beta
    \]
    up to a phase, where every nonidentity $f_{A,j}$ is supported within one vertex set or on one gauge qubit, $(t,f_{A,j})$ is Abelian, and 
    \[
    \supp{f_{A,j}}\subseteq \supp{\aS{i}}.
    \] 
\end{lemma}
\begin{proof}
    For a Pauli $P$ on $\vertexset{v}$, let us define its orbit product by 
    \[
    o(P) := P^\dagger \T_v P \T_v^\dagger.
    \]
    We refer to any Pauli operator that can be written as the above form as an \emph{orbit product}. We now show that $o(P)$ is an Abelian fault.
    Since $\T_v$ is Clifford, $o(P)$ is a Pauli on $\vertexset{v}$, and $\T_v^2 = \id$ gives $\T_v o(P) \T_v^\dagger = \pm o(P)$. 
    Suppose $\T_v M \T_v ^\dagger=\pm M$, and $PM = \epsilon MP$ for $\epsilon\in\{\pm 1\}$.
    Conjugating the latter relation by $\T_v$ gives 
    \[
    (\T_v P \T_v^\dagger) M = \epsilon M (\T_v P \T_v ^\dagger ).
    \]
    The two $\epsilon$ commutation signs cancel for $o(P) = (\T_v P \T_v^\dagger) \cdot P$, so 
    \[
    o(P)M = M o(P).
    \]

    First we show that $o(P)$ commutes with all checks in $\cN$.
    Let $M$ be the restriction to $\vertexset{v}$ of a stabilizer $N \in \cC$, where $N\T N \T = \id$. Since $N$ is invariant under $\T$, $N$ must satisfy $\T_v N \T_v^\dagger = \pm M$.
    Using our calculation above gives $[o(P),M]=0$, and therefore $[o(P),N]=0$. 
    Moreover, $o(P)$ acts only on matter qubits, so we must have $[o(P), \mathbb{N}] = 0$ as well. Similarly, $o(P)$ is disjoint from every $\bp$ check, and so commutes trivially. The remaining checks in $\cN$ are just products of gauged invariant checks $\mathbb{N}$ and $\bp$, so $[o(P),\nS{j}]$ for all $\nS{j}\in \cN$. 

    To verify the second Abelian condition in Eq.~\eqref{eq:abelian-condition-2}, recall the two forms of checks in $\cA$. 
    The vertex check is 
    \[\av = \T_v\prod_{e\in \delta(v)}\cX_e.
    \]
    If $u\neq v$, then $o(P)$ is disjoint from $\av$, so they commute. If $u=v$, then $\T_vo(P) \T_v^\dagger = \pm o(P)$ means $o(P)\av = \pm \av o(P)$.
    For a decorated check $\gaugedS$ associated with $S$ and with root vertex $v_0$, Thm~\ref{thm:z2-gauged-stabilizers} gives
    \[
    \gaugedS = S 
    \prod_{\substack{v\in \lambda (S)\\v\ne v_0}}
    \left(
    \mathsf{S}_{p(v,v_0)}^{\mu_v}
    \prod_{e\in p(v,v_0)}
    \mathrm{C_e}\bigl({N_{S}}_v\bigr)
    \right).
    \]
    Here, the $\mathsf{S}_{p(v,v_0)}^{\mu_v}$ acts only on the gauge qubits in $p(v,v_0)$ and is diagonal in the $\cZ$ basis, while 
    $\mathrm{C_e}\bigl( {N_{S}}_v\bigr)$ has edge control $e$ and matter target ${N_{S}}_v$. Moreover
    \[
    i^{\mu_v}(N_{S})_v = \grpcom{S_{v}}{\T_v},
    \]
    so each $(N_{S})_v$ is an orbit product up to a phase.

    Our earlier calculation shows that $o(P)$ commutes with every $({N_S})_u$, and therefore with every control gate $\mathrm{C}_e({N_S})_u$. Moreover, $o(P)$ is disjoint from $\mathsf{S}_{p(v,v_0)}^{\mu_v}$ so they commute, and since $S \in \cC$ is Pauli, the commutator with $S$ is a phase. Therefore, $o(P)\gaugedS = \pm \gaugedS o(P)$, and thus 
    \[
    o(P)^\dagger \aS{i} o(P) \aS{i}^\dagger\in \{\pm \id\},\qquad \aS{i}\in \cA,
    \]
    so each $(t, o(P))$ is Abelian. 
    Each gauge qubit $\cZ_e$ is also an elementary Abelian fault: it commutes with every check in $\cN$ and with every factor in $\gaugedS$, and only anti-commutes with overlapping $\av$ checks.

    Now, we compute the operators $f_{A,1} f_{A,2} \cdots f_{A,m}$ in the claim. It suffices to consider an arbitrary elementary spatial fault $f$, which includes any possible elementary non-Abelian spatial fault.
    First, let $\aS{i}=\av$.
    If $f$ is supported on $\vertexset{v}$, then
    \[
    f^\dagger \av f \av ^\dagger = f^\dagger \T_v f \T_v^\dagger = o(f),
    \]
    up to a phase. 
    A matter fault $f$ on another vertex set is disjoint from $\av$, while a gauge qubit overlaps only with the Pauli $\cX$ part of $\av$, so 
    \[
    f^\dagger \av f \av^\dagger\in\{\pm\id\}.
    \]
    Hence, the claim holds for $\av$ with $m\le 1$.

    Now let $\aS{i}=\gaugedS$.
    If $f$ is a matter fault, its commutator with $S$ is just a phase, and it is disjoint from $\mathsf{S}_{p(v,v_0)}^{\mu_v}$, and we just need to worry about the target of the controlled-Pauli decorations.
    Indeed,
    \begin{equation}
        f^\dagger \bigl(C_e(N_{S})_v\bigr) f \bigl(C_e(N_{S})_v\bigr)^\dagger = \begin{cases}
            \id,\qquad & f({N_{S}})_v = ({N_{S}})_v f,\\
            \cZ_e,\qquad & f({N_{S}})_v = -({N_{S}})_v f.
        \end{cases}
    \end{equation}
    In the second case, conjugation by $f$ replaces $(N_S)_v$ by $-(N_S)_v$, and 
    \[
    \mathrm C_e\bigl(-(N_S)_v\bigr)
      =\cZ_e\,\mathrm C_e\bigl((N_S)_v\bigr).
    \]
    Thus, $f^\dagger\gaugedS f \gaugedS^\dagger$ is, up to a phase, at most $O(1)$ gauge $\cZ$ operators.  

    Finally, let $f$ be a single-qubit fault on a gauge qubit. 
    A gauge qubit fault $f=\cZ_e$ commutes with every factor in $\gaugedS$.
    For $f=\cX_e$,
    \[
    \cX_e\mathrm{C}_e ({N_S})_v \cX_e\mathrm{C}_e ({N_S})_v ^\dagger = ({N_S})_v,
    \]
    so an overlapping controlled-Pauli decoration contributes $({N_S})_v$, which is an elementary Abelian fault because it is an orbit product.
    For a path $p$, set
    $\cZ_p:=\prod_{e'\in p}\cZ_{e'}$.
    If $e\notin p$, then $\cX_e$ commutes with $\mathsf{S}_p^\mu$. If $e\in p$, then 
    \[
    \cX_e\mathsf S_p^\mu\cX_e
    \bigl(\mathsf S_p^\mu\bigr)^\dagger
      =i^\mu\cZ_p^\mu.
    \]
    Up to a phase, this is either the identity or a product of the single gauge qubit $\cZ$ operators. 
    We know there are $O(1)$ of these Paulis, because of the bounded number and length of the paths decorating $\gaugedS$.
    The same conclusion holds for $\mathcal{Y}_e = i \cX_e \cZ_e$ because $\cZ_e$ commutes with $\gaugedS$.

    The $\cZ_e$ and  $({N_S})_v$ factors we computed for different cases of $f^\dagger \aS{i} f \aS{i}^\dagger$ are all elementary Abelian faults supported in $\supp{\aS{i}}$.
    Bounded check and decoration weights bound the total number of these factors to be $m\le \beta = O(1)$.
\end{proof}

A measurement event $(Q,t)$ is the measurement of the operator $Q$ at time $t$ in Def.~\ref{def:detailed-gauging-procedure}.
For any quantum trajectory of this gauging procedure, the measurement at $(Q,t)$ projects the state by some projector $\Pi_{r(Q,t)}$ onto the $r(Q,t)$-eigenspace of $Q$, where $r(Q,t)\in \lambda(Q)$.
We let $r$ denote the function on the measurement events that assigns eigenvalues to each measurement event.

We now define the noisy version of the $r$ that is affected by temporal faults in a fault configuration $F$.
The measurement outcome history is
\[
r_F(Q,t) :=\begin{cases}
    \pi(r(Q,t)),\qquad &(t,Q,\pi)\in F,\\
    r(Q,t),\qquad &\mathrm{otherwise}.
\end{cases} 
\]
Since there is only one measurement fault per measurement event $(Q,t)$, $r_F(Q,t)$ is well-defined.

For $t\in \{1,\dots, 3d\}$, define
\[
E_F(t) := \prod_{f:(t,f)\in F} f,
\]
with the product in a fixed order, and let $E_F(t) = \id$ if the set is empty.

Write $\ket{0}_\edgeE:=\bigotimes_{e\in \edgeE}\ket{0}_e$. 
For the ungauging round, let $r(\cZ_e, 2d+1) = (-1)^{z_e}$ be the projection applied when measuring gauge qubit $e$.
Let $\bra{z}_\edgeE:=\bigotimes_{e\in \edgeE}\bra{z_e}_e$.
Define $M_t$, which are the projective measurements performed in each round $t$, and when applicable, also the state preparation and Clifford feedforward operations applied to the state. 
To be explicit, we list all $M_t(r,r_F)$ below according to the procedure in Def.~\ref{def:detailed-gauging-procedure}. We remark that the dependence on the noisy measurement $r_F$ is due to the Clifford feedforward correction, $U_{c(z)}$.
\begin{align}
\label{eq:cases-M}
    M_t(r,r_F):=\begin{cases}
    \prod_{i}\Pi_{r(S_i,t)}, & 1\le t < d,\\
    \prod_{i}\Pi_{r(S_i,t)} \bigl(
    \id \otimes \ket{0}_\edgeE
    \bigr), & t = d,\\
    \bigl( \prod_{\mathbb{C}\in \cA_\kappa \backslash \{\av:v\in \vertexV \}} \Pi_{r(\mathbb{C},t)}\bigr)\dots
    \bigl( \prod_{\mathbb{C}\in \cA_1 \backslash \{\av:v\in \vertexV \}} \Pi_{r(\mathbb{C},t)}\bigr)\\
    \cdot \bigl(\prod_{\mathbb{N}\in \cN}\Pi_{r(\mathbb{N},t)}\bigr)
    \bigl(
    \prod_{v\in\vertexV}\Pi_{r(\av,t)}   
    \bigr)
    , & d+1 \le t \le 2d,\\
    \bigl(\prod_{i}\Pi_{r(S_i,t)}
    \bigr)
    U_{c(z)} \otimes
    \bra{z}_\edgeE, & t=2d+1\\
    \prod_{i}\Pi_{r(S_i,t)}, & 2d+2\le t\le 3d
\end{cases}
\end{align}

\begin{definition}[(Valid) quantum trajectories]    
    A \emph{trajectory} is denoted by the pair $(F,r)$.
    The corresponding \emph{trajectory Kraus operator} is
    \[
    K(F,r) := M_{3d}(r,r_F)E_F(3d)\dots M_1(r,r_F) E_F(1).
    \]
    For an input state $\rho$, the gauging procedure in Def.~\ref{def:detailed-gauging-procedure} outputs the unnormalized conditional state $K(F,r)\rho K(F,r)^\dagger$.
    Let $\Pi_{\cC}$ be the projector onto the codespace of $\cC$. 
    We say a trajectory $(F,r)$ is \emph{valid} if $K(F,r)\Pi_\cC\neq 0$, and \emph{invalid} otherwise.
\end{definition}

In other words, an invalid trajectory has zero probability for every encoded input state.
Thus, it suffices to examine the fault weights of only valid trajectories for the fault distance.

\begin{fact}[Moving faults through trajectories]
\label{fact:moving-nonabelian}
Define the following products of Abelian faults:
\[
B_{\rightarrow}(f_N, \mathbb{C}) = \mathbb{C}^\dagger f_N \mathbb{C} f_N^\dagger, \quad B_{\leftarrow}(f_N, \mathbb{C}) = f_N^\dagger \mathbb{C} f_N \mathbb{C}^\dagger 
\]
These are Abelian faults since they are related to the commutator in Lemma \ref{lem:byproducts} by conjugating with $\mathbb{C}$, which only adds a phase.

\[
f_N \Pi_{r(\mathbb{C}, t)} = \frac{1}{2} \sum_{\epsilon = \pm 1} \Pi_{\epsilon r(\mathbb{C}, t)} \bigl ( \id + \epsilon B_{\rightarrow} \bigr ) f_N = \frac{1}{2} \sum_{\epsilon = \pm 1} \bigl ( \id + \epsilon \mathbb{C} B_{\rightarrow}\mathbb{C}^\dagger \bigr ) \Pi_{\epsilon r(\mathbb{C}, t)} f_N
\]

\[
\Pi_{r(\mathbb{C}, t)} f_N = f_N  \frac{1}{2}  \sum_{\epsilon = \pm 1} \bigl ( \id + \epsilon B_\leftarrow  \bigr ) \Pi_{\epsilon r(\mathbb{C},t)}  = f_N  \frac{1}{2} \sum_{\epsilon = \pm 1} \Pi_{\epsilon r(\mathbb{C},t)} \bigl ( \id + \epsilon \mathbb{C}^\dagger B_\leftarrow \mathbb{C} \bigr )
\]
    
\end{fact}
\begin{proof}
    For the backwards move:
    \begin{align}
        f_N \Pi_{r(\mathbb{C}, t)} &= \frac{f_N + r(\mathbb{C}, t) f_N \mathbb{C}}{2} \notag \\
        &= \frac{f_N + r(\mathbb{C}, t) \mathbb{C} B_{\rightarrow}(f_N, \mathbb{C}) f_N }{2} \notag\\
        &= \frac{\id + r(\mathbb{C}, t) \mathbb{C} B_{\rightarrow}(f_N, \mathbb{C})}{2} f_N \notag\\
        &= \left (\frac{\id + r(\mathbb{C}, t) \mathbb{C}}{2} \frac{\id + B_{\rightarrow}}{2} + \frac{\id - r(\mathbb{C}, t) \mathbb{C}}{2} \frac{\id - B_{\rightarrow}}{2}  \right ) f_N \notag \\
        &= \frac{1}{2} \sum_{\epsilon = \pm 1} \Pi_{\epsilon r(\mathbb{C}, t)} \bigl ( \id + \epsilon B_{\rightarrow} \bigr ) f_N
    \end{align}
    To move the byproducts back to the other side, we conjugate $B_{\rightarrow}$ by $\mathbb{C}$:
    \begin{align}
        f_N \Pi_{r(\mathbb{C}, t)} &= \frac{f_N + r(\mathbb{C}, t) f_N \mathbb{C}}{2} \notag \\
        &= \frac{f_N + r(\mathbb{C}, t) \mathbb{C} B_{\rightarrow}(f_N, \mathbb{C}) \mathbb{C}^\dagger \mathbb{C} f_N }{2} \notag\\
        &= \frac{1}{2} \sum_{\epsilon = \pm 1} \bigl ( \id + \epsilon \mathbb{C} B_{\rightarrow}\mathbb{C}^\dagger \bigr ) \Pi_{\epsilon r(\mathbb{C}, t)} f_N \notag
    \end{align}
    The derivation of the forward moves follow the same steps.
\end{proof}

To decide whether a valid faulty trajectory $(F,r)$ with $F\neq \emptyset$ fails, we compare it with a valid fault-free trajectory, which we define below.

\begin{definition}[Fault-free trajectories]
    An \emph{fault-free trajectory} is a valid trajectory $(\emptyset, r')$. Here, $\emptyset$ indicates that the fault configuration $F$ is empty, so no spatial or temporal faults occur, and
    $r'$ assigns an eigenvalue $r'(Q,t)\in \lambda(Q)$ on every measurement event $(Q,t)$. Since there are no temporal faults, the recorded outcomes $r'_{\emptyset}(Q,t)$ are the same as $r'(Q,t)$ for all $Q,t$. 
\end{definition}
Generically, the same gauging protocol and input code $\cC$ may admit multiple valid, fault-free assignments $r'$. For example, even in the absence of faults, $\av$ measurements at $t=d+1$ are random, so clearly multiple $r'$ outcomes are possible.
The next lemma shows that two valid fault-free trajectories $(\emptyset,r')$ and $(\emptyset,r'')$ have the same action on the codespace when the reported $\T$ outcome is the same.

Let 
\[
m_\T(r_F) := \prod_{v\in \vertexV} r_F(\av,2d).
\]

\begin{lemma}[Valid fault-free trajectories have the same logical action]
\label{lem:fault-free}
    Let $(\emptyset, r')$ and $(\emptyset, r'')$ be valid fault-free trajectories, where $r'(Q,t),r''(Q,t)\in \lambda(Q)$.
    If 
    \[
    m_\T(r') = m_\T(r''),
    \]
    then
    \[
    K(\emptyset, r')\Pi_\cC \propto K(\emptyset,r'')\Pi_\cC.
    \]
\end{lemma}
\begin{proof}
    Observe that 
    \[
    E_\emptyset(t) = \id
    \]
    so the Kraus operator for a fault-free trajectory is given by
    \[
    K(\emptyset, r) = \prod_{t=1}^{3d} M_t(r, r_\emptyset) .
    \]
    Assume at $t=0$ we have $S_i = +1$. Note at the end of the procedure we project onto the $\cC$ codespace. Then, the first $d$ timesteps of the fault-free Kraus operators are all equivalent to the projector onto the codespace of $\cC$, as $r(S_i, t) = +1$ for $1 \leq t \leq d$.
    \[
    K(\emptyset, r) \Pi_{\cC} = \left (\prod_{t=d+1}^{3d} M_t (r, r_\emptyset) \prod_{t=1}^d \Pi_{\cC} \right ) \Pi_{\cC} = \prod_{t=d+1}^{3d} M_t (r, r_\emptyset)
    \]
    At $t = d+1$, we begin measuring the gauged code stabilizers. The outcomes for the $\av$ stabilizers are random, with parity $m_\T(r') = m_\T(r'')$. As there are no faults, the measurement outcomes for the stabilizers $\mathbb{N} \in \cN \backslash \{\bp \}$ and $\mathbb{C} \in \cA \backslash \{\av\}$ remain $+1$. Additionally, since we have initialized all edges in $\ket{0}$ without any faults, $r(\bp, d+1) = +1$. Let $\Pi_{\cN}$ be the projector onto all $+1$ outcomes for $\mathbb{N} \in \cN$, and $\Pi_{\cA \backslash \{\av\}}$ be the same for all Abelian stabilizers excluding the $\av$. Additionally, since the procedure stays in the $\Pi_{\cN}$ subspace while we are in the gauged code, all stabilizers commute with one another. 
    \[
    M_{d+1}(r, r_\emptyset) = \Pi_{\cN} \Pi_{\cA \backslash \{\av\}}\left ( \prod_{v \in \vertexV} \Pi_{r(\av, d+1)} \right ) \left ( \id \otimes \ket{0}_{\edgeE} \right )
    \] 
    Similarly, for $d+2 \leq t \leq 2d$, 
    \[
    M_{t}(r, r_\emptyset) = \Pi_{\cN} \Pi_{\cA \backslash \{\av\}} \left ( \prod_{v \in \vertexV} \Pi_{r(\av, 2d)} \right )
    \]
    Combining these two, we get the collapsed operator: 
    \[
    \prod_{t = d+1}^{2d} M_{t}(r, r_\emptyset) = \Pi_{\cN} \Pi_{\cA \backslash \{\av\}} \left ( \prod_{v \in \vertexV} \Pi_{r(\av, 2d)} \right ) \left ( \id \otimes \ket{0}_{\edgeE} \right )
    \] 
    Note we have enforced the measurement outcomes for $\av$ at $t = d+1$ and $t = 2d$ are equal, as this is a fault-free trajectory.
    Next, consider $t = 2d +1$.
    \[
    M_{2d +1}(r, r_\emptyset) = \left ( \prod_i \Pi_{r(S_i, 2d + 1)} \right ) U_{c(z)} \bra{z}_{\edgeE}
    \]
    Consider the operator
    \[
    \prod_{t = d+1}^{2d+1} M_{t}(r, r_\emptyset) = \left ( \prod_i \Pi_{r(S_i, 2d + 1)} \right ) U_{c(z)} \bra{z}_{\edgeE} \Pi_{\cN} \Pi_{\cA \backslash \{\av\}} \left ( \prod_{v \in \vertexV} \Pi_{r(\av, 2d)} \right ) \left ( \id \otimes \ket{0}_{\edgeE} \right )
    \]
    Note that $\Pi_{\cN}$ includes the projector onto the $\bp = +1$ subspace; the measurement outcomes $z$ for the edge ancilla must be consistent with this for a nonzero trajectory. 
    Since $\graphG$ is homologically trivial, and we have a $\bp$ for each cycle of $\graphG$, the support of $-1$ measurement outcomes must form a \emph{coboundary} $\delta_1 \sigma$ of some subset of vertices $\sigma \subseteq \vertexV$. 
    The $\Xp$ support of the product $\prod_{v \in \sigma} \av$ is exactly this coboundary. 
    Thus, we have 
    \[
    \bra{z}_{\edgeE} = \bra{0}_{\edgeE} \prod_{v \in \sigma} \av \T_\sigma
    \] 
    where $\T_\sigma = \prod_{v \in \sigma} \T_v$. We sum over all consistent choices of subset $\sigma$ to retain all possible trajectories at the end. Note $\sigma$ is a strict subset since the product of all $\av$ is trivial on the edge ancilla. Note the correction step $U_{c(z)} = \T_\sigma$.
    \[
    \prod_{t = d+1}^{2d+1} M_{t}(r, r_\emptyset) =  \left ( \prod_i \Pi_{r(S_i, 2d + 1)} \right ) \bra{0}_{\edgeE} \Pi_{\cN} \Pi_{\cA \backslash \{\av\}} \left (\prod_{v \in \sigma} \av \right ) \left ( \prod_{v \in \vertexV} \Pi_{r(\av, 2d)} \right ) \ket{0}_{\edgeE}
    \] 
    Letting $m_{\T}(r_F, \sigma) = \prod_{v \in \sigma} r_F(\av, 2d)$, we find:
    \[
     \prod_{t = d+1}^{2d+1} M_{t}(r, r_\emptyset) =  m_{\T}(r_\emptyset, \sigma) \left ( \prod_i \Pi_{r(S_i, 2d + 1)} \right ) \bra{0}_{\edgeE} \Pi_{\cN} \Pi_{\cA \backslash \{\av\}} \left ( \prod_{v \in \vertexV} \Pi_{r(\av, 2d)} \right ) \ket{0}_{\edgeE} 
    \]
    As discussed in Sec. \ref{sec:ungauging}, projecting all edge ancilla to the $\ket{0}$ state maps the decorated stabilizers onto the original code stabilizers, and kills the $\bp$ stabilizers. It also preserves their eigenvalues as $+1$.
    \begin{align*}
        \prod_{t = d+1}^{2d+1} M_{t}(r, r_\emptyset) &= m_{\T}(r_\emptyset, \sigma) \left ( \prod_i \Pi_{r(S_i, 2d + 1)} \right ) \left (\prod_{i} \Pi_{r(S_i, 0)}  \right ) \bra{0}_{\edgeE} \prod_{v \in \vertexV} \Pi_{r(\av, 2d)} \ket{0}_{\edgeE} \\
        &=  m_{\T}(r_\emptyset, \sigma) \Pi_{\cC} \bra{0}_{\edgeE} \prod_{v \in \vertexV} \Pi_{r(\av, 2d)} \ket{0}_{\edgeE} 
    \end{align*}
    Consider 
    \begin{align*}
        \bra{0}_{\edgeE} \prod_{v \in \vertexV} \Pi_{r(\av, 2d)} \ket{0}_{\edgeE} &= \bra{0}_{\edgeE} \frac{1}{2^{|\vertexV|}}\sum_{\{\phi\}: \phi \subseteq \vertexV} \prod_{v \in \phi} r(\av, 2d) \av \ket{0}_{\edgeE} \\
        &= \frac{1}{2^{|\vertexV|}}\sum_{\{\phi\}: \phi \subseteq \vertexV} \prod_{v \in \phi} r(\av, 2d) \bra{0}_{\edgeE}\prod_{v \in \phi} \av \ket{0}_{\edgeE} \\
        &= \frac{1}{2^{|\vertexV|}}\sum_{\{\phi\}: \phi \subseteq \vertexV} \prod_{v \in \phi} r(\av, 2d) \T_v \bra{0}_{\edgeE} \ket{\delta_1 \phi}_{\edgeE} \\
        &= \frac{1}{2^{|\vertexV|}} \left (\id + m_{\T}(r_\emptyset) \T \right )
    \end{align*}
    where $\ket{\delta_1 \phi}_\edgeE$ indicates the state with $\ket{1}$ in the support of $\delta_1 \phi$ and $\ket{0}$ elsewhere. We can see that only terms with $\delta_1 \phi = 0$ survive, and thus the only remaining terms are $\phi = \emptyset$, and $\phi = \vertexV$. 

    So, the Kraus operator for the trajectory corresponding to $\sigma$ satisfies
    \[
    K_\sigma(\emptyset, r)\Pi_{\cC} = \frac{1}{2^{|\vertexV|-1}} m_{\T}(r_\emptyset, \sigma) \frac{1}{2} \left (\id + m_{\T}(r_\emptyset) \T \right ) \Pi_{\cC}
    \]
    The normalization reflects the uniform probability to measure the edges in the state $\ket{\delta_1 \sigma}$. Note that the $\sigma$ dependence only comes in as a global phase. Since $m_\T(r_\emptyset, \sigma_r)$ is just a phase, we see that the action of a fault-free trajectory on the codespace of $\cC$ is purely dependent on $m_{\T}(r_\emptyset) = m_{\T}(r)$. Thus, if $r' = r''$ for two fault-free trajectories, $K(\emptyset, r') \Pi_{\cC} \propto K(\emptyset, r'') \Pi_{\cC}$.
    
\end{proof} 

In our proof of Lemma \ref{lem:fault-free} above, as an intermediate result, we have demonstrated that the logical action of the gauging protocol, in the absence of noise, is to project the initial codespace onto an eigenstate of $\T$.

\begin{corollary}[Logical action of the gauging protocol] \label{cor:gauging-logical-action}
    For a fault-free trajectory $(\emptyset, r')$, the gauging protocol in Def. \ref{def:gauging-measurement-protocol} implements the logical projection 
    \begin{equation}
        \Pi_m = \frac{\id + m \T}{2}, \quad m = m_\T(r').
    \end{equation}
\end{corollary}
\begin{proof}
    See the proof of Lemma \ref{lem:fault-free}.
\end{proof}

\begin{definition}[Logical failure]\label{def:logical-failure}
    A valid trajectory $(F,r)$ is a \emph{logical success} if there exists a valid fault-free trajectory $(\emptyset, r')$ and a scalar $\alpha\in \mathbb{C}\setminus \{0\}$
    such that
    \begin{enumerate}
        \item \label{cond:logical-success-1} $m_\T(r') = m_\T (r_F)$,
        \item \label{cond:logical-success-2} $K(F,r)\Pi_\cC = \alpha K(\emptyset , r')\Pi_\cC$.
    \end{enumerate}
    
    It is a \emph{logical failure} otherwise.
\end{definition}
To interpret this condition, let $\rho =\Pi_{\cC} \rho \Pi_\cC$ be any encoded input density operator.
For a fixed fault configuration $F$, define
$p(r|F,\rho):=\Tr( K(F,r) \rho K(F,r)^\dagger )$.
This is the probability of obtaining an output state with the projected eigenspaces specified by $r$, conditional on $F$ and $\rho$.
Since $\rho$ is in the codespace, condition~\ref{cond:logical-success-2} gives
\[
K(F,r) \rho K(F,r)^\dagger = |\alpha|^2 K(\emptyset,r') \rho K(\emptyset,r')^\dagger.
\] 
Taking the trace of both sides gives
\[
p(r|F,p) = |\alpha|^2 p (r'|\emptyset,\rho).
\]
Thus, when $p(r|F,\rho)>0$, the faulty and fault-free give the same normalized conditional output state,
\[
    \frac{K(F,r)\rho K(F,r)^\dagger}{p(r|F,p)} = \frac{K(\emptyset,r')\rho K(\emptyset, r')^\dagger}{p(r'|\emptyset,\rho)}.
\]
The same scalar $\alpha$ applies to the entire input codespace, so this equality holds for every encoded state, including logical superpositions. Consequently, every output measurement statistic, and the expectation values of every logical operator in the faulty execution agree with the corresponding fault-free trajectory.

One may worry that condition~\ref{cond:logical-success-1} only equates the reported $m_\T$ outcomes, which include any temporal faults, without requiring the output state of the faulty $(F,r)$ trajectory to lie in the reported $m_\T$-eigenspace. However, in condition~\ref{cond:logical-success-2} rules this out, as we now show.
Let $m_\T(r) = m_\T(r_F)=m$. Cor.~\ref{cor:gauging-logical-action} shows $K(\emptyset,r')\Pi_\cC \propto \Pi_{m}\Pi_{\cC}$, so
$\Pi_{-m} K(\emptyset, r')\Pi_{\cC} = 0$.
Condition~\ref{cond:logical-success-2} implies that 
\[
\Pi_{-m} K(F, r)\Pi_{\cC}  = \Pi_{-m} K(\emptyset, r')\Pi_{\cC} =0.
\]
Hence, a faulty trajectory reporting $m$, has no overlap with the $-m$-eigenspace of $\T$.

\subsection{Detectors}

We now describe the detectors appearing in the Clifford gauging protocol. 

\begin{definition}[Detectors and dropped detectors]
   A \emph{detector} is a function $D$ of the recorded outcomes whose values are roots of unity and such that $D(r')=1$ for every valid fault-free trajectory $(\emptyset, r')$. 

    Let the last measurement round on which $D$ depends be $t^*$. 
    A detector $D$ is \emph{dropped} by a fault configuration $F$ if there are valid trajectories $(F,r)$ and $(F,\tilde{r})$ such that 
    \begin{align}
        \tilde{r}(Q,t)=r(Q,t),\qquad  \forall (Q,t),\ t<t^*
    \end{align}
    and 
    \[
    D(\tilde{r}_F)\neq D(r_F),
    \]
    where $\tilde{r}_F$ and $r_F$ are the measurement outcome histories of $(F,\tilde{r})$ and $(F,r)$ trajectories respectively.
    Call the set of all dropped detectors given $F$, $\mathrm{Drop}(F)$. For a valid trajectory $(F,r)$, a detector $D\notin \mathrm{Drop}(F)$ is \emph{satisfied} if $D(r_F)=1$ and \emph{unsatisfied} if $D(r_F)\neq 1$.
    \label{def:detectors}
\end{definition}

\emph{Bulk detectors} compare consecutive measurements of an operator $Q$ at times $t$ and $t+1$.
We use 
\[
D^t_Q(r_F) := r_F(Q,t)^*\ r_F(Q,t+1),\qquad \mathrm{for}\ \begin{cases}
    Q\in \{S_i\}_i\quad &t\in \{1,\dots,d-1\}\cup \{2d+1,\dots, 3d-1\}\\
    Q\in \cN\cup \cA,\quad &t\in \{d+1,\dots 2d-1\}.
\end{cases}
\]

These bulk detectors cover only consecutive rounds in which the same check is measured. 
We also need detectors that express the relationship between the known eigenvalues of the input codestate and the first measurements at $t=1$, as well as between rounds $d$ and $d+1$, and between rounds $2d$ and $2d+1$.
We call the detector functions expressing these relations \emph{boundary detectors} and list them below.
If $Q$ is measured at exactly one of the two times $t$ and $t+1$, we replace the missing outcome of $Q$ in $D^t_Q$ by the corresponding eigenvalue fixed by state preparation or inferred from other measurement outcomes.

Because the input state before $t=1$ lies in the codespace of $\cC$, every base code stabilizer $S_i$ has eigenvalue $+1$.
Its boundary detector at $t=1$ is thus
\[
D^0_{S_i}(r_F) := r_F(S_i,1),\qquad \forall S_i.
\]

Let $\gaugedS_i=\mathfrak{C}[S_i]$ be a decorated stabilizer first measured at the gauging time boundary $t=d$.
Its Clifford decorations act trivially on the initialized gauge qubits in $\ket{0}_\edgeE$, so gauging preserves the $S_i$ eigenvalue.
Thus, their boundary detectors at $t=d$ are 
\[
D^{d}_{\gaugedS_i}(r_F) := r_F(S_i, d)^* r_F(\gaugedS_i, d+1),\qquad \forall \gaugedS_i.
\]

To derive the detector across the ungauging boundary, $t=2d$, consider a fault-free trajectory $(\emptyset,r')$. 
Let $\ket{\Phi}$ be the state immediately before ungauging, where $\gaugedS_i\ket{\Phi}=\lambda\ket{\Phi}$.
Let $z=(z_e)_{e\in\edgeE}\in \mathbb{F}_2^{|\edgeE|}$ be the fault-free gauge qubit measurement string defined by $r'(\cZ_e, 2d+1) = (-1)^{z_e}$. 
For this outcome, the ungauging operation in step~\ref{step:ungauging} of Def.~\ref{def:detailed-gauging-procedure} is 
\[
W_z := U_{c(z)} \bra{z}_\edgeE.
\]
One can show that  
\[
S_iW_z\ket{\Phi}= W_z \gaugedS_i \ket{\Phi} = \lambda W_z\ket{\Phi},
\]
so the first $S_i$ outcome after ungauging is equal to the final outcome of $\gaugedS_i$ before ungauging.
Hence, every fault-free trajectory satisfies $r'(\gaugedS_i,2d) = r'(S_i,2d+1)$, so the boundary detector at $2d$ is
\[
D_{\gaugedS_i}^{2d}(r_F):=r_F (\gaugedS_i,2d)^* r_F(S_i,2d+1),\qquad \forall S_i.
\]
We refer to detectors consisting of non-Abelian (Abelian) stabilizer outcomes as \emph{non-Abelian (Abelian) detectors}.

\begin{lemma}[Only Abelian detectors are dropped]
    \label{lem:abelian-dropped-only}
    Every detector in $\mathrm{Drop}(F)$ is an Abelian detector.
\end{lemma}

\begin{proof}
    By Def.~\ref{def:detectors}, dropping a detector $D$ requires two valid trajectories with fault configuration $F$ that diverge only at $t=t_1$, where $t_1$ is the last round on which the detector $D$ depends. Suppose we did drop a non-Abelian detector $D_{\mathbb{N}}(r_F) = r_F(Q, t_1-1)^*r_F(Q, t_1)$ corresponding to some $\mathbb{N} \in \cN$. Recall that the non-Abelian stabilizers are always Pauli (see Def. \ref{thm:z2-gauged-stabilizers}). Our elementary faults are also Pauli, and so must commute or anti-commute with a given non-Abelian stabilizer. 
    
    Consider the Kraus operators corresponding to two histories $r$ and $\tilde{r}$ that agree for all $t < t_1$. Let their shared Kraus operator up until this point be $K_{< t_1}(F) $. Suppose some spatial fault $f_1$ occurs at $t = t_1$, with $f_1 \mathbb{N} f_1^\dagger \mathbb{N}^\dagger = \epsilon \id$, $\epsilon \in \{\pm 1\}$. In the case of a temporal fault, we have essentially the same expression, as $\mathbb{N}$ has two eigenspaces to toggle between: either the measurement fault does nothing, or it flips the eigenvalue. 
    \[
    K(F, r) = M_{t_1}(r, r_F)f_1 K_{< t_1}(F) = f_1 \Pi_{\epsilon r(\mathbb{N}, t_1)} K_{< t_1}(F) 
    \]
    Similarly, we have
    \[
    K(F, \tilde{r}) = M_{t_1}(\tilde{r}, \tilde{r}_F)f_1 K_{< t_1}(F) = f_1 \Pi_{\epsilon \tilde{r}(\mathbb{N}, t_1)} K_{< t_1}(F)
    \]
    Let the measurement outcome $r(\mathbb{N}, t_1-1) = \tilde{r}(\mathbb{N}, t_1-1) = m$. For either trajectory to have nonzero amplitude, we must have
    \[
    \epsilon r(\mathbb{N}, t_1) = m = \epsilon \tilde{r}(\mathbb{N}, t_1) \implies  r(\mathbb{N}, t_1) = \tilde{r}(\mathbb{N}, t_1) = \epsilon m
    \]
    But this is a contradiction; both trajectories agree on the entirety of detector $D_{\mathbb{N}}$, and so it cannot be dropped.
 
\end{proof}

To define our notion of fault distance, we introduce \emph{loop-like} faults, which violate no deterministic detectors. 

\begin{definition}[Loop-like fault]
\label{def:loop-like-fault}
    A fault configuration $F$ is \emph{loop-like} if every detector $D\notin \mathrm{Drop}(F)$ is satisfied on every valid trajectory $(F,r)$.
    Equivalently,
    \[
    D(r_F) = 1\qquad \text{for every valid $(F,r)$, and every $D\notin \mathrm{Drop}(F)$.}
    \]
\end{definition}

We can then explicitly write down what we mean by a nontrivial space time fault:
\begin{definition}[Nontrivial space time faults and fault distance]
\label{def:fault-distance}
    A loop-like fault $F$ is a \emph{nontrivial} logical fault if any associated valid trajectory $(F, r)$ is a logical failure by Def. \ref{def:logical-failure}. Let the set of all such logical failures be denoted $L$.

    The \emph{fault distance} of the gauging measurement protocol is the minimum distance of any nontrivial logical fault.
    \[
    d_f = \text{min}_{F \in L} |F|
    \]
\end{definition}

Next, we find it useful to situate detectors and faults on a fault-location hypergraph. This allows us to reason about the range over which a given elementary fault may impact detectors.

\begin{definition}[Fault-location hypergraph]
\label{def:fault-location-hypergraph}
    Let
    \begin{align}
        \mathsf{L}_{\mathrm{s}}&:=\{(t,\vertexset{v}):1\le t\le 3d,v\in\vertexV\}\\
        &\qquad \sqcup \{(t,\{e\}):d+2\le t\le 2d+1,\ e\in\edgeE\},\notag\\
        \mathsf{L}_\mathrm{t}&:=\{(Q,t): Q\text{ measured in round }t\}.
\end{align}
Set 
\begin{align}
    V_\mathcal{H}:= \mathsf{L}_{\rm{s}} \sqcup \mathsf{L}_{\rm{t}}.\notag
\end{align}
    For a fault configuration $F$, define its support by
    \begin{align}
        \supp{F}:=&\{(t,l)\in \mathsf{L}_{\rm{s}}:\exists f,\ (t,f)\in F, \supp{f}\subseteq l\}\notag\\
        &\cup \{ (Q,t) \in \mathsf{L}_\mathrm{t}: \exists\pi, (t,Q,\pi)\in F\}.\notag
    \end{align}
    
    For each detector $D^t_Q$, let $\mathsf{I}(D^t_Q)\subseteq V_F$ be its hyperedge. It contains the temporal locations of every recorded outcome appearing in the defining expression of $D^t_{Q}$. It also contains every spatial location $(t+1,l)$ for which $l$ intersects the support of an operator whose eigenvalue appears in $D^t_Q$. 
    
    Set 
    $E_\mathcal{H}=\{\mathsf{I}(D): D\text{ is a detector}\}$.
    The \emph{fault location hypergraph} is $\mathcal{H}=(V_\mathcal{H}, E_\mathcal{H})$. 
\end{definition}

As an example, for a bulk detector $D_Q^t$,
\[
\mathsf{I}(D_Q^t):=\{ (Q,t), (Q,t+1)\}\cup \{(t+1,l)\in \mathsf{L}_\mathrm{s}:l\cap \supp{Q}\neq \emptyset\}.
\]

Two vertices in $\mathcal{H}$ are adjacent when they belong to a common hyperedge. $\mathrm{dist}_\mathcal{H}$ denotes the corresponding shortest path distance.
Each hyperedge has $O(1)$ vertices and each vertex belongs to $O(1)$ hyperedges, because each outcome appears in $O(1)$ detectors, and the measured checks have bounded weight, qubit degree, and vertex set size. 

We define the neighbors of a detector $D_Q^t$ within a radius $\xi$ as
\begin{align}
\label{eq:detector-neighborhood}
    \mathsf{N}_\xi (D) := \{ x\in V_F: \min_{y\in \mathsf{I}(D)}\mathrm{dist}_{\mathcal{H}}(x,y)\le \xi\}.
\end{align}
The minimum is $+\infty$ if $x$ and $\mathsf{I}(D)$ lie in different connected components.

We now argue that the set of detectors dropped by a given loop-like fault must be localized to the support of the fault on the fault-location hypergraph.

\begin{lemma}[Localization of dropped detectors around a non-Abelian loop] \label{lem:dropped-detector-localization}
There is an integer $\upsilon=O(1)$ such that, for every loop-like fault configuration $F$, all dropped bulk detectors $D_{\aS{i}}^t$ with $\aS{i}\in \cA$ and $d+1\leq t\leq 2d-1$ lie within hypergraph distance $\upsilon$ of $\supp{F_N}$.
Equivalently,
\[
\mathsf{N}_{\upsilon}(D_{\aS{i}}^t)\cap \supp{F_N}\neq \emptyset
\]
for every such dropped detector.
Likewise, all dropped boundary detectors lie within hypergraph distance $\upsilon$ of $\supp{F}$:
\[
\mathsf{N}_\upsilon(D_Q^t)\cap \supp{F}\neq \emptyset.
\]
Consequently,
\[
|\mathrm{Drop}(F)|=O(\Wt(F)).
\]
\end{lemma}

\begin{proof}

Away from the gauging boundaries, we know that only non-Abelian faults drop detectors. Anywhere in spacetime, only Abelian detectors are dropped, by Lemma \ref{lem:abelian-dropped-only}.

Consider some valid trajectory $(F, r)$ for the loop-like fault $F$. Consider the trajectory Kraus operator 
\[
K(F, r) = M_{3d}(r, r_F) E_F(3d) \cdots M_1(r, r_F) E_F(1)
\] Let $t_0$ be the earliest time on which $F$ has support, and let $f_N \in F$ be a non-Abelian spatial fault. Each non-Abelian elementary fault $f_N$ commutes with an Abelian stabilizer up to some constant number of Abelian faults, by Lemma \ref{lem:byproducts}. By Fact \ref{fact:moving-nonabelian}, 
\[
\Pi_{r(\mathbb{C}_i, t_0+1)} f_N \Pi_{r(\mathbb{C}_i, t_0)} = f_N \frac{1}{2} \left ( \sum_{\epsilon = \pm 1} \bigl (\id + \epsilon B_\leftarrow \bigr ) \Pi_{\epsilon r(\mathbb{C},t_0 + 1)}\Pi_{r(\mathbb{C}, t_0)} \right )
\]
We see that the trajectory has split into two valid options: one with the detector $D_{\mathbb{C}}^{t_0}$ violated, and one with the detector satisfied. Both of these are valid for the fault $F$, and agree up until the time $t_0 + 1$, and so this detector must be dropped. Note that any subsequent detectors are \emph{not} dropped, since pushing $f_N$ past them adds an $\epsilon$, which then cancels with the $\epsilon$ on the prior outcomes.
\[
\Pi_{r(\mathbb{C}, t_0+2)}f_N  \left ( \frac{1}{2}\sum_{\epsilon = \pm 1} \bigl ( \id + \epsilon B_{\leftarrow} \bigr ) \Pi_{\epsilon r(\mathbb{C},t_0 + 1)}\Pi_{r(\mathbb{C}, t_0)} \right ) 
\]
\[
= f_N \frac{1}{4} \sum_{\epsilon, \epsilon'}  \bigl (\id + \epsilon' B_\leftarrow \bigr) \Pi_{\epsilon' r(\mathbb{C},t_0 + 2)} \bigl ( \id + \epsilon B_{\leftarrow} \bigr ) \Pi_{\epsilon r(\mathbb{C},t_0 + 1)}\Pi_{r(\mathbb{C}, t_0)}
\]
\[
= f_N \frac{1}{4} \sum_{\epsilon, \epsilon'}  \bigl (\id + \epsilon' B_\leftarrow \bigr) \bigl ( \id + \epsilon B_{\leftarrow} \bigr )\Pi_{\alpha \epsilon' r(\mathbb{C},t_0 + 2)}  \Pi_{\epsilon r(\mathbb{C},t_0 + 1)}\Pi_{r(\mathbb{C}, t_0)}
\]
\[
= f_N \frac{1}{4} \sum_{\epsilon}  \bigl (\id + \alpha \epsilon B_\leftarrow \bigr) \bigl ( \id + \epsilon B_{\leftarrow} \bigr )\Pi_{\epsilon r(\mathbb{C},t_0 + 2)}  \Pi_{\epsilon r(\mathbb{C},t_0 + 1)}\Pi_{r(\mathbb{C}, t_0)}
\]
\[
\neq 0 \implies \epsilon r(\mathbb{C}, t_0) = r(\mathbb{C}, t_0 + 1) = r(\mathbb{C}, t_0 + 2)
\]

In the third to the fourth line, we have used the fact that $B_\leftarrow$ is a collection of Abelian Pauli faults, and so must commute up to $\alpha = \pm 1$ with $\mathbb{C}$. We see in the last line only the detector at $t_0$ can be violated (in the case $\epsilon = -1$).

Because the gauged code is LDPC, we know that the number of Abelian stabilizers with nontrivial commutation with $f_N$ is $O(1)$, and thus the number of dropped detectors is $O(1)$. Furthermore, by our definition of the fault-location hypergraph, the hyper-edges corresponding to these dropped detectors must touch the vertex corresponding to $f_N$. So all such dropped detectors are in a radius of $1$ around a given non-Abelian fault. If we consider all spatial faults belonging to the full configuration $F$, each of them produces some $O(1)$ number of dropped detectors via the same mechanism. 

There is another mechanism whereby non-Abelian faults may produce dropped detectors. This is indirectly, by flipping the outcome of a non-Abelian stabilizer. Suppose we have fixed some ordering for measuring the Abelian stabilizers, so they are partitioned into sets $\{\cA_r\}$. Stabilizers belonging to different sets only commute up to some non-Abelian stabilizer. If a non-Abelian stabilizer is violated, when we measure the Abelian stabilizers in our fixed ordering, measuring one will randomize the other. 

Consider another spatial fault $f_N$, and suppose there are Abelian stabilizers $\mathbb{C}_i, \mathbb{C}_j$ such that
\[
[\mathbb{C}_i, \mathbb{C}_j] = \mathbb{N}, \quad \grpcom{f_N}{\mathbb{N}} = -1
\]
Assume in our fixed schedule, $\mathbb{C}_i$ is measured before $\mathbb{C}_j$. Also, we take $f_N$ to commute with both $\mathbb{C}_i$ and $\mathbb{C}_j$ up to phases $\alpha_i, \alpha_j$ (this is fine, since we know they would be dropped otherwise). In the $\mathbb{N}= -1$ subspace:
\[
\Pi_{r(\mathbb{C}_i, t_1)}\Pi_{r(\mathbb{C}_j, t_2)} = 
    \frac{1}{2} \sum_{\epsilon = \pm 1} \Pi_{r(\mathbb{C}_j, t_2)} \Pi_{\epsilon r(\mathbb{C}_i, t_1)} + \epsilon \Pi_{-r(\mathbb{C}_j, t_2)} \Pi_{\epsilon r(\mathbb{C}_i, t_1)}
\]
Whatever the measured outcome of $\mathbb{C}_j$, there is an equal chance of either outcome for $\mathbb{C}_i$, no matter what the initial state was. As long as we remain in the $\mathbb{N} = -1$ subspace, we always reset the outcomes for $\mathbb{C}_j$ when we measure $\mathbb{C}_i$, and vice versa. Thus, the detectors for both $\mathbb{C}_i$ and $\mathbb{C}_j$ are dropped. For a given $f_N$, it may anti-commute with an $O(1)$ number of non-Abelian stabilizers, which in turn can only share support with an $O(1)$ number of Abelian stabilizers. So, all in all, only an $O(1)$ number of pairs of Abelian stabilizers may begin to anti-commute as a result of a given elementary non-Abelian fault. 

Note that this time the detectors are dropped because the \emph{absolute} eigenvalue of $\mathbb{N}$ is $-1$--- not the detector value, which may be $+1$! So in general these dropped detectors persist in the system for as long as the excitation does, and are not localized to the fault $f_N$. However, we can argue that they are localized in the case of a \emph{loop-like} fault. 

A detector may be $+1$ even in the presence of a physical violation of a stabilizer if a measurement error means we readout the violated stabilizer incorrectly. Consider the earliest time a generic loop-like fault $F$ has non-Abelian support. As we start in the codespace, these non-Abelian faults at the earliest time $t_0$ must \emph{create} $-1$ stabilizer outcomes for some $\mathbb{N} \in \cN$. To prevent any non-Abelian detectors from being be violated, we must have measurement errors at every location a $\mathbb{N}$ stabilizer is flipped. The dropped detectors for $\mathbb{C}_i$ and $\mathbb{C}_j$ persist as long as the measurement errors do. When the measurement errors stop at some later time $t_1$, we must have had another physical error that got rid of the $\mathbb{N} = -1$ stabilizer violations, otherwise we would have a detector violation. At this point, the detectors for $\mathbb{C}_i, \mathbb{C}_j$ are restored, as $\mathbb{C}_i, \mathbb{C}_j$ no longer anti-commute.

We now turn to the question of whether these dropped detectors are localized on the graph around the full fault configuration $F$. 
At the time $t_0$ of the initial physical faults, the first set of dropped detectors is localized, since $D_{\mathbb{C}_i}^{t_0}, D_{\mathbb{C}_j}^{t_0}$ must share support with $D_{\mathbb{N}}^{t_0}$, and so must all share a vertex. 
As the vertex corresponding to $f_N$ also touches $D_{\mathbb{N}}^{t_0}$, there is a length-2 path connecting them. 
The same argument holds for the final time $t_1$ with the second set of physical faults. For the intermediate times $t_0 < t < t_1$, there must be measurement faults preventing $D_{\mathbb{N}}^t$ from being violated, and these faults correspond to vertices that have a length-2 path to the dropped detectors $D_{\mathbb{C}_i}^{t}, D_{\mathbb{C}_j}^t$. 

Thus, for a loop-like fault, we have demonstrated all dropped detectors lie in a bounded region around the support of the fault $F$. Furthermore, due to the LDPC stabilizer structure of the gauged code, the number of dropped detectors is proportional to the weight of the fault $F$.

We note that all of the above arguments hold for detectors living at the boundary of the gauged code; we simply broaden the definition of a fault that can cause a dropped detector, since we are comparing gauged and ungauged stabilizers.

\end{proof}

\begin{remark}[Two types of dropped detectors]
As discussed in the proof of Lemma \ref{lem:dropped-detector-localization}, there are two mechanisms behind dropped detectors: (i) non-Abelian faults creating byproducts when commuting through the Abelian projectors; (ii) non-commuting Abelian projectors resulting from a violated non-Abelian stabilizer. 

\vspace{6pt}
In the section below, when we describe how to clean errors through spacetime, we deal with the first kind of dropped detector by expanding the Kraus operator into all possible trajectories for affected detectors.

\vspace{6pt}
We show that the second kind of dropped detectors are still localized to a loop-like fault when after cleaning. Furthermore, we can also eventually transform the Kraus operator of a loop-like fault (subject to some additional conditions) such that there are no dropped detectors at all. 
    
\end{remark}

We now consider fault configurations that are too small to span the timelike direction of the Clifford gauging protocol. We show that such fault configurations can be ``cut'' into disjoint pieces. 

Let 
\[
I_\mathrm{pre} :=\{1,\dots, d\},\qquad I_\mathrm{g}:=\{d+1,\dots,2d\},\qquad I_\mathrm{post} :=\{2d+1,\dots,3d\}.
\]
Let $\mathcal{H}[I] = (V_{\mathcal{H}[I]},E_{\mathcal{H}[I]})$ be the subhypergraph induced by the vertices whose time coordinate lies in $I$. 
Let $\partial_-(I)$ and $\partial_+(I)$ be its vertices at times $\min(I)$ and $\max(I)$ respectively.

\begin{definition}[Cutting condition] \label{def:cutting-condition}
An interval $I$ satisfies the \emph{cutting condition} for a fault configuration $F$ if there are disjoint sets of vertices $P_-$, $C$, and $P_+$ such that
\[
V_{\mathcal{H}[I]} = P_-\sqcup C \sqcup P_+,
\]
where
\[
\partial_-(I)\subseteq P_-,\qquad \partial_{+}(I)\subseteq P_+,
\]
and no hyperedge of $\mathcal{H}[I]$ intersects both $P_-$ and $P_+$.
Further, $C$ must satisfy 
\[
C\cap \left( \supp{F} \cup \bigcup_{D\in \mathrm{Drop}(F)}\mathsf{N}_{\upsilon}(D)\right) = \emptyset.
\]
Then, we call $C$ a \emph{clean cut}.
\end{definition}

We can show that any fault $F$ with suitably low weight must satisfy this cutting condition: in particular, there is a clean cut supported on each of the following regions: (i) the initial $d$ rounds spent in the base code, (ii) the $d$ rounds spent in the gauged code, and (iii) the final $d$ rounds in the base code. 

\begin{figure}
    \centering
    \includegraphics[width=0.75\linewidth]{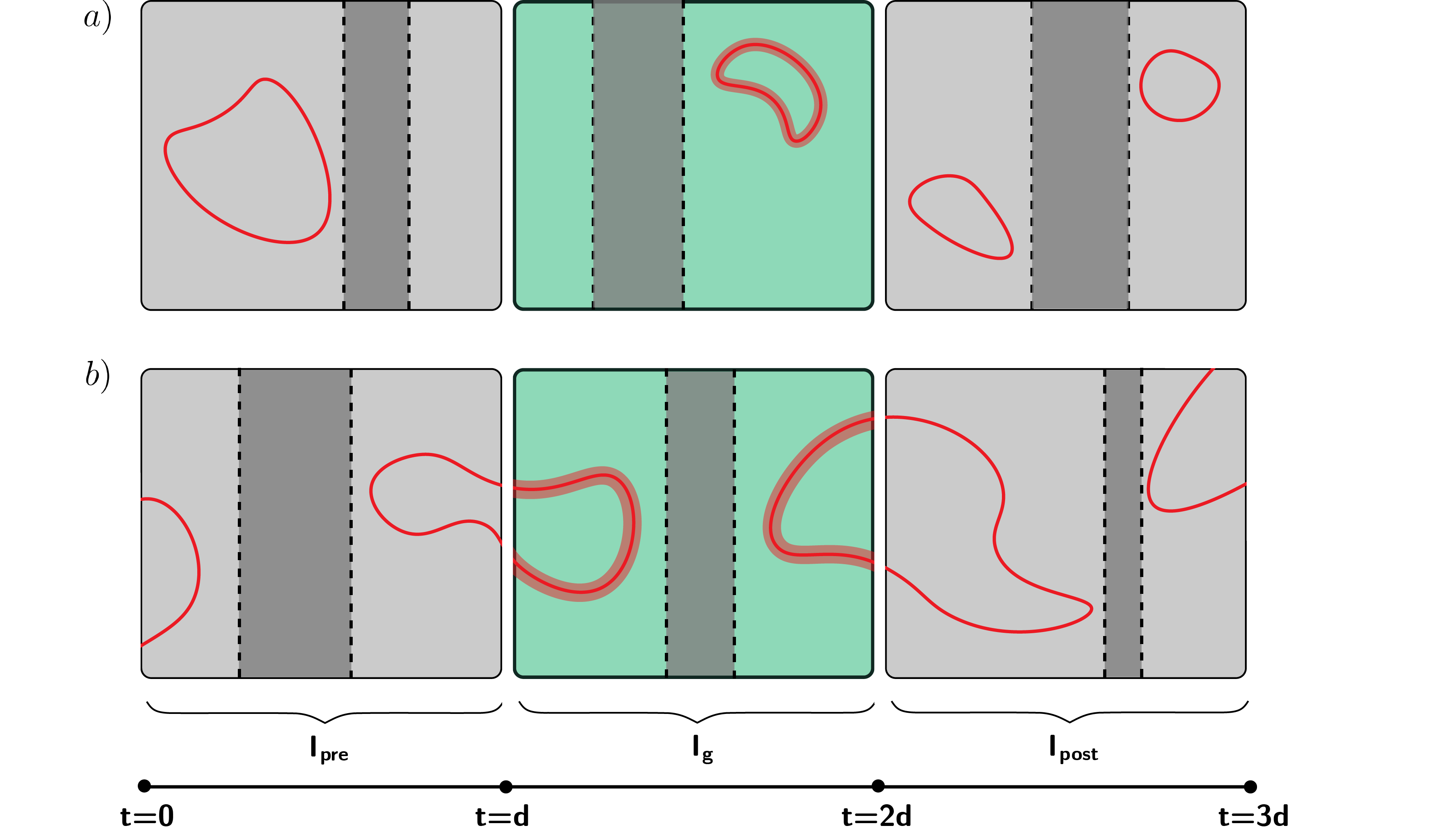}
    \caption{Two different loop-like faults in spacetime that satisfy the cutting condition (Def. \ref{def:cutting-condition}). a) $P_\pm$ are the unshaded regions, while $C$ is the shaded region of spacetime where the faults have no support. In each of the three segments of the measurement protocol, pre-, during, and post-gauging, we can find a suitable region $C$. b) The cutting condition can hold for faults spanning gauging boundaries as well.}
    \label{fig:cutting-condition}
\end{figure}

Intuitively, if the total weight of a fault is less than $d$, there are not enough elementary faults to span the time direction, which consists of $d$ time steps in each stage of the gauging protocol. Thus, we must be able to find some set of time steps on which the fault $F$ has no support. If we ``cut'' along these time steps, we partition spacetime into disjoint chunks without touching the support of the fault. See Fig. \ref{fig:cutting-condition} for an illustration. 

We now formalize this intuition.

\begin{lemma}[Existence of a cut]
\label{lem:cutting-condition-existence}
For some $1 < c_\mathrm{cut}=O(1)$, every loop-like fault configuration $F$ with $\Wt(F)<d/c_\mathrm{cut}$ has a clean cut for each of $I_\mathrm{pre}, I_\mathrm{g}, I_\mathrm{post}$.
\end{lemma}
\begin{proof}
Our strategy to prove the existence of a clean cut is to first break a given fault into connected components on the detector hypergraph. Given a large enough $c_{\rm cut}$, we are able to show that there must always be some interval of size $4 \upsilon$ between some pair of connected components. This interval can then define our clean cut. 

We know a loop-like fault must consist of some union of connected components--- other loop-like faults--- because it does not violate any (deterministic) detectors, and so must have no boundary on $\mathcal{H}$. Let the weight of a loop-like fault $F$ be $|F| \leq d/c_{\rm cut}$. We know that since $I_{\rm pre}, I_{\rm g}, I_{\rm post}$ all consist of $d$ time steps, there must be at least $d - d/c_{\rm cut}$ time steps \emph{without any faults}. 

We still need to demonstrate that there is a single interval between connected components of $F$ that has suitably large weight, say $4 \upsilon$. This ensures that the dropped detectors localized to a connected component also do not touch. 

In the worst case, the set of $d( 1- 1/c_{\rm cut})$ time steps without faults are spread out between many small connected components. In particular, say we have $d/c_{\rm cut}$ connected components, equally spaced. Then the number of time steps between each connected component is:
\[
\frac{d - d/c_{\rm cut}}{d/c_{\rm cut}} = c_{\rm cut} - 1 \geq 4 \upsilon \implies c_{\rm cut} \geq 4\upsilon + 1
\]
As $\upsilon = O(1)$, we can also find a suitable $O(1)$ $c_{\rm cut}$. Note that $c_{\rm cut} > 1$ always as well. 

We can then pick one interval in each of $I_{\rm pre}, I_{\rm g}, I_{\rm post}$ to perform the cut.

\end{proof}

If a fault satisfies the cutting condition, we are able to ``clean'' each portion of the fault on either side of the cut onto one of the temporal boundaries. 

\subsection{Cleaning} \label{sec:cleaning}

We now describe a process that can be applied to ``clean'' loop-like fault configurations that do not span the time direction of the protocol. Cleaning such a fault results in an equivalent fault that is supported completely at the gauging and ungauging time steps. 

More precisely, Lemma~\ref{lem:rewriting-abelian} and~\ref{lem:rewriting-non-abelian}, show how to rewrite Abelian and non-Abelian faults across one gauged round.
Def.~\ref{def:two-stage} combines these rewrite rules into a two-stage cleaning procedure to move all faults from one time step to the next, while Lemma~\ref{lem:cleaning-equivalence} proves the cleaning procedure is an exact re-writing of any given trajectory Kraus operator.
Def.~\ref{def:iter-cleaning} defines an iterated cleaning procedure to move all faults to one of the boundary rounds $t = 1, d+1, 2d+1, 3d$.

For a fault configuration $F$ and $d+2\le t\le 2d$, let
\[
F_t:=\{(t,f)\in F : (t,f) \text{ is a spatial fault}\}=F_t^A \sqcup F_t^N,
\]
where $F_t^A$ and $F_t^N$ are its Abelian and non-Abelian faults. 

Fix an order of the spatial faults at each time $t$ and let
\begin{align}
E_t(F) := \eta_t N_tA_t,\qquad A_t :=\prod_{(t,f)\in F_t^A}f_A,\qquad 
N_t:=\prod_{(t,f)\in F_t^N}f_N,
\label{eq:Et-At-Nt}
\end{align}
where $\eta_t \in \{\pm 1\}$ is the phase produced by this reordering.

\begin{lemma}[Rewriting Abelian faults across a gauged round]
\label{lem:rewriting-abelian}
    Let $(F,r)$ be a trajectory, 
    let $d+2\le t\le 2d$ and let $f_A$ be the Pauli operator of an elementary Abelian fault at time $t$ or $t+1$.
    For each $Q\in \cN \cup \cA$, define $\epsilon(f_A,Q)\in \{\pm 1\}$ by
    \[
        Qf_A = \epsilon(f_A,Q) f_A Q.
    \]
    In particular, $\epsilon(f_A,Q)=1$ for all $\nS{j}\in \cN$.
    Define 
    \[
    r^{(t,f_A)}(Q,t^*) :=
    \begin{cases}
        \epsilon(f_A,Q) r(Q,t),\qquad &t^*=t,\\
        r(Q,t^*),\qquad &t^*\neq t.
    \end{cases}
    \]
    Then,
    \begin{align}
        \label{eq:push-abelian-forward}
        M_t(r,r_F)f_A = f_A M_t(r^{(t,f_A)},r_F),\\
        \label{eq:push-abelian-backward}
        f_A M_t(r,r_F) = M_t(r^{(t,f_A)},r_F) f_A.
    \end{align}
    We refer to Eq.~\eqref{eq:push-abelian-forward} and Eq.~\eqref{eq:push-abelian-backward} as \emph{forward} and \emph{backward Abelian rewrites}, respectively.
\end{lemma}
\begin{proof}
    
    For every $Q\in \cN\cup \cA$, the definition of $\epsilon(f_A,Q)$ directly gives
    \begin{align}
    \label{eq:abelian-projector-commutator}
        \Pi_{r(Q,t)}f_A = f_A\Pi_{\epsilon(f_A,Q) r(Q,t)}.
    \end{align}
    Recall 
    \[
    M_t(r,r_F) = \bigl(\prod_{\mathbb{C}\in \cA_\kappa}\Pi_{r(\mathbb{C},t)}
    \bigr)\dots \bigl(\prod_{\mathbb{C}\in \cA_1}\Pi_{r(\mathbb{C},t)}
    \bigr)
    \bigl(\prod_{\mathbb{N}\in \cN} \Pi_{r(\mathbb{N},t)}
    \bigr)
    \]
    from Eq.~\eqref{eq:cases-M}.
    Using Eq.~\eqref{eq:abelian-projector-commutator}, compute 
    \begin{align}
        M_t(r,r_F)f_A&= \left(\prod_{C\in\cA_\kappa}\Pi_{r(C,t)}\right)\cdots\left(\prod_{C\in\cA_1}\Pi_{r(C,t)}\right)\left(\prod_{N\in\cN}\Pi_{r(N,t)}\right)f_A\\
        &=f_A\left(\prod_{C\in\cA_\kappa}\Pi_{r^{(t,f_A)}(\aS{},t)}\right)
        \cdots\left(\prod_{C\in\cA_1}\Pi_{r^{(t,f_A)}(\aS{},t)}\right)
        \left(\prod_{N\in\cN}\Pi_{r^{(t,f_A)}(\nS{},t)}\right)\\
        &=f_A M_t\bigl(r^{(t,f_A)},r_F\bigr).
    \end{align}
    Eq.~\eqref{eq:push-abelian-backward} follows from Eq.~\eqref{eq:push-abelian-forward}: since $\epsilon_{f_A,Q}^2 = 1$,  $(r^{(f_A,t)})^{(f_A,t)}=r$. Relabeling $r$ with $r^{(f_A,t)}$ and $r^{(f_A,t)}$ with $r$ in equation~\ref{eq:non-abelian-forward} gives Eq.~\eqref{eq:push-abelian-backward}.
\end{proof}

Thus, $f_A$ can be rewritten from round $t$ to $t+1$ (Eq.~\eqref{eq:push-abelian-forward}) or from round $t+1$ to round $t$ (Eq.~\eqref{eq:push-abelian-backward}) without changing its Pauli support.
During this move, the \textit{projected} history $r(Q,t)$ changes to $r^{(t,f_A)}$, while the \textit{recorded} history remains $r_F$.
In order for measured outcomes $r_F$ to remain the same despite the changed eigenspace projections at round $t$, the fault configuration needs to change. In particular, measurement errors are added at round $t$. Changing these measurement errors while pushing these Abelian faults between rounds is analogous to multiplying by Pauli spacetime stabilizers in standard spacetime cleaning methods in Pauli stabilizer codes.

We later compose these rewrites iteratively over many rounds. This requires us to bound the number of spatial and temporal faults changed by each rewrite, since the final cleaned configuration need not have $O(\Wt(F))$ if we are not careful.

\begin{corollary}[Fault support and weight after Abelian rewrites]
\label{cor:fault-support-abelian-rewrite}
    
    Each Abelian rewrite in in Lemma~\ref{lem:rewriting-abelian} does not change the support of $f_A$.
    Further, recorded history remains $r_F$. 
\end{corollary}
\begin{proof}
    The same Pauli operator $f_A$ appears on both sides of equations~\eqref{eq:push-abelian-forward} and~\eqref{eq:push-abelian-backward}. Only the time at which it occurs changes. Thus, the number of elementary spatial faults and the qubit support of the spatial fault is unchanged after the rewrite.
    
    Now we show that the Abelian rewrite can keep the $r_F$ measurement outcome history the same by changing only $O(1)$ measurement events in round $t$.
    It suffices to only examine $\cA$ checks because $f_A$ rewrites do not change the projected eigenspaces of $\cN$ checks.
    For each $\aS{i}\in \cA$, let $\pi_{i}$ be the temporal fault permutation at $(\aS{i},t)$. Take $\pi_{\aS{i}}=\id$ when no temporal fault occurs. 
    Replace the temporal fault permutation $\pi_i$ by the permutation $\pi_i'$ defined by
    \[
    \pi_{i}'(\epsilon(f_A,\aS{i})\lambda ):= \pi_i {\lambda},\qquad \lambda\in \lambda(\aS{i}).
    \]
    This defines a permutation where multiplication by $\epsilon(f_A,\aS{i})$ permutes $\lambda(\aS{i})$.
    Then,
     \begin{align}
         \pi_i'(r^{(t,f_A)}(\aS{i},t)) &= \pi_i'(\epsilon_i r(\aS{i},t))\\
        &=\pi_i( r(\aS{i},t))\\
        &=r_F(\aS{i},t).
     \end{align}
    Thus the recorded history $r_F$ is unchanged.
    Moreover, $\pi_i=\pi_i'$ when $\epsilon_i=1$. Since $f_A$ is local, and the measured checks have bounded qubit degree, these permutations change at only $O(1)$ measurement events.    
\end{proof}

Moving an elementary non-Abelian fault through a measurement projector (see Fact.~\ref{fact:moving-nonabelian}) can change the projected eigenvalue and produce a linear combination of Abelian byproducts.
To track the operators produced by compositions of these rewrites, we introduce the following error space containing all possible products and linear combinations of a set of elementary spatial faults.
Below, we apply this definition to the Abelian byproducts generated by the non-Abelian rewrites (Lemma~\ref{lem:rewriting-non-abelian}).

\begin{definition}[Error space generated by faults]
    For a finite set $R$ of elementary spatial faults, define 
    \[
    \mathrm{Err}(R) := \rm{span}\Biggl\{\prod_{(t,f)\in R'} f:R'\subseteq R\Biggr\}
    \]
    where every product is taken in a fixed order of the spatial faults $f$, and an empty product is $\id$.
    Thus, $\mathrm{Err}(R)$ contains linear combinations of products of the faults in $R$.

    We use the same notation $\mathrm{Err}(S)$ for a set of spatial locations.
    If $S\subseteq \mathsf{L}_\mathrm{s}$, whose elements are the spatial locations $(t,\vertexset{v})$ and $(t,\{e\})$ from Def.~\ref{def:fault-location-hypergraph}, 
    define
    \[
    \mathrm{Err}(S) := \mathrm{Err}(\{(t,f)\in \Lambda: \exists (t,l)\in S, \supp{f}\subseteq l\}).
    \]
\end{definition}

For every $(t,f_N)\in F_t^N$, $\aS{}\in \cA$, and $\diamond\in \{\leftarrow, \rightarrow\}$, fix the ordered factorization in Lemma~\ref{lem:byproducts},
\begin{align}
\label{eq:b-arrow-factorization}
    B_\diamond(f_N,\aS{}) = f_{A,1}^\diamond(f_N,\aS{})\dots f_{A,m}^\diamond(f_N,\aS{})\ \text{up to a phase},\qquad m\le \beta,
\end{align}
where each $f_{A,j}^\diamond(f_N,\aS{})$ is the Pauli operator of an elementary Abelian fault at time $t$.

Define 
\[
\cB_t:=\{(t,f_{A,j}^{\diamond}(f_N, \aS{})):(t,f_N)\in F_t^N,\ \aS{}\in \cA,\  \diamond\in \{\rightarrow, \leftarrow\},\ 1\le j\le m,\ f_{A,j}^\diamond(f_N, \aS{})\neq \id\}.
\]
This is the set of all nonidentity elementary Abelian byproducts that can arise when non-Abelian faults in $F_t^N$ are moved, in either direction, through the $\cA$-projectors measured in round $t$.

\begin{lemma}[Rewriting non-Abelian faults across a round in the gauged code]
\label{lem:rewriting-non-abelian}
Let $(F,r)$ be a trajectory. For $d+2 \le t \le 2d$, moving $N_t$ forward to round $t+1$ across $M_t$ gives 
\begin{align}
    M_t(r,r_F) N_t &= N_t \left(\sum_j M_t(r_j, r_F) A_j\right),\qquad A_j\in \mathrm{Err}(\cB_t).\label{eq:non-abelian-forward}
\end{align}
For $d+2 \le t \le 2d-1$, moving $N_{t+1}$ backwards to round $t$ across $M_t$ gives
\begin{align}
    N_{t+1} M_t(r,r_F) &= \left(\sum_j A_j' M_t (r_j',r_F)\right) N_{t+1},\qquad A_j'\in \mathrm{Err}(\cB_{t+1}).\label{eq:non-abelian-backward}
\end{align}
Every $r_j(Q,t^*)$ and $r_j'(Q,t^*)$ is an outcome function that agrees with $r(Q,t^*)$ for $t^*\neq t$.
Moreover, every term has the same recorded term $r_F$.
We refer to Eq.~\eqref{eq:non-abelian-forward} and Eq.~\eqref{eq:non-abelian-backward} as \emph{forward} and \emph{backward non-Abelian rewrites}, respectively.
\end{lemma}
\begin{proof}
    Write
    \[
    N_t = f_{N,s}\dots f_{N,1},\qquad M_t =P_{l}\dots P_1
    \]
    using a fixed fault order and a fixed order of the projectors at round $t$ (Eq.~\eqref{eq:cases-M}). We use $P_\ell$ with subscript $1\le \ell \le l$ to enumerate the projectors in $M_t$.
    
    Suppose $P_\ell=\Pi_{\mu}$ projects onto an 
    $\nS{}\in \cN$ eigenspace with eigenvalue $\mu$, 
    then the fact that both $\nS{}$ and $f_{N,k}$ ($1\le k\le s$) are Pauli means that $\nS{}f_{N,k}=\epsilon f_{N,k}\nS{}$ for some $\epsilon\in \{\pm 1\}$,
    and thus
    \begin{align}
    \label{eq:move-non-abelian-projector}
        \Pi_{\mu} f_{N,k} = f_{N,k}\Pi_{\epsilon \mu}.
    \end{align}
    Now suppose that $P_\ell=\Pi_{\mu}$ instead projects onto a $\aS{}\in \cA$ eigenspace. Abbreviate $O_\leftarrow:= \aS{}B_\leftarrow(f_{N,k},\aS{}) \aS{}^\dagger$ and $O_\rightarrow:=B_\rightarrow (f_{N,k},\aS{})$.
    Fact~\ref{fact:moving-nonabelian} directly gives forward and backward rewrites for an elementary non-Abelian fault across an Abelian projector: 
    \begin{align}
        \Pi_{\mu}f_{N,k} = f_{N,k}\sum_{\epsilon=\pm1} \Pi_{\epsilon \mu} 
        \frac{\id + \epsilon O_\leftarrow}{2}
        \label{eq:forward-elementary-non-abelian}\\
        f_{N,k}\Pi_{\mu}=
        \sum_{\epsilon=\pm1}
        \Pi_{\epsilon \mu}
        \frac{\id + \epsilon O_\rightarrow}{2}f_{N,k}.
        \label{eq:backward-elementary-non-abelian}
    \end{align}

    First we observe that $O_\leftarrow,O_\rightarrow\in\mathrm{Err}(\cB_t)$. To see this, recall that each $B_\leftarrow (f_{N,k},\aS{})$ has a fixed factorization (Eq.~\eqref{eq:b-arrow-factorization})
    \[
    B_\leftarrow (f_{N,k},\aS{}) = \omega_{\leftarrow}\prod_if_{A,i}^{\leftarrow},
    \]
    where we have abbreviated $f_{A,i}^\leftarrow = f_{A,i}^\leftarrow(f_{N,k},\aS{})$.
    $(t,f_{A,i}^\leftarrow)\in \cB_t$ and $\omega_\leftarrow$ is a phase. 
    Every $f_{A,i}^\leftarrow$ is an elementary Abelian spatial fault, $\aS{}f_{A,i}^{\leftarrow} \aS{}^{\dagger}=\eta_i f_{A,i}^{\leftarrow}$ for some $\eta_i\in \{\pm 1\}$.
    Therefore,
    \[
    O_\leftarrow = \omega_\leftarrow \bigl(\prod_{i}\eta_i \bigr) \prod_i f_{A,i}^\leftarrow \in \mathrm{Err}(\cB_t).
    \]
    The same calculation with $B_\rightarrow$ gives $O_\rightarrow\in \mathrm{Err}(\cB_t)$.

    After one forward rewrite, Eq.\eqref{eq:forward-elementary-non-abelian} when expanded is
    \[
    \Pi_\mu f_{N,k} = \frac{1}{2}f_{N,k} (\Pi_\mu + \Pi_{-\mu} + \Pi_\mu O_\leftarrow - \Pi_{-\mu} O_\leftarrow).
    \]
    Moving an elementary non-Abelian fault through the $\aS{}$ measurement produces trajectories with either projected history $\mu$ or $-\mu$. Additionally, this move results in a linear combination of Abelian byproducts to the right of the projector $\Pi_{\epsilon \mu}$, $(\id\pm \epsilon O_\leftarrow)$, which depends on the projected eigenspace.

    Now apply Eqs.~\eqref{eq:move-non-abelian-projector},~\eqref{eq:forward-elementary-non-abelian} and~\eqref{eq:backward-elementary-non-abelian} successively to the factors of $N_t = f_{N,s}\dots f_{N,1}$, and sequentially move each $f_{N,k}$ to the left of $M_t$ from $f_{N,s}$ to $f_{N,1}$.
    When moving $f_{N,k}$ past a non-Abelian projector, we change the sign of the projected eigenspace, as in Eq.~\eqref{eq:move-non-abelian-projector}. When moving $f_{N,k}$ past an Abelian projector, we use Eq.~\eqref{eq:forward-elementary-non-abelian} and~\eqref{eq:backward-elementary-non-abelian}.
    Each time a non-Abelian fault is moved past an Abelian projector, it replaces the current term in the sum with four terms: the projected eigenspace is either $\mu$ or $-\mu$, and the factor to the right of the projector is either $\id$, $O_\leftarrow$, or $-O_\leftarrow$. 
    We continue until every factor of $N_t$ has been moved to the left of every projector in $M_t$. Let $j$ label one non-zero term in this finite sum. By definition, once a $j$ is fixed, so are all projectors $P_{\ell,j}$ ($1\le \ell \le l$, $\ell\in \mathbb{N}$) and Abelian faults between these projectors $A_{\ell,j}$. We include any factors of $1/2$ and other commutation signs in $A_{\ell,j}$.
    Presently, we have
    \[
    M_t(r,r_F)N_t = N_t \sum_{j} P_{l,j}  A_{l,j}\dots P_{1,j}A_{1,j},\qquad A_{\ell,j}\in \mathrm{Err}(\cB_t).
    \]
    Every $A_{\ell,j}\in \mathrm{Err}(\cB_t)$ because it is a scalar multiple of a product of generated elementary Abelian byproducts.

    We apply the backward Abelian rewrite (Lemma~\ref{lem:rewriting-abelian}) to each term labeled by $j$ to put all $A_{\ell,j}$ to the right of $M_t$:
    \[
    M_t(r,r_F)N_t = N_t \sum_{j} P_{l,j}  A_{l,j}\dots P_{1,j}A_{1,j}= 
    N_t\sum_j M_t(r_j,r_F)A_j
    \qquad A_j\in \mathrm{Err}(\cB_t) 
    \]
    Thus we have exactly Eq.~\eqref{eq:non-abelian-forward} in the claim. Starting instead with Eq.~\eqref{eq:backward-elementary-non-abelian} and following the same argument with $O_\rightarrow$ instead of $O_\leftarrow$ gives Eq.~\eqref{eq:non-abelian-backward}.
\end{proof}

\begin{corollary}[Fault support and weight after non-Abelian rewrites]
\label{cor:fault-support-non-abelian-rewrites}
    In the forward (backward) non-Abelian rewrite in Eq.~\eqref{eq:non-abelian-forward} (Eq.~\eqref{eq:non-abelian-backward}), every $A_j$ \emph{($A_j'$)} is a scalar multiple of a product of at most $O(|F_t^N|)$ \emph{($O(|F_{t+1}^N|)$)} elementary Abelian spatial faults at time $t$ \emph{($t+1$)}.
\end{corollary}
\begin{proof}
    Each elementary non-Abelian fault intersects $O(1)$ measured checks.
    When we sequentially move each non-Abelian elementary fault in $F_t^N$ ($F_{t+1}^N$) during the forward (backward) rewrite, the elementary fault non-trivially overlaps with $O(1)$ projectors. 
    
    Thus, we invoke Fact~\ref{fact:moving-nonabelian} only $O(|F_t^N|)$ ($O(|F_{t+1}^N|)$) times.
    Each use of Fact~\ref{fact:moving-nonabelian} produces at most $\beta = O(1)$ elementary Abelian factors by Lemma~\ref{lem:byproducts}. 
    We then must have that the total number of elementary Abelian byproducts produced by moving $F_t^N$ ($F_{t+1}^N$) across $M_t$ is at most $O(|F_t^N|)$ ($O(|F_{t+1}^N|)$).
    
    Using the Abelian rewrite rule (Lemma~\ref{lem:rewriting-abelian}) to move all Abelian byproducts to the right (left) of $M_t$ keeps the same spatial support and number of elementary Abelian faults by Lemma~\ref{cor:fault-support-abelian-rewrite}. 
    Therefore, every product $A_j$ ($A_j'$) contains at most $O(|F_t^N|)$ ($O(|F_{t+1}^N|)$) elementary Abelian faults at time $t$ ($t+1$).
    
\end{proof}

\begin{definition}[Two-stage cleaning across one time step]
\label{def:two-stage}
    Let $(F,r)$ be a trajectory, and let $M_t$ be the stabilizer measurements in round $t$.

    Let $d+2\le t\le 2d$.
    To clean spatial faults at time $t$ forward to time $t+1$,
    \begin{enumerate}
        \item Apply the forward non-Abelian rewrite in Lemma~\ref{lem:rewriting-non-abelian}. This moves $F_t^N$ from time $t$ to $t+1$, and leaves Abelian byproducts at time $t$. The trajectory $(F,r)$ is expanded into a sum of many trajectories, $(F_j,r_j)$, each with the same noisy outcome history $r_F$.
        \item For each $(F_j,r_j)$ trajectory in the expansion, use the forward rewrite in Lemma~\ref{lem:rewriting-abelian} to move every original fault in $F_t^A$ and all Abelian byproducts across $M_t$ to round $t+1$.
    \end{enumerate}
    We refer to the above procedure as \emph{forward two-stage cleaning}.

    Let $d+2\le t\le 2d-1$.
    To clean spatial faults at time $t+1$ backwards to time $t$,
    \begin{enumerate}
        \item Apply the backward non-Abelian rewrite in Lemma~\ref{lem:rewriting-non-abelian}. This moves $F_{t+1}^N$ from time $t+1$ to $t$, and leaves Abelian byproducts at time $t+1$. The trajectory $(F,r)$ is expanded into a sum of many trajectories, $(F_j,r_j)$, each with the same noisy outcome history $r_F$.
        \item For each $(F_j,r_j)$ trajectory in the expansion, use the backwards rewrite in Lemma~\ref{lem:rewriting-abelian} to move every original fault in $F_{t+1}^A$ and all Abelian byproducts across $M_t$ to round $t$.
    \end{enumerate}
    We refer to the above procedure as \emph{backward two-stage cleaning}.
\end{definition}

\begin{lemma}[Cleaning equivalence]
\label{lem:cleaning-equivalence}
    Let $(F,r)$ be a trajectory, and perform a forward two-stage cleaning or a backwards
    two-stage cleaning (Def.~\ref{def:two-stage}). The resulting sum of trajectories $(F_j,r_j)$ and coefficients $\alpha_j\in \mathbb{C}$ satisfy
    \[
    K(F,r) = \sum_j \alpha_j  K(F_j,r_j),\ \qquad (r_j)_{F_j} = r_F
    \]
    for every non-zero term. $(r_j)_{F_j}$ denotes the recorded history computed from $r_j$ and $F_j$.
    Thus, the two-stage cleaning is an exact rewrite of the Kraus operator corresponding to the trajectory $(F,r)$ and it does not change the recorded outcome history.
\end{lemma}
\begin{proof}
    We first prove the claim for forward two-stage cleaning by applying the exact operator identities in lemmas~\ref{lem:rewriting-abelian} and~\ref{lem:rewriting-non-abelian}.
    Let us write the Kraus operator for the $(F,r)$ trajectory as
    \[
        K(F,r) = K_{>t}(F,r)M_t(r,r_F) E_t(F) K_{<t}(F,r),
    \]
    where 
    \[
    K_{>t}(F,r):= M_{3d}E_{3d}\dots M_{t+1}E_{t+1},\qquad
    K_{<t}(F,r):= M_{t-1}E_{t-1}\dots M_1E_1,
    \]
    we have dropped the $(F,r)$ arguments for readability.
    Using Eq.~\eqref{eq:Et-At-Nt} and the forward non-Abelian rewrite, 
    \begin{align}
        K(F,r) &= \eta_t K_{>t} M_t(r,r_F)N_t A_t K_{<t}\notag\\
        &= \eta_t K_{>t} N_t \bigl( \sum_{j}M_{t}(r_j,r_F)A_j\bigr) A_t K_{<t}.
    \end{align}
    Expand each $A_j\in \mathrm{Err}(\cB_t)$ as $A_j\in \sum_a c_{j,a}G_{j,a}$ where $G_{j,a}$ is a product of elementary Abelian byproducts.
    Applying the forward Abelian rewrite in Lemma~\ref{lem:rewriting-abelian} to the factors of $G_{j,a}A_t$ alters the projected outcome,
    \[
    M_t(r_j,r_F) G_{j,a} A_t = G_{j,a} A_t M_t(r_{j,a}, r_F).
    \]
    Thus, we presently have
    \[
    K(F,r) = \sum_{j,a} \eta_t c_{j,a} K_{>t} N_t G_{j,a} A_t M_{t}(r_{j,a},r_F) K_{<t}.
    \]
    In each term in the sum, the spatial faults that were formerly at time $t$, and all their spatial Abelian byproducts, now occur at time $t+1$.
    We collect them all and the rest of the faults in $F$ into a fault configuration $F_{j,a}$, and absorb the scalar coefficients into $\alpha_{j,a}$.
    Finally, we have 
    \[
    K(F,r) = \sum_{j,a} \alpha_{j,a} K(F_{j,a},r_{j,a}).
    \]

    For the backward rewrite, it is convenient to first reorder the Abelian and non-Abelian factors in $E_{t+1}(F)$ to $E_{t+1}(F) =\eta_t' A_{t+1}N_{t+1}$, so that $N_{t+1}$ is adjacent to $M_t$.
    Since both operators are Pauli, this only introduces an overall sign absorbed into the scalar coefficient.
    The backward non-Abelian and Abelian rewrites then give the same expansion in the claim, with all spatial faults moved to time $t$.

    The projected outcome function changes twice during this calculation: the non-Abelian move replaces $r$ by $r_j$ and the subsequent Abelian moves replace $r_j$ by $r_{j,a}$. Both changes only occur at measurement events in round $t$.
    To keep the recorded outcome history, $r_F$, the same for all trajectories $(F_{j,a},r_{j,a})$, replace the temporal fault permutation at each changed measurement event $(Q,t)$ by a permutation $\pi_{j,a,Q}$ that satisfies
    \[
    \pi_{j,a,Q}\bigl(r_{j,a}(Q,t)\bigr) = r_F(Q,t).
    \]
    At every unchanged measurement event, we keep the same distribution of temporal faults as $F$.
    It follows that \[
    (r_{j,a})_{F_{j,a}} = r_F
    \]
    at every measurement event, for all trajectories $(F_{j,a},r_{j,a})$.
    Replacing the double index $j,a$ with a single index $j$ shows the claimed result.    
\end{proof}

\begin{remark}[Cleaned trajectories remain loop-like]
    Note, since the observed outcomes $r_{F_j}=r_F$ are preserved under cleaning, if we begin with a loop-like fault $F$, all resulting trajectories $F_j$ will be loop-like as well.
\end{remark}

\begin{corollary}[Fault support and weight after two-stage cleaning]
\label{cor:two-stage-fault-support}
For every nonzero term $\alpha_jK(F_j,r_j)$ in the expansion produced by forward two-stage cleaning across $M_t$,
\begin{enumerate}
    \item No spatial faults remain at time $t$, so $(F_j)_t=\emptyset$;
    \item the spatial faults at the destination time $t+1$ satisfy \[
    |(F_j)_{t+1}| \le |F_{t+1}| + c_{\rm{clean}}|F_t|
    \]
    for a constant $c_{\rm{clean}}=O(1)$;
    \item $F_j$ and $F$ have the same spatial faults at every time
    $t^*\notin\{t,t+1\}$; and
    \item $F$ and $F_j$ differ by at most $O(|F_t|)$ measurement events in round $t$.
\end{enumerate}
These claims hold for backward two-stage cleaning after interchanging $t$ and $t+1$ and replacing $F_t$ with $F_{t+1}$.
\end{corollary}
\begin{proof}
    Consider forward cleaning.
    By Lemma~\ref{cor:fault-support-non-abelian-rewrites}, the forward non-Abelian rewrite moves all faults in $F_t^N$ across $M_t$ and generates at most $c_N|F_t^N|$ elementary Abelian byproducts in every resulting trajectory, for a constant $c_N=O(1)$.
    The subsequent forward Abelian rewrites move these byproducts and every fault in $F_t^A$ across $M_t$. Hence, every spatial fault originally at time $t$, which is the union $F_t^N\cup F_t^A$, and any byproducts generated by moving these faults occur at time $t+1$ after two-stage cleaning. This proves that $(F_j)_t=\emptyset$.
    
    Now we consider all spatial faults at time $t+1$ after two-stage cleaning.
    The faults newly placed at time $t+1$ consist of the original $|F_t|$ faults and at most $c_N|F_t^N|$ byproducts. Together with the faults that were already at $F_{t+1}$ from the original fault configuration $F$,
    \begin{align}
        |(F_j)_{t+1}| &\le |F_{t+1}| + |F_t| + c_N|F_t^N|\notag\\
        &\le |F_{t+1}| + (1 + c_N)|F_t| .
    \end{align}
    Thus the second claim holds for $c_{\mathrm{clean}}:=(1+c_N)=O(1)$. Any overlapping faults in $F_{t+1}$ that happen to cancel out on their support only decrease the left hand side of this inequality.

    Both stages in the two-stage cleaning only rewrite the faults at times $t$ and $t+1$. They therefore leave the spatial faults at every time $t^*\notin \{t,t+1\}$ unchanged from $F$, which is the third claim.

    Finally, the forward non-Abelian rewrite changes temporal faults at $O(|F_t^N|)$ measurement events at round $t$, by Lemma~\ref{cor:fault-support-non-abelian-rewrites}.
    There are at most $|F_t^A|+c_N|F_t^N| = O(|F_t|)$ Abelian faults and byproducts that are moved in the second stage of the two-stage cleaning. Each forward elementary Abelian rewrite changes temporal faults at $O(1)$ round $t$ measurement events by Lemma~\ref{cor:fault-support-abelian-rewrite}; moving $O(|F_t|)$ elementary Abelian faults in $t$ changes $ O(|F_t|)$ round $t$ measurement events.
    Thus, $F_j$ differs from $F$ at only $O(|F_t|)$ temporal locations, all in round $t$.

    For backwards cleaning, the same arguments apply to all faults in $F_{t+1}$ by interchanging $t$ and $t+1$ in the first three claims. 
\end{proof}

\begin{remark}[Weight after iterating two-stage cleaning]
    Note that the weight of a cleaned fault does not grow exponentially with the number of rounds we have cleaned it through; the byproducts that may be produced by some non-Abelian elementary fault are always the same, since the stabilizers being measured each round are the same. Thus the spatial support of byproducts remains restricted to a bounded neighborhood around the fault as we move the fault through time, and thus the total byproduct weight remains bounded as well.
\end{remark}

For the next step of the cleaning, we need to consider what happens when we move faults through the gauging boundaries at $t=d+1$ and $t=2d+2$. We will first clean everything near the ungauging boundaries into the gauged code to produce a loop-like fault near $t=d+2$ and $t=2d+1$. Then, after proving the bounded weight of these remnant loop-like excitations, it will be most convenient to move them back into the base code to map the whole trajectory onto a completely deterministic set of detectors. 

When we are considering the boundary rewrites into the gauged code, there are three kinds of errors to consider: (i) Abelian matter faults, (ii) non-Abelian matter faults, and (iii) initialization/measurement errors on the edge ancilla. The following lemma describes the re-write rules across the gauging and ungauging boundaries:

\begin{lemma}[Boundary rewrites into the gauged code]
\label{lem:boundary-rewrites-into-gauged-code}
The boundary re-writes we need to consider are as follows.
\begin{itemize}
    \item Abelian matter faults: Abelian rewrites (Lemma \ref{lem:rewriting-abelian}) are unchanged across the boundary from the base code to the ungauged code as compared to the bulk of the gauged code.
    \begin{equation}
        M_{d+1}(r, r_F) A_{d+1} = A_{d+1} M_{d+1}(r^{(A_{d+1}, d+1)}, r_F)
    \end{equation}
    \begin{equation}
        A_{2d+2} M_{2d+1}(r, r_F) = \alpha(z) M_{2d+1}(r^{(f_A, 2d+1)}, r_F) A_{2d+2}
    \end{equation}
    Here, $\alpha(z) = A_{2d+2} U_{c(z)} A_{2d+2}^\dagger U_{c(z)}^\dagger$.
    \item non-Abelian matter faults: as with the bulk rewrites, non-Abelian faults generate Abelian byproduct operators when they pass from the base code to the gauged code.
    \begin{equation}
       M_{d+1}(r,r_F) N_{d+1} = N_{d+1} \left(\sum_j M_{d+1}(r_j, r_F) A_j\right),\qquad A_j\in \mathrm{Err}(\cB_{d+1})
    \end{equation}
    \begin{equation}
        N_{2d+2} M_{2d+1}(r, r_F) = M_{2d+1}(r^{(2d+1, N)}, r_F) B_{\rm bd.} N_{2d+2}
    \end{equation}
    Here, $B_{\rm bd.} = U_{c(z)}^\dagger N_{2d+2} U_{c(z)} N_{2d+2}^\dagger$.
    \item Initialization/readout errors on edge qubits: Ancilla initialization errors into non-Abelian faults when we push them onto the gauged code. 
    \begin{equation}
        M_{d+1}(r, r_F) \Xp_{d+1} = \Xp_{d+1} \left(\sum_j M_{d+1}(r_j, r_F) A_j\right),\qquad A_j\in \mathrm{Err}(\cB_{d+1}).
    \end{equation}
    Readout errors are already equivalent to non-Abelian errors on the gauged code, so we don't need to do anything to them.

    All boundary rewrites preserve the recorded history $r_F$.

\end{itemize}
    
\end{lemma}
\begin{proof}
    The two boundary times are $t = d+1$ and $t = 2d +1$. We want to push faults occurring just after we measure the base code stabilizers for the final time at $t=d+1$, onto the next slice where we have begun measuring the gauged code stabilizers at $t = d+2$. Likewise, we want to push any faults on the first time slice after returning to the base code, $t = 2d+2$, onto the time slice $t = 2d+1$ right after we have measured the gauged code stabilizers for the last time.

    Consider a fault configuration $F$, and let $A_{d+1}$ be the Abelian faults of $F$ acting on the base code at time $t=d+1$. By definition, each elementary Abelian Pauli fault must commute or anti-commute with the \emph{gauged} Abelian stabilizers. There is a stabilizer of the original code for every stabilizer $\mathbb{C} \in \cA \backslash \{\av\}$. The only difference between the decorated stabilizers $\mathbb{C}$ and the equivalent base code stabilizers is $\Zp$ support on the gauge qubits, thus an Abelian error supported on the matter qubits has the same commutation relations with $\mathbb{C}$ and its representative on the gauged code. Thus, the re-write rule across the boundary is identical to the usual Abelian re-write inside the gauged code.
    \begin{align}
        M_{d+1}(r, r_F) A_{d+1} = A_{d+1} M_{d+1}(r^{(A_{d+1}, d+1)}, r_F) \notag
    \end{align}
    Any projector values $r^{(A_{d+1}, d+1)}$ that are changed by passing $A_{d+1}$ through should have their assigned permutation flipped to ensure $r_F$ remains the same (as in Corollary \ref{cor:fault-support-abelian-rewrite}).
    
    Now consider all Abelian faults $A_{2d+2}$ at time $t = 2d+2$. The faults must be pushed backwards through the feedback circuit $U_{c(z)}$. Note that the action of $U_{c(z)}$ is identical to some product of the $\av$ operators when restricted to the matter qubits. As the $\av$ are Abelian stabilizers, $A_{2d+2}$ may only commute or anti-commute with them. Commuting $A_{2d+2}$ through $U_{c(z)}$ thus produces a global phase $\alpha(z)$. We retrieve the usual Abelian rewrite rule again, but up to this phase $\alpha(z)$.
    \begin{align}
        A_{2d+2} M_{2d+1}(r, r_F) = \alpha(z) M_{2d+1}(r^{(f_A, 2d+2)}, r_F) A_{2d+2} \notag 
    \end{align}
    Again, we also perform the necessary changes to the measurement faults to ensure $r_F$ is unchanged.

    Now consider all non-Abelian faults $N_{d+1}$ (supported on the matter degrees of freedom) on the base code at time $t=d+1$. We simply start measuring the gauged code stabilizers at $t=d+1$, so the re-write rules are exactly the same as the ones in Lemma \ref{lem:rewriting-non-abelian}.
    \[
    M_{d+1}(r,r_F) N_{d+1} = N_{d+1} \left(\sum_j M_{d+1}(r_j, r_F) A_j\right),\qquad A_j\in \mathrm{Err}(\cB_{d+1}).
    \]
    Here, $\cB_{d+1}$ should be understood as the set of all Abelian byproducts that arise when $N_{d+1}$ is pushed through the Abelian stabilizer measurements in time step $d+1$.
    
    For all non-Abelian faults $N_{2d+2}$ (supported on the matter degrees of freedom) on the base code, we must push them through the last round of base code syndrome measurements and the feedback circuit $U_{c(z)}$. Restricting any non-Abelian stabilizer of the gauged code onto the matter qubits yields either identity ($\bp$ stabilizers) or a stabilizer of the base code (other $\mathbb{N} \in \cN$). $N_{2d+2}$ may anti-commute with some of these ``non-Abelian'' base code stabilizers, as with the rewrites in the gauged code itself. 
    
    For the feedback circuit, we know that the action of $U_{c(z)}$ looks like a product of $\av$ stabilizers restricted to the matter qubits. As non-Abelian faults may produce Abelian byproducts as they pass through a set of Abelian stabilizers, we find: 
    \[
    N_{2d+2} M_{2d+1}(r, r_F) = M_{2d+1}(r^{(2d+1, N)}, r_F) B_{\rm bd.} N_{2d+2}
    \]
    where we define $B_{\rm bd.} = U_{c(z)}^\dagger N_{2d+2} U_{c(z)} N_{2d+2}^\dagger$, and $r^{(2d+1, N)}$ captures any outcomes that were flipped due to anti-commutation with $N_{2d+2}$. For both non-Abelian rewrites, use Corollary \ref{cor:fault-support-non-abelian-rewrites} to ensure $r_F$ remains unchanged.

    Finally, consider initialization and measurement errors on the gauge qubits. At $t = d+1$, we can have initialization errors that cause some ancilla to start in $\ket{1}$ instead of $\ket{0}$. $\Xp$ is a non-Abelian fault; it anti-commutes with some $\bp$ and creates Abelian byproducts when we push it through the gauged code. Let $\Xp_{d+1}$ denote the product of all initialization errors in the fault configuration.
    \[
    M_{d+1}(r, r_F) \Xp_{d+1} = \Xp_{d+1} \left(\sum_j M_{d+1}(r_j, r_F) A_j\right),\qquad A_j\in \mathrm{Err}(\cB_{d+1}).
    \]
    Here, $\cB_{d+1}$ should be understood as the set of all Abelian byproducts that arise when $\Xp_{d+1}$ is pushed through the Abelian stabilizer measurements in time step $d+1$. Again, apply Corollary \ref{cor:fault-support-non-abelian-rewrites} to ensure $r_F$ is unchanged. 
    
    We don't need to do anything to a readout error at $t = 2d+1$, as it already acts on the gauged code. In particular, the set of all readout errors $\Xp_{2d+1}$ creates some $\bp$ violations on the gauged code. If we did want to push $\Xp_{2d+1}$ further back in time, it would create byproducts. However, it is not necessary here.
    
\end{proof}

\begin{corollary}[Bounded support after boundary rewrites]
\label{cor:boundary-rewrite-fault-support}
    There is a constant $c_{\mathrm{bd}}=O(1)$ such that every valid trajectory $(F_j,r_j)$ produced by the boundary rewrites in Lemma~\ref{lem:boundary-rewrites-into-gauged-code} satisfy the following.
    \begin{enumerate}
        \item Moving all faults in $F_{d+1}$ (including initialization errors on edge qubits) forward into the gauged code leaves no spatial fault at time $d+1$ and gives 
            \begin{equation}
                |(F_j)_{d+2}| \le |F_{d+2}|+ c_{\rm{bd}} |F_{d+1}|.
            \end{equation}
        \item Moving all faults in $F_{2d+2}$ (including readout errors on edge qubits) backward into the gauged code leaves no spatial faults at time $2d+2$ and gives 
        \begin{equation}
            |(F_j)_{2d+1}|\le |F_{2d+1}| + c_{\rm{bd}} |F_{2d+2}|.
        \end{equation}
    \end{enumerate}
    Consequently, the union over all valid trajectories in the rewrite has  $O(|F_{d+1}|)$ faults at $d+2$, and $O(|F_{2d+2}|)$ faults at $2d+1$ in addition to the faults already present at those times.
\end{corollary}
\begin{proof}
    This can be seen by examining the faults after rewrites case by case in Lemma~\ref{lem:boundary-rewrites-into-gauged-code}. An Abelian fault is unchanged when it crosses a gauging or ungauging boundary. Each elementary non-Abelian fault overlaps only $O(1)$ checks, and therefore only produces $O(1)$ Abelian byproducts by Lemma~\ref{lem:byproducts}.  
    
    As $U_{c(z)}$ is transversal, $B_{\mathrm{bd.}}$ can only have support on vertex sets intersecting with $F_{2d+2}$; this must be proportional to $|F_{2d+2}|$. Thus, the number of Abelian byproducts created when moving through the ungauging boundary is $O(|F_{2d+2}|)$. 
    
    Initialization errors on gauge qubits follow the same bound as elementary non-Abelian faults.
    Thus, if we have $|F_{d+1}|$ ($|F_{2d+2}|$) faults before the forward (backward) boundary rewrite, then after the rewrite, there are at most $c_{\rm{bd}}|F_{d+1}|$ ($c_{\rm{bd}}|F_{2d+2}|$) additional faults at times $d+2$ ($2d+1$) for some $O(1)$ constant $c_{\rm bd}$.
\end{proof}

\begin{definition}[Iterative cleaning procedure]
\label{def:iter-cleaning}
Fix a trajectory $(F,r)$ and suppose that $F$ satisfies the cutting condition in Def.~\ref{def:cutting-condition} in intervals $I_\mathrm{pre}$, $I_\mathrm{g}$, and $I_\mathrm{post}$, such that 
\[
V_\cH (I_\alpha) = P_{\alpha,-} \sqcup C_\alpha \sqcup P_{\alpha,+},\qquad \alpha\in \{\mathrm{pre,\ g,\ post}\}.
\]

The \emph{iterative cleaning procedure} is a sequence of rewrites on $(F,r)$ that yields an expansion of $K(F,r)$,
\begin{align}
    K(F,r) = \sum_{j}\alpha_j K(F_j,r_j),\qquad (r_j)_{F_j}=r_F,
\end{align}
such that spatial faults in every trajectory, $K(F_j,r_j)$ lie solely in the support of four rounds, $t\in \{1,d+2,2d+1,3d\}$.

All steps in this iterative cleaning procedure are applied to every trajectory $(F_{j},r_j)$ in the sum.
Fix one such trajectory and let $F_{t,j}$ be its spatial faults at time $t$. Let $P_{\alpha,\pm }\cap F_{t,j}$ denote spatial faults whose location vertices lie in $P_{\alpha,\pm}$.

To clean $P_{\alpha,+}$ ($P_{\alpha,-}$) forwards (backwards) to a time $s$, repeatedly choose the smallest (largest) $t<s$ ($t>s$) with $P_{\alpha,+}\cap F_{t,j}\neq \emptyset$ ($P_{\alpha,-}\cap F_{t,j}\neq \emptyset$) and apply forwards (backwards) cleaning across $M_{t}$ ($M_{t-1}$). Repeat until there is no such $t$ where $P_{\alpha,+}\cap F_{t,j}\neq \emptyset$ ($P_{\alpha,-}\cap F_{t,j}\neq \emptyset$).

For $I_{\mathrm{g}}$, let the forwards or backwards cleaning be the two-stage cleaning procedure in Def.~\ref{def:two-stage}.
For $I_{\mathrm{pre}}, I_{\mathrm{post}}$, use $\Pi_{\lambda}f  = f\Pi_{\pm\lambda}$ to rewrite spatial faults $(t,f)$.

\end{definition}

Written out explicitly, the iterative cleaning procedure is:
\begin{enumerate}
    \item In $I_\mathrm{pre}$, clean $P_{\text{pre},-}$ backwards to time $t=1$.
    \item In $I_\mathrm{pre}$, clean $P_{\text{pre},+}$ forwards to time $t=d+1$.
    Using Lemma~\ref{lem:boundary-rewrites-into-gauged-code}, move the faults in $F_{d+1,j}$ into the gauged code to time $t=d+2$.
    \item In $I_\mathrm{g}$, clean $P_{\mathrm{g},-}$ backwards to time $d+2$.
    \item In $I_\mathrm{g}$, clean $P_{\mathrm{g},+}$ forwards to time $2d+1$.
    \item In $I_{\mathrm{post}}$, clean $P_{\mathrm{post},-}$ backwards to time $2d+2$. Then, use Lemma~\ref{lem:boundary-rewrites-into-gauged-code} to move faults in $F_{2d+2,j}$ into the gauged code to time $t=2d+1$.
    \item In $I_{\mathrm{post}}$, clean $P_{\mathrm{post},+}$ forwards to time $t=3d$.
\end{enumerate}

Now we implement the iterative cleaning. When we apply the iterative cleaning procedure above to a loop-like fault that satisfies the cutting condition, we are able to show that the resulting rewritten Kraus operator has support solely on the rounds $t \in \{1, d+2, 2d+2, 3d\}$. We use the bound on the initial weight of the fault $F$, the bounded growth in the weight due to the rewrites, and the space distance of both the gauged and base codes to show any trajectory in the cleaned Kraus operator must be logically trivial.

\begin{lemma}[Cleaned trajectory after iterative cleaning]
\label{lem:cleaned-trajectory}
    Let $(F,r)$ be a valid trajectory of a loop-like fault configuration with 
    \[
    |F| \leq \frac{d}{c_{\rm cut}}
    \]
    such that it possesses clean cuts in $I_\mathrm{pre}$, $I_\mathrm{g}$, and $I_\mathrm{post}$.
    Every valid trajectory $(F_j,r_j)$ produced by iterative cleaning has no temporal faults, and all of its spatial faults occur at 
    \[
    T_\mathrm{res.} = \{1,d+2,2d+1,3d\}.
    \]
    We refer to the spatial faults of a trajectory $(F_j,r_j)$ after cleaning as \emph{residual faults}.
    
    The only dropped detectors in each trajectory $(F_j, r_j)$ are localized to the time steps $t = d+1$ and $2d$, in the neighborhood of the corresponding residual spatial faults at $T_{res.}$.

    For $t\in T_\mathrm{res.}$, define the total residual support, which is the set of all residual spatial locations after cleaning,
    \[
    \mathsf{R}_{t}:=\{(t,\ell)\in \mathsf{L}_s: \exists j,\ \exists (t,f)\in F_j, \text{ s.t. } \supp{f}\subseteq \ell\}
    \]
    Then, 
    \begin{align}
    \label{eq:iter-bounded-fault}
        \sum_{t\in T_\mathrm{res.}}|\mathsf{R}_t|\le c\Wt(F)< \frac{c}{c_{\rm cut}}d.
    \end{align}
    for an $O(1)$ factor $c$.

\end{lemma}
\begin{proof}

    We focus on cleaning a fault $F$ supported purely on either side of the cuts $C_{\rm g}$ and $C_{\rm pre}$ to the boundary at $d+2$, or cleaning faults from between $C_{\rm g}$ and $C_{\rm post}$ to the boundary at $2d+1$. The cleaning to $t=1$ and $t=3d$ occur purely within the base code, and thus standard Pauli cleaning arguments apply. 

    First, consider any trajectory in the vicinity of the clean cut $C_{\rm g} \in I_{\rm g}$ inside the gauged code. As there are no faults or dropped detectors supported on the cut by Def. \ref{def:cutting-condition}, we must have all non-Abelian stabilizers projected to their $+1$ eigenspace at the cut. Otherwise, we would have some dropped detectors created by the non-Abelian excitations. Furthermore, since there can be no measurement faults on the cut, the projected outcomes and recorded outcomes agree:
    \[
    r_F(Q, t) = r(Q, t) \quad (Q, t) \in C_{\rm g}
    \]

    Now consider our trajectory of interest, $(F, r)$. Let the time step separating the cut $C_{\rm g}$ from the first fault or dropped detector in $F$ be $t^*$. Consider cleaning downwards away from $t^*$ towards $t = d+1$ (the logic is the same for the upwards cleaning to $t = 2d+2$ from the upper edge of the cut). 

    By the above arguments, the form of the projectors at $t^*$ must be:
    \[
    M(t^*) = \prod_{\mathbb{N} \in \cN} \Pi_{r(\mathbb{N}, t^*) = +1} \prod_{\mathbb{C} \in \cA} \Pi_{r(\mathbb{C}, t^*)} := M_{\rm clean}(t^*)
    \]
    Thus the Kraus operator for a valid trajectory $(F, r)$ in the vicinity of the cut looks like 
    \[
    K(F, r) = \cdots M_{\rm clean}(t^*) M(t^* - 1) F_{t^*-1} M(t^*-2) \cdots 
    \]
    \[
    = \cdots  M_{\rm clean}(t^*) M_{\rm clean}(t^* - 1) F_{t^*-1} M(t^*-2) \cdots 
    \]
    As we clean $F$ downwards by one step to $t^*-2$ to generate new trajectories $F_j$, by Corollary \ref{cor:boundary-rewrite-fault-support}, no spatial faults remain at time $t^*-1$, so we have:
    \[
    K(F, r) = \cdots  M_{\rm clean}(t^*) M_{\rm clean}(t^* - 1) \bigl ( \sum_j M_{t^* - 2}(r_j, r_F) A_j N_{t^*-2}  \bigr )
    \]
    \[
    = \cdots  M_{\rm clean}(t^*) M_{\rm clean}(t^* - 1) M_{\rm clean}(t^* - 2) \bigl ( \sum_j A_j N_{t^*-2} \bigr ) \cdots 
    \]
    Here, $A_j$ combines the initial Abelian support $A_{t^*-1}$ we cleaned backwards, the byproducts produced by moving $N_{t^*-1}$, and any Abelian support already at $t^*-2$. Similarly, $N_{t^*-2}$ now contains both the support originally at $t^*-2$ and the moved non-Abelian faults. We repeat the two-stage cleaning until we reach $t=d+2$. At this point, the Kraus operator looks like
    \[
    K(F, r) = \cdots M_{\rm clean}(d+3) M_{\rm clean}(d+2) \bigl ( \sum_j A_j N_{d+2} \bigr ) M_{d+1}(r, r_F) \cdots 
    \]

    At the same time as we have been cleaning inside the gauged code downwards from the clean cut $C_{\rm g}$, we can clean faults on the interval $I_{\rm pre}$ upwards from the cut $C_{\rm pre}$. Using the same arguments, there is a set of clean measurements on the base code, $M'_{\rm clean}$, that gets pushed up as we clean away from the cut, which eventually meet with the clean projectors on the gauged code:
    \[
    K(F, r) = \cdots M_{\rm clean}(d+2) \bigl ( \sum_j A_j N_{d+2} \bigr ) M_{d+1}(r, r_F) N_{d+1} A_{d+1} \bigl ( \id \otimes \ket{0}_{\edgeE}  \bigr ) M'_{\rm clean}(d) \cdots 
    \]
    \[
    = \cdots M_{\rm clean}(d+2) \bigl ( \sum_j A'_j N'_{d+2} M_{d+1}(r_j, r_F) \bigr ) \bigl ( \id \otimes \ket{0}_{\edgeE}  \bigr ) M'_{\rm clean}(d) \cdots 
    \]
    where $N', A'$ represent the combined support of the non-Abelian and Abelian faults once we have moved over $N_{d+1}$, its byproducts, and $A_{d+1}$ from the base code. The projector $M_{d+1}(r_j, r_F)$ inherits its $r_j$ (and thus $r_F$) from the projector $M'_{\rm clean}$ as each base code stabilizer corresponds to some gauged code stabilizer. The $\bp$ terms must be $+1$ in $M_{d+1}$ as the clean code projector at $d$ includes the initial edge state $\ket{0}_\edgeE$. Finally, the $\av$ terms are randomized to some values $\bm a_v$.
    \[
    K(F, r) = \cdots M_{\rm clean}(d+2) \bigl ( \sum_j A'_j N'_{d+2} \bigr ) M'_{\rm clean}(d+1) \bigl ( \id \otimes \ket{0}_{\edgeE}  \bigr ) M'_{\rm clean}(d) \cdots = \sum_j K_j(F_j, r_j) 
    \]
    The spatial faults now occur solely at $t = d+2$ and we have converted all measurement projectors in the vicinity of the gauging boundary $t=d+1$ to clean projectors where the projected outcomes equal the recorded outcomes. Thus, we have $r_j = r_F$ for all times $t$ and trajectories in the expansion, so there are no measurement faults anymore. 

    Additionally, since the clean projectors must have $\mathbb{N} = +1$ for all non-Abelian stabilizers, we cannot have any dropped detectors due to non-commutativity of Abelian stabilizers. The only remaining non-deterministic detectors are ones at $t=d+1$ from the remaining non-Abelian fault support.

    If we apply these arguments to the other boundary time slices, we have shown the first set of statements: (1) we can clean the fault $F$ such that all support is spatial and lies in $t \in \{1, d+2, 2d+1, 3d\}$, and (2) the only dropped detectors are the result of non-Abelian faults not commuting with the detectors at the boundary, and so are localized to the boundary time slices and the spatial locations of the residual faults. 

    Now, we argue that the weight of the residual faults must be bounded. This is a simple application of Corollaries \ref{cor:two-stage-fault-support} and \ref{cor:boundary-rewrite-fault-support}. Corollary \ref{cor:two-stage-fault-support} tells us the errors cleaned through the gauge code remain related to their original weight by a constant factor, while Corollary \ref{cor:boundary-rewrite-fault-support} tells us that any faults on the base code pushed onto the gauged code also only increase in weight by an $O(1)$ factor. 

    As we have $|F| \leq d/c_{\rm cut}$, 
    \begin{equation}
        \sum_{t \in T_{res}} |\mathrm{R}_t| \leq c |F| \leq \frac{c}{c_{\rm cut}} d
    \end{equation}

    \end{proof}

Our final step is to get rid of all the remaining dropped detectors localized to the residual spatial faults in the gauged code by pushing these loop-like faults \emph{back} into the base code in either direction. This will allow us to decompose the Kraus operator for the trajectory $(F, r)$ into a set of fully deterministic projectors.

\begin{lemma}[Ungauging cleaning onto the base code]
\label{lem:ungauging-cleaning}
    Let $(F, r)$ be a valid trajectory of a loop-like fault configuration with 
    \[
    |F| \leq \frac{d}{c_{\rm cut}}
    \]
    Take the set of trajectories $(F_j, r_j)$ produced by iterative cleaning, and push the residual faults on $t = d+2$ and $d = 2d+1$ to $t=d$ and $t=2d+2$, respectively. 

    Then the new set of resultant trajectories have no dropped detectors, and the residual errors on $T'_{res} = \{1, d, 2d+2, 3d\}$ are still bounded in weight:
    \[
    \sum_{t \in T'_{res}} |\mathrm{R}_t| \leq \frac{c'}{c_{\rm cut}} d
    \]
\end{lemma}
\begin{proof}
    
    First, consider pushing the residual fault at $t= d+2$ for trajectory $(F_j, r_j)$ two time steps backwards onto the base code at $t = d$. We have:
    \[
    K(F_j, r_j) =\cdots  M_{\rm clean}(d+2) A_j N_{d+2} M'_{\rm clean}(d+1) \bigl ( \id \otimes \ket{0}_{\edgeE}  \bigr ) M'_{\rm clean}(d) \cdots 
    \]
    Recall $M_{\rm clean}$ and $M'_{\rm clean}$ represent the projectors inherited from the clean cuts in the gauged code and the base code pre-gauging. The projector $M'_{\rm clean}$ is still the projector onto the gauged code, so we apply another round of two-stage cleaning, to arrive at:
    \[
    K(F_j, r_j) = \cdots  M_{\rm clean}(d+2) \sum_{k} M_{d+1}(r_k, r_{F}) A_k N'_{d+1} \bigl ( \id \otimes \ket{0}_{\edgeE}  \bigr ) M'_{\rm clean}(d) \cdots 
    \]
    \[
    = \cdots  M_{\rm clean}(d+2) M_{\rm clean}(d+1) \sum_{k} A_k N'_{d+1} \bigl ( \id \otimes \ket{0}_{\edgeE}  \bigr ) M'_{\rm clean}(d) \cdots 
    \]
    Note that we have removed all dropped detectors: the non-Abelian faults now act on the base code, and so return to acting like regular Pauli faults.
    
    We want to push the faults $A_k N'_{d+1}$ onto the base code. Separate $A_k = A^m_k A^g_k$ and $N'_{d+1} = N^m N^g$ into matter (m) and gauge (g) support. We know that $A^g_k$ is some configuration of $\Zp$ on the edge ancilla, while $N^g$ is some configuration of $\Xp$ on the edge ancilla. Furthermore, since the fault $A_k N'_{d+1}$ must be loop-like, we know that the support of $N^g$ must form a coboundary on the gauging graph, which we will label $z_0 = \delta_1 \bm{s}$. 

    Thus, we can push the gauge support through onto $\ket{0}_{\edgeE}$:
    \[
    A_k N'_{d+1} \bigl ( \id \otimes \ket{0}_{\edgeE}  \bigr ) = A^m N^m \bigl ( \id \otimes \ket{\delta_1 s}\bigr )
    \]
    Thus, we have
    \[
    K(F_k, r_k) = \cdots M_{\rm clean}(d+1) \bigl ( \id \otimes \ket{\delta_1 s}_{\edgeE}  \bigr ) A^m N^m M_{\rm clean}(d) \cdots 
    \]
    As a final step: 
    \[
    K(F_k, r_k) = \cdots M_{\rm clean}(d+1) \bigl (\prod_{v \in \bm{s}} \av \bigr ) \T^{\bm s}  \ket{0}_{\edgeE}  \bigr ) A^m N^m M_{\rm clean}(d) \cdots  
    \]
    \[
    K(F_k, r_k) = \cdots \alpha M_{\rm clean}(d+1) \ket{0}_{\edgeE}  \T^{\bm s} A^m N^m M_{\rm clean}(d) \cdots 
    \]
    \[
    = \cdots \alpha M_{\rm clean}(d+1) \ket{0}_{\edgeE} F_{d+1} M_{\rm clean}(d) \cdots, \quad F_{d+1} = \T^{\bm s} A^m N^m
    \]
    Using the expansion properties of the gauging graph, we know that since $|\delta_1 s| = |N^g| \leq (cd)/c_{\rm cut}$, the number of vertices in $\bm{s}$ or $\bm{s} + \id$ (take whichever is minimal) is bounded:
    \[
    |\bm{s}| \leq \frac{|\delta_1 s|}{h} \leq \frac{c}{c_{\rm cut} h} d \implies |\T^{\bm s}| \leq  \frac{c \nu}{c_{\rm cut} h} d
    \]
    Recall $\nu$ is the maximum size of a vertex set. So the final residual fault pushed onto the base code at $t = d+1$ is also bounded in weight:
    \[
    |\T^{\bm s} A^m N^m| \leq \bigl ( 1 + \frac{\nu}{h} \bigr )\frac{c}{c_{\rm cut}} d
    \]
    
    Now, consider pushing the residual fault at $t = 2d+1$ for $(F_j, r_j)$ one step forwards onto the base code at $t = 2d+2$. This time, however, we need not perform a last non-Abelian re-write, since we already cleaned past the last non-Abelian projector. Initially, we have:
    \[
    K(F_j, r_j) = \cdots M''_{\rm clean}\bigl ( U_{c(z)} \otimes \bra{z}_\edgeE \bigr ) N_{2d+1} A_j M_{\rm clean}(2d) \cdots 
    \]
    Here $M''_{\rm clean}$ is the projector inherited from the clean cut in the post-gauging base code. Again, separate $N_{2d+1}$ and $A_j$ into their support on the matter and gauge qubits. Since this residual spatial fault is still a loop-like fault, the support of $N^g$ is again a coboundary, denoted $z_1 = \delta_1 \bm{w}$. $A^g$ is a set of $\Zp$ operators, and is thus absorbed into $\bra{z}_\edgeE$ with some global phase. Since there are no measurement faults at the ungauging boundary from our iterative cleaning, we also know $z$ is a coboundary for some other set of vertices $\bm{r}$. This is exactly the local Clifford correction. Thus we have:
    \[
    K(F_j, r_j) = \cdots M''_{\rm clean}\bigl ( U_{c(z)} \otimes \bra{z + \delta_1 \bm{w}}_\edgeE \bigr ) N^m A^m M_{\rm clean}(2d) \cdots
    \]
    \[
    = \cdots M''_{\rm clean} \bigl ( U_{c(z)} N^m A^m U_{c(z)}^\dagger \bigr ) \bigl ( \T^{\bm w} \otimes \bra{0}_\edgeE \bigr )  M_{\rm clean}(2d) \cdots 
    \]
    \[
    = \cdots M''_{\rm clean} F'_{2d+2} \bra{0}_{\edgeE} M_{\rm clean}(2d) \cdots , \quad F'_{2d+2} =  U_{c(z)} N^m A^m U_{c(z)}^\dagger \T^{\bm w}
    \]
    Again, we can argue that $\T^{\bm w}$ is bounded. Since $U_{c(z)}$ is also just some application of the local symmetry operators, conjugating $N^m A^m$ by it will not increase their weight by more than a factor of $\nu$. So in all, we have:
    \[
    |F'_{2d+2}| \leq \bigl( 1 + \frac{1}{h} \bigr) \frac{\nu c}{c_{\rm cut}} d
    \]

    Thus, the final Kraus operator for the fully deterministic trajectories $(F_k, r_k)$ are:
    \begin{align}
        K(F_k, r_k) &= M'''_{\rm clean}(3d) F_{3d} M''_{\rm clean}(3d-1) \cdots M''_{\rm clean}(2d+3) F'_{2d+2} \notag \\
        &\cdot \bra{0}_{\edgeE} M_{\rm clean}(2d) \cdots M_{\rm clean}(d+1) \ket{0}_{\edgeE} F'_{d+1} M'_{\rm clean}(d) \cdots M'_{\rm clean}(1) F_1 \notag
    \end{align}

    With 
    \[
    |F_{3d}| + |F'_{2d+3}| + |F'_{d+1}| + |F_1| \leq \bigl ( 1 + \frac{1}{h} \bigr ) \frac{4 \nu c}{c_{\rm cut}} d \leq \frac{c'}{c_{\rm cut}} d, \quad c' = 4 \nu c ( 1 + \frac{1}{h} \bigr ) 
    \]
    
\end{proof}

\begin{lemma}[Logical success condition~\ref{cond:logical-success-1}]
\label{lem:logical-success-condition-1}
    Let $(F,r)$ be a valid trajectory, and let $(F_j,r_j)$ be any valid trajectory obtained by iterative cleaning followed by ungauging cleaning.
    Let $r_j'$ be the candidate fault-free outcome history obtained by removing residual spatial faults from $(F_j,r_j)$ while leaving every measurement outcome unchanged. Then,
    $(\emptyset,r')$ with
    \begin{equation}
        m_\T(r'_j)=m_\T\bigl((r_j)_{F_j}\bigr) = m_\T (r_F).
        \label{eq:cleaning-preserves-recorded-outcome}
    \end{equation}
\end{lemma}
\begin{proof}
    Every rewrite used in the iterative cleaning and ungauging cleaning preserves the complete recorded outcome history. Therefore,
    \[
    (r_j)_{F_j}(Q,t) = r_F(Q,t)\qquad \text{for every measurement event }(Q,t)
    \]
    After ungauging cleaning, all residual spatial faults occur at times $\{1,d,2d+2,3d\}$ by construction.
    In particular, no residual fault occurs at an $\av$ measurement event in the gauged region. 
    The fault-free candidate's recorded outcome history is defined by removing all such residual spatial faults without changing the measurement labels.
    Consequently,
    \[
    r_j'(\av,2d) = (r_j)_{F_j} (\av, 2d) = r_F (\av, 2d)
    \]
    for every $v\in \vertexV$.
    Taking the product over all vertices gives
    \[
    m_\T(r'_j) = \prod_{v\in \vertexV}r_j'(\av, 2d) = \prod_{v\in \vertexV}(r_j)_{F_j}(\av,2d) = \prod_{v\in \vertexV}(r)_{F}(\av,2d) = m_\T (r_F).
    \]
    This proves Eq.~\eqref{eq:cleaning-preserves-recorded-outcome}.
\end{proof}

\begin{lemma}[Applying the space distance proof]
\label{lem:applying-space-distance}
Let $\Pi_{\cC}$ be the projector onto the codespace of $\cC$, whose distance is $d$.
Let $R$ be a set of spatial locations in one round of $\cC$.
If $|R|<d/\nu$ then, for every $E\in \mathrm{Err}(R)$
\[
\Pi_\cC E \Pi_\cC = c_E \Pi_{\cC},
\]
for some scalar $c_E\in \mathbb{C}$.
\end{lemma}

\begin{proof}
    Each spatial location contains at most, $\nu$ physical qubits. 
    Therefore, every Pauli operator generater by the locations in $R$ has weight at most $\nu |R|<d$.

    Let $P$ be such a Pauli operator. If $P$ does not preserve the codespace, then it maps the codespace into an orthogonal syndrome sector, and hence, $\Pi_{\cC }P \Pi_{\cC}=0$.

    Instead if $P$ preserves the codespace, then $P$ belongs to the normalizer of the stabilizer group. Since $\Wt(P)<d$, then it cannot be a nontrivial logical operator. 
    Therefore its restriction to the codespace is proportional to the identity.
    \[
    \Pi_{\cC}P\Pi_{\cC} = c_P \Pi_{\cC}.
    \]
    
    Every operator in $\mathrm{Err}(R)$ is a linear combination of such Pauli operators, and thus by linearity, 
    \[
    \Pi_{\cC}E\Pi_{\cC}
      =\sum_P e_P\Pi_{\cC}P\Pi_{\cC}
      =\left(\sum_P e_Pc_P\right)\Pi_{\cC}
      =c_E\Pi_{\cC}.
    \]
\end{proof}

\begin{lemma}[Logical success condition~\ref{cond:logical-success-2}]
\label{lem:logical-success-condition-2}
    Let $(F,r)$ be a valid trajectory of a loop-like fault configuration
    with clean cuts in $I_{\mathrm{pre}}, I_\mathrm{g},$ and $ I_\mathrm{post}$. 
    Perform iterative cleaning from Def.~\ref{def:iter-cleaning}, followed by the ungauging cleaning of Lemma~\ref{lem:ungauging-cleaning}, obtaining the expansion
    \begin{align}
        K(F,r)\Pi_\cC =  \sum_{j} \alpha_j K(F_j,r_j) \Pi_\cC.
        \label{cleaned-trajectory}
    \end{align}
    Let $\{\mathsf{R}_t:t\in T_{\mathrm{res.}}\}$ be the common residual support produced by this cleaning, and suppose that 
    \begin{equation}
        \sum_{t\in T_\mathrm{res.}} |\mathsf{R}_t|< \frac{d}{\nu}.
        \label{eq:residual-support-assumption}
    \end{equation}
    
    Then, for each valid $(F_j,r_j)$ trajectory, there exists a valid, fault-free trajectory $(\emptyset , r'_j)$ and a scalar $c_j \neq 0$ such that
    \begin{equation}
    \label{eq:logical-success-j-proportional}
        K(F_j,r_j)\Pi_\cC = c_j K(\emptyset , r'_j)\Pi_\cC.
    \end{equation}

    Furthermore, this implies that
    \begin{align}
    \label{eq:logical-success-proportional}
        K(F,r) \Pi_{\cC} = \alpha K(\emptyset, r') \Pi_{\cC},\qquad \alpha\neq 0.
    \end{align}
\end{lemma}

\begin{proof}

    By construction, the only residual spatial faults after iterative cleaning followed by ungauging cleaning are in $T_{\mathrm{res.}} = \{1, d, 2d+2, 3d\}$. 
    For each $t\in T_{\mathrm{res.}}$ let $\mathsf{R}_t$ denote the common set of spatial locations supporting the residual spatial faults at time $t$, over all terms in the expansion over $j$.
    Thus, the product of the residual spatial faults, $R_{t,j}\in \mathrm{Err}(\mathsf{R}_t)$, for every $t\in T_{\mathrm{res.}}$

    By Eq.~\eqref{eq:residual-support-assumption}, 
    since every $|\mathsf{R}_t|$ is nonnegative, for each $t\in T_{\mathrm{res.}}$, 
    \[
    |\mathsf{R}_t|<d/\nu.
    \]
    Thus, Lemma~\ref{lem:applying-space-distance} applies to every residual operator $R_{t,j}\in \mathrm{Err}(\mathsf{R}_t)$.

    Fix a valid, cleaned trajectory $(F_j,r_j)$. 
    Let $M_{t_1:t_2} := M_{t_2} \dots M_{t_1}$.
    Thus, 
    \[
    K(F_j,r_j)\Pi_{\cC} =  M_{3d}
    R_{3d,j}
    M_{2d+2:3d-1}
    R_{2d+2,j}
    M_{d:2d+1}
    R_{d,j}
    M_{1:d-1}
    R_{1,j}\Pi_{\cC}.
    \]
    Because all rewrites in iterative cleaning and ungauging cleaning preserve the outcomes history $r_F$, if $(F,r)$ is a loop-like fault, then each $(F_j,r_j)$ is loop-like as well.
    In particular, by Lemma~\ref{lem:cleaned-trajectory} Lemma~\ref{lem:ungauging-cleaning}, each trajectory has no temporal faults and every detector is not dropped. By virtue of $F_j$ being a loop-like fault, every detector is satisfied. 

    We first consider the measurements before gauging. Since the input lies in the codespace, the initial boundary detector for each stabilizer $S_i$ gives $r_j(S_i,1) = +1$.
    The bulk detectors between rounds $t$ and $t+1$ are 
    \[
    D_{S_i}^t (r_j) = r_j(S_i,t)^* r_{j}(S_i,t+1).
    \]
    Since this detector is satisfied, it must be that $r_j(S_i,t+1) = r_j(S_i,t)$. Induction therefore gives
    \begin{align}
        r_j(S_i,t) = +1,\qquad 1\le t\le d,\qquad \forall S_i.
        \label{eq:all-plus-one-stabilizers}
    \end{align}
    Consequently, the product of all measurement projectors in each of these rounds is the codespace projector $\Pi_\cC$. 
    Since $\Pi_{\cC}^2 = \Pi_\cC$, repeated rounds become one codespace projector, 
    \[
    M_{1:d-1}(r_j) \Pi_\cC = \Pi_{\cC}.
    \]

    Let $V\ket{\psi} :=\ket{\psi}\otimes\ket{0}_{\edgeE}$. 
    Recall that 
    \[ M_d = \left(\prod_i \Pi_{r_j (S_i, d)}\otimes\id \right)\ket{0}_\edgeE.
    \]    
    By Eq.~\eqref{eq:all-plus-one-stabilizers}, $M_d(r_j) = V \Pi_\cC$. 
    It follows that 
    \[
    M_d R_{d,j} M_{1:d-1} R_{1,j} \Pi_\cC = V \Pi_{cC} R_{d,j} \Pi_{\cC} R_{1,j} \Pi_{\cC}.
    \]
    Now we consider the rounds after ungauging. 
    First we derive that
    \[
    r_j(S_i,t),\qquad 2d+1 \le t\le 3d.
    \]
    For every base code stabilizer $S_i$, let $\mathbb{S}_i$ denote its corresponding gauged code decorated stabilizer. Since $r_j(S_i,d) = +1$, satisfaction of the gauging boundary detector $D_{S_i}^d:=r_j(S_i,d)^* r_j(\mathbb{S}_i,d+1)$ implies that
    \[
    r_{j}(\mathbb{S}_i,d+1) = 1.
    \] 
    Then the satisfied bulk detectors $D_{\mathbb{S_i}}^t:=r_j(\mathbb{S}_i,t)^* r_j(\mathbb{S}_i,t+1)$ for $d+1\le t\le 2d$ then give 
    \[
        r_j(\mathbb{S}_i,t) = 1, d+1\le t\le 2d.
    \]
    Finally, the satisfaction of the ungauging boundary detector $D_{\mathbb{S}_i}^{2d} :=r_{j}(\mathbb{S}_i, 2d)^* r_j(S_i,2d+1)$ implies that 
    \[
    r_j(S_i, 2d+1)=1,\qquad \forall S_i,
    \]
    and by satisfies bulk detectors after gauging,
    \[
    r_j(S_i,t),\qquad 2d+1 \le t\le 3d.
    \]
    
    The fault-free ungauging operation associated with the edge measurement string $\bm z_j$ is 
    \[
    W_{\bm z_j}:= U_{c(\bm z_j)}\bra {\bm{z}_j}.
    \]
    The measurement round immediately after ungauging is thus
    \[
    M_{2d+1} :=  \left(\prod_i \Pi_{r_j (S_i, 2d+1)}\otimes\id \right) U_{c(\bm z_j)}\bra {\bm{z}_j}=  \Pi_{\cC}W_{\bm{z}_j}.
    \]
    Thus, 
    \[
    M_{3d}R_{3d,j} M_{2d+2:3d-1} R_{2d+2,j} M_{2d+1} 
    = \Pi_{\cC} R_{3d,j} \Pi_{\cC} R_{2d+2,j} \Pi_{\cC} W_{\bm{z}_j}
    \]

    Since there are no residual spatial or temporal faults in the rounds $d+1,\dots, 2d$, the corresponding outcomes constitute a fault-free gauged code history.
    More explicitly, the gauging boundary detectors fix the outcomes of the checks in $\cN$ and $\cA \setminus \{\av\}_v$ to their fault-free values. The first $\{\av\}_v$ measurement determines some assignment $\bm a_{j}= (a_{v,j})_v$, and satisfaction of all bulk $\av$ detectors implies that this assignment is unchanged in every subsequent gauged round. 
    Thus, $M_{d+1:2d}V\Pi_{\cC}$ is the gauged portion of a fault-free trajectory, with $\av$-outcome assignment $\bm a_j$. The outcomes $\bm z_j$ are also compatible with this fault-free gauged trajectory.

    Therefore we have,
    \[
    K(F_j,r_j) \Pi_\cC = \left( \Pi_{\cC} R_{3d,j} \Pi_{\cC}\right) \left( \Pi_{\cC} R_{2d+2,j} \Pi_{\cC}\right) \Pi_{\cC} W_{\bm{z}_j}M_{d+1,2d}V \Pi_{\cC} \left( \Pi_{\cC} R_{d,j} \Pi_{\cC}\right) \left( \Pi_{\cC} R_{1,j} \Pi_{\cC}\right) .
    \]
    Now apply Lemma~\ref{lem:applying-space-distance}. Since $|\mathsf{R}_t|<d/\nu$ for every $t\in T_{\mathrm{res}}$, there is a scalar $c_{t,j}$ such that 
    \[
    \Pi_{\cC}R_{t,j} \Pi_{\cC} = c_{t,j} \Pi_{\cC}.
    \]
    So,
    \begin{align}
    \label{eq:final-cleaned-trajectory}
        K(F_j,r_j) \Pi_\cC =c_{3d} c_{2d+2} \Pi_{\cC} W_{\bm{z}_j}M_{d+1,2d}V \Pi_{\cC} c_{d} c_{1}
    \end{align}
    Define $r_j'$ to be the outcome history obtained from $r_j$ by removing the four residual spatial faults and leaving all measurement outcomes unchanged. 
    Consequently, 
    \begin{align}
    \label{eq:final-fault-free}
        K(\emptyset, r_j') \Pi_{\cC} = \Pi_{\cC} W_{\bm{z}_j}M_{d+1,2d}V \Pi_{\cC}.
    \end{align}
    Equation~\eqref{eq:final-cleaned-trajectory} therefore becomes
    \begin{align}
    \label{eq:logical-success}
        K(F_j,r_j) \Pi_\cC = c_j K(\emptyset, r_j') \Pi_{\cC}, \qquad c_j :=c_{3d} c_{2d+2} c_{d} c_{1}.
    \end{align}
    Since $(F_j,r_j)$ is a valid trajectory, $ K(F_j,r_j) \Pi_\cC\neq 0$.
    Then Eq.~\eqref{eq:logical-success} implies $c_j \neq 0$ and $K(\emptyset, r_j') \Pi_{\cC}\neq 0$. Thus $(\emptyset,r_j')$ is a valid fault-free trajectory, and the claimed equation~\eqref{eq:logical-success-j-proportional} holds.

    Now we show Eq.~\eqref{eq:logical-success-proportional}. Because $(F,r)$ is valid, there must be at least one non-zero term in the expansion in Eq.~\eqref{cleaned-trajectory} since $K(F,r)\Pi_{\cC}\neq 0$. Fix one such term, labeled by $j_0$, and let $r':=r'_{j_0}$.
    Using Lemma~\ref{eq:cleaning-preserves-recorded-outcome}, every surviving term satisfies 
    \[
    m_{\T}(r'_j) =  m_{\T}(r_F).
    \]
    In particular, 
    \[
    m_{\T}(r'_j) = m_{\T}(r')
    \]
    for every surviving $j$.

    Since $(\emptyset, r'_j)$ and $(\emptyset, r')$ are both valid fault-free trajectories with the same reported $\T$ outcome, Lemma~\ref{lem:fault-free} says that there is a scalar $\beta_j\neq 0$ such that 
    \[
    K(\emptyset,r'_j)\Pi_{\cC} = \beta_{j} K(\emptyset,r') \Pi_\cC.
    \]

    In summary, if we omit any zero terms from the cleaned expansion in Eq.~\eqref{cleaned-trajectory} and apply Eq.~\eqref{eq:logical-success-j-proportional}, we have
    \begin{align}
        K(F,r)\Pi_{\cC} &=
        \sum_j \alpha_j K(F_j,r_j)\Pi_{\cC}\notag\\
        &=\sum_j \alpha_j c_j  K(\emptyset,r_j')\Pi_{\cC}\notag\\
        &=\sum_j \alpha_j c_j \beta_j  K(\emptyset,r')\Pi_{\cC}\notag\\
        &=\left(\sum_{j}\alpha_j c_j \beta_j  \right) K(\emptyset,r')\Pi_{\cC}.
    \end{align}
    If we define $\alpha:=\sum_{j}\alpha_j c_j \beta_j $
    then the scalar $\alpha$ must be nonzero, else it would imply that  $K(F,r)\Pi_{\cC}=0$, which is a contradiction because $(F,r)$ is a valid trajectory.
    Thus,
    \[
        K(F,r)\Pi_{\cC}=\alpha K(\emptyset,r')\Pi_{\cC},
        \qquad \alpha\neq0,
    \]
    proving Eq.~\eqref{eq:logical-success-proportional}.

\end{proof}

\begin{theorem}[Fault distance of the gauging procedure (formal)] \label{thm:spacetime-distance}
    Let $\cC$ be a qubit LDPC stabilizer code of distance $d$ with a transversal (Def.~\ref{def:transversal-gate}) Clifford symmetry $\T$ satisfying $\T^2=\id$. Let $\graphG$ satisfy the desiderata of Def.~\ref{def:graph-desiderata}.
    
    Consider the fault-tolerant gauging procedure in Def.~\ref{def:detailed-gauging-procedure} with the elementary fault model of Def.~\ref{def:elementary-faults}.
    Let logical failure be defined as in Def.~\ref{def:logical-failure}, and let $d_{\mathrm{f}}$ be the resulting adversarial spacetime fault distance from Def.~\ref{def:fault-distance}.

    There is a constant $\gamma = O(1)$, such that  $d_\mathrm{f}\ge d/\gamma$.
\end{theorem}
\begin{proof}
    We will choose $\gamma$ to satisfy two properties. First, a fault of weight less than $d/\gamma$ must admit three cuts needed for the iterative cleaning. Second, after cleaning, its total residual spatial support must be smaller than the spatial distance $d/\nu$.
    
    Let $c_{\mathrm{cut}}=O(1)$ be the constant from Lemma~\ref{lem:cutting-condition-existence}, and let $c_{\mathrm{res.}}=O(1)$ be the constant in the residual support bound,
    \[
    \sum_{t\in T_{\mathrm{res.}}}|\mathsf{R}_t|\leq c_{\rm{res.}}\Wt(F).
    \]
    Choose 
    \begin{equation}
        \gamma:= \max(c_{\mathrm{cut}}, \nu c_{\mathrm{res.}}).
        \label{def:gamma}
    \end{equation}
    Because $c_\mathrm{cut}$, $c_{\mathrm{res.}}$, and $\nu$ are all $O(1)$, we have $\gamma = O(1)$.

    We now show that every loop-like fault below weight $d/\gamma$ is logically successful.
    Let $F$ be an arbitrary loop-like fault configuration satisfying 
    \[
    \Wt(F)<\frac{d}{\gamma},
    \]
    and let $(F,r)$ be any valid trajectory of $F$. 

    We first verify that the clean cuts exist.
    Since $\gamma \geq c_{\mathrm{cut}}$, 
    \[\Wt(F)\le d/\gamma \le d/c_{\mathrm{cut}}.
    \]
    Lemma~\ref{lem:cutting-condition-existence} thus guarantees that there are three clean cuts supported on the intervals $I_{\rm pre} = [1, d]$, $I_g = [d+1, 2d]$, $I_{\rm post} = [2d+1, 3d]$.

    We next use these cuts to move faults onto the reduced time slices.
    Perform iterative cleaning from Def.~\ref{def:iter-cleaning}, followed by the ungauging cleaning in Lemma~\ref{lem:ungauging-cleaning}. Since these are exact rewrites of the Kraus operator, we have
    \[
    K(F,r)\Pi_{\cC} = \sum_{j}\alpha_j K(F_j,r_j) \Pi_\cC.
    \]
    The total residual support bound is
    $\sum_{t\in T_{\mathrm{res.}}} |\mathsf{R}_t|\le c_{\mathrm{res.}}\Wt(F)$. We now verify that this residual support is less than $d/\nu$, so we can use the distance of the base code, as in the proof of Lemma~\ref{cond:logical-success-2}.
    Because $\gamma \ge \nu c_{\mathrm{res.}}$, 
    \[
    \sum_{t\in T_{\mathrm{res.}}} |\mathsf{R}_t|\le c_{\mathrm{res.}}\Wt(F) < \frac{c_{\mathrm{res.}}d}{\gamma} \le \frac{d}{\nu}.
    \]
    Thus, the assumption in Eq.~\eqref{eq:residual-support-assumption} in Lemma~\ref{lem:logical-success-condition-2} is satisfied.
    Now that the assumption is valid, we can use Lemma~\ref{lem:logical-success-condition-2}, which gives a valid, fault-free trajectory $(\emptyset, r')$ and a scalar $\alpha\neq 0$ such that $K(F,r)\Pi_{\cC}= \alpha K(\emptyset, r')\Pi_{\cC}$. 
    This shows the logical success condition~\ref{cond:logical-success-2} in Def.~\ref{def:logical-failure}.

    It remains to show the other condition for logical success, which is that a valid fault-free trajectory $(\emptyset, r')$ selected above has the same measured symmetry value as the original faulty trajectory.
    The trajectory $(\emptyset, r')$ selected in the proof of Lemma~\ref{lem:logical-success-condition-2} is one of the fault-free candidate trajectories to which Lemma~\ref{lem:logical-success-condition-1} applies.
    Therefore, $m_{\T}(r') = m_{\T}(r_F) $,
    proving condition~\ref{cond:logical-success-1} and thus, logical success of $(F,r)$.

    Finally, $F$ and its valid trajectory were arbitrary, thus every valid loop-like fault configuration satisfying
    \[
    \Wt(F) <\frac{d}{\gamma}
    \]
    is logically successful. Thus, no loop-like logical failure has weight below $d/\gamma$.
    By the definition of the adversarial spacetime fault distance,
    \[
    d_{\rm{f}}\ge \frac{d}{\gamma}.
    \]
\end{proof}

\ifdefined\ismainfile
\else
    \end{document}
\fi

\section{Examples} \label{sec:examples}\ifdefined\ismainfile
\else
    \documentclass[11pt]{article}

    \input{preamble.tex}
    
    \begin{document}
    \title{Examples}
    \date{\today}
    \bibliographystyle{abbrv}

    \maketitle
\fi

In this section we present a range of examples that demonstrate the flexibility of the general Clifford gauging measurement protocol. These examples include local topological codes as well as more general QLDPC codes that are not local in any fixed spatial dimension. We focus our discussion on finding the deformed gauged codes that occur during the gauging measurement protocol in each example. The logical action of the gauging measurement protocol in each case is simply to measure the relevant transversal Clifford gate on the initial stabilizer code. We discuss applications of such logical measurements in the following section, below. 

\subsection{Gauging XS on the color code}
We start with Clifford gauging on the color code as a pedagogical example, before moving to higher-rate examples. First, as the color code is self-dual, it possesses a transversal $XS$ gate. If we gauge this symmetry, we can obtain $\ket{\CZbar}$ states on the color code. 

We begin by partitioning the sites of the honeycomb lattice into vertex sets. There are many choices of such a partition, with three convenient ones being (i) one site per vertex set, (ii) one unit cell (two sites) per vertex set, and (iii) disjoint hexagons forming vertex sets. For this example, we will consider all three choices to give a sense for how the resulting models differ. We note that a similar protocol to $(i)$ was considered in~\cite{Tantivasadakarn23}. 

The precise form of the $XS$ symmetry on the honeycomb lattice is \begin{equation}
    \T = \prod_{A} W \prod_B \tilde{W}, \quad W = T^\dagger X T = e^{-i\pi/4}XS, \quad \tilde{W} = TXT^\dagger = e^{+i\pi/4}XS^\dagger
\end{equation} As the honeycomb lattice is bipartite, it can be separated into two sublattices $A, B$. Going forward, we will denote $A$ sites with black circles and $B$ sites with white circles.

We use the following commutation relations when performing the gauging procedure: 
\begin{equation}
    \begin{aligned}
    CW_{c \rightarrow t} X_c [CW_{c \rightarrow t}]^\dagger &= X_c \otimes W_t \\
    C\tilde{W}_{c \rightarrow t} X_c [C\tilde{W}_{c \rightarrow t}]^\dagger &= X_c \otimes \tilde{W}_t \\
    CW_{c \rightarrow t} X_t [CW_{c \rightarrow t}]^\dagger &= X_t S_c^\dagger \CZ_{c \to t} \\
    C\tilde{W}_{c \rightarrow t} X_t [C\tilde{W}_{c \rightarrow t}]^\dagger &= X_t S_c \CZ_{c \to t} \\
    \end{aligned}
\end{equation}
\begin{equation}
\begin{aligned}
    CW_{c \rightarrow t} Z_t [CW_{c \rightarrow t}]^\dagger =
    C\tilde{W}_{c \rightarrow t} Z_t [C\tilde{W}_{c \rightarrow t}]^\dagger = Z_c \otimes Z_t
\end{aligned}
\end{equation}

\subsubsection{Qubit Per Vertex Set}
As our first example, we take vertex sets to be single sites of the honeycomb lattice. We introduce a new ancilla qubit for every vertex set, initialized in the $\ket{+}$ state, and we entangle it with its vertex set by applying a $CW$ or $C\tilde{W}$ gate depending on the sublattice. The ancilla acts as the control, and the code qubit as the target.

The new stabilizers after this step are as follows: 
\begin{equation}
    \raisebox{-35pt}{
    \begin{tikzpicture}

        \node[name=g, regular polygon, regular polygon sides = 6, draw, inner sep = 0.5cm, fill=gray, fill opacity=0.25, thick] at (-4, 0) {};
        
        \foreach \anchor/\placement/\n/\fill/\text in
        {corner 1/below/1/white/black, corner 2/below/2/black/white, corner 3/right/3/white/black, corner 4/above/4/black/white, corner 5/above/5/white/black, corner 6/left/6/black/white}
        {
        \filldraw[shift=(g.\anchor), fill=red!50!white] (\n*60:0.4) circle (5pt);
        \node[inner sep = 1pt] at ($(g.\anchor) + (\n*60:0.4)$) {\tiny{$Z$}};
        \node[draw, circle, fill=\fill, inner sep = 1pt, text=\text] at (g.\anchor) {\tiny{$Z$}};
        }

        \node[name=g, regular polygon, regular polygon sides = 6, draw, inner sep = 0.5cm, fill=gray, fill opacity=0.25, thick] at (0, 0) {};
        
        \foreach \anchor/\placement/\n/\fill/\text in
        {corner 1/below/1/white/black, corner 2/below/2/black/white, corner 3/right/3/white/black, corner 4/above/4/black/white, corner 5/above/5/white/black, corner 6/left/6/black/white}
        {
        \draw[-, red, thick] ($(g.\anchor) + (\n*60:0.5)$) to (g.\anchor);
        \filldraw[shift=(g.\anchor), fill=red!50!white] (\n*60:0.5) circle (5pt);
        \node[draw, circle, fill=\fill, inner sep = 1pt, text=\text] at (g.\anchor) {\tiny{$X$}};
        }

        \foreach \n/\anchor in
        {1/corner 1,3/corner 3,5/corner 5}
        {
        \node[inner sep = 1pt] at ($(g.\anchor) + (\n*60:0.5)$) {\tiny{$S$}};
        }
        \foreach \n/\anchor in
        {2/corner 2,4/corner 4,6/corner 6}
        {
        \node[inner sep = 1pt] at ($(g.\anchor) + (\n*60:0.5)$) {\tiny{$S$}};
        }
        
        \node[draw, circle, fill=black, inner sep = 1pt, text=white] at (4, 0) {\tiny{$W$}};
        \node[draw, circle, fill=red!50!white, inner sep = 2pt] at (4.5, 0) {\tiny{$X$}};

        \node[draw, circle, fill=white, inner sep = 0.5pt, text=black] at (6, 0) {\tiny{$W^\dagger$}};
        \node[draw, circle, fill=red!50!white, inner sep = 2pt] at (6.5, 0) {\tiny{$X$}};
        
    \end{tikzpicture}
    }\label{eq:honeycomb_X_term_with_ancilla_vertices}
\end{equation}
Here we have displaced the red circles to avoid overlap for visual clarity; the red nodes also sit on vertices of the honeycomb lattice. A $\CZ$ gate is denoted by a red line between both degrees of freedom acted on by the gate.

We now introduce edges between the vertex set ancillas using the following rule: there must be a path connecting two ancillas if there exists a stabilizer with support on both the vertex sets associated with the ancillas. Since every vertex set is a single site, one way to satisfy this rule is to simply place the ancillas on the sites of another honeycomb lattice, sitting on top of the original one. The purple node and the red node correspond to the ancilla qubits on the edges and the vertices of the gauging graph, respectively.
We then add a new qubit on each edge of this new lattice (called the gauging graph). The new qubits are initialized in the $\ket{0}$ state. We entangle the new ancillas and the original vertex set ancillas using a cluster state entangler. The new stabilizers are as follows:
\begin{equation}
\raisebox{-30pt}{
    \begin{tikzpicture}
        \filldraw[] (0,0) circle (2pt) node[left] {$W$};
        \draw[-, thick] (0, 0) -- (1.5, 0);
        \draw[-, thick] (0, 0) -- (120:1.5);
        \draw[-, thick] (0, 0) -- (240:1.5);

        \filldraw[fill=blue!50!white, thick] (0:0.75) circle (3pt) node[above] {$X$};
        \filldraw[fill=blue!50!white, thick] (120:0.75) circle (3pt) node[left] {$X$};
        \filldraw[fill=blue!50!white, thick] (240:0.75) circle (3pt) node[left] {$X$};
        
        \filldraw[fill=red!50!white, thick] (0.2,0) circle (3pt) node[above] {$X$};

        \draw[-, thick] (4, 0) -- (2.5, 0);
        \draw[-, thick] (4, 0) -- ($(4, 0) + (60:1.5)$);
        \draw[-, thick] (4, 0) -- ($(4, 0) + (300:1.5)$);
        \filldraw[fill=white] (4,0) circle (3pt) node[right] {$W^\dagger$};

        \filldraw[fill=blue!50!white, thick] ($(4, 0) + (180:0.75)$) circle (3pt) node[above] {$X$};
        \filldraw[fill=blue!50!white, thick] ($(4, 0) + (60:0.75)$) circle (3pt) node[left] {$X$};
        \filldraw[fill=blue!50!white, thick] ($(4, 0) + (300:0.75)$) circle (3pt) node[left] {$X$};
        
        \filldraw[fill=red!50!white, thick] (3.8, 0) circle (3pt) node[above] {$X$};

        \draw[-, thick] (0+6, 0) -- (1.5+6, 0);
        \filldraw[fill=red!50!white, thick] (0+6,0) circle (3pt) node[above] {$Z$};
        \filldraw[fill=red!50!white, thick] (1.5+6, 0) circle (3pt) node[above] {$Z$};
        \filldraw[fill=blue!50!white, thick] (0.75+6, 0) circle (3pt) node[above] {$Z$};

    \end{tikzpicture}
    }
\end{equation} Note that the hexagonal stabilizers originating from the color code itself are unchanged by this step.

Finally, we project the vertex set ancillas onto the $\ket{+}$ state (in a real protocol, we would measure in the $X$ basis and then apply feedback to correct undesired $-1$ outcomes). The surviving stabilizers are those that commute with $X$ on the ancilla. Some are straightforward to determine--- the vertex stabilizers clearly commute with $X$, as they have only $X$ support on the vertex ancilla. Additionally, the product of the vertex-edge-vertex ancilla stabilizers $Z_{v} Z_{e} Z_{v'}$ around any hexagon will have no support on the vertex ancilla, and so trivially commutes.  
\begin{equation}
    \raisebox{-30pt}{
    \begin{tikzpicture}
        \filldraw[] (0,0) circle (2pt) node[left] {$W$};
        \draw[-, thick] (0, 0) -- (1.5, 0);
        \draw[-, thick] (0, 0) -- (120:1.5);
        \draw[-, thick] (0, 0) -- (240:1.5);

        \filldraw[fill=blue!50!white, thick] (0:0.75) circle (3pt) node[above] {$X$};
        \filldraw[fill=blue!50!white, thick] (120:0.75) circle (3pt) node[left] {$X$};
        \filldraw[fill=blue!50!white, thick] (240:0.75) circle (3pt) node[left] {$X$};

        \draw[-, thick] (4, 0) -- (2.5, 0);
        \draw[-, thick] (4, 0) -- ($(4, 0) + (60:1.5)$);
        \draw[-, thick] (4, 0) -- ($(4, 0) + (300:1.5)$);
        \filldraw[fill=white] (4,0) circle (3pt) node[right] {$W^\dagger$};

        \filldraw[fill=blue!50!white, thick] ($(4, 0) + (180:0.75)$) circle (3pt) node[above] {$X$};
        \filldraw[fill=blue!50!white, thick] ($(4, 0) + (60:0.75)$) circle (3pt) node[left] {$X$};
        \filldraw[fill=blue!50!white, thick] ($(4, 0) + (300:0.75)$) circle (3pt) node[left] {$X$};

        \node[name = h, regular polygon, regular polygon sides = 6, draw, inner sep = 0.5cm, fill=gray!50!white, fill opacity=0.25, thick] at (7, 0) {};
        \foreach \anchor/\placement in
        {side 1/above, side 2/left, side 3/left, side 4/below, side 5/right, side 6/right}
        {
        \filldraw[shift=(h.\anchor), fill=blue!50!white] (0,0) circle (3pt) node[\placement] {$Z$};
        }
    \end{tikzpicture}
    }
\end{equation} These are the $A_v$ and $B_p$ terms, to use the notation of the previous sections.

Now we consider the original hexagon stabilizers of the color code, which have been deformed by the gauging procedure so far. The $Z$ hexagon does not commute with the measurement due to the $Z$ support it picked up on the vertex qubits; we can get rid of this support by multiplying with the three-body cluster state stabilizer $Z_v Z_e Z_{v'}$.
\begin{equation}
    \raisebox{-35pt}{
    \begin{tikzpicture}

        \node[name=g, regular polygon, regular polygon sides = 6, draw, inner sep = 0.5cm, fill=gray, fill opacity=0.25, thick] at (-4, 0) {};
        
        \foreach \anchor/\placement/\n/\fill/\text in
        {corner 1/below/1/white/black, corner 2/below/2/black/white, corner 3/right/3/white/black, corner 4/above/4/black/white, corner 5/above/5/white/black, corner 6/left/6/black/white}
        {
        \node[draw, circle, fill=\fill, inner sep = 1pt, text=\text] at (g.\anchor) {\tiny{$Z$}};
        }

        \foreach \anchor/\placement in {side 1/above, side 3/left, side 5/right}
        \filldraw[shift=(g.\anchor), fill=blue!50!white] (0,0) circle (3pt) node[\placement] {\small{$Z$}};
        
    \end{tikzpicture}
    }
\end{equation} The $X$ hexagon is a little trickier--- we will again use the three-body cluster state stabilizer. We will use the following identity:
\begin{equation}
\begin{aligned}
    \CZ_{a \rightarrow b} &= \ket{0}\bra{0}_a \otimes \mathbb{I}_b + \ket{1}\bra{1}_a \otimes Z_b \\
    &= \frac{1 + Z_a}{2} \otimes \mathbb{I}_b + \frac{1- Z_a}{2} \otimes Z_b \\
    &\sim \frac{1 + Z_{e(a, a')} Z_{a'}}{2} \otimes \mathbb{I}_b + \frac{1 - Z_{e(a, a')} Z_{a'}}{2} \otimes Z_b \\
    &= \mathbb{I}_{e(a, a')} \mathbb{I}_{a'} \otimes \frac{1 + Z_b}{2} + Z_{e(a, a')} Z_{a'} \otimes \frac{1 - Z_b}{2} \\
    &= \CZ_{b \rightarrow e(a, a')} \CZ_{b \rightarrow a'}
\end{aligned}
\label{eq:CZ-deformation}
\end{equation} 
We have used the three-body cluster state stabilizer $Z_a Z_{e(a, a')} Z_{a'}$ in the third line. This allows us to shift the support of a control gate away from a given vertex, at the cost of introducing more control gates elsewhere. Additionally, we require the following identity:
\begin{equation}
    \begin{aligned}
        S_a S_{a'}^\dagger &= (\dyad{0}_a + i \dyad{1}_a)(\dyad{0}_{a'} - i \dyad{1}_{a'}) \\
        &= \left(\frac{1+Z_a}{2} + i \frac{1-Z_a}{2} \right) \left(\frac{1+Z_{a'}}{2} - i \frac{1-Z_{a'}}{2} \right) \\
        &\sim \left ( \frac{1+Z_{e(a, a')}Z_{a'}}{2} + i \frac{1-Z_{e(a,a')}Z_{a'}}{2} \right ) \left (\frac{1+Z_{a'}}{2} - i \frac{1-Z_{a'}}{2} \right ) \\
        &= \frac{1 + Z_{e(a, a')}}{2} +  \frac{1- Z_{e(a, a')}}{2} (i Z_{a'}) \\
        &= S_{e(a, a')} \CZ_{e(a,a') \rightarrow a'}
    \end{aligned}
    \label{eq:S-deformation}
\end{equation}
Using Eq.~\ref{eq:S-deformation}, the $X$ hexagon can be modified in the following way:
\begin{equation}
    \raisebox{-35pt}{
    \begin{tikzpicture}
        
        \node[name=g, regular polygon, regular polygon sides = 6, draw, inner sep = 0.5cm, fill=gray, fill opacity=0.25, thick] at (0, 0) {};

        \foreach \anchor/\target/\n in
        {corner 2/side 1/2, corner 4/side 3/4, corner 6/side 5/6}
        {
        \draw[-, red, thick] ($(g.\anchor) + (\n*60:0.5)$) to (g.\target);
        }
        
        \foreach \anchor/\placement/\n/\fill/\text in
        {corner 1/below/1/white/black, corner 2/below/2/black/white, corner 3/right/3/white/black, corner 4/above/4/black/white, corner 5/above/5/white/black, corner 6/left/6/black/white}
        {
        \draw[-, red, thick] ($(g.\anchor) + (\n*60:0.5)$) to (g.\anchor);
        \filldraw[shift=(g.\anchor), fill=red!50!white] (\n*60:0.5) circle (3pt);
        \node[draw, circle, fill=\fill, inner sep = 1pt, text=\text] at (g.\anchor) {\tiny{$X$}};
        }

        \foreach \anchor/\placement in {side 1/above, side 3/left, side 5/right}
        \filldraw[shift=(g.\anchor), fill=blue!50!white] (0,0) circle (3pt) node[\placement] {\small{$S$}};
        
    \end{tikzpicture}
    }
\end{equation}
Here, the red lines joining two degrees of freedom indicate a $CZ$ gate.

Now all of the $S$ support has been moved away from the vertex ancilla onto the edge ancilla, but we have added some new $CZ$ gates.
Using Eq.~\ref{eq:CZ-deformation}, we can push all support on the vertex qubits to a single, root vertex, labeled ``a''. For each vertex ancilla $b \neq a$ touched by a CZ, we find a path $p(a, b)$ between $a$ and $b$. The first identity tells us we can trade a $CZ$ on $b$ for a $CZ$ on $a$ and all the edges along the path $p(a, b)$. To illustrate, let $a$ be the top left vertex ancilla, and $b$ the leftmost ancilla. Then Eq.~\ref{eq:CZ-deformation} gives: 
\begin{equation}
    \raisebox{-45pt}{
    \begin{tikzpicture}[scale=1]
        
        \node[name=g, regular polygon, regular polygon sides = 6, draw, inner sep = 0.75cm, fill=gray, fill opacity=0.25, thick] at (0, 0) {};

       \foreach \anchor/\target/\n in
        {corner 2/side 1/2, corner 4/side 3/4, corner 6/side 5/6}
        {
        \draw[-, red, thick] ($(g.\anchor) + (\n*60:0.5)$) to (g.\target);
        }
        
        \foreach \anchor/\placement/\n/\fill/\text in
        {corner 1/below/1/white/black, corner 2/below/2/black/white, corner 3/right/3/white/black, corner 4/above/4/black/white, corner 5/above/5/white/black, corner 6/left/6/black/white}
        {
        \draw[-, red, thick] ($(g.\anchor) + (\n*60:0.5)$) to (g.\anchor);
        \filldraw[shift=(g.\anchor), fill=red!50!white] (\n*60:0.5) circle (3pt);
        \node[draw, circle, fill=\fill, inner sep = 1pt, text=\text] at (g.\anchor) {\tiny{$X$}};
        }

        \filldraw[shift=(g.corner 2), fill=white!50!red, thick] (2*60:0.5) circle (3.5pt) node[name=v] {\tiny {$a$}};

        \filldraw[shift=(g.corner 3), fill=white!50!red, thick] (3*60:0.5) circle (3.5pt) node[name=v] {\tiny {$b$}};

        \foreach \anchor/\placement in {side 1/above, side 3/left, side 5/right}
        \filldraw[shift=(g.\anchor), fill=blue!50!white] (0,0) circle (3pt) node[\placement] {\small{$S$}};

        \node at (2.5, 0) {$=$};

        \node[name=g, regular polygon, regular polygon sides = 6, draw, inner sep = 0.75cm, fill=gray, fill opacity=0.25, thick] at (5, 0) {};

        \foreach \anchor/\target/\n in
        {corner 2/side 1/2, corner 4/side 3/4, corner 6/side 5/6}
        {
        \draw[-, red, thick] ($(g.\anchor) + (\n*60:0.5)$) to (g.\target);
        }

        \foreach \anchor/\placement/\n/\fill/\text in
        {corner 1/below/1/white/black, corner 2/below/2/black/white, corner 4/above/4/black/white, corner 5/above/5/white/black, corner 6/left/6/black/white}
        {
        \draw[-, red, thick] ($(g.\anchor) + (\n*60:0.5)$) to (g.\anchor);
        }

        \draw[-, red, thick] ($(g.corner 2) + (2*60:0.5)$) to[in=90, out = -120] (g.corner 3);
        \draw[-, red, thick] ($(g.corner 3)$) to[in=180, out=90] (g.side 2);
        
        \foreach \anchor/\placement/\n/\fill/\text in
        {corner 1/below/1/white/black, corner 2/below/2/black/white, corner 3/right/3/white/black, corner 4/above/4/black/white, corner 5/above/5/white/black, corner 6/left/6/black/white}
        {
        \filldraw[shift=(g.\anchor), fill=red!50!white] (\n*60:0.5) circle (3pt);
        \node[draw, circle, fill=\fill, inner sep = 1pt, text=\text] at (g.\anchor) {\tiny{$X$}};
        }

        \filldraw[shift=(g.corner 2), fill=white!50!red, thick] (2*60:0.5) circle (3.5pt) node[name=v] {\tiny {$a$}};

        \filldraw[shift=(g.corner 3), fill=white!50!red, thick] (3*60:0.5) circle (3.5pt) node[name=v] {\tiny {$b$}};

        \foreach \anchor/\placement in {side 1/above, side 3/left, side 5/right}
        \filldraw[shift=(g.\anchor), fill=blue!50!white] (0,0) circle (3pt) node[\placement] {\small{$S$}};

        \foreach \anchor/\placement in {side 2/above, side 4/left, side 6/right}
        \filldraw[shift=(g.\anchor), fill=blue!50!white] (0,0) circle (3pt) node[\placement] {};
        
    \end{tikzpicture}
    }
\end{equation}
We have removed $Z$ support on vertex ancilla $b$ from the stabilizer. We can repeat this process for all the other vertex ancillas (excluding the root, $a$). We obtain the following deformed stabilizer:
\begin{equation}
    \raisebox{-35pt}{
    \begin{tikzpicture}

        \node[name=g, regular polygon, regular polygon sides = 6, draw, inner sep = 0.75cm, fill=gray, fill opacity=0.25, thick] at (0, 0) { };

        \filldraw[shift=(g.corner 2), fill=white!50!red, thick] (-0.5, 0.5) circle (3.5pt) node[name=v] {\tiny {$a$} };

        \foreach \anchor/\placement/\n/\fill/\text/\name in
        {corner 1/below/1/white/black/A, corner 2/below/2/black/white/B, corner 3/right/3/white/black/C, corner 4/above/4/black/white/D, corner 5/above/5/white/black/E, corner 6/left/6/black/white/F}
        {
        \node[draw, circle, fill=\fill, inner sep = 1pt, text=\text, name=\name] at (g.\anchor) {\tiny{$X$}};
        }

        \foreach \anchor/\placement/\name in
        {side 2/left/b, side 4/below/d, side 6/right/f}
        {
        \filldraw[shift=(g.\anchor), fill=blue!50!white, fill opacity = 1, thick] (0,0) circle (4pt);
        \node[name=\name] at (g.\anchor) { }; 
        }

        \foreach \anchor/\placement/\name in
        {side 1/above/a, side 3/left/c, side 5/right/e}
        {
        \filldraw[shift=(g.\anchor), fill=blue!50!white, fill opacity = 1, thick] (0,0) circle (4pt);
        \node[name=\name] at (g.\anchor) {\tiny{$S$}}; 
        }

        \draw[-, thick, red] (a) to[in=-100, out=-90] (A);
        \draw[-, thick, red] (a) to[in=180, out=-90] (F);
        \draw[-, thick, red] (a) to[in=120, out=-90] (E);

        \draw[-, thick, red] (f) to[in=180, out=-130] (F);
        \draw[-, thick, red] (f) to[in=120, out=-130] (E);

        \draw[-, thick, red] (e) to[in=120, out=170] (E);

        \draw[-, thick, red] (c) to[in=50, out=10] (D);
        
        \draw[-, thick, red] (b) to[in=50, out=-30] (D);
        \draw[-, thick, red] (b) to[in=0, out=-70] (C);

        \draw[-, red, thick] (v) to[in=130, out=10] (A);
        \draw[-, red, thick] (v) to[in=120, out=-30] (B);
        \draw[-, red, thick] (v) to[in=100, out=-110] (C);
        \draw[-, red, thick] (v) to[in=180, out=-160] (D);
        \draw[-, red, thick] (v) to[in=140, out=-90] (E);
        \draw[-, red, thick] (v) to[in=60, out=50] (F);
        \draw[-, red, thick] (v) to[in=90, out=5] (a);
        \draw[-, red, thick] (v) to[in=170, out=220] (c);
        \draw[-, red, thick] (v) to[in=150, out=-70] (e);

        \draw[-, red, thick] (a) to[in=200, out=-90] (f);
        \draw[-, red, thick] (a) to[in=10, out=-90] (c);

        \draw[-, red, thick] (c) to[in=90, out=10] (d);
        \draw[-, red, thick] (c) to[in=150, out=10] (e);
        \draw[-, red, thick] (f) to[in=10, out=-130] (c);
        
        \draw[-, red, thick] (f) to[in=120, out=-130] (e);

        \node at (2.5, 0) {$=$};

        \node[name=g, regular polygon, regular polygon sides = 6, draw, inner sep = 0.75cm, fill=gray, fill opacity=0.25, thick] at (5, 0) { };
        
        \foreach \anchor/\placement/\n/\fill/\text/\name in
        {corner 1/below/1/white/black/A, corner 2/below/2/black/white/B, corner 3/right/3/white/black/C, corner 4/above/4/black/white/D, corner 5/above/5/white/black/E, corner 6/left/6/black/white/F}
        {
        \node[draw, circle, fill=\fill, inner sep = 1pt, text=\text, name=\name] at (g.\anchor) {\tiny{$X$}};
        }

        \foreach \anchor/\placement/\name in
        {side 2/left/b, side 4/below/d, side 6/right/f}
        {
        \filldraw[shift=(g.\anchor), fill=blue!50!white, fill opacity = 1, thick] (0,0) circle (4pt);
        \node[name=\name] at (g.\anchor) { }; 
        }

        \foreach \anchor/\placement/\name in
        {side 1/above/a, side 3/left/c, side 5/right/e}
        {
        \filldraw[shift=(g.\anchor), fill=blue!50!white, fill opacity = 1, thick] (0,0) circle (4pt);
        \node[name=\name] at (g.\anchor) {\tiny{$S$}}; 
        }
        
        \draw[-, thick, red] (a) to[in=-100, out=-90] (A);
        \draw[-, thick, red] (a) to[in=180, out=-90] (F);
        \draw[-, thick, red] (a) to[in=120, out=-90] (E);

        \draw[-, thick, red] (f) to[in=180, out=-130] (F);
        \draw[-, thick, red] (f) to[in=120, out=-130] (E);

        \draw[-, thick, red] (e) to[in=120, out=170] (E);

        \draw[-, thick, red] (c) to[in=50, out=10] (D);
        
        \draw[-, thick, red] (b) to[in=50, out=-30] (D);
        \draw[-, thick, red] (b) to[in=0, out=-70] (C);

        \draw[-, red, thick] (a) to[in=200, out=-90] (f);
        \draw[-, red, thick] (a) to[in=10, out=-90] (c);

        \draw[-, red, thick] (c) to[in=90, out=10] (d);
        \draw[-, red, thick] (c) to[in=150, out=10] (e);
        \draw[-, red, thick] (f) to[in=10, out=-130] (c);
        
        \draw[-, red, thick] (f) to[in=120, out=-130] (e);

        \node at (7, 0) {$\times$};
        \node at (7.5, 0) {$C_a$};
        \draw [decorate, decoration = {calligraphic brace}, thick] (8, -1.25) -- (8, 1.25);
        \draw [decorate, decoration = {calligraphic brace, mirror}, thick] (11.5, -1.25) -- (11.5, 1.25);

        \node[name=g, regular polygon, regular polygon sides = 6, draw, inner sep = 0.75cm, fill=gray, fill opacity=0.25, thick] at (9.75, 0) { };

        \foreach \anchor/\placement/\n/\fill/\text/\name in
        {corner 1/below/1/white/black/A, corner 2/below/2/black/white/B, corner 3/right/3/white/black/C, corner 4/above/4/black/white/D, corner 5/above/5/white/black/E, corner 6/left/6/black/white/F}
        {
        \node[draw, circle, fill=\fill, inner sep = 1pt, text=\text, name=\name] at (g.\anchor) {\tiny{$Z$}};
        }

        \foreach \anchor/\placement/\name in
        {side 1/above/a, side 3/left/c, side 5/right/e}
        {
        \filldraw[shift=(g.\anchor), fill=blue!50!white, fill opacity = 1, thick] (0,0) circle (4pt); 
        \node[name=\name] at (g.\anchor) {\tiny{$Z$}};
        }
    \end{tikzpicture}
    }
\end{equation} 
The controlled-stabilizer is trivial on the codespace, and so we can ignore it (with the caveat that the resulting stabilizers will only strictly commute \emph{in the codespace}). Thus, we have removed all support on the vertex qubits that does not commute with the $X$ measurement. The final stabilizers are:
\begin{equation}
    \raisebox{-30pt}{
    \begin{tikzpicture}
        \filldraw[] (0,0) circle (2pt) node[left] {$W$};
        \draw[-, thick] (0, 0) -- (1.5, 0);
        \draw[-, thick] (0, 0) -- (120:1.5);
        \draw[-, thick] (0, 0) -- (240:1.5);

        \filldraw[fill=blue!50!white, thick] (0:0.75) circle (3pt) node[above] {$X$};
        \filldraw[fill=blue!50!white, thick] (120:0.75) circle (3pt) node[left] {$X$};
        \filldraw[fill=blue!50!white, thick] (240:0.75) circle (3pt) node[left] {$X$};

        \draw[-, thick] (4, 0) -- (2.5, 0);
        \draw[-, thick] (4, 0) -- ($(4, 0) + (60:1.5)$);
        \draw[-, thick] (4, 0) -- ($(4, 0) + (300:1.5)$);
        \filldraw[fill=white] (4,0) circle (3pt) node[right] {$W^\dagger$};

        \filldraw[fill=blue!50!white, thick] ($(4, 0) + (180:0.75)$) circle (3pt) node[above] {$X$};
        \filldraw[fill=blue!50!white, thick] ($(4, 0) + (60:0.75)$) circle (3pt) node[left] {$X$};
        \filldraw[fill=blue!50!white, thick] ($(4, 0) + (300:0.75)$) circle (3pt) node[left] {$X$};

        \node[name = h, regular polygon, regular polygon sides = 6, draw, inner sep = 0.5cm, fill=gray!50!white, fill opacity=0.25, thick] at (7, 0) {};
        \foreach \anchor/\placement in
        {side 1/above, side 2/left, side 3/left, side 4/below, side 5/right, side 6/right}
        {
        \filldraw[shift=(h.\anchor), fill=blue!50!white] (0,0) circle (3pt) node[\placement] {$Z$};
        }
    \end{tikzpicture}
    }
\end{equation}
\begin{equation}
    \raisebox{-30pt}{
    \begin{tikzpicture}
        \node[name=g, regular polygon, regular polygon sides = 6, draw, inner sep = 0.75cm, fill=gray, fill opacity=0.25, thick] at (0, 0) { };
        
        \foreach \anchor/\placement/\n/\fill/\text/\name in
        {corner 1/below/1/white/black/A, corner 2/below/2/black/white/B, corner 3/right/3/white/black/C, corner 4/above/4/black/white/D, corner 5/above/5/white/black/E, corner 6/left/6/black/white/F}
        {
        \node[draw, circle, fill=\fill, inner sep = 1pt, text=\text, name=\name] at (g.\anchor) {\tiny{$X$}};
        }

        \foreach \anchor/\placement/\name in
        {side 2/left/b, side 4/below/d, side 6/right/f}
        {
        \filldraw[shift=(g.\anchor), fill=blue!50!white, fill opacity = 1, thick] (0,0) circle (4pt);
        \node[name=\name] at (g.\anchor) { }; 
        }

        \foreach \anchor/\placement/\name in
        {side 1/above/a, side 3/left/c, side 5/right/e}
        {
        \filldraw[shift=(g.\anchor), fill=blue!50!white, fill opacity = 1, thick] (0,0) circle (4pt);
        \node[name=\name] at (g.\anchor) {\tiny{$S$}}; 
        }
        
        \draw[-, thick, red] (a) to[in=-100, out=-90] (A);
        \draw[-, thick, red] (a) to[in=180, out=-90] (F);
        \draw[-, thick, red] (a) to[in=120, out=-90] (E);

        \draw[-, thick, red] (f) to[in=180, out=-130] (F);
        \draw[-, thick, red] (f) to[in=120, out=-130] (E);

        \draw[-, thick, red] (e) to[in=120, out=170] (E);

        \draw[-, thick, red] (c) to[in=50, out=10] (D);
        
        \draw[-, thick, red] (b) to[in=50, out=-30] (D);
        \draw[-, thick, red] (b) to[in=0, out=-70] (C);

        \draw[-, red, thick] (a) to[in=200, out=-90] (f);
        \draw[-, red, thick] (a) to[in=10, out=-90] (c);

        \draw[-, red, thick] (c) to[in=90, out=10] (d);
        \draw[-, red, thick] (c) to[in=150, out=10] (e);
        \draw[-, red, thick] (f) to[in=10, out=-130] (c);
        
        \draw[-, red, thick] (f) to[in=120, out=-130] (e);

        \node[name=g, regular polygon, regular polygon sides = 6, draw, inner sep = 0.75cm, fill=gray, fill opacity=0.25, thick] at (5, 0) {};
        
        \foreach \anchor/\placement/\n/\fill/\text in
        {corner 1/below/1/white/black, corner 2/below/2/black/white, corner 3/right/3/white/black, corner 4/above/4/black/white, corner 5/above/5/white/black, corner 6/left/6/black/white}
        {
        \node[draw, circle, fill=\fill, inner sep = 1pt, text=\text] at (g.\anchor) {\tiny{$Z$}};
        }

        \foreach \anchor/\placement in {side 1/above, side 3/left, side 5/right}
        \filldraw[shift=(g.\anchor), fill=blue!50!white] (0,0) circle (3pt) node[\placement] {\small{$Z$}};
        
    \end{tikzpicture}
    }
\end{equation}
We note that the deformed $X$ hexagon has a large number of $CZ$ gates, and so would be rather impractical to measure in an actual device. Different choices of vertex set lead to different decorations on the original code stabilizers, which may be more or less tractable; thus the choice of vertex set is an important choice that should be taken into consideration when designing a Clifford measurement scheme for practical use.

\subsubsection{Unit Cell per Vertex Set}
The procedure to generate the stabilizers is the same as the previous example, but now we partition the transversal $XS$ symmetry by unit cells of the honeycomb lattice, rather than by individual sites. A vertex set  consists of two neighboring $A, B$ sites--- we choose sites connected by a flat, horizontal edge.

After the symmetry enrichment, the stabilizers are given by:
\begin{equation}
    \raisebox{-45pt}{
    \begin{tikzpicture}

        \node[name=g, regular polygon, regular polygon sides = 6, draw, inner sep = 0.5cm, fill=gray, fill opacity=0.25, thick] at (-4, 0) {};

        \filldraw[shift=(g.corner 3), fill=red!50!white] (180:0.55) circle (5pt);
        \node[inner sep = 1pt] at ($(g.corner 3) + (180:0.55)$) {\tiny{$Z$}};

        \filldraw[shift=(g.corner 6), fill=red!50!white] (0:0.55) circle (5pt);
        \node[inner sep = 1pt] at ($(g.corner 6) + (0:0.55)$) {\tiny{$Z$}};
        
        \foreach \anchor/\placement/\n/\fill/\text in
        {corner 1/below/1/white/black, corner 2/below/2/black/white, corner 3/right/3/white/black, corner 4/above/4/black/white, corner 5/above/5/white/black, corner 6/left/6/black/white}
        {
        \node[draw, circle, fill=\fill, inner sep = 1pt, text=\text] at (g.\anchor) {\tiny{$Z$}};
        }

        \node[name=g, regular polygon, regular polygon sides = 6, draw, inner sep = 0.5cm, fill=gray, fill opacity=0.25, thick] at (0, 0) {};

        \draw[-, red, thick] ($(g.corner 3) + (180:0.55)$) to (g.corner 3);
        \filldraw[shift=(g.corner 3), fill=red!50!white] (180:0.55) circle (5pt);
        \node[inner sep = 1pt] at ($(g.corner 3) + (180:0.55)$) {\tiny{$S$}};

        \draw[-, red, thick] ($(g.corner 6) + (0:0.55)$) to (g.corner 6);
        \filldraw[shift=(g.corner 6), fill=red!50!white] (0:0.55) circle (5pt);
        \node[inner sep = 1pt] at ($(g.corner 6) + (0:0.55)$) {\tiny{$S^\dagger$}};

        \draw[-, red, thick] (g.corner 1) to ($(g.side 1) + (0, 0.65)$);
        \draw[-, red, thick] (g.corner 2) to ($(g.side 1) + (0, 0.65)$);
        \filldraw[shift=(g.side 1), fill=red!50!white] (0, 0.65) circle (3pt);

        \draw[-, red, thick] (g.corner 4) to ($(g.side 4) + (0, -0.65)$);
        \draw[-, red, thick] (g.corner 5) to ($(g.side 4) + (0, -0.65)$);
        \filldraw[shift=(g.side 4), fill=red!50!white] (0, -0.65) circle (3pt);
        
        \foreach \anchor/\placement/\n/\fill/\text in
        {corner 1/below/1/white/black, corner 2/below/2/black/white, corner 3/right/3/white/black, corner 4/above/4/black/white, corner 5/above/5/white/black, corner 6/left/6/black/white}
        {
        \node[draw, circle, fill=\fill, inner sep = 1pt, text=\text] at (g.\anchor) {\tiny{$X$}};
        }

        \node[shape=circle, fill=red!50!white, draw, thick, inner sep=1pt] at (5, 0) {\small{$X$}};
        \node[shape=circle, fill=black, text=white, draw, inner sep=1pt] at (4, 0) {\small{$W$}};
        \node[shape=circle, fill=white, text=black, draw, inner sep=1pt] at (6, 0) {\small{$W^\dagger$}};
        
    \end{tikzpicture}
    }
\end{equation}
Again, we introduce edge ancilla and entangle with the vertex ancilla in a cluster state. Here, the convenient choice is a square lattice defined by our choice of unit cells. The resultant stabilizers are
\begin{equation}
    \raisebox{-35pt}{
    \begin{tikzpicture}[scale=0.75]

        \draw[-, dashed, thick, blue] ($(5,0) + (-45:2)$) -- ($(5, 0) + (135:2)$);
        \draw[-, dashed, thick, blue] ($(5,0) + (45:2)$) -- ($(5, 0) + (225:2)$);

        \node[shape=circle, fill=blue!50!white, draw, thick, inner sep=1pt] at ($(5,0) + (45:1.25)$) {\tiny{$X$}};
        \node[shape=circle, fill=blue!50!white, draw, thick, inner sep=1pt] at ($(5,0) + (-45:1.25)$) {\tiny{$X$}};
        \node[shape=circle, fill=blue!50!white, draw, thick, inner sep=1pt] at ($(5,0) + (135:1.25)$) {\tiny{$X$}};
        \node[shape=circle, fill=blue!50!white, draw, thick, inner sep=1pt] at ($(5,0) + (225:1.25)$) {\tiny{$X$}};
        
        \node[shape=circle, fill=red!50!white, draw, thick, inner sep=1pt] at (5, 0) {\small{$X$}};
        \node[shape=circle, fill=black, text=white, draw, inner sep=1pt] at (4, 0) {\small{$W$}};
        \node[shape=circle, fill=white, text=black, draw, inner sep=1pt] at (6, 0) {\small{$W^\dagger$}};

        \draw[-, dashed, thick, blue] ($(10,0) + (-45:1.5)$) -- ($(10, 0) + (135:1.5)$);
        \node[shape=circle, fill=blue!50!white, draw, thick, inner sep=1pt] at (10, 0) {\tiny{$Z$}};
        
        \node[shape=circle, fill=red!50!white, draw, thick, inner sep=1pt] at ($(10, 0) + (-45:1.5)$) {\tiny{$Z$}};
        \node[shape=circle, fill=red!50!white, draw, thick, inner sep=1pt] at ($(10, 0) + (135:1.5)$) {\tiny{$Z$}};

        \draw[-, dashed, thick, blue] ($(14,0) + (45:1.5)$) -- ($(14, 0) + (-135:1.5)$);
        \node[shape=circle, fill=blue!50!white, draw, thick, inner sep=1pt] at (14, 0) {\tiny{$Z$}};
        
        \node[shape=circle, fill=red!50!white, draw, thick, inner sep=1pt] at ($(14, 0) + (45:1.5)$) {\tiny{$Z$}};
        \node[shape=circle, fill=red!50!white, draw, thick, inner sep=1pt] at ($(14, 0) + (-135:1.5)$) {\tiny{$Z$}};
        
    \end{tikzpicture}
    }
\end{equation}

Again, we need to project the vertices onto the $X$ basis, and so need to get rid of any support on the vertices that does not commute with this measurement. We use the same identities as the previous example. The $Z$ hexagon, $A_v$, and $B_p$ terms are straightforward:

\begin{equation}
    \raisebox{-45pt}{
    \begin{tikzpicture}

        \node[name=g, regular polygon, regular polygon sides = 6, draw, inner sep = 0.5cm, fill=gray, fill opacity=0.25, thick] at (-4, 0) {};

        \draw[-, dashed, blue, thick] ($(g.corner 3) + (180:0.65)$) -- ($(g.side 4) + (0, -0.85)$);
        \draw[-, dashed, blue, thick] ($(g.corner 6) + (0:0.65)$) -- ($(g.side 4) + (0, -0.85)$);
        \node[shape=circle, fill=blue!50!white, draw, thick, inner sep=1pt] at ($(g.side 4) + (-0.9, 0.1)$) {\tiny{$Z$}};
        \node[shape=circle, fill=blue!50!white, draw, thick, inner sep=1pt] at ($(g.side 4) + (0.9, 0.1)$) {\tiny{$Z$}};

        \foreach \anchor/\placement/\n/\fill/\text in
        {corner 1/below/1/white/black, corner 2/below/2/black/white, corner 3/right/3/white/black, corner 4/above/4/black/white, corner 5/above/5/white/black, corner 6/left/6/black/white}
        {
        \node[draw, circle, fill=\fill, inner sep = 1pt, text=\text] at (g.\anchor) {\tiny{$Z$}};
        }

        \draw[-, dashed, thick, blue] ($(0,0) + (-45:2)$) -- ($(0, 0) + (135:2)$);
        \draw[-, dashed, thick, blue] ($(0,0) + (45:2)$) -- ($(0, 0) + (225:2)$);

        \node[shape=circle, fill=blue!50!white, draw, thick, inner sep=1pt] at ($(0,0) + (45:1.25)$) {\tiny{$X$}};
        \node[shape=circle, fill=blue!50!white, draw, thick, inner sep=1pt] at ($(0,0) + (-45:1.25)$) {\tiny{$X$}};
        \node[shape=circle, fill=blue!50!white, draw, thick, inner sep=1pt] at ($(0,0) + (135:1.25)$) {\tiny{$X$}};
        \node[shape=circle, fill=blue!50!white, draw, thick, inner sep=1pt] at ($(0,0) + (225:1.25)$) {\tiny{$X$}};
        
        \node[shape=circle, fill=black, text=white, draw, inner sep=1pt] at (-1, 0) {\small{$W$}};
        \node[shape=circle, fill=white, text=black, draw, inner sep=1pt] at (1, 0) {\small{$W^\dagger$}};

        \node[name= g, regular polygon, regular polygon sides = 4, inner sep = 0.5cm, rotate=45, draw, blue, thick, dashed] at (4, 0) {};

        \foreach \anchor in 
        {side 1, side 2, side 3, side 4}
        {
        \node[shape=circle, fill=blue!50!white, draw, thick, inner sep=1pt] at (g.\anchor) {\tiny{$Z$}};
        }
        
    \end{tikzpicture}
    }
\end{equation}

The $X$ hexagon is modified as follows (again using Eqs.~\ref{eq:CZ-deformation} and \ref{eq:S-deformation}, with the topmost vertex as the root):
\begin{equation}
    \raisebox{-45pt}{
    \begin{tikzpicture}

        \node[name=g, regular polygon, regular polygon sides = 6, draw, inner sep = 0.5cm, fill=gray, fill opacity=0.25, thick] at (-4, 0) {};

        \node[name=s, regular polygon, regular polygon sides = 4, rotate= 45, inner sep = 0.85cm, draw, blue, dashed, thick] at (-4, 0) {};

        \draw[-, red, thick] (s.side 1) to[in=90, out=-90] (g.corner 3);
        \draw[-, red, thick] (s.side 1) to[in=180, out=230] (g.corner 4);

        \draw[-, red, thick] (s.side 4) to[in=0, out=-50] (g.corner 5);
        \draw[-, red, thick] (s.side 4) to[in=90, out=-90] (g.corner 6);

        \draw[-, red, thick] (s.side 2) to[in=220, out=-90] (g.corner 4);

        \draw[-, red, thick] (s.side 3) to[in=220, out=-90] (g.corner 5);
        
        \foreach \anchor/\placement/\n/\fill/\text in
        {corner 1/below/1/white/black, corner 2/below/2/black/white, corner 3/right/3/white/black, corner 4/above/4/black/white, corner 5/above/5/white/black, corner 6/left/6/black/white}
        {
        \node[draw, circle, fill=\fill, inner sep = 1pt, text=\text] at (g.\anchor) {\tiny{$X$}};
        }

        \node[shape=circle, fill=blue!50!white, draw, thick, inner sep=2.5pt] at (s.side 1) {\tiny{$S^\dagger$}};
        \node[shape=circle, fill=blue!50!white, draw, thick, inner sep=1pt] at (s.side 4) {\tiny{$S$}};
        \node[shape=circle, fill=blue!50!white, draw, thick, inner sep=3pt] at (s.side 2) {};
        \node[shape=circle, fill=blue!50!white, draw, thick, inner sep=3pt] at (s.side 3) {};

    \end{tikzpicture}
    }
\end{equation}

\subsubsection{Hexagon per Vertex Set}

For the last part of this example, we assign a three-coloring to the honeycomb lattice. Each hexagonal face of the lattice is assigned red, blue, or green, and no hexagon of the same color may share a site. Hexagons of a single color (say, green) thus form a partition of the lattice, and a convenient choice of vertex set. 

We introduce a new ancilla qubit for every vertex set, initialized in the $\ket{+}$ state, and we entangle it with its vertex set by applying $CW$/$C\tilde{W}$ gates with the ancilla as the control. The new stabilizers of the model are as follows: 
\begin{equation}
    \raisebox{-35pt}{
    \begin{tikzpicture}

        \node[name=g, regular polygon, regular polygon sides = 6, draw, inner sep = 0.5cm, fill=green, fill opacity=0.25, thick] at (0, 0) {};
        \node[shape=circle, fill=gray!50!white, draw, thick, inner sep=2pt, name=ga] at (g.center) { };
        
        \foreach \anchor/\placement/\n/\fill/\text/\op in
        {corner 1/below/1/white/black, corner 2/below/2/black/white, corner 3/right/3/white/black, corner 4/above/4/black/white, corner 5/above/5/white/black, corner 6/left/6/black/white}
        {
        \draw[-, red, thick] (ga) to (g.\anchor);
        \node[draw, circle, fill=\fill, inner sep = 1pt, text=\text] at (g.\anchor) {\tiny{$X$}};
        }

        \node[name= b, regular polygon, regular polygon sides = 6, draw, inner sep = 0.5cm, fill=blue, fill opacity=0.25, thick] at (4, 0) {};
        
        \node at (b.center) [shape=circle, fill=gray!50!white, draw, thick, inner sep=2pt, name=ba,  above=1.25cm] { };
        \node [shape=circle, fill=gray!50!white, draw, thick, inner sep=2pt, name=bb] at ($ (1.299, -0.75) + (b.center) $) { };
        \node [shape=circle, fill=gray!50!white, draw, thick, inner sep=2pt, name=bc] at ($ (-1.299, -0.75) + (b.center) $) { };

        \draw[-, red, thick] (ba) to (b.corner 1);
        \draw[-, red, thick] (ba) to (b.corner 2);

        \draw[-, red, thick] (bb) to (b.corner 6);
        \draw[-, red, thick] (bb) to (b.corner 5);

        \draw[-, red, thick] (bc) to (b.corner 4);
        \draw[-, red, thick] (bc) to (b.corner 3);

       \foreach \anchor/\placement/\n/\fill/\text/\op in
        {corner 1/below/1/white/black, corner 2/below/2/black/white, corner 3/right/3/white/black, corner 4/above/4/black/white, corner 5/above/5/white/black, corner 6/left/6/black/white}
        {
        \node[draw, circle, fill=\fill, inner sep = 1pt, text=\text] at (b.\anchor) {\tiny{$X$}};
        }

        \node[name = r, regular polygon, regular polygon sides = 6, draw, inner sep = 0.5cm, fill=red, fill opacity=0.25, thick] at (8, 0.5) {};
        \node at (r.center) [shape=circle, fill=gray!50!white, draw, thick, inner sep=2pt, name=ra,  below=1.25cm] { };
        \node [shape=circle, fill=gray!50!white, draw, thick, inner sep=2pt, name=rb] at ($ (-1.299, 0.75) + (r.center) $) { };
        \node [shape=circle, fill=gray!50!white, draw, thick, inner sep=2pt, name=rc] at ($ (1.299, 0.75) + (r.center) $) { };

        \draw[-, red, thick] (ra) to (r.corner 5);
        \draw[-, red, thick] (ra) to (r.corner 4);

        \draw[-, red, thick] (rb) to (r.corner 3);
        \draw[-, red, thick] (rb) to (r.corner 2);

        \draw[-, red, thick] (rc) to (r.corner 6);
        \draw[-, red, thick] (rc) to (r.corner 1);
        
        \foreach \anchor/\placement/\n/\fill/\text/\op in
        {corner 1/below/1/white/black, corner 2/below/2/black/white, corner 3/right/3/white/black, corner 4/above/4/black/white, corner 5/above/5/white/black, corner 6/left/6/black/white}
        {
        \node[draw, circle, fill=\fill, inner sep = 1pt, text=\text] at (r.\anchor) {\tiny{$X$}};
        }

        \node[name = g2, regular polygon, regular polygon sides = 6, draw, inner sep = 0.5cm, fill=green, fill opacity=0.25, thick] at (11.5, 0) {};
        \node[shape=circle, fill=gray!50!white, draw, thick, inner sep=1.5pt, name=g2a] at (g2.center) {\tiny{$X_a$ }};
        
        \foreach \anchor/\placement/\n/\fill/\text/\op in
        {corner 1/below/1/white/black/$W^\dagger$, corner 2/below/2/black/white/$W$, corner 3/right/3/white/black/$W^\dagger$, corner 4/above/4/black/white/$W$, corner 5/above/5/white/black/$W^\dagger$, corner 6/left/6/black/white/$W$}
        {
        \node[draw, circle, fill=\fill, inner sep = 1pt, text=\text] at (g2.\anchor) {\tiny{\op}};
        }
        
    \end{tikzpicture}
    }
\end{equation}
Here, the vertex ancillas are pictured in grey.

We now introduce edge ancillas, which in this case are placed on all green edges (the edges forming the triangular sublattice defined by the vertex set ancillas). The new edge ancillas are entangled with the vertex ancillas using the CSS cluster entangler.
\begin{equation}
    \raisebox{-45pt}{
    \begin{tikzpicture}
        \node[name = g2, regular polygon, regular polygon sides = 6, draw, inner sep = 0.5cm, fill=green, fill opacity=0.25, thick] at (0, 0) {};
        \node[shape=circle, fill=white, draw, thick, inner sep=2.5pt, name=g2a] at (g2.center) { };
        \node[right, outer sep=1pt] at (g2.center) {\scriptsize\texttt{$X_a$}};

        \foreach \n/\placement in {0/right, 1/above right, 2/above left, 3/left, 4/below left, 5/below right}
        {
        \draw[dashed, thick, green!80!black] (g2a) to (\n*60:1.5);
        \filldraw[shift=(g2.center), fill=white, draw = green!80!black, thick] (\n*60:1.5) circle (3pt) node[\placement] {\scriptsize\texttt{$X$}};
        }
        
        \foreach \anchor/\placement/\n/\fill/\text/\op in
        {corner 1/below/1/white/black/$W^\dagger$, corner 2/below/2/black/white/$W$, corner 3/right/3/white/black/$W^\dagger$, corner 4/above/4/black/white/$W$, corner 5/above/5/white/black/$W^\dagger$, corner 6/left/6/black/white/$W$}
        {
        \node[draw, circle, fill=\fill, inner sep = 1pt, text=\text] at (g2.\anchor) {\tiny{\op}};
        }

        \node[name = g0, regular polygon, regular polygon sides = 6, draw, inner sep = 0.5cm, fill=green, fill opacity=0.25, thick] at (4, 0) {};
        \node[shape=circle, fill=gray!50!white, draw, thick, inner sep=2.5pt, name=g0a] at (g0.center) { };
        \node[above, outer sep=1pt] at (g0.center) {\scriptsize\texttt{$Z_a$}};

        \node[name = g1, regular polygon, regular polygon sides = 6, draw, inner sep = 0.5cm, fill=green, fill opacity=0.25, thick] at (8.5-2, 0) {};
        \node[shape=circle, fill=gray!50!white, draw, thick, inner sep=2.5pt, name=g1a] at (g1.center) { };
        \node[above, outer sep=1pt] at (g1.center) {\scriptsize\texttt{$Z_a$}};

        \draw[dashed, thick, green!80!black] (g0a) to (g1a);
        \filldraw[shift=(g0a.center), fill=white, draw = green!80!black, thick] (0:1.25) circle (3pt) node[above] {\scriptsize\texttt{$Z$}};

    \end{tikzpicture}
    }
\end{equation}

We can again pull a controlled-stabilizer out of the deformed stabilizers we currently have:
\begin{equation}
    \raisebox{-75pt}{
    \begin{tikzpicture}
        \node[name=g, regular polygon, regular polygon sides = 6, draw, inner sep = 0.5cm, fill=green, fill opacity=0.25, thick] at (0, 0) {};
        \node[shape=circle, fill=gray!50!white, draw, thick, inner sep=1.5pt, name=ga] at (g.center) { \tiny{$Z_a$}};
        
        \foreach \anchor/\placement/\n/\fill/\text/\op in
        {corner 1/below/1/white/black, corner 2/below/2/black/white, corner 3/right/3/white/black, corner 4/above/4/black/white, corner 5/above/5/white/black, corner 6/left/6/black/white}
        {
        \draw[-, red, thick] (ga) to (g.\anchor);
        \node[draw, circle, fill=\fill, inner sep = 1pt, text=\text] at (g.\anchor) {\tiny{$X$}};
        }

        \node at (2, 0) {$=$};

        \node[name=g2, regular polygon, regular polygon sides = 6, draw, inner sep = 0.5cm, fill=green, fill opacity=0.25, thick] at (4, 0) {};
        \node[shape=circle, fill=gray!50!white, draw, thick, inner sep=1.5pt, name=g2a] at (g2.center) { \tiny{$a$}};
        
        \foreach \anchor/\placement/\n/\fill/\text/\op in
        {corner 1/below/1/white/black, corner 2/below/2/black/white, corner 3/right/3/white/black, corner 4/above/4/black/white, corner 5/above/5/white/black, corner 6/left/6/black/white}
        {
        \node[draw, circle, fill=\fill, inner sep = 1pt, text=\text] at (g2.\anchor) {\tiny{$X$}};
        }
        
        \node at (5.75, 0) {$\times$};
        \node at (6.5, 0) {$C_a$};

        \draw [decorate, decoration = {calligraphic brace}, thick] (7, -1.25) -- (7, 1.25);
        \draw [decorate, decoration = {calligraphic brace, mirror}, thick] (10, -1.25) -- (10, 1.25);

        \node[name=g3, regular polygon, regular polygon sides = 6, draw, inner sep = 0.5cm, fill=green, fill opacity=0.25, thick] at (8.5, 0) {};
        \node[shape=circle, fill=gray!50!white, draw, thick, inner sep=2.5pt, name=g3a] at (g3.center) { \tiny{$a$}};
        \foreach \anchor/\placement/\n/\fill/\text/\op in
        {corner 1/below/1/white/black, corner 2/below/2/black/white, corner 3/right/3/white/black, corner 4/above/4/black/white, corner 5/above/5/white/black, corner 6/left/6/black/white}
        {
        \node[draw, circle, fill=\fill, inner sep = 1pt, text=\text] at (g3.\anchor) {\tiny{$Z$}};
        }

        \node at (2, -3) {$=$};

        \node[name=g2, regular polygon, regular polygon sides = 6, draw, inner sep = 0.5cm, fill=green, fill opacity=0.25, thick] at (4, -3) {};
        \node[shape=circle, fill=gray!50!white, draw, thick, inner sep=2.5pt, name=g2a] at (g2.center) { \tiny{$a$}};
        
        \foreach \anchor/\placement/\n/\fill/\text/\op in
        {corner 1/below/1/white/black, corner 2/below/2/black/white, corner 3/right/3/white/black, corner 4/above/4/black/white, corner 5/above/5/white/black, corner 6/left/6/black/white}
        {
        \node[draw, circle, fill=\fill, inner sep = 1pt, text=\text] at (g2.\anchor) {\tiny{$X$}};
        }

        \node at (9, -3) {(in the color code ground state manifold)};
        
    \end{tikzpicture}
    }
\end{equation}

\begin{equation}
    \raisebox{-90pt}{
    \begin{tikzpicture}
        
        \node[name= b, regular polygon, regular polygon sides = 6, draw, inner sep = 0.5cm, fill=blue, fill opacity=0.25, thick] at (0, 0) {};
        \draw[-, green!80!black, very thick] (b.corner 2) -- (b.corner 3);
        \draw[-, green!80!black, very thick] (b.corner 4) -- (b.corner 5);
        \draw[-, green!80!black, very thick] (b.corner 6) -- (b.corner 1);

        \filldraw[fill=white, draw = green!80!black, thick] (b.side 2) circle (3pt);
        \filldraw[fill=white, draw = green!80!black, thick] (b.side 4) circle (3pt);
        \filldraw[fill=white, draw = green!80!black, thick] (b.side 6) circle (3pt);
        
        \node at (b.center) [shape=circle, fill=white!50!gray, draw, thick, inner sep=1.5pt, name=ba,  above=1.25cm] {\tiny{$Z_a$}};
        \node [shape=circle, fill=white!50!gray, draw, thick, inner sep=1.5pt, name=bb] at ($ (1.299, -0.75) + (b.center) $) {\tiny{$Z_b$}};
        \node [shape=circle, fill=white!50!gray, draw, thick, inner sep=1.5pt, name=bc] at ($ (-1.299, -0.75) + (b.center) $) {\tiny{$Z_c$}};

        \draw[-, red, thick] (ba) to (b.corner 1);
        \draw[-, red, thick] (ba) to (b.corner 2);

        \draw[-, red, thick] (bb) to (b.corner 6);
        \draw[-, red, thick] (bb) to (b.corner 5);

        \draw[-, red, thick] (bc) to (b.corner 4);
        \draw[-, red, thick] (bc) to (b.corner 3);

        \foreach \anchor/\placement/\n/\fill/\text/\op in
        {corner 1/below/1/white/black, corner 2/below/2/black/white, corner 3/right/3/white/black, corner 4/above/4/black/white, corner 5/above/5/white/black, corner 6/left/6/black/white}
        {
        \node[draw, circle, fill=\fill, inner sep = 1pt, text=\text] at (b.\anchor) {\tiny{$X$}};
        }

        \node at (2.25, 0) {$=$};

        \node[name= b2, regular polygon, regular polygon sides = 6, draw, inner sep = 0.5cm, fill=blue, fill opacity=0.25, thick] at (4.75, 0) {};

        \node at (b2.center) [shape=circle, fill=white!50!gray, draw, thick, inner sep=1.5pt, name=b2a,  above=1.25cm] {\tiny{$Z_a$}};
        \node [shape=circle, fill=white!50!gray, draw, thick, inner sep=1.5pt, name=b2b] at ($ (1.299, -0.75) + (b2.center) $) {\tiny{$Z_b$}};
        \node [shape=circle, fill=white!50!gray, draw, thick, inner sep=1.5pt, name=b2c] at ($ (-1.299, -0.75) + (b2.center) $) {\tiny{$Z_c$}};

        \draw[-, red, thick] (b2a) to (b2.corner 1);
        \draw[-, red, thick] (b2a) to (b2.corner 2);
        \draw[-, red, thick] (b2a) to[out=180, in=180] (b2.corner 3);
        \draw[-, red, thick] (b2a) to[out=180, in=180] (b2.corner 4);
        \draw[-, red, thick] (b2a) to[out=0, in=0] (b2.corner 5);
        \draw[-, red, thick] (b2a) to[out=0, in=0] (b2.corner 6);

        \draw[-, red, thick] (b2.side 6) to[out=180, in=200] (b2.corner 6);
        \draw[-, red, thick] (b2.side 6) to[out=180, in=120] (b2.corner 5);

        \draw[-, red, thick] (b2.side 2) to[out=0, in=0] (b2.corner 3);
        \draw[-, red, thick] (b2.side 2) to[out=0, in=60] (b2.corner 4);

        \draw[-, green!80!black, very thick] (b2.corner 2) -- (b2.corner 3);
        \draw[-, green!80!black, very thick] (b2.corner 4) -- (b2.corner 5);
        \draw[-, green!80!black, very thick] (b2.corner 6) -- (b2.corner 1);

        \filldraw[fill=white, draw = green!80!black, thick] (b2.side 2) circle (3pt);
        \filldraw[fill=white, draw = green!80!black, thick] (b2.side 4) circle (3pt);
        \filldraw[fill=white, draw = green!80!black, thick] (b2.side 6) circle (3pt);

        \foreach \anchor/\placement/\n/\fill/\text/\op in
        {corner 1/below/1/white/black, corner 2/below/2/black/white, corner 3/right/3/white/black, corner 4/above/4/black/white, corner 5/above/5/white/black, corner 6/left/6/black/white}
        {
        \node[draw, circle, fill=\fill, inner sep = 1pt, text=\text] at (b2.\anchor) {\tiny{$X$}};
        }

        \node at (0, -3) {$=$};

        \node[name= b3, regular polygon, regular polygon sides = 6, draw, inner sep = 0.5cm, fill=blue, fill opacity=0.25, thick] at (3.75-2, -3) {};

        \draw[-, red, thick] (b3.side 6) to[out=180, in=200] (b3.corner 6);
        \draw[-, red, thick] (b3.side 6) to[out=180, in=120] (b3.corner 5);

        \draw[-, red, thick] (b3.side 2) to[out=0, in=0] (b3.corner 3);
        \draw[-, red, thick] (b3.side 2) to[out=0, in=60] (b3.corner 4);
        
        \draw[-, green!80!black, very thick] (b3.corner 2) -- (b3.corner 3);
        \draw[-, green!80!black, very thick] (b3.corner 4) -- (b3.corner 5);
        \draw[-, green!80!black, very thick] (b3.corner 6) -- (b3.corner 1);

        \filldraw[fill=white, draw = green!80!black, thick] (b3.side 2) circle (3pt) node {\tiny{$\beta$}};
        \filldraw[fill=white, draw = green!80!black, thick] (b3.side 4) circle (3pt) node {\tiny{$\alpha$}};
        \filldraw[fill=white, draw = green!80!black, thick] (b3.side 6) circle (3pt) node {\tiny{$\gamma$}};

        \foreach \anchor/\placement/\n/\fill/\text/\op in
        {corner 1/below/1/white/black, corner 2/below/2/black/white, corner 3/right/3/white/black, corner 4/above/4/black/white, corner 5/above/5/white/black, corner 6/left/6/black/white}
        {
        \node[draw, circle, fill=\fill, inner sep = 1pt, text=\text] at (b3.\anchor) {\tiny{$X$}};
        }

        \node at (b3.center) [shape=circle, fill=white!50!gray, draw, thick, inner sep=1.5pt, name=b3a,  above=1.25cm] {\tiny{$a$}};
        \node [shape=circle, fill=white!50!gray, draw, thick, inner sep=1.5pt, name=b3b] at ($ (1.299, -0.75) + (b3.center) $) {\tiny{$b$}};
        \node [shape=circle, fill=white!50!gray, draw, thick, inner sep=1.5pt, name=b3c] at ($ (-1.299, -0.75) + (b3.center) $) {\tiny{$c$}};

        \node at (5.25-2, -3) {$\times$};
        \node at (5.75-2, -3) {$C_a$};

        \draw [decorate, decoration = {calligraphic brace}, thick] (6.25-2, -1.25-3) -- (6.25-2, 1.25-3);
        \draw [decorate, decoration = {calligraphic brace, mirror}, thick] (7.75, -1.25-3) -- (7.75, 1.25-3);

        \node[name= b4, regular polygon, regular polygon sides = 6, draw, inner sep = 0.5cm, fill=blue, fill opacity=0.25, thick] at (6, -3) { };

        \foreach \anchor/\placement/\n/\fill/\text/\op in
        {corner 1/below/1/white/black, corner 2/below/2/black/white, corner 3/right/3/white/black, corner 4/above/4/black/white, corner 5/above/5/white/black, corner 6/left/6/black/white}
        {
        \node[draw, circle, fill=\fill, inner sep = 1pt, text=\text] at (b4.\anchor) {\tiny{$Z$}};
        }

        \node at (b4.center) [shape=circle, fill=white!50!gray, draw, thick, inner sep=1.5pt, name=b4a,  above=1.25cm] {\tiny{$a$}};
        \node [shape=circle, fill=white!50!gray, draw, thick, inner sep=1.5pt, name=b4b] at ($ (1.299, -0.75) + (b4.center) $) {\tiny{$b$}};
        \node [shape=circle, fill=white!50!gray, draw, thick, inner sep=1.5pt, name=b4c] at ($ (-1.299, -0.75) + (b4.center) $) {\tiny{$c$}};

        \node at (8.75, -3) {$=$};

        \node[name= b5, regular polygon, regular polygon sides = 6, draw, inner sep = 0.5cm, fill=blue, fill opacity=0.25, thick] at (10.75, -3) {};

        \draw[-, red, thick] (b5.side 6) to[out=180, in=200] (b5.corner 6);
        \draw[-, red, thick] (b5.side 6) to[out=180, in=120] (b5.corner 5);

        \draw[-, red, thick] (b5.side 2) to[out=0, in=0] (b5.corner 3);
        \draw[-, red, thick] (b5.side 2) to[out=0, in=60] (b5.corner 4);
        
        \draw[-, green!80!black, very thick] (b5.corner 2) -- (b5.corner 3);
        \draw[-, green!80!black, very thick] (b5.corner 4) -- (b5.corner 5);
        \draw[-, green!80!black, very thick] (b5.corner 6) -- (b5.corner 1);

        \filldraw[fill=white, draw = green!80!black, thick] (b5.side 2) circle (3pt) node {\tiny{$\beta$}};
        \filldraw[fill=white, draw = green!80!black, thick] (b5.side 4) circle (3pt) node {\tiny{$\alpha$}};
        \filldraw[fill=white, draw = green!80!black, thick] (b5.side 6) circle (3pt) node {\tiny{$\gamma$}};

        \foreach \anchor/\placement/\n/\fill/\text/\op in
        {corner 1/below/1/white/black, corner 2/below/2/black/white, corner 3/right/3/white/black, corner 4/above/4/black/white, corner 5/above/5/white/black, corner 6/left/6/black/white}
        {
        \node[draw, circle, fill=\fill, inner sep = 1pt, text=\text] at (b5.\anchor) {\tiny{$X$}};
        }

        \node at (b5.center) [shape=circle, fill=white!50!gray, draw, thick, inner sep=1.5pt, name=b5a,  above=1.25cm] {\tiny{$a$}};
        \node [shape=circle, fill=white!50!gray, draw, thick, inner sep=1.5pt, name=b5b] at ($ (1.299, -0.75) + (b5.center) $) {\tiny{$b$}};
        \node [shape=circle, fill=white!50!gray, draw, thick, inner sep=1.5pt, name=b5c] at ($ (-1.299, -0.75) + (b5.center) $) {\tiny{$c$}};

    \end{tikzpicture}
    }
\end{equation}
where $\alpha,\beta,\gamma$ are the gauge ancillas on  edges of three types of orientation respectively, and the red line is the $\CZ$ gate.
Note that the last equality again follows from the fact we are acting within the ground state manifold. There is a similar identity for the red hexagons we omit for brevity. 

After projecting the vertex ancilla in the $X$ basis, the final stabilizers take the following form: 
\begin{equation}
    \raisebox{-50pt}{
    \begin{tikzpicture}
        \node[name = g2, regular polygon, regular polygon sides = 6, draw, inner sep = 0.5cm, fill=green, fill opacity=0.25, thick] at (0, 0) {};
        \node[name=g2a] at (g2.center) { };

        \foreach \n/\placement in {0/right, 1/above right, 2/above left, 3/left, 4/below left, 5/below right}
        {
        \draw[dashed, thick, green!80!black] (g2a) to (\n*60:1.5);
        \filldraw[shift=(g2.center), fill=white, draw = green!80!black, thick] (\n*60:1.5) circle (3pt) node[\placement] {\scriptsize\texttt{$X$}};
        }
        
        \foreach \anchor/\placement/\n/\fill/\text/\op in
        {corner 1/below/1/white/black/$W^\dagger$, corner 2/below/2/black/white/$W$, corner 3/right/3/white/black/$W^\dagger$, corner 4/above/4/black/white/$W$, corner 5/above/5/white/black/$W^\dagger$, corner 6/left/6/black/white/$W$}
        {
        \node[draw, circle, fill=\fill, inner sep = 1pt, text=\text] at (g2.\anchor) {\tiny{\op}};
        }

        \node[name=v, regular polygon, regular polygon sides = 3, draw =green!80!black, dashed, thick, inner sep = 0.4cm] at (4, 0) { };
        
        \filldraw[shift=(v.center), fill=white, draw = green!80!black, thick] (v.side 1) circle (3pt) node[left, inner sep = 4pt] {\scriptsize\texttt{$Z$}};
        \filldraw[shift=(v.center), fill=white, draw = green!80!black, thick] (v.side 2) circle (3pt) node[below, inner sep = 4pt] {\scriptsize\texttt{$Z$}};
        \filldraw[shift=(v.center), fill=white, draw = green!80!black, thick] (v.side 3) circle (3pt) node[right, inner sep = 4pt] {\scriptsize\texttt{$Z$}};

    \end{tikzpicture}
    }
\end{equation}
\begin{equation}
    \raisebox{-30pt}{
    \begin{tikzpicture}
    
        \node[name= g, regular polygon, regular polygon sides = 6, draw, inner sep = 0.5cm, fill=green, fill opacity=0.25, thick] at (0, 0) { };

        \foreach \anchor/\placement/\n/\fill/\text/\op in
        {corner 1/below/1/white/black, corner 2/below/2/black/white, corner 3/right/3/white/black, corner 4/above/4/black/white, corner 5/above/5/white/black, corner 6/left/6/black/white}
        {
        \node[draw, circle, fill=\fill, inner sep = 1pt, text=\text] at (g.\anchor) {\tiny{$X$}};
        }

        \node[name= b5, regular polygon, regular polygon sides = 6, draw, inner sep = 0.5cm, fill=blue, fill opacity=0.25, thick] at (4, 0) {};

        \draw[-, red, thick] (b5.side 6) to[out=180, in=200] (b5.corner 6);
        \draw[-, red, thick] (b5.side 6) to[out=180, in=120] (b5.corner 5);

        \draw[-, red, thick] (b5.side 2) to[out=0, in=0] (b5.corner 3);
        \draw[-, red, thick] (b5.side 2) to[out=0, in=60] (b5.corner 4);
        
        \draw[-, green!80!black, very thick] (b5.corner 2) -- (b5.corner 3);
        \draw[-, green!80!black, very thick] (b5.corner 4) -- (b5.corner 5);
        \draw[-, green!80!black, very thick] (b5.corner 6) -- (b5.corner 1);

        \filldraw[fill=white, draw = green!80!black, thick] (b5.side 2) circle (3pt) node {\tiny{$\beta$}};
        \filldraw[fill=white, draw = green!80!black, thick] (b5.side 4) circle (3pt) node {\tiny{$\alpha$}};
        \filldraw[fill=white, draw = green!80!black, thick] (b5.side 6) circle (3pt) node {\tiny{$\gamma$}};

        \foreach \anchor/\placement/\n/\fill/\text/\op in
        {corner 1/below/1/white/black, corner 2/below/2/black/white, corner 3/right/3/white/black, corner 4/above/4/black/white, corner 5/above/5/white/black, corner 6/left/6/black/white}
        {
        \node[draw, circle, fill=\fill, inner sep = 1pt, text=\text] at (b5.\anchor) {\tiny{$X$}};
        }

        \node[name= r, regular polygon, regular polygon sides = 6, draw, inner sep = 0.5cm, fill=red, fill opacity=0.25, thick] at (8, 0) {};

        \draw[-, red, thick] (r.side 5) to[out=180, in=200] (r.corner 1);
        \draw[-, red, thick] (r.side 5) to[out=180, in=120] (r.corner 6);

        \draw[-, red, thick] (r.side 3) to[out=0, in=0] (r.corner 2);
        \draw[-, red, thick] (r.side 3) to[out=0, in=60] (r.corner 3);
        
        \draw[-, green!80!black, very thick] (r.corner 1) -- (r.corner 2);
        \draw[-, green!80!black, very thick] (r.corner 3) -- (r.corner 4);
        \draw[-, green!80!black, very thick] (r.corner 5) -- (r.corner 6);

        \filldraw[fill=white, draw = green!80!black, thick] (r.side 1) circle (3pt) node {\tiny{$\alpha$}};
        \filldraw[fill=white, draw = green!80!black, thick] (r.side 3) circle (3pt) node {\tiny{$\gamma$}};
        \filldraw[fill=white, draw = green!80!black, thick] (r.side 5) circle (3pt) node {\tiny{$\beta$}};

        \foreach \anchor/\placement/\n/\fill/\text/\op in
        {corner 1/below/1/white/black, corner 2/below/2/black/white, corner 3/right/3/white/black, corner 4/above/4/black/white, corner 5/above/5/white/black, corner 6/left/6/black/white}
        {
        \node[draw, circle, fill=\fill, inner sep = 1pt, text=\text] at (r.\anchor) {\tiny{$X$}};
        }
        
    \end{tikzpicture}
    }
\end{equation}
We have omitted the $Z$ hexagon stabilizers, which remain unchanged throughout the gauging procedure.

\subsection{Gauging bivariate bicycle codes}
We now turn to some higher rate examples. In particular, we focus on the $[[144, 12, 12]]$ (Gross) code \cite{Bravyi2024faulttolerant}, which has received much attention recently, as it trades a modest amount of non-local connectivity for a large boost in encoding rate.

\subsubsection{Construction of self-dual bivariate bicycle codes}

In this example, we gauge a transversal symmetry on a \textit{self-dual bivariate bicycle} code with $k$ logical qubits. The transversal gate that we gauge is
\begin{equation}
    \T = \bar{W} = \prod_{A} W \prod_B \tilde{W}, \quad W = XS, \quad \tilde{W} = XS^\dagger
\end{equation} where, as with the color code, $A, B$ denote sublattices of a bipartite graph. 

Several papers have already considered self-dual versions of the bivariate bicycle code \cite{LiangChen2025Bivariate, Xu2025batched}.  We work with a different construction which is a generalization of the octagonal/square color code \cite{Bombin2006TopologicalQuantumDistillation}. The procedure we use to construct a self-dual code is generic and maps any initial CSS code to a new self-dual CSS code \cite{Criger2016NoiseThresholds422ToricCode, Bravyi2010MajoranaFermionCodes}; the action of $\bar{XS}$ on the new self-dual code is equivalent to the transversal $\CNOT$ between the original codeblocks (we work through the $\CNOT$ example below). 

Let the starting CSS code $C$ be an $[[n,k,d]]$ code with $X$ stabilizers $S_{X,1}, \cdots, S_{X, m_x}$ and $Z$ stabilizers $S_{Z,1}, \cdots, S_{Z, m_z}$.  Denote by $C_H$ the same code except with all stabilizers mapped to $S_{w,m_i} \rightarrow H^{\otimes n} S_{w,m_i} H^{\otimes n}$.  We also consider a subcode $C_{422}$ which is the self-dual $[[4,2,2]]$ code with stabilizers $X_1 X_2 X_3 X_4$ and $Z_1 Z_2 Z_3 Z_4$.  The logical operators of $C_{422}$ are $X_{L,1} = X_1 X_2$, $Z_{L,1} = Z_1 Z_3$, $X_{L,2} = X_1 X_3$, and $Z_{L,2} = Z_1 Z_2$.  

We form the self-dual code by pairing corresponding qubits on codeblocks $C$ and $C_H$, then encoding this pair of qubits into $C_{422}$, such that the first logical of $C_{422}$ corresponds to qubits of $C$, while the second logical corresponds to qubits of $C_H$. In other words, the new set of stabilizers contains
\begin{equation}
    \left \{ X_{j,1}X_{j,2}X_{j,3}X_{j,4}, ~~ Z_{j,1}Z_{j,2}Z_{j,3}Z_{j,4}\right \}_{j\in [n]}
\end{equation}
\begin{equation}
\begin{aligned}
    &\left \{\prod_{j\in \mathrm{supp}(S_{X,i})} X_{L,1,j} \right \}_i
    \quad \left \{\prod_{j\in \mathrm{supp}(S_{Z,i})} Z_{L,1,j}\right \}_i  \\
   &\left \{\prod_{j\in \mathrm{supp}(S_{X,i})} Z_{L,2,j}\right \}_i
   \quad \left \{\prod_{j\in \mathrm{supp}(S_{Z,i})} X_{L,2,j}\right \}_i 
\end{aligned}
\end{equation}
where $X_{L,1,j}$ corresponds to the first logical of the $[[4,2,2]]$ code that took the place of qubit $j$ in the original code (similarly for Z).  $X_{j,1}$ ($Z_{j,1}$) corresponds to the X (Z) operator on the first qubit in the $[[4,2,2]]$ code at location $j$.

We can quickly see that all of these stabilizers commute. The first and second group commute because the X and Z operators of the original code commute and this is preserved for the concatenated code since the two rows reference the same logical on the subcode (similarly for the third and fourth row).  The first and third group commute because they address different logicals in the subcode (same for second and fourth row).  The last row commutes with everything because the stabilizers of the subcode commute with logicals of the subcode.  The code is clearly self-dual because the first group is the Hadamard of the third (same for second and fourth) and the $[[4,2,2]]$ code is self-dual on its own.  

The code realized by this mapping takes an initial $[[n,k,d]]$ CSS code to a self-dual $[[4n, 2k, 2d]]$ CSS code.  The distance is doubled because each qubit in the support of an original logical is replaced by two qubits in the $[[4,2,2]]$ code.

\begin{figure}[htbp]
    \centering
    \includegraphics[width=0.7\linewidth]{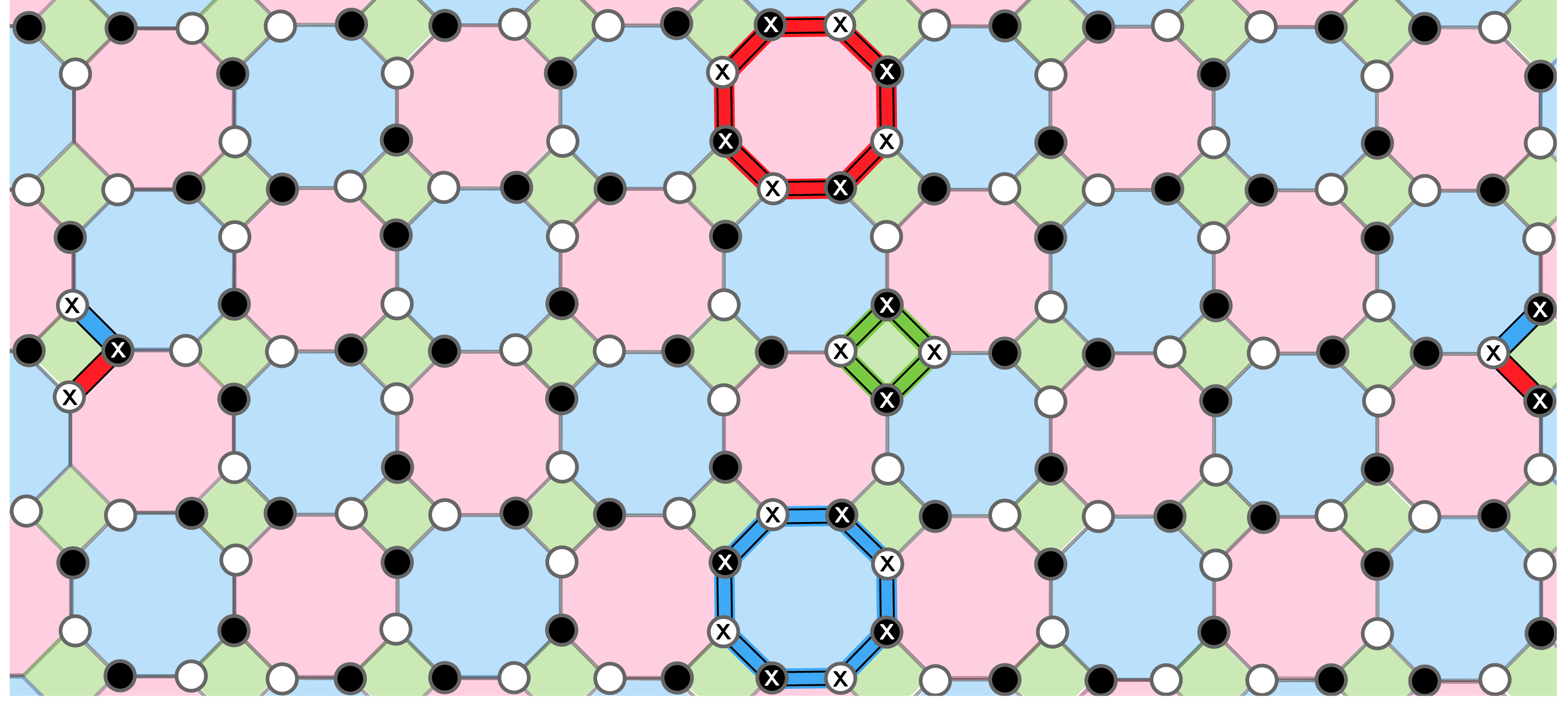}
    \caption{\textbf{Self-dual bivariate bicycle code}:  Lattice of the self-dual bivariate bicycle code, with the $A$ sublattice denoted by black circles, and the $B$ sublattice by white circles.  The red and blue highlights show the support of the stabilizers.  All stabilizers are present as both X and Z type. }
    \label{fig:bivariate_self_dual}
\end{figure}

If this construction is applied to the toric code, one generates the octagonal/square color code.  To realize a higher rate quantum LDPC code we choose to apply this general construction to the $[[144,12,12]]$ code of \cite{Bravyi2024faulttolerant} realizing a self-dual $[[576, 24, 24]]$ code. The stabilizers of the code are shown in Fig.~\ref{fig:bivariate_self_dual}: the code is CSS with two kinds of weight-12 checks, and one kind of weight-4 check. 

Next we characterize the logical action of the transversal gate $\bar{W}$. The $X$ logicals of the self-dual code have a representative which is computed from the original $X$ logical $\bar{X}_L = \prod_{j\in \mathrm{supp}(\bar{X}_L)} X_j$ as $\bar{X}_{L,1} = \prod_{j\in \mathrm{supp}(\bar{X}_L)} X_{L,1,j}$ and $\bar{X}_{L,2} = \prod_{j\in \mathrm{supp}(\bar{Z}_L)} X_{L,2,j}$.  The same type of relation holds for the logical $Z$ operators.  As a consequence, the transversal $W$ has the following action:
\begin{align}
    \prod_{j=1}^n \CZ_{(L,1,j), (L,2,j)},
\quad \text{equivalent to}\quad
    \prod_{l=1}^k \CZ_{(\bar{1},l),(\bar{2},l)},
\end{align}
a product of logical $\CZ$ gates of the self-dual code. This set of logical $\CZ$ gates can be directly mapped back onto $C$ and $C_H$, as logicals $(\bar{1}, \cdot)$ correspond to logicals of $C$, and logicals $(\bar{2}, \cdot)$ correspond to $C_H$. As $C_H$ is related to $C$ by a Hadamard, the action is also equivalent to $\CNOT$ between two copies of $C$.

\subsubsection{Gauging \texorpdfstring{$XS$}{XS} on a self-dual BB code} 
Below, we separate the $X$($Z$) stabilizers based on the color of the face on which they are supported:
\begin{equation}
    \green{S^{(g)}_{i, X(Z)}} \in \green{\mathcal{S}^{(g)}_{X(Z)}}, \quad \red{S^{(r)}_{i, X(Z)}} \in \red{\mathcal{S}^{(r)}_{X(Z)}}, \quad \blue{S^{(b)}_{i,X(Z)}}\in \blue{\mathcal{S}^{(b)}_{X(Z)}}
\end{equation}

Starting from the original code we apply the circuit-based gauging procedure.  As a first step we identify the vertex sets of the gauged code. Similar to the color code example above, we partition the qubits by green faces: i.e. the support of a $\green{S^{(g)}_{i, X}}$ stabilizer forms our vertex set. To begin the circuit-based gauging procedure we add a vertex ancilla $v$, initialized in the state $|+\rangle$, to each vertex set. We will denote this set of vertex ancillas by $V$. Since these vertex ancillas are in the state $|+\rangle$ they are stabilized by the single-qubit $X$ operator.  Hence the stabilizers at this point are those of the original code along with $\mathcal{S}_Q = \{X_i \mid i \in V\}$.   

Next we apply $CW$ (and $C\tilde{W}$) gates from each vertex ancilla $v \in V$ to the code qubits belonging to the corresponding vertex set.  Making use of the following relations
\begin{align}
    CW_{a\rightarrow b}\,X_b\,CW_{a\rightarrow b}^\dagger &= C(i Z)_{a\rightarrow b}\, X_b, \\
    C\tilde{W}_{a\rightarrow b}\,X_b\,C\tilde{W}_{a\rightarrow b}^\dagger &= C(-i Z)_{a\rightarrow b}\, X_b
\end{align}
we determine the stabilizers after the application of these gates.  The $Z$ stabilizers of the original code, $\green{\mathcal{S}^{(g)}_Z}, \red{\mathcal{S}^{(r)}_Z}, \blue{\mathcal{S}^{(b)}_Z}$, are unaffected by this circuit transformation.  A new set of stabilizers corresponding to each stabilizer in $\green{\mathcal{S}^{(g)}_X}$ is given by \begin{equation}
    \left \{\prod_{j \in \mathrm{supp}(\green{S^{(g)}_{X,i}})} X_j \, \CZ_{\vertexset{\green{S^{(g)}_{X,i}}}\rightarrow j} \mathrel{\bigg|} i \in V \right \}
\end{equation} where $\vertexset{\green{S^{(g)}_{X,i}}}$ returns the vertex ancilla qubit for a given green stabilizer.  We also add the stabilizers 
\begin{equation}
    \green{A_V'} = \left \{X_i \prod_{j \in \mathrm{supp}(\green{S^{(g)}_{X,i}}) \cap A} W_{j}  \prod_{k \in \mathrm{supp}(\green{S^{(g)}_{X,i}}) \cap B} \tilde{W}_{k} \mathrel{\bigg|} i \in V \right \}
\end{equation}
Here, $A$ is the sublattice where the transversal gate acts as $W$, while $B$ is the sublattice where it acts as $\tilde{W}$. Lastly we need to compute the updated $X$ stabilizers for the octagonal (red, blue) stabilizers.  These are described by 
\begin{equation}
    \left \{ \prod_{j \in \mathrm{supp}(\red{S^{(r)}_{X,i}})} X_j \CZ_{\Sigma^{-1}(j)\rightarrow j} \right \}, \quad \left \{ \prod_{j \in \mathrm{supp}(\blue{S^{(b)}_{X,i}})} X_j \CZ_{\Sigma^{-1}(j)\rightarrow j} \right \}
\end{equation}
In other words, we dress the $X$ stabilizers with $\CZ$ gates supported on each qubit and its corresponding vertex ancilla.  

\begin{figure}[htbp]
    \centering
    \includegraphics[width=0.9\linewidth]{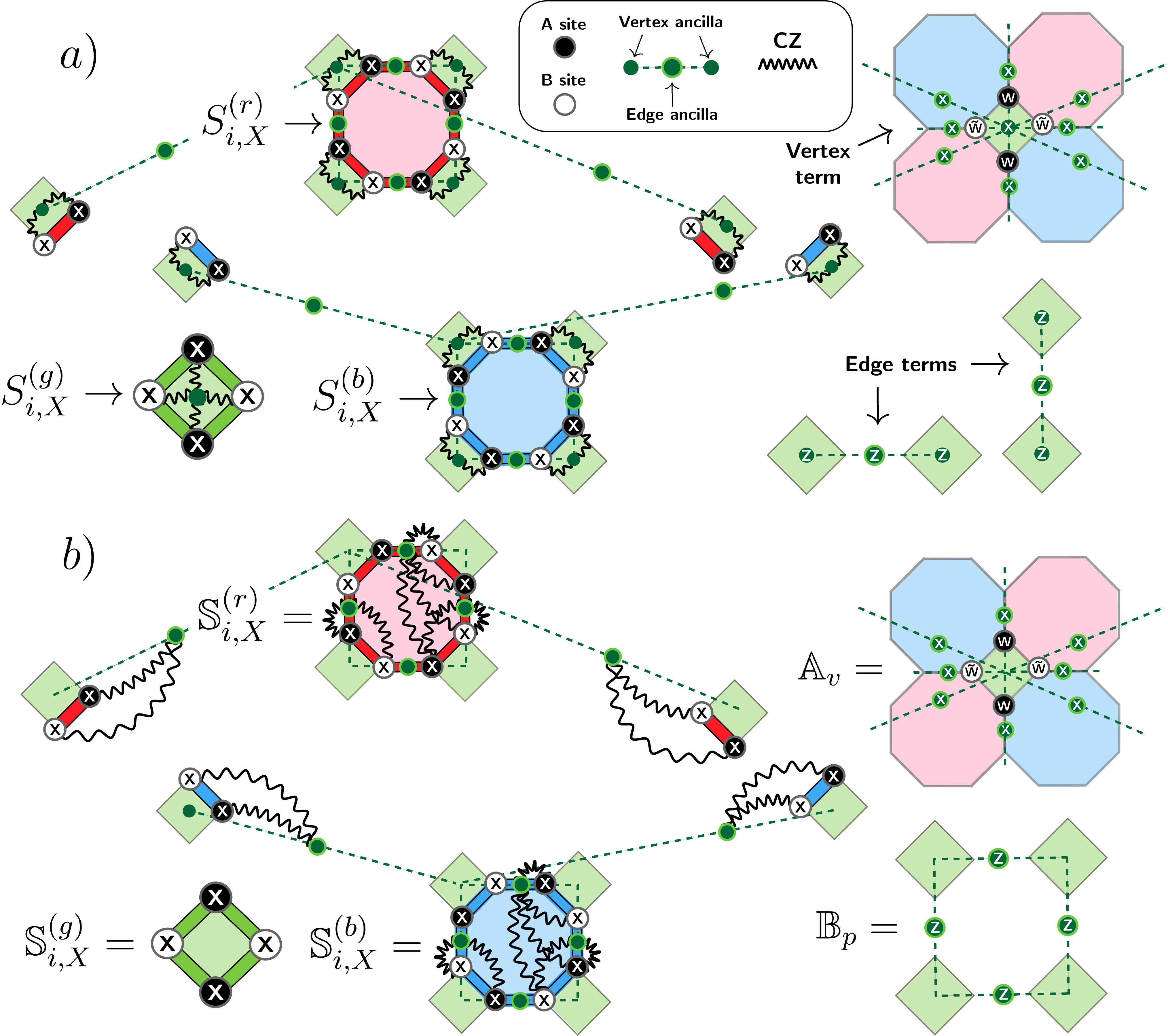}
    \caption{\textbf{Gauged self-dual bivariate bicycle code}:  We show selected stabilizers at different stages of the gauging procedure. Physical qubits are placed at the vertices of the underlying 4.8.8 tiling. (a) Intermediate step where we have added vertex ancillas, performed the entangling gates between vertex ancillas and neighboring qubits, added new edges and gauge qubits, and performed the cluster state entangler. (b) Final gauged code. Squigly lines represent $\CZ$ gates that result from dressing.  }
    \label{fig:gauged_bivariate_self_dual}
\end{figure}

Next we add gauge qubits into the system on the outgoing edges from the vertex sets.  The choice of edges here is up to the convention and ours is shown in Fig.~\ref{fig:gauged_bivariate_self_dual}; we have set the gauging graph to be the square sublattice naturally formed by the green faces with additional long range edges set by the connectivity of the octagonal stabilizers. Let the set of gauge qubits be $G$ and $\partial$ the mapping from gauge qubits to the neighboring vertex ancillas in $V$. The gauge qubits are initialized in the state $|0\rangle$. We entangle all of the gauge qubits with the neighboring vertex ancillas using the CSS cluster entangler. In addition to adding the cluster state edge stabilizers given by 
\begin{equation}
    \left \{Z_i \prod_{j\in \delta(i)} Z_{j}\mid i \in G \right \},
\end{equation}
the $\green{A_V'}$ are modified to the form 
\begin{equation}
    \green{A_V''} = \left \{X_i \prod_{\ell \mid i \in \delta(\ell)} X_{\ell} \prod_{j \in \mathrm{supp}(\green{S^{(g)}_{X,i}}) \cap A} W_{j}  \prod_{k \in \mathrm{supp}(\green{S^{(g)}_{X,i}}) \cap B} \tilde{W}_{k} \mathrel{\bigg|} i \in V \right \}
\end{equation}
The other stabilizers are untouched by the cluster state entangler. The stabilizers at this point are shown in Fig.~\ref{fig:gauged_bivariate_self_dual}a, with the omission of the $S_Z$ stabilizers which have not changed.   

The last step of constructing the gauged code is to measure all of the vertex ancillas in the $X$ basis.  We aim to find new stabilizers which commute with this measurement operation (perhaps up to other stabilizers). The initial $Z$ stabilizers of the code trivially commute with this measurement, as they were untouched by the gauging circuit. The $A_V''$ simply have their $X$ support on the vertex ancilla removed: 
\begin{equation}
    \green{A_V''} \to \green{\av} = \left \{ \prod_{\ell \mid i \in \delta(\ell)} X_{\ell} \prod_{j \in \mathrm{supp}(\green{S^{(g)}_{X,i}}) \cap A} W_{j}  \prod_{k \in \mathrm{supp}(\green{S^{(g)}_{X,i}}) \cap B} \tilde{W}_{k} \mathrel{\bigg|} i \in V \right \}
\end{equation}
Additionally we have loop stabilizers formed from products of the edge stabilizers around every cycle of the gauging graph: 
\begin{equation}
    \bp = \prod_{i \mid \partial(i) = 0} Z_i
\end{equation}
To determine the final form of the other stabilizers, we will again take advantage of Eq.~\ref{eq:CZ-deformation}. We can pull a controlled-$\green{S^{(g)}_{Z, i}}$ stabilizer out of each $\green{S^{(g)}_{X, i}}$: 
\begin{equation}
\begin{aligned}
    \left \{\prod_{j \in \mathrm{supp}(\green{S^{(g)}_{X,i}})} X_j \, \CZ_{\vertexset{\green{S^{(g)}_{X,i}}}\rightarrow j} \mathrel{\bigg|} i \in V \right \} &\rightarrow \left \{\prod_{j \in \mathrm{supp}(\green{S^{(g)}_{X,i}})} X_j \, \mathsf{C}_i\green{S^{(g)}_{Z, i}}\mathrel{\bigg|} i \in V \right \}\\
    &\xrightarrow{\green{S^{(g)}_{Z, i}} = +1} \left \{\prod_{j \in \mathrm{supp}(\green{S^{(g)}_{X,i}})} X_j \mathrel{\bigg|} i \in V \right \}
\end{aligned}
\end{equation}
Similarly, we can pull a controlled-$\red{S^{(r)}_{Z, i}} (\blue{S^{(b)}_{Z, i}})$ out of the deformed $\red{S^{(r)}_{X, i}} (\blue{S^{(b)}_{X, i}})$. However, since these stabilizers are supported between multiple vertex sets, when we pull out the controlled-stabilizers, we will leave behind some $\CZ$ gates between the original qubits and the edge ancilla. 

The final set of gauged stabilizers are pictured in Fig.~\ref{fig:gauged_bivariate_self_dual}b).

As an aside, we note that gauging the $XS$ symmetry of the color code yields a topological code with the same properties as the quantum double of $D_4$. This is a consequence of the fact $D_4 = \mathbb{Z}_2 \times \mathbb{Z}_2 \rtimes \mathbb{Z}_2$ .\footnote{Usually, we write $D_4 = \mathbb{Z}_4 \rtimes \mathbb{Z}_2$, but it can equivalently be expressed as $\mathbb{Z}_2 \times \mathbb{Z}_2 \rtimes \mathbb{Z}_2$, as both $\mathbb{Z}_4$ and $\mathbb{Z}_2 \times \mathbb{Z}_2$ are normal subgroups.} In essence, when we gauge, we implement this semi-direct product \cite{Tantivasadakarn_2023}. The color code is equivalent to two copies of the $\mathbb{Z}_2$ quantum double, more commonly known as the toric code--- gauging a layer-swap $\mathbb{Z}_2$ symmetry on this model leads to a $D_4$ anyon model. 

Recent work~\cite{LiangChen2025Bivariate} has shown that placing the stabilizers of the $\llbracket 144, 12, 12\rrbracket$ code on an infinite square lattice yields a code equivalent to 8 copies of the toric code. As we used two copies of the input BB code, we should expect the self-dual code considered here to be equivalent to 16 copies of the toric code. Thus, we can conjecture that the $XS$-gauged model has topological order described by $\mathbb{Z}_2^{16} \rtimes \mathbb{Z}_2$.

\subsubsection{Gauging \texorpdfstring{$\CNOT$}{CNOT} on bivariate bicycle codes}
Here we gauge the transversal $\CNOT$ between a pair of bivariate bicycle codes\footnote{We note that this example was also discussed in Ref.~\cite{christos2026nonabelianquantumlowdensityparity}.}. The bivariate bicycle code is a CSS code and hence admits a transversal $\CNOT$ gate.  By gauging the transversal $\CNOT$ symmetry we are measuring the operator $\CNOT^{\otimes n}$; this does not allow us to individually resolve the value of the $\CNOT$ measurement for each logical qubit--- this is not an addressable measurement.

\begin{figure}[htbp]
    \centering
    \includegraphics[width=0.75\linewidth]{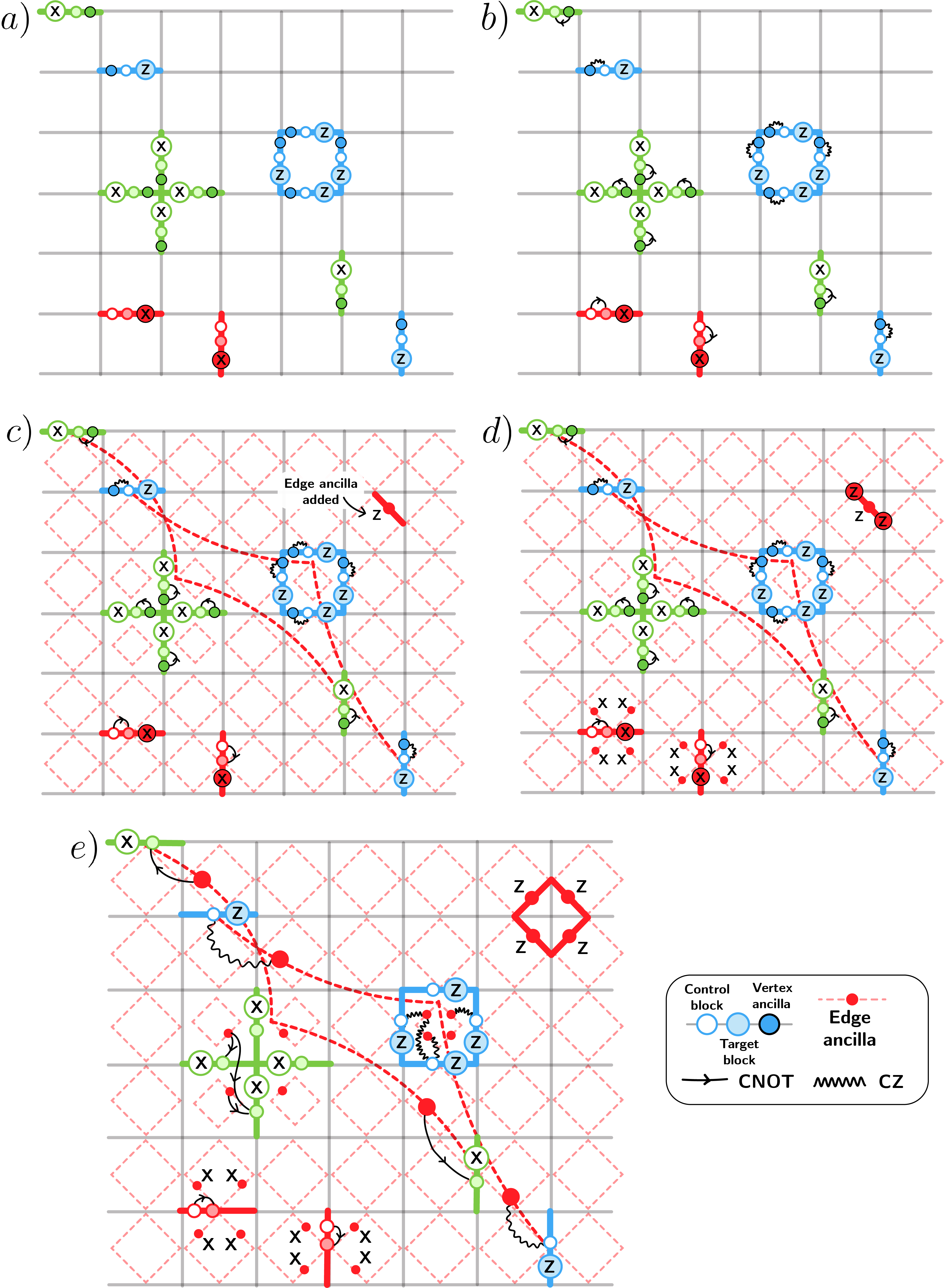}
    \caption{\textbf{Gauging $\CNOT$ between bivariate bicycle codes: } We show all the steps for gauging a $\CNOT$ between two bivariate bicycle codes.  In this example a gauging vertex is associated to each pair of qubits in the support of a $\CNOT$. (a) We show control-block $X$ stabilizers (green), target-block $Z$ stabilizers (blue), and vertex ancilla (red). (b) Stabilizers after performing the $CCNOT$ operation between vertex ancillas and the qubits in both codeblocks. (c) The gauging graph is pictured in red dashed lines (only some of the long-range connections are pictured and the lattice is slightly deformed around the support of the stabilizers purely for visual clarity), edge ancillas initialized in $\ket{0}$. (d) Stabilizers after the cluster state entangler. (e) Final stabilizers after projecting vertex ancilla onto $\ket{+}$.}
    \label{fig:bb_cnot_gauging}
\end{figure}

We pick vertex sets consisting of pairs of corresponding qubits in the two codeblocks; the gauging operation is shown step-by-step in Fig.~\ref{fig:bb_cnot_gauging}, with Fig.~\ref{fig:bb_cnot_gauging}e) depicting the stabilizers of the gauged code.

\subsection{Gauging qudit symmetries}
Here, to exemplify the general gauging procedure, we consider examples of  non-$\mathbb{Z}_2$ Clifford symmetries on qubit and qudit codes. 

\subsubsection{Gauging \texorpdfstring{$HS$}{HS} on color codes and BB codes}
The color code, being self-dual, possesses a transversal facet gate $HS$. This is an order 3 permutation that takes $X \to Y \to Z$. As with the $XS$ gate, the physical representation of the logical facet gate $F = HS$ has different action on the $A$ and $B$ sublattices: $\T = \prod_{A} F \prod_B \tilde{F}$, where $\tilde{F} = HS^\dagger$. The facet gate is a $\mathbb{Z}_3$ symmetry: $(HS)^3 = (HS^\dagger)^3 = \id$. Thus, we use \emph{qutrit} ancillas to perform the gauging. We write qutrit operators as $\mathcal{X}, \mathcal{Z}$: $\mathcal{Z}\mathcal{X} = e^{2\pi i /3} \mathcal{XZ}$. As we are gauging a $\mathbb{Z}_3$ symmetry on top of a code with associated group $\mathbb{Z}_2 \times \mathbb{Z}_2$ (being related to two copies of the $\mathbb{Z}_2$ toric code), we should expect the resulting model to be described by $(\mathbb{Z}_2 \times \mathbb{Z}_2) \rtimes \mathbb{Z}_3 \cong A_4$.

We make the same choice of vertex set as the prior example: each green hexagon. The control gate we use for the gauging is a mixed qutrit-qubit gate, with the qutrit acting as the control and the qubit as the target:
\begin{equation}
    \mathcal{C}_i F_j = \sum_{a = 0}^2 \dyad{a}_i \otimes F^a_j
\end{equation}

After gauging, we find the following checks:
\begin{equation}
    \raisebox{-50pt}{
    \begin{tikzpicture}
        \node[name = g2, regular polygon, regular polygon sides = 6, draw, inner sep = 0.5cm, fill=green, fill opacity=0.25, thick] at (0, 0) {};
        \node[name=g2a] at (g2.center) { };

        \foreach \n/\placement in {0/right, 1/above right, 2/above left}
        {
        \draw[dashed, thick, green!80!black] (g2a) to (\n*60:1.5);
        \filldraw[shift=(g2.center), fill=white, draw = green!80!black, thick] (\n*60:1.5) circle (3pt) node[\placement] {\scriptsize\texttt{$\mathcal{X}$}};
        }

        \foreach \n/\placement in {3/left, 4/below left, 5/below right}
        {
        \draw[dashed, thick, green!80!black] (g2a) to (\n*60:1.5);
        \filldraw[shift=(g2.center), fill=white, draw = green!80!black, thick] (\n*60:1.5) circle (3pt) node[\placement] {\scriptsize\texttt{$\mathcal{X}^\dagger$}};
        }
        
        \foreach \anchor/\placement/\n/\fill/\text/\op in
        {corner 1/below/1/white/black/$\tilde{F}$, corner 2/below/2/black/white/$F$, corner 3/right/3/white/black/$\tilde{F}$, corner 4/above/4/black/white/$F$, corner 5/above/5/white/black/$\tilde{F}$, corner 6/left/6/black/white/$F$}
        {
        \node[draw, circle, fill=\fill, inner sep = 1pt, text=\text] at (g2.\anchor) {\tiny{\op}};
        }

        \node[name=v, regular polygon, regular polygon sides = 3, draw =green!80!black, dashed, thick, inner sep = 0.4cm] at (4, 0) { };
        
        \filldraw[shift=(v.center), fill=white, draw = green!80!black, thick] (v.side 1) circle (3pt) node[left, inner sep = 4pt] {\scriptsize\texttt{$\mathcal{Z}^\dagger$}};
        \filldraw[shift=(v.center), fill=white, draw = green!80!black, thick] (v.side 2) circle (3pt) node[below, inner sep = 4pt] {\scriptsize\texttt{$\mathcal{Z}$}};
        \filldraw[shift=(v.center), fill=white, draw = green!80!black, thick] (v.side 3) circle (3pt) node[right, inner sep = 4pt] {\scriptsize\texttt{$\mathcal{Z}$}};

        \node[name=v, regular polygon, regular polygon sides = 3, draw =green!80!black, dashed, thick, inner sep = 0.4cm, rotate=180] at (7, 0.5) { };
        
        \filldraw[shift=(v.center), fill=white, draw = green!80!black, thick] (v.side 1) circle (3pt) node[right, inner sep = 4pt] {\scriptsize\texttt{$\mathcal{Z}^\dagger$}};
        \filldraw[shift=(v.center), fill=white, draw = green!80!black, thick] (v.side 2) circle (3pt) node[above, inner sep = 4pt] {\scriptsize\texttt{$\mathcal{Z}$}};
        \filldraw[shift=(v.center), fill=white, draw = green!80!black, thick] (v.side 3) circle (3pt) node[left, inner sep = 4pt] {\scriptsize\texttt{$\mathcal{Z}$}};

    \end{tikzpicture}
    }
\end{equation}
\begin{equation}
    \raisebox{-30pt}{
    \begin{tikzpicture}
        \node[name= g, regular polygon, regular polygon sides = 6, draw, inner sep = 0.5cm, fill=green, fill opacity=0.25, thick] at (0, 0) { };

        \foreach \anchor/\placement/\n/\fill/\text/\op in
        {corner 1/below/1/white/black, corner 2/below/2/black/white, corner 3/right/3/white/black, corner 4/above/4/black/white, corner 5/above/5/white/black, corner 6/left/6/black/white}
        {
        \node[draw, circle, fill=\fill, inner sep = 1pt, text=\text] at (g.\anchor) {\tiny{$X$}};
        }

        \node at (1.5, 0) {$+$};

        \node[name= g, regular polygon, regular polygon sides = 6, draw, inner sep = 0.5cm, fill=green, fill opacity=0.25, thick] at (3, 0) { };

        \foreach \anchor/\placement/\n/\fill/\text/\op in
        {corner 1/below/1/white/black, corner 2/below/2/black/white, corner 3/right/3/white/black, corner 4/above/4/black/white, corner 5/above/5/white/black, corner 6/left/6/black/white}
        {
        \node[draw, circle, fill=\fill, inner sep = 1pt, text=\text] at (g.\anchor) {\tiny{$Y$}};
        }

        \node at (4.5, 0) {$+$};

        \node[name= g, regular polygon, regular polygon sides = 6, draw, inner sep = 0.5cm, fill=green, fill opacity=0.25, thick] at (6, 0) { };

        \foreach \anchor/\placement/\n/\fill/\text/\op in
        {corner 1/below/1/white/black, corner 2/below/2/black/white, corner 3/right/3/white/black, corner 4/above/4/black/white, corner 5/above/5/white/black, corner 6/left/6/black/white}
        {
        \node[draw, circle, fill=\fill, inner sep = 1pt, text=\text] at (g.\anchor) {\tiny{$Z$}};
        }
    \end{tikzpicture}
    }
\end{equation}
\begin{equation}
    \raisebox{-30pt}{
    \begin{tikzpicture}
        
        \node[name= b5, regular polygon, regular polygon sides = 6, draw, inner sep = 0.5cm, fill=blue, fill opacity=0.25, thick] at (0, 0) {};

        \draw[-, red, thick] (b5.side 6) to[out=180, in=200] (b5.corner 6);
        \draw[-, red, thick] (b5.side 6) to[out=180, in=120] (b5.corner 5);

        \draw[-, red, thick] (b5.side 2) to[out=0, in=0] (b5.corner 3);
        \draw[-, red, thick] (b5.side 2) to[out=0, in=60] (b5.corner 4);
        
        \draw[-, green!80!black, very thick] (b5.corner 2) -- (b5.corner 3);
        \draw[-, green!80!black, very thick] (b5.corner 4) -- (b5.corner 5);
        \draw[-, green!80!black, very thick] (b5.corner 6) -- (b5.corner 1);

        \filldraw[fill=white, draw = green!80!black, thick] (b5.side 2) circle (3pt) node {\tiny{$\beta$}};
        \filldraw[fill=white, draw = green!80!black, thick] (b5.side 4) circle (3pt) node {\tiny{$\alpha$}};
        \filldraw[fill=white, draw = green!80!black, thick] (b5.side 6) circle (3pt) node {\tiny{$\gamma$}};

        \foreach \anchor/\placement/\n/\fill/\text/\op in
        {corner 1/below/1/white/black, corner 2/below/2/black/white, corner 3/right/3/white/black, corner 4/above/4/black/white, corner 5/above/5/white/black, corner 6/left/6/black/white}
        {
        \node[draw, circle, fill=\fill, inner sep = 1pt, text=\text] at (b5.\anchor) {\tiny{$X$}};
        }

        \node at (1.5, 0) {$+$};

        \node[name= b5, regular polygon, regular polygon sides = 6, draw, inner sep = 0.5cm, fill=blue, fill opacity=0.25, thick] at (3, 0) {};

        \draw[-, green, thick] (b5.side 6) to[out=180, in=200] (b5.corner 6);
        \draw[-, green, thick] (b5.side 6) to[out=180, in=120] (b5.corner 5);

        \draw[-, green, thick] (b5.side 2) to[out=0, in=0] (b5.corner 3);
        \draw[-, green, thick] (b5.side 2) to[out=0, in=60] (b5.corner 4);
        
        \draw[-, green!80!black, very thick] (b5.corner 2) -- (b5.corner 3);
        \draw[-, green!80!black, very thick] (b5.corner 4) -- (b5.corner 5);
        \draw[-, green!80!black, very thick] (b5.corner 6) -- (b5.corner 1);

        \filldraw[fill=white, draw = green!80!black, thick] (b5.side 2) circle (3pt) node {\tiny{$\beta$}};
        \filldraw[fill=white, draw = green!80!black, thick] (b5.side 4) circle (3pt) node {\tiny{$\alpha$}};
        \filldraw[fill=white, draw = green!80!black, thick] (b5.side 6) circle (3pt) node {\tiny{$\gamma$}};

        \foreach \anchor/\placement/\n/\fill/\text/\op in
        {corner 1/below/1/white/black, corner 2/below/2/black/white, corner 3/right/3/white/black, corner 4/above/4/black/white, corner 5/above/5/white/black, corner 6/left/6/black/white}
        {
        \node[draw, circle, fill=\fill, inner sep = 1pt, text=\text] at (b5.\anchor) {\tiny{$Y$}};
        }

        \node at (4.5, 0) {$+$};

        \node[name= b5, regular polygon, regular polygon sides = 6, draw, inner sep = 0.5cm, fill=blue, fill opacity=0.25, thick] at (6, 0) {};

        \draw[-, blue, thick] (b5.side 6) to[out=180, in=200] (b5.corner 6);
        \draw[-, blue, thick] (b5.side 6) to[out=180, in=120] (b5.corner 5);

        \draw[-, blue, thick] (b5.side 2) to[out=0, in=0] (b5.corner 3);
        \draw[-, blue, thick] (b5.side 2) to[out=0, in=60] (b5.corner 4);
        
        \draw[-, green!80!black, very thick] (b5.corner 2) -- (b5.corner 3);
        \draw[-, green!80!black, very thick] (b5.corner 4) -- (b5.corner 5);
        \draw[-, green!80!black, very thick] (b5.corner 6) -- (b5.corner 1);

        \filldraw[fill=white, draw = green!80!black, thick] (b5.side 2) circle (3pt) node {\tiny{$\beta$}};
        \filldraw[fill=white, draw = green!80!black, thick] (b5.side 4) circle (3pt) node {\tiny{$\alpha$}};
        \filldraw[fill=white, draw = green!80!black, thick] (b5.side 6) circle (3pt) node {\tiny{$\gamma$}};

        \foreach \anchor/\placement/\n/\fill/\text/\op in
        {corner 1/below/1/white/black, corner 2/below/2/black/white, corner 3/right/3/white/black, corner 4/above/4/black/white, corner 5/above/5/white/black, corner 6/left/6/black/white}
        {
        \node[draw, circle, fill=\fill, inner sep = 1pt, text=\text] at (b5.\anchor) {\tiny{$Z$}};
        }

    \end{tikzpicture}
    }
\end{equation}
\begin{equation}
    \raisebox{-30pt}{
    \begin{tikzpicture}

        \node[name= r, regular polygon, regular polygon sides = 6, draw, inner sep = 0.5cm, fill=red, fill opacity=0.25, thick] at (0, 0) {};

        \draw[-, red, thick] (r.side 5) to[out=180, in=200] (r.corner 1);
        \draw[-, red, thick] (r.side 5) to[out=180, in=120] (r.corner 6);

        \draw[-, red, thick] (r.side 3) to[out=0, in=0] (r.corner 2);
        \draw[-, red, thick] (r.side 3) to[out=0, in=60] (r.corner 3);
        
        \draw[-, green!80!black, very thick] (r.corner 1) -- (r.corner 2);
        \draw[-, green!80!black, very thick] (r.corner 3) -- (r.corner 4);
        \draw[-, green!80!black, very thick] (r.corner 5) -- (r.corner 6);

        \filldraw[fill=white, draw = green!80!black, thick] (r.side 1) circle (3pt) node {\tiny{$\alpha$}};
        \filldraw[fill=white, draw = green!80!black, thick] (r.side 3) circle (3pt) node {\tiny{$\gamma$}};
        \filldraw[fill=white, draw = green!80!black, thick] (r.side 5) circle (3pt) node {\tiny{$\beta$}};

        \foreach \anchor/\placement/\n/\fill/\text/\op in
        {corner 1/below/1/white/black, corner 2/below/2/black/white, corner 3/right/3/white/black, corner 4/above/4/black/white, corner 5/above/5/white/black, corner 6/left/6/black/white}
        {
        \node[draw, circle, fill=\fill, inner sep = 1pt, text=\text] at (r.\anchor) {\tiny{$X$}};
        }

        \node at (1.5, 0) {$+$};

        \node[name= r, regular polygon, regular polygon sides = 6, draw, inner sep = 0.5cm, fill=red, fill opacity=0.25, thick] at (3, 0) {};

        \draw[->, green, thick] (r.side 5) to[out=180, in=200] (r.corner 1);
        \draw[->, green, thick] (r.side 5) to[out=180, in=120] (r.corner 6);

        \draw[->, green, thick] (r.side 3) to[out=0, in=0] (r.corner 2);
        \draw[->, green, thick] (r.side 3) to[out=0, in=60] (r.corner 3);
        
        \draw[-, green!80!black, very thick] (r.corner 1) -- (r.corner 2);
        \draw[-, green!80!black, very thick] (r.corner 3) -- (r.corner 4);
        \draw[-, green!80!black, very thick] (r.corner 5) -- (r.corner 6);

        \filldraw[fill=white, draw = green!80!black, thick] (r.side 1) circle (3pt) node {\tiny{$\alpha$}};
        \filldraw[fill=white, draw = green!80!black, thick] (r.side 3) circle (3pt) node {\tiny{$\gamma$}};
        \filldraw[fill=white, draw = green!80!black, thick] (r.side 5) circle (3pt) node {\tiny{$\beta$}};

        \foreach \anchor/\placement/\n/\fill/\text/\op in
        {corner 1/below/1/white/black, corner 2/below/2/black/white, corner 3/right/3/white/black, corner 4/above/4/black/white, corner 5/above/5/white/black, corner 6/left/6/black/white}
        {
        \node[draw, circle, fill=\fill, inner sep = 1pt, text=\text] at (r.\anchor) {\tiny{$Y$}};
        }

        \node[name= r, regular polygon, regular polygon sides = 6, draw, inner sep = 0.5cm, fill=red, fill opacity=0.25, thick] at (6, 0) {};

        \draw[->, blue, thick] (r.side 5) to[out=180, in=200] (r.corner 1);
        \draw[->, blue, thick] (r.side 5) to[out=180, in=120] (r.corner 6);

        \draw[->, blue, thick] (r.side 3) to[out=0, in=0] (r.corner 2);
        \draw[->, blue, thick] (r.side 3) to[out=0, in=60] (r.corner 3);
        
        \draw[-, green!80!black, very thick] (r.corner 1) -- (r.corner 2);
        \draw[-, green!80!black, very thick] (r.corner 3) -- (r.corner 4);
        \draw[-, green!80!black, very thick] (r.corner 5) -- (r.corner 6);

        \filldraw[fill=white, draw = green!80!black, thick] (r.side 1) circle (3pt) node {\tiny{$\alpha$}};
        \filldraw[fill=white, draw = green!80!black, thick] (r.side 3) circle (3pt) node {\tiny{$\gamma$}};
        \filldraw[fill=white, draw = green!80!black, thick] (r.side 5) circle (3pt) node {\tiny{$\beta$}};

        \foreach \anchor/\placement/\n/\fill/\text/\op in
        {corner 1/below/1/white/black, corner 2/below/2/black/white, corner 3/right/3/white/black, corner 4/above/4/black/white, corner 5/above/5/white/black, corner 6/left/6/black/white}
        {
        \node[draw, circle, fill=\fill, inner sep = 1pt, text=\text] at (r.\anchor) {\tiny{$Z$}};
        }

        \node at (4.5, 0) {$+$};
        
    \end{tikzpicture}
    }
\end{equation}

The decorations correspond to the following controlled gates (here, $a$ is a qutrit edge ancilla, $b$ is a qubit of the original code):
\begin{equation}
\raisebox{-30pt}{
    \begin{tikzpicture}
        \draw[->, red, thick] (0, 0) to (1, 0);
        \node[red] at (-0.25, 0) {$a$};
        \node[red] at (1.25, 0) {$b$};
        \node at (5, 0) {$= \dyad{0}_a \otimes X_b + \dyad{1}_a \otimes Y_b + \dyad{2}_a \otimes Z_b$};

        \draw[->, green, thick] (0, -1) to (1, -1);
        \node[green] at (-0.25, -1) {$a$};
        \node[green] at (1.25, -1) {$b$};
        \node at (5, -1) {$= \dyad{0}_a \otimes Y_b + \dyad{1}_a \otimes Z_b + \dyad{2}_a \otimes X_b$};

        \draw[->, blue, thick] (0, -2) to (1, -2);
        \node[blue] at (-0.25, -2) {$a$};
        \node[blue] at (1.25, -2) {$b$};
        \node at (5, -2) {$= \dyad{0}_a \otimes Z_b + \dyad{1}_a \otimes X_b + \dyad{2}_a \otimes Y_b$};
    \end{tikzpicture}
    }
    \label{eq:facet-commutators}
\end{equation}
Note that the gauged stabilizers are $\emph{not}$ unitary--- the qudit gauging procedure does not generically return unitary stabilizers. 

\begin{figure}[htbp]
    \centering
    \includegraphics[width=1\linewidth]{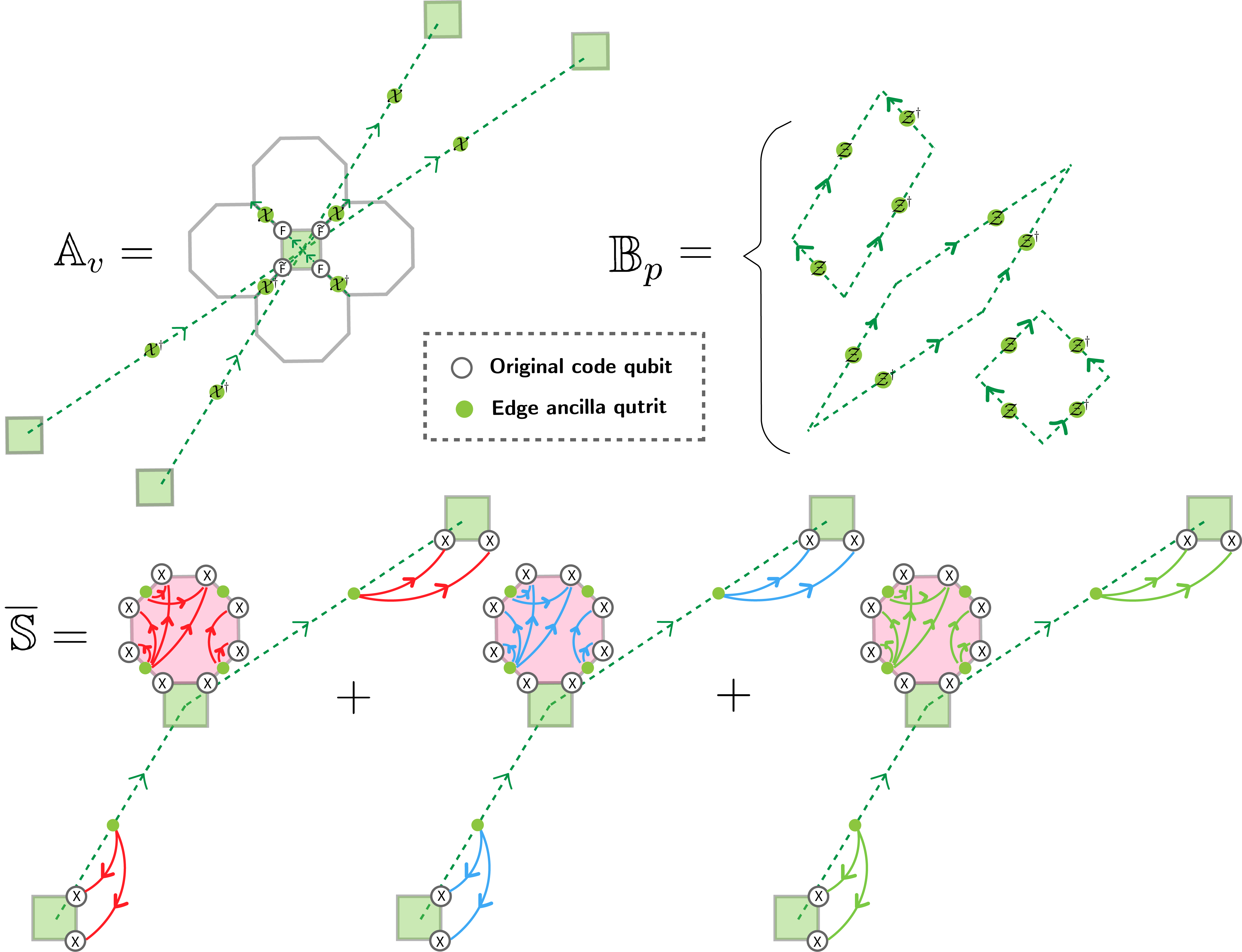}
    \caption{\textbf{Gauging $HS$ on a Self-Dual BB Code:} We illustrate some of the stabilizers that result from gauging $HS$ on top of the self-dual BB code introduced at the beginning of the example section. The gauging graph is indicated by green dashed lines, while the grey lines denote the original lattice. The control gates used in $\gaugedS$ are defined in Eq.~\ref{eq:facet-commutators}.}
    \label{fig:facet-BB-stabilizers}
\end{figure}

The self-dual version of the BB code introduced in examples above possesses the same transversal facet gate, although the logical action is more complicated, given the larger number of logical qubits. The physical implementation is the same however, and so the gauging procedure is very similar, up to modifying the connectivity of the gauging graph. The resultant stabilizers are pictured in Fig.~\ref{fig:facet-BB-stabilizers}. Again, since this self-dual BB code should be describable as a $\mathbb{Z}_2^{16}$ topological order, we conjecture this gauged model has $\mathbb{Z}_2^{16} \rtimes \mathbb{Z}_3$ topological order.

\subsubsection{Charge conjugation on a qutrit BB code}
\begin{figure}[htbp]
    \centering
    \includegraphics[width=1\linewidth]{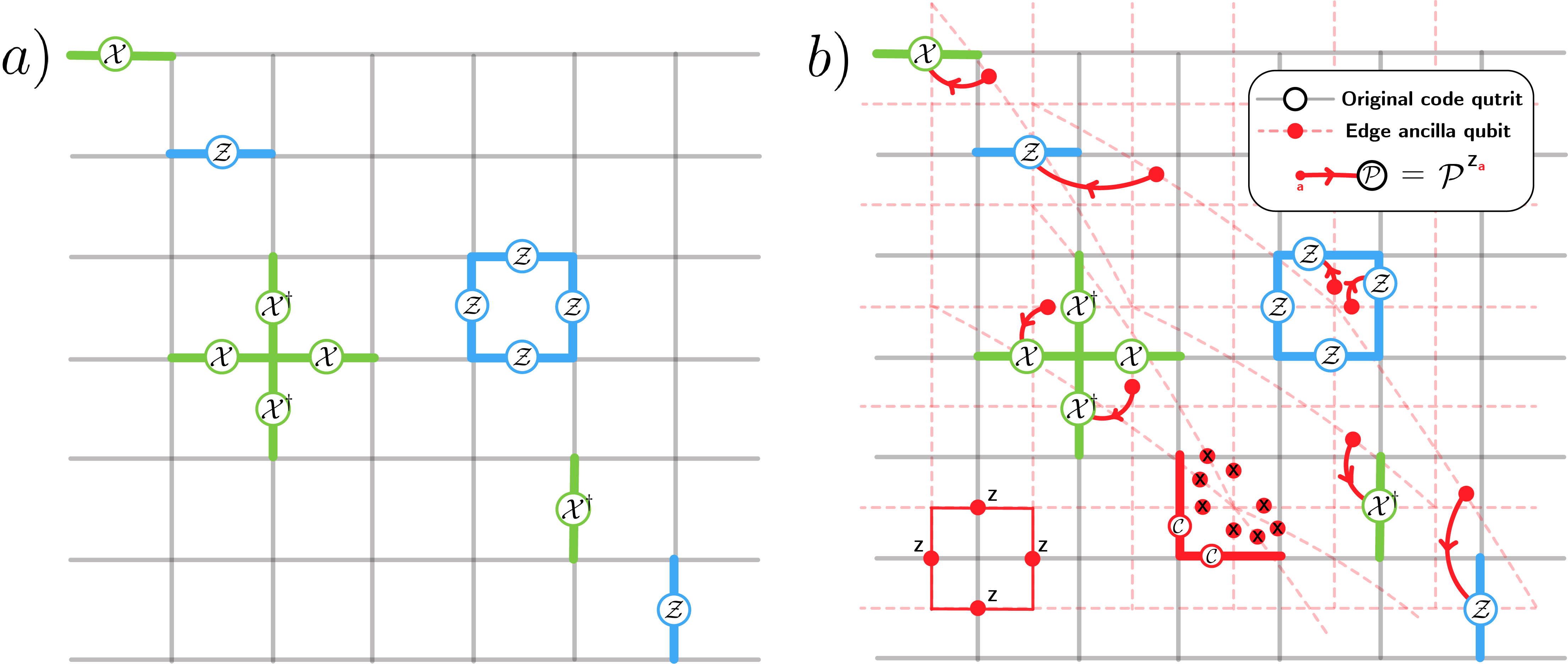}
    \caption{\textbf{Gauging charge conjugation on the qutrit BB code:} a) Stabilizers for the initial qutrit code. b) Stabilizers for the code after gauging the $\mathbb{Z}_2$ symmetry. The gate $\mathcal{P}_i^{Z_j} = \dyad{0}_j \otimes \mathcal{P}_i + \dyad{1}_j \otimes \mathcal{P}^\dagger_i$ is a control gate with a \emph{qubit} control and a \emph{qutrit} target. Here, $\mathcal{P}$ is a tensor product of clock and shift operators. We have pictured unitary components of the decorated $\mathcal{X}$ and $\mathcal{Z}$ stabilizers (green and blue, respectively)--- note that the true stabilizers will be the sum of this component with its Hermitian conjugate. We also depict the vertex stabilizer $\av$ (bottom center) and an example cycle check $\bp$ (bottom left). }
    \label{fig:S3-BB-stabilizers}
\end{figure}

As an example starting with a non-qubit stabilizer code, consider the qutrit version of the $[[144, 12, 12]]$ Gross code. The Pauli operators are replaced with clock and shift operators: 
\begin{equation}
    \mathcal{ZX} = e^{2\pi i /3}\mathcal{XZ}, \mathcal{X}^3 = \mathcal{Z}^3 = \id
\end{equation}

The stabilizers are pictured in Fig.~\ref{fig:S3-BB-stabilizers}a. This model possesses a transversal charge conjugation gate, $\T = \prod \mathcal{C}$, where $\mathcal{C}$ acts as $\mathcal{C} \mathcal{X}^a \mathcal{Z}^b \mathcal{C} = \mathcal{X}^{-a} \mathcal{Z}^{-b}$. This is a $\mathbb{Z}_2$ symmetry. Choosing unit cells of the square lattice as our vertex set, the resultant stabilizers are pictured in Fig.~\ref{fig:S3-BB-stabilizers}b. 

Note that the logical action of $\T = \prod \mathcal{C}$ is to conjugate \emph{all} logical qutrits.

\subsubsection{Gauging qudit \texorpdfstring{$\CNOT$}{CNOT} on qudit CSS codes}
While the qudit gauging procedure generally returns non-unitary stabilizers, there are some special cases where we can find unitary stabilizers akin to the $\mathbb{Z}_2$ case. The structure of the resulting code is \emph{nilpotent}: in terms of the syndromes non-Abelian stabilizer violations always fuse to Abelian stabilizer violations. In terms of the stabilizers, we may always split the stabilizer group into two parts, one corresponding to (Abelian) stabilizers that only commute on the codespace, and the other to (non-Abelian) stabilizers that control those commutation relations. Essentially, the structure will always have the form of Corollary \ref{cor:z2-commutation-relations} in Sec. \ref{sec:gauging-procedure}.

\begin{figure}[htbp]
    \centering
    \includegraphics[width=1\linewidth]{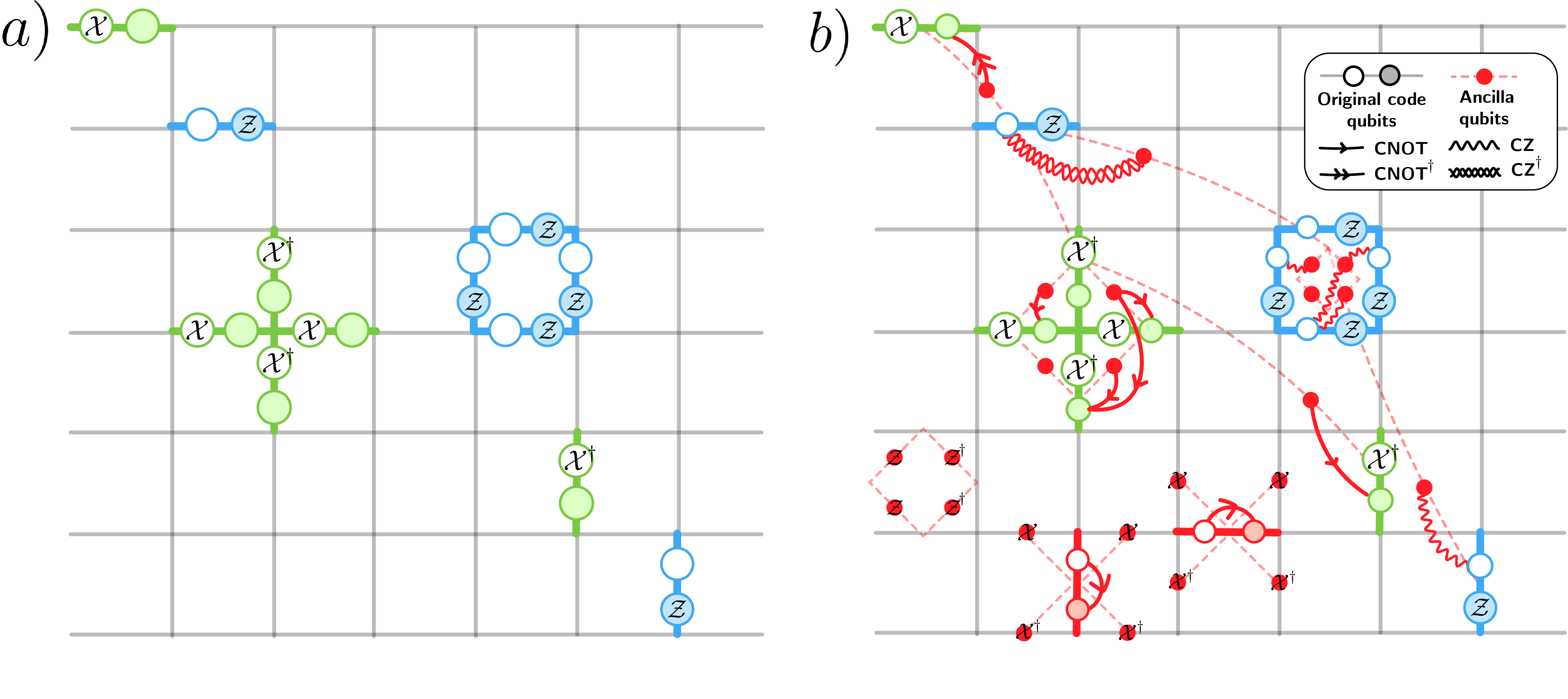}
    \caption{\textbf{Gauging $\CNOT$ on the qutrit BB code:} a) Stabilizers for the initial qutrit code. We have only pictured the stabilizers that are affected by the gauging procedure. b) Stabilizers for the code after gauging transversal qutrit $\CNOT$. We have taken the pair of qutrits belonging to the same edge in the two codeblocks as a vertex set. We have omitted depicting the full gauging graph for visual clarity, but it has the same connectivity as in Fig.~\ref{fig:bb_cnot_gauging}.}
    \label{fig:bb-qudit-CNOT}
\end{figure}

Groups built from $\mathbb{Z}_2$ extensions of $\mathbb{Z}_2$ are always nilpotent, and so the qubit code/qubit symmetry case yields unitary stabilizers. Another class of nilpotent groups is $\mathbb{Z}_p^2 \rtimes \mathbb Z_p$; such a group would correspond to gauging a $\mathbb{Z}_q$ symmetry on two copies of a $q$-dimensional qudit code. A natural example of such a $\mathbb{Z}_p$ operator is qudit $\CNOT$ on two CSS qudit codes: 
\begin{equation}
    \CNOT_{a \to b} \ket{a} \ket{b} = \ket{a} \ket{b + a \text{ mod } q}
\end{equation}
The procedure to gauge this $\CNOT$ is the same as the qubit case, and results in much the same stabilizers, up to some daggers. See Fig.~\ref{fig:bb-qudit-CNOT} for the result of gauging qutrit $\CNOT$ on the qutrit BB code, using a single edge of the square lattice as a vertex set. It is instructive to compare this example to the earlier qubit BB code.

\subsection{Fold-transversal gates on hypergraph product codes}

In this example we gauge a fold-transversal $\mathbb{Z}_2$ symmetry on a generic square hypergraph product (HGP) code \cite{Breuckmann2024, Quintavalle_2023}. We define a choice of vertex sets compatible with the structure of the fold transversal gates defined in Ref.~\cite{Breuckmann2024}, and derive the resulting gauged code.

Recall the hypergraph product of two classical codes $\mathcal{C}_1: C_1^{(1)}\overset{H_1}{\rightarrow} C_0^{(1)}$ and $\mathcal{C}_2:C_1^{(2)}\overset{H_2}{\rightarrow}C_0^{(2)}$ is the quantum code associated to the tensor product chain complex: 
\begin{equation}
    \mathcal{C} = \mathcal{C}_1 \otimes \mathcal{C}_2: C_1^{(1)} \otimes C_1^{(2)} \overset{H_X}{\rightarrow} C_0^{(1)} \otimes C_1^{(2)} \oplus C_1^{(1)} \otimes C_0^{(2)} \overset{H^T_Z}{\rightarrow} C_0^{(1)} \otimes C_0^{(2)}
\end{equation}
\begin{align}
    H_X = (H_1\otimes I_{n_2}\ |\ I_{n_1}\otimes H_2)\\
    H_Z = (I_{m_1}\otimes H^T_2\ |\ H_1^T\otimes I_{m_2})
\end{align}
If $\mathcal{C}_2 = \mathcal{C}_1^T$ is the dual classical code of $\mathcal{C}_1$, then the quantum code $\mathcal{C}$ is a \emph{square} HGP code. If the input classical code is self-dual, then $\mathcal{C}$ is a \emph{symmetric} HGP code.

Square codes may possess transversal Clifford gates of the following form \cite{Breuckmann2024}: 
\begin{equation}
    \T_\tau = \bigotimes_{\substack{i = 1, \dots, n \\ i < \tau(i)}} \text{SWAP}_{i, \tau(i)} \bigotimes_{i = 1}^n H_i
\end{equation}
where $\tau$ is a permutation on the labels of physical qubits. We assume $\tau^2 = \id$, which allows us to break the circuit of $\text{SWAP}$ gates into disjoint local actions, and ensures that $\T^2 = \id$. For a general permutation $\tau$, we must include the entire orbit of a given qubit in a vertex set, and the order of the gate $\T$ becomes larger than 2, necessitating gauge qudits.

\begin{figure}
    \centering \includegraphics[width=0.75\linewidth]{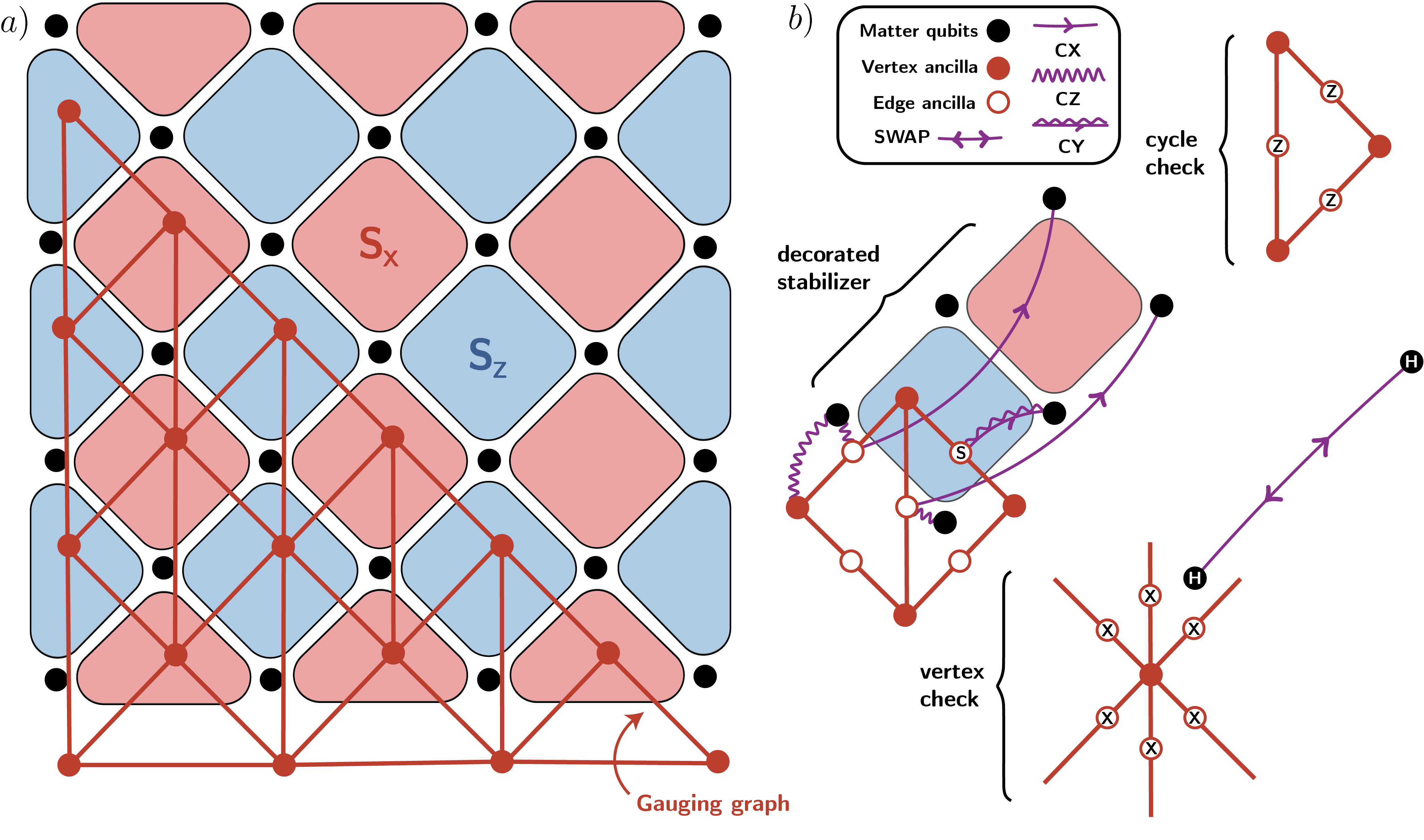}
    \caption{a) Consider the fold transversal Hadamard gate on the surface code, which exchanges $S_X$ (pink) and $S_Z$ (blue) stabilizers along the diagonal running from the top left to the bottom right.  A convenient choice of gauging graph is given in the figure (red lines). b) The resulting gauged stabilizers. }
    \label{fig:fold-transversal-gauging}
\end{figure}

We define vertex sets $\invvertexset{v_i} = \{i, \tau(i)\}$ for $i \leq \tau(i)$. The associated local action of $\T_\tau$ is 
\begin{equation}
    (\T_\tau)_{v_i} \equiv (\T_\tau)_i = \text{SWAP}_{i, \tau(i)} H_i H_{\tau(i)}, \quad \T_\tau = \bigotimes_{i < \tau(i)} (\T_\tau)_i.
\end{equation}
As $\tau^2 = \id$, we have at most order-2 orbits for each qubit, and also at most order-2 orbits for the stabilizers. In particular, each $Z$ stabilizer $S_Z$ is paired with an $X$ stabilizer $S_X = \tau(S_Z)$ to form the orbit $O(S_Z) = \{S_Z, \tau(S_Z)\}$. Thus, the commutation relation between any stabilizer and the local symmetry action is:
\begin{equation}
\ComGrp{S_Z}{(\T_\tau)_i} = \left [(\tau(S_Z))_i \prod_{j \neq i}(S_Z)_j \right ] S_Z = [\tau(S_Z)]_i (S_Z)_i 
\end{equation}
We require a short path connecting any two vertex ancillas whose vertex sets have support on the same stabilizer. Assuming we have constructed the gauging graph with such short paths, the gauged stabilizers are:\footnote{Due to the LDPCness of the initial HGP code, we can always construct this gauging graph while preserving bounded degree by simply adding an edge whenever a short path is missing.}
\begin{equation}
    \av = (\T_\tau)_i \prod_{e \in \delta v_i} X_e, \quad \bp = \prod_{\substack{e \in f \\ \partial f = 0}} Z_e,
\end{equation}
\begin{equation}
    \gaugedS = S \cdot \left ( \prod_{i \neq i_0} \prod_{e \in p(i, i_0)} C_e (\tau(S))_i C_e S_i \right ).
\end{equation}
See Fig.~\ref{fig:fold-transversal-gauging} for an illustration of these stabilizers in the simplest case of the fold transversal gate on the surface code \cite{Moussa_2016}.
We emphasize that this procedure applies to any square HGP code, but we choose to illustrate the surface code for the sake of clarity in the presentation of the figure.

\ifdefined\ismainfile
\else
    \bibliography{bib}
    \end{document}
\fi

\section{Applications}\label{sec:applications}\ifdefined\ismainfile
\else
    \documentclass[11pt]{article}

    \input{preamble.tex}
    
    \begin{document}
    \title{Applications}
    \date{\today}
    \bibliographystyle{abbrv}

    \maketitle
\fi

In this section we describe applications of the gauging measurement protocol. The logical action of a Clifford gate on a high-rate QLDPC code may be complicated and consequently the gauging measurement of that Clifford gate may lead to a logical magic state that is challenging to use. Below, we describe several approaches to achieve useful logical primitives using gauging measurements on QLDPC codes. 

\subsection{Preparing $\ket{\mathsf{Tof}}$ states in CSS codes by gauging}
\label{sec:applicationsA}

First, we present a scheme to prepare multiple $\tof$ states on a single CSS QLDPC codeblock using our fault-tolerant gauging protocol. We highlight that this scheme can be applied to the transversal $\CX$ gate in a general CSS QLDPC code with no further structure required. 
Unless otherwise specified, all states and operators in this section are at the logical level.

Often, QLDPC codes admit a transversal logical Clifford operator of the form $\T^{\otimes k}$, where $k$ is the number of logical qubits in one codeblock.
In the context of magic state preparation, it is well-motivated to prepare a logical product state such as $\ket{+\T}^{\otimes k}$, where $\ket{+\T}$ is a logical state in the $+1$ eigenspace of $\T$.
Unfortunately, naively measuring the logical transversal operator $\T^{\otimes k}$ is not sufficient to obtain a tensor product of $\ket{+\T}$ states. 

Let $\ket{+\T}=\ket{(-1)^{0}\T}$ denote a state in the $+1$ eigenspace of $\T$, and $\ket{-\T}=\ket{(-1)^1\T}$ denote a state in the $-1$ eigenspace of $\T$. 
Let the entries of $\vec{v} = (v_1, \dots, v_k) \in\{0,1\}^k$, be used to denote which eigenspace of $\T$ logical qubit $i$ is in.

Suppose for the sake of simplicity that we begin with $k$ logical qubits in a single codeblock in the stabilizer state such that each logical qubit $i$ has equal overlap with the $\ket{+\T}_i$ and $\ket{-\T}_i$ states.
If we measure the operator $\T^{\otimes k}$ and obtain measurement outcome $m_\T$, the codestate is projected into a state proportional to
\begin{align}
    \sum_{\vec{v}:\ \prod_i(-1)^{v_i}=m_\T}\ket{(-1)^{v_1}\T}_1
    \otimes\ket{(-1)^{v_2}\T}_2
    \otimes ...\otimes \ket{(-1)^{v_k}\T}_k.
    \label{eq:entangled-magic}
\end{align}
This is technically a magic state, although its computational utility is unclear. It is more desirable to prepare individual $\ket{+\T}$ states that can be consumed to perform gate teleportation or some similar implementation of a non-Clifford gate.

Certainly, if one had transversal addressable logical Clifford operators $\T_i$ for all $i\in [k]$, then we could measure $\T_i$ one at a time (or in parallel if the operators on the physical Hilbert space are disjoint) to obtain the product state we desire.
However, this places a strict requirement on the QLDPC code that we use, because it must come equipped with many addressable logical Clifford operators $\T_i,\ i\in[k]$ that we can measure.
A natural question to ask is: can we measure a symmetry $\T$ that is present in a broad family of QLDPC codes, and still yields known computationally-useful magic states?

To answer this question, we propose a protocol that only requires a transversal $\CX^{\otimes k}$ between two QLDPC code blocks.
The existence of this transversal operation is guaranteed for any CSS code, which means that this protocol is applicable to a very large family of QLDPC codes.
Algorithm~\ref{alg:cx-states} presents our protocol to produce $\ket{+\CX}^{\otimes k}$ between two QLDPC codeblocks, where $$\ket{+\CX}=\frac{1}{\sqrt{3}}(\ket{0+}+\ket{0-}+\ket{1+}).$$
Importantly, $\ket{+\CX}^{\otimes k}$ is a product state of many magic states unlike Eq.~\eqref{eq:entangled-magic}.
This means that the magic is easily separable, and so more broadly useful for algorithmic purposes. 

We briefly remark on how these $\ket{+\CX}$ states can be consumed to perform non-Clifford logic. 
There is a known probabilistic conversion that takes two copies of $\ket{+\CX}$  and creates one $$\tof=\ket{000}+\ket{010}+\ket{100}+\ket{111}$$ state~\cite{dennis2001toward,gupta2024encoding} using just Pauli measurements and Clifford operations, succeeding with probability $4/9$.
This means that running our protocol once over two QLDPC code blocks with $k$ logical qubits each yields on expectation $2k/9$ $\tof$ states.
In particular, these $\tof$ states are resource magic states that can be used to deterministically perform the Toffoli gate using just Clifford gates, Pauli corrections, and Pauli measurements~\cite{dennis2001toward}.
We remark that Ref.~\cite{vaknin2026efficientmagicstatecultivation} considers measurement of the $\CX$ operator using post-selection and a large GHZ state. 
There, the authors considered codes with $k=1$, so preparing a product state $\ket{+\CX}^{\otimes k}$ is straightforward. 

\begin{example}
    Given two code blocks of the same $[[n,k,d]]$ CSS code, we can prepare $\ket{+\CX}^{\otimes k}$ between these two code blocks with $\Omega(k)$ measurements of $\CX^{\otimes k}$, Clifford operations, and Pauli measurements.
\end{example}

We now introduce the notation used in Alg.~\ref{alg:cx-states}.
This protocol uses two $[[n,k,d]]$ QLDPC code blocks. 
Let the \textit{control code block}, denoted by subscript $\mathrm{c}$, be the code block on which the control of $\CX^{\otimes k}$ is acted on.
Let the \textit{target code block}, denoted by the subscript $\mathrm{t}$, be the code block on which the target of $\CX^{\otimes k}$ is acted on. 
The $i$th logical qubit in the control and target codeblock constitute the $i$th \textit{logical qubit pair}.
A state or operator of the $i$th logical qubit pair is denoted by the subscript $i$.

Our algorithm makes use of some facts about the structure of the $\pm1$ eigenspace of a controlled-Pauli operator.
Let $\ket{P_{\pm}}$ denote the $\pm1$ eigenstates of the Pauli operator $P$.
Then the $+1$ eigenspace of a controlled-Pauli operator $\mathrm{CP}_i$ is spanned by $\{\ket{00}_i,\ket{01}_i,\ket{1P_+}_i\}$ and the $-1$ eigenspace is spanned by $\{\ket{1P_-}_i\}$. 
From this basis of the eigenspaces, we observe that the measurement of the Pauli operators $Z_\mathrm{c}$ and $P_{\mathrm{t}}$ collapses a state to either the $+1$ or $-1$ eigenspace of $\mathrm{CP}$.
For example, if the measurement of $Z_\mathrm{c}$ is $m_{Z_\mathrm{c}}=+1$, then we know that the post-measurement state collapsed to the $+1$ eigenspace of $\mathrm{CP}_i$.

Finally, we remark that the successful preparation of a magic state relies on the initial state of the code prior to the Clifford measurement. 
For example, measuring $\CX_i$ on the state $\ket{00}_i$ leaves the state invariant; the resulting state is a stabilizer state. 
Clearly, our choice of state prior to Clifford measurement is crucial to preparing the right magic state.  
What state should we initialize prior to $\CX$ measurement?
The initial state, $\ket{+}_c\ket{0}_t$ suffices because measuring $m_{\CX_i}=+1$ projects to the following post measurement state
\begin{align}
    \frac{I_\mathrm{c}I_\mathrm{t}+\CX_i}{4} \left(\ket{0+}+\ket{0-}+\ket{1+}+\ket{1-}\right)
    &=\frac{1}{4}\left(2\ket{0+}+2\ket{0-}+2\ket{1+}\right)\\
    &\propto \frac{1}{\sqrt{3}}\left(\ket{0+}+\ket{0-}+\ket{1+}\right)=:\ket{+\CX},
\end{align}
which is proportional to the $\ket{+\CX}$ we desired.
Let $c_+$ be a constant from normalization, $$\ket{+\CX}=\frac{1}{c_+} \frac{1}{2}(\ket{0+}+\ket{0-}+\ket{1+}),$$ thus $c_+=\frac{\sqrt{3}}{2}$.
If we measure $\CX^{\otimes k}$ and obtain $m_{\CX_i}=-1$, then the post-measurement state is a stabilizer state,
\begin{align}
    \frac{I_\mathrm{c}I_\mathrm{t}-\CX_i}{4} \left(\ket{0+}+\ket{0-}+\ket{1+}+\ket{1-}\right) \propto \ket{1-}=:\ket{-\CX}.
\end{align}
Again, let $c_-=\frac{1}{2}$ be a constant from normalization.

\begin{algorithm}
    \caption{Preparing $k$ copies of $\ket{+\CX}$ states.}
    \label{alg:cx-states}
    \KwIn{Two copies of the same $[[n,k,d]]$ CSS QLDPC code.}
    \KwOut{$k$ copies of $\ket{+\CX}$ states across two $[[n,k,d]]$ codes.}
    \vspace{0.5cm}
    Initialize the control codeblock $C$ into $\ket{+}^{\otimes k}$ and the target codeblock $T$ into $\ket{0}^{\otimes k}$ OR $C$ into $\ket{1}^{\otimes k}$ and $T$ into $\ket{-}^{\otimes k}$.

    $m_{\CX^{\otimes k}}\leftarrow$ Measure $\CX^{\otimes k}$. For the second initialization option, instead set $m_{\CX^{\otimes k}}\leftarrow(-1)^k$.

    \For{$i\in[1,k]$}{
        
        $m_{Z_{\mathrm{c}_i}},m_{X_{\mathrm{t}_i}}$ $\leftarrow$ Measure $Z_{\mathrm{c}_i}$ and $X_{\mathrm{t}_i}$

        \If{$(m_{Z_{\mathrm{c}_i}},m_{X_{\mathrm{t}_i}})=(+1,+1)$}{
            $\lambda_i\leftarrow+1$ \quad \quad \quad\quad \quad \tcp{Store the parity of pair $i$ as $+1$}

            Apply $H_{\mathrm{c}_i}H_{\mathrm{t}_i}$ \quad\quad\quad \tcp{Apply Clifford correction to reset to $\ket{+0}_{i}$}
        }

        \If{$(m_{Z_{\mathrm{c}_i}},m_{X_{\mathrm{t}_i}})=(+1,-1)$}{
            
            $\lambda_i\leftarrow+1$ \quad\quad \quad\quad\quad  \tcp{Store the parity of pair $i$ as $+1$}

            Apply $H_{\mathrm{c}_i}H_{\mathrm{t}_i}Z_{\mathrm{t}_i}$ \quad\quad\quad \tcp{Apply Clifford correction to reset to $\ket{+0}_{i}$}
        }

        \If{$(m_{Z_{\mathrm{c}_i}},m_{X_{\mathrm{t}_i}})=(-1,+1)$}{
            $\lambda_{i}\leftarrow+1$ \quad\quad \quad \quad\quad  \tcp{Store the parity of pair $i$ as $+1$}

            Apply $H_{\mathrm{c}_i}H_{\mathrm{t}_i}X_{\mathrm{c}_i}$ \quad\tcp{Apply Clifford correction to reset to $\ket{+0}_{i}$}
        }

        \If{$(m_{Z_{\mathrm{c}_i}},m_{X_{\mathrm{t}_i}})=(-1,-1)$}{
            $\lambda_{i}\leftarrow-1$ \quad \quad \quad\quad \quad \tcp{Store the parity of pair $i$ as $-1$}

            Apply $H_{\mathrm{c}_i}H_{\mathrm{t}_i}X_{\mathrm{c}_i}Z_{\mathrm{t}_i}$ \quad\quad\quad \tcp{Apply Clifford correction to reset to $\ket{+0}_{i}$}
        }
        $\lambda_{[k]\setminus i} \leftarrow\lambda_i\cdot m_{\CX^{\otimes k}}$ \tcp{The parity of all $\CX$ pairs except for pair $i$}

        $\lambda'_i\leftarrow -1$ \tcp{Initialization to enter the while loop}

        \While{$\lambda'_i\neq +1$}{
            $m'_{\CX^{\otimes k}}\leftarrow$ Measure $\CX^{\otimes k}$.  \tcp{Measure the global operator again}

            $\lambda'_i\leftarrow\lambda_{[k]\setminus i}\cdot m'_{\CX^{\otimes k}}$ \tcp{Get the eigenvalue of pair $i$}

            \If{$\lambda'_i\neq +1$}{
                Apply $H_{\mathrm{c}_i}H_{\mathrm{t}_i}X_{\mathrm{c}_i}Z_{\mathrm{t}_i}$ \tcp{Reset pair $i$ from $\ket{1-}_i$ back to $\ket{+0}_{i}$}
            }
        }

        $m_{\CX^{\otimes k}}\leftarrow m'_{\CX^{\otimes k}}$

    }
\end{algorithm}

The working principle of Alg.~\ref{alg:cx-states} is as follows.
First, we initialize the control code block in $\ket{+}^{\otimes k}$ and the target code block in $\ket{0}^{\otimes k}$.
Then we measure $\CX^{\otimes k}$ on the control and target blocks.
Let $n_+$ ($n_-$) be the number of $0$s ($1$s) in $\vec{v}\in\{0,1\}^{k}$.
Then, the post-measurement state $\ket{\Psi_0}$ dependent on the measurement outcome, $m_{\CX^{\otimes k}}^{(0)}$ is the following
\begin{align}
    \ket{\Psi_0}=\sum_{\vec{v},\ (-1)^{n_-}=m_{\CX^{\otimes k}}^{(0)}} c_+^{n_+}{c_-}^{n_-}\left( \bigotimes_{i\in [k]}\ket{(-1)^{v_i}\CX}_i\right).
    \label{eq:psi0}
\end{align}
This is a superposition over all possible combinations of logical pairs $i$ in states $\ket{\pm\CX}$ with the total number of $\ket{-\CX}$s being even (odd) when the measurement outcome $m_{\CX^{\otimes k}}=+1$($=-1$).

The following steps narrow the number of terms in this superposition until we are left with a product state over all logical pairs. First, consider how the first logical pair of qubits can be disentangled from the rest:

\begin{enumerate}
    \item We measure logical Paulis $Z_{\mathrm{c}_1}$ and $X_{\mathrm{t}_1}$ on the first pair of logical qubits and store the outcomes of these measurements in $(m_{Z_{\mathrm{c}_1}},m_{X_{\mathrm{t}_1}})$.
    \begin{itemize}
        \item If $(m_{Z_{\mathrm{c}_1}},m_{X_{\mathrm{t}_1}})\in \{ (+1,+1),(+1,-1),(-1,+1)\}$, then we know that logical pair $1$ has collapsed to the $+1$ eigenspace of $\CX_1$.
    Let us call this post-measurement state $\ket{\Psi_1}$. Since $\CX^{\otimes k}$ commutes with $Z_{\mathrm{c}_1}$ and with $X_{\mathrm{t}_1}$, $\ket{\Psi_1}$ must be a simultaneous eigenstate of all three operators. We can use the previous measurement result of $\CX^{\otimes k}$ to deduce that logical pairs $2,\dots, k$ are in the $m_{\CX^{\otimes k}}$ eigenspace of $\CX_2\otimes \dots \otimes\CX_k$. 
    \item Conversely, if we had measured $(m_{Z_{\mathrm{c}_1}},m_{X_{\mathrm{t}_1}})=(-1,-1)$, then logical pair $1$ collapsed to the $-1$ eigenspace of $\CX_1$.
    Using our previously obtained eigenvalue of $\CX^{\otimes k}$, we deduce that logical pairs from $2$ through $k$ are in the $-m_{\CX^{\otimes k}}$ eigenspace of $\CX_2\otimes ...\otimes\CX_k$. 
    \end{itemize}
    
    \item Regardless of which eigenvalue of $\CX_2\otimes ...\otimes\CX_k$ $\ket{\Psi_1}$ has, we reset logical pair $1$ back to $\ket{+0}_1$ by applying the Clifford operation $$H_{\mathrm{c}_1}H_{\mathrm{t}_1}X_{\mathrm{c}_1}^{(1-m_{Z_{\mathrm{c}_1}})/2}Z_{\mathrm{t}_1}^{(1-m_{X_{\mathrm{t}_1}})/2}$$
    \item Finally, we measure $\CX^{\otimes k}$ again.
    Crucially, measuring the global operator $\CX^{\otimes k}$ at this point does not change the eigenvalue of $\CX_2\otimes ...\otimes\CX_k$, so we know whether the state of logical pair $1$ is $\ket{+\CX}_1$ or $\ket{-\CX}_1$. 
    \item With probability $3/4$, we have produced the magic state $\ket{+\CX}_{1}$. If we fail and project to $\ket{-\CX}_{1}$, we can repeat until success by resetting pair $1$ into $\ket{+0}_{1}$ via Clifford operations and measuring the global operator $\CX^{\otimes k}$ again.
\end{enumerate}

We indicate the eigenvalue of $\CX_2\otimes ...\otimes\CX_k$ with $\lambda_{[k]\setminus 1}=\pm 1$. Let $\vec{v}\in \{0,1\}^{k-1}$ and index its entries by integers $i\in [2,k]$.
We define $n_+$ and $n_-$ in the same way as above.
After success with the first pair, we have the state 
\begin{align}
    \ket{\Psi_2}=\ket{+\CX}_{1}\otimes \sum_{\vec{v},\ (-1)^{n_-}=\lambda_{[k]\setminus 1}} c_+^{n_+}c_-^{n_-}\bigotimes_{i\in [2,k]} \ket{(-1)^{v_i}\CX}_i
\end{align}

The rest of the algorithm repeats this process for all other pairs of logical qubits one by one. Once a logical pair is in the $\ket{+\CX}_i$ state, subsequent repetitions of these steps do not disturb or change the state of pair $i$. This eventually results in $k$ copies of $\ket{+\CX}$ states. 

As an alternative initialization step, one can instead prepare the control and target code blocks in logical $\ket{1}$ and $\ket{-}$ states, respectively, to fix the initial eigenvalue of the codestate under $\CX^{\otimes k}$. This avoids the requirement for the initial $\CX^{\otimes k}$ measurement. The remainder of the algorithm proceeds as above.  

We now discuss how many operations, aside from the initial global measurement, are required to fix a single logical pair into the $\ket{+\CX}_i$ state. 
The procedure requires at least two Pauli measurements, one Clifford operation, and one global $\CX^{\otimes k}$ measurement. 
If the inferred eigenvalue $\lambda'_i$ is $-1$, we reset pair $i$ to $\ket{+0}_i$ with the Clifford correction and repeat the global $\CX^{\otimes k}$ measurement.
Fortunately, the expected number of repetitions until success is a small constant, $\langle \mathrm{Rep.}\rangle=4/3\simeq 1.33$. 
Thus, the expected number of $\CX^{\otimes k}$ measurements, Pauli measurements, and Clifford operations our algorithm requires to prepare $\ket{+\CX}^{\otimes k}$ is $\Theta(k)$.

One can similarly prepare a comparable number of $\tof$ states with only half of the measurements required in Alg.~\ref{alg:cx-states}. 
To do this, we prepare a product of $$\ket{+\CX}\ket{+\CX}+\ket{-\CX}\ket{-\CX}$$ states instead; these can still be turned into $\tof$ states. We only require half of the measurements because we only need to determine the parity of neighboring logical pairs, rather than each individual pair's eigenvalue. We also note that it is possible to get $> 2k/9\ \tof$ states if one is willing to repeat Alg.~\ref{alg:cx-states} on the $\ket{\Log{\CZ}}$-to-$\tof$ conversions that fail.

\subsection{Performing structured controlled-Clifford gates in parallel}
\label{sec:ctrl-t-gadget}

In this section, we introduce a gadget that uses a single logical Clifford measurement to deterministically perform a logical controlled-Clifford gate, $\CT$, where $\T$ is a $\mathbb{Z}_2$ Clifford symmetry of a QLDPC code.
When $\T$ is a symmetry that acts on many logical qubits, for example $\T = \bigotimes_{i}\T_i$, this gadget performs many non-Clifford gates $\CT_i$ in parallel.
Note that the control qubit of these operations is the same, which limits the addressability of $\CT_i$ operations. However, this one-control many-target circuit structure is readily applicable to subroutines in quantum state preparation, block encoding of second-quantized Hamiltonians,  unitary synthesis, and the hidden subgroup problem~\cite{Low_2024,Zhang_2022,liu2025blockencodinglowgate,bouland2024statehiddensubgroupproblem}. We discuss how our gadget may be applied to these instances and the savings obtained as compared to existing circuits below in Section~\ref{sec:applications-with-cswap}.

\begin{table}
    \centering
    \begin{tabular}{|c|c|c|}
        \hline
        \textbf{Algorithm}& \textbf{Existing Protocols: $T$-count} & \textbf{$\CT$ Gadget: \# Clifford meas.} \\
        \hline
         $\SWAP$ network & $\mathcal{O}(bK)$ & $\mathcal{O}(\log K)$\\
         Alg. 1 (non-adaptive) \cite{bouland2024statehiddensubgroupproblem}& $\mathcal{O}(K^3/\epsilon^2)$ & $O(K^2)$\\
         Alg. 2 (adaptive) \cite{bouland2024statehiddensubgroupproblem} & $\mathcal{O}(K^2/\epsilon^2)$ & $O(K)$\\
         \hline
    \end{tabular}
    \caption{Comparison between $T$-count for various existing algorithms/algorithm primitives and our protocols using the $\CT$ gadget. We list the number of $T$ states consumed by existing protocols vs. the number of Clifford measurements needed for the $\CT$ gadget. In all cases, $K$ is the number of logical qubits in the target codeblock. For the $\SWAP$ network, $b$ gives the size of registers in our lookup table. In the last two algorithms for the hidden cut problem, $\epsilon \in [0, 1]$ is a parameter indicating how far away the two factor states are from product states.}
    \label{tab:placeholder}
\end{table}

\subsubsection{$\CT$ gadget}
First, we introduce the gadget itself. Let $c$ denote the logical qubit designated as the control, and $t$ denote the logical qubit(s) being targeted. Additionally, we need $a$, a logical ancilla qubit. 

The control qubit before the gadget is in a state $(\alpha_c\ket{0}_{c}  + \beta_c\ket{1}_c )$. 
As $\T^2=1$, $\T$ has eigenvalues $\pm1$. 
We can decompose a general state on the target logical qubit(s) in terms of the $\pm1$ eigenspaces of $\T$, written $(\gamma_t\ket{+\T}_t+\delta_t\ket{-\T}_t)$. 
The most general input state we consider is thus
\begin{align}
    \ket{\Psi_0} = (\alpha_c\ket{0}_{c}  + \beta_c\ket{1}_c )\otimes (\gamma_t\ket{+\T}_t+\delta_t\ket{-\T}_t)\otimes \ket{0}_a.
\end{align}
Below, we present a gadget $\CT$ that performs the following linear transformation:
\begin{align}
    \CT: \ket{\Psi_0} \to
    \ket{0}_c \otimes\biggl( (\gamma_t\ket{+\T}_t +\delta_t\ket{-\T}_t)\otimes \alpha_c \ket{0}_a + (\gamma_t\ket{+\T}_t-\delta_t\ket{-\T}_t)\otimes \beta_c\ket{1}_a\biggr)
\end{align}
To this end, we first measure the operator $O_1 = X_c\T_tX_a$. 
We remark that $O_1$ must still be a transversal Clifford operator. $X_c$ and $X_a$ must have transversal, Pauli representatives in the QLDPC codes hosting $c$ and $a$, and so multiplying $\T$ by $X_c$ and $X_a$ leads to an operator that is both Clifford and transversal.
Furthermore, the eigenvalues of $O_1$ are still $\pm 1$ because $$O_1^2 = (X_c\T_tX_a)(X_c\T_tX_a) = X_c^2 \T_t^2 X_a^2 = \id.$$
Since $O_1$ is still a $\mathbb{Z}_2$, transversal Clifford gate, we can use Clifford gauging to measure it fault tolerantly. We remark that if the QLDPC code block, or blocks, hosting $X_a$ and $X_c$ are disjoint from the block hosting the $\T_t$ gate, one can build an ancilla system for the joint gauging measurement from the individual ancilla systems joined together via adapters~\cite{swaroop2025universaladaptersquantumldpc}. We write $\ket{\pm m_j}_{cta}$ for the post-measurement state of the registers $c$, $t$, $a$ after the $j$th measurement of the gadget returns the outcome $\pm 1$.
Up to a global normalization constant, the state after measuring $O_1$ with outcome $+1$ is
\begin{align}
    \ket{+m_1}_{cta}=(\alpha_c+\beta_c)\gamma_t \ket{+}_c\ket{+\T}_t \ket{+}_a +(\alpha_c + \beta_c)\delta_t \ket{+}_c\ket{-\T}_t \ket{-}_a \notag\\
    +(\alpha_c - \beta_c)\delta_t \ket{-}_c\ket{-\T}_t\ket{+}_a +(\alpha_c -\beta_c)\gamma_t \ket{-}_c\ket{+\T}_t\ket{-}_a.
    \label{eq:o1-plus-one}
\end{align}
The state after measuring $O_1$ with outcome $-1$ is
\begin{align}
    \ket{-m_1}_{cta}&=
    (\alpha_c+\beta_c)\gamma_t \ket{+}_c\ket{+\T}_t \ket{-}_a +(\alpha_c + \beta_c)\delta_t \ket{+}_c\ket{-\T}_t \ket{+}_a \label{eq:o1-minus-one}\\
    &+(\alpha_c - \beta_c)\delta_t \ket{-}_c\ket{-\T}_t\ket{-}_a +(\alpha_c -\beta_c)\gamma_t \ket{-}_c\ket{+\T}_t\ket{+}_a
    \notag\\&=Z_a\ket{+m_1}.
\end{align}
Whenever measurement of $O_1$ returns $-1$, we apply $Z_a$ to deterministically prepare the state in Eq.~\eqref{eq:o1-plus-one}.

Next, we measure $Z_c$. Expanding $\ket{+m_1}$ from Eq.~\eqref{eq:o1-plus-one} in the $Z_c$ basis, we have
\begin{align}
    \ket{+m_1}_{cta}&\propto\ket{0}_c \big [(\alpha_c+\beta_c)\gamma_t\ket{+\T}_t\ket{+}_a+(\alpha_c+\beta_c)\delta_t\ket{-\T}_t\ket{-}_a\notag\\
    &+(\alpha_c-\beta_c)\delta_t\ket{-\T}_t\ket{+}_a+(\alpha_c-\beta_c)\gamma_t\ket{+\T}_t\ket{-}_a \big]\notag\\
    &+\ket{1}_c \big [(\alpha_c+\beta_c)\gamma_t\ket{+\T}_t\ket{+}_a+(\alpha_c+\beta_c)\delta_t\ket{-\T}_t\ket{-}_a\notag\\
    &-(\alpha_c-\beta_c)\delta_t\ket{-\T}_t\ket{+}_a-(\alpha_c-\beta_c)\gamma_t\ket{+\T}_t\ket{-}_a \big].
\end{align}
The states after measuring $Z_c$ with outcomes $+1$ and $-1$ respectively are
\begin{align}
    \ket{+m_2}_{cta} &\propto \ket{0}\bra{0}_c \ket{+m_1}\label{eq:plus-m2}\\
    &\propto\ket{0}_c[(\alpha_c+\beta_c)\gamma_t\ket{+\T}_t\ket{+}_a+(\alpha_c+\beta_c)\delta_t\ket{-\T}_t\ket{-}_a\notag\\
    &+(\alpha_c-\beta_c)\delta_t\ket{-\T}_t\ket{+}_a+(\alpha_c-\beta_c)\gamma_t\ket{+\T}_t\ket{-}_a]\notag\\
    \ket{-m_2}_{cta}&\propto\ket{1}\bra{1}_c\ket{+m_1}\notag\\
    &\propto\ket{1}_c[(\alpha_c+\beta_c)\gamma_t\ket{+\T}_t\ket{+}_a+(\alpha_c+\beta_c)\delta_t\ket{-\T}_t\ket{-}_a\notag\\
    &-(\alpha_c-\beta_c)\delta_t\ket{-\T}_t\ket{+}_a-(\alpha_c-\beta_c)\gamma_t\ket{+\T}_t\ket{-}_a]\\
    &=X_c\T_t X_a\ket{+m_2}
\end{align}
If we measure $Z_c$ and get the outcome $-1$, we apply $X_c\T_tX_a$ to deterministically prepare the state $\ket{+m_2}$ in Eq.~\eqref{eq:plus-m2}.

Now, we show that $\ket{+m_2}$ is the state following the desired linear transformation $\CT$.
Up to normalization constants, we take equation~\eqref{eq:plus-m2} and expand in the $Z_a$ basis,
{\small
\begin{align}
    \ket{+m_2}_{cta}&\propto\ket{0}_c\otimes \biggl( \biggl[ (\alpha_c+\beta_c)\gamma_t\ket{+\T}_t+(\alpha_c+\beta_c)\delta_t\ket{-\T}_t + (\alpha_c-\beta_c)\delta_t\ket{-\T}_t
    + (\alpha_c-\beta_c)\gamma_t\ket{+\T}_t\biggr]\otimes \ket{0}_a\notag\\
    &+ \biggl[(\alpha_c+\beta_c)\gamma_t\ket{+\T}_t
    - (\alpha_c+\beta_c)\delta_t\ket{-\T}_t + (\alpha_c-\beta_c)\delta_t\ket{-\T}_t
    - (\alpha_c-\beta_c)\gamma_t\ket{+\T}_t\biggr]\otimes \ket{1}_a\biggr)\notag\\
    &\propto \ket{0}_c\otimes  \biggl((\gamma_t\ket{+\T}_t +\delta_t\ket{-\T}_t) \otimes \alpha_c \ket{0}_a + (\gamma_t\ket{+\T}_t -\delta_t\ket{-\T}_t) \beta_c\otimes \ket{1}_a\biggr)\\
    &=\ket{0}_c (\CT)_{at} \biggl( (\gamma_t\ket{+\T}_t + \delta_t \ket{-\T}_t) (\alpha_c\ket{0}_a+ \beta_c\ket{1}_a) \biggr) = \ket{\Psi_f}
\end{align}}
By inspection, we can see that this has performed the $\CT$ operation with the control on the input state of $c$, and the target on $t$.
Furthermore, the state of the control qubit has been teleported to the ancilla qubit $a$.

The $\CT$ gadget is summarized in Algorithm \ref{alg:cu-gadget}.

\begin{algorithm}
    \caption{Implementing $\CT$}
    \label{alg:cu-gadget}

    \KwIn{A control codeblock with a designated control logical $c$, target codeblock $t$ with a transversal Clifford $\T$, and an ancilla codeblock with a designated ancilla logical $a$.}
    \KwOut{$\CT$ applied to the target codeblock, and the state of the control teleported to the ancilla.}
    \vspace{0.5cm}

     $\ket{\Psi} \leftarrow \ket{\Psi_0} = \ket{\psi}_c \ket{\psi'}_t \ket{0}_a$ \quad \quad \quad \tcp{Initialize the ancilla in $\ket{0}_a$. $\ket{\psi}_c \ket{\psi'}_t$ are generic.}

    $m_1 \leftarrow$ Measure $O_1 = X_c \T_t X_a$.
    
    \If{$m_1 = +1$}{
    $\ket{\Psi} \leftarrow \ket{+m_1}_{cta}$
    
    }

    \If{$m_1 = -1$}{
    $\ket{\Psi} \leftarrow \ket{-m_1}_{cta}$
    
    Apply $Z_a$ \quad\quad\quad \tcp{Corrects back to $\ket{+m_1}$.}
    
    }

    $m_2 \leftarrow$ Measure $Z_c$.

    \If{$m_2 = +1$}{
    $\ket{\Psi} \leftarrow \ket{+m_2}_{cta}$
    }

    \If{$m_2 = -1$}{
    $\ket{\Psi} \leftarrow \ket{-m_2}_{cta}$

    Apply $X_c \T_t X_a$ \quad \quad \quad \tcp{Corrects back to $\ket{+m_2}$.}
    }

\end{algorithm}

\subsubsection{Performing $\CSWAP^{\otimes k}$ in parallel in SWAP networks}
\label{sec:applications-with-cswap}

Now we consider two practical applications of the $\CT$ gadget. The first is a parallel implementation of $\CSWAP^{\otimes k}$ between many logical qubits, with the same control qubit. This structure of $\CSWAP$ operations is common in certain information processing tasks, in particular, $\SWAP$ networks.
$\SWAP$ networks in data-lookup oracles are an important subroutine in arbitrary quantum state preparation, unitary synthesis, and block encoding of Hamiltonians~\cite{Low_2024,liu2025blockencodinglowgate}.

Given two copies of the same $[[n,k,d]]$ stabilizer code, let $i_\mathrm{A (B)}\in[n]$ and $j_\mathrm{A (B)}\in [k]$ enumerate the physical and logical qubits in code block A (B).
Then the transversal operation $\prod_{i=1}^n\SWAP _{i_\mathrm{A}i_{\mathrm{B}}}$ implements a logical $\SWAP$ between logical qubits in A and B, $ \prod_{j=1}^k\SWAP_{j_\mathrm{A}j_{\mathrm{B}}}$. 
Taking $\T = \prod_{i=1}^n\SWAP _{i_\mathrm{A}i_{\mathrm{B}}}$ we can use the $\CT$ gadget in Section~\ref{sec:ctrl-t-gadget} to implement the following logical operations in parallel:
\begin{align}
    \prod_{j=1}^k C_{\mathrm{c}}\SWAP _{j_\mathrm{A}j_\mathrm{B}}.
\end{align}
In other words, a single logical qubit $c$ acts as the control for the $\SWAP$ of every $j_\mathrm{A},j_{\mathrm{B}}$ logical qubit pair.
Importantly, this means that an arbitrary number of $\CSWAP$ operations sharing a common control qubit can be performed in parallel. This is achieved by encoding each pair of logical qubits to be swapped in the logical qubits $j_\mathrm{A}$ and $j_\mathrm{B}$ of two separate code blocks.

We can show that this implementation significantly reduces the amount of non-Clifford operations needed to implement the $\SWAP$ network.
Suppose there are $K$ registers $\{\ket{\phi_x} \}, x \in \{0, \dots, K-1\}$, where each $\ket{\phi_x}$ is a $b$-qubit state. We discuss how these registers are distributed across codeblocks $A, B$ below; we require that $A$ and $B$ each encode at least $b \lfloor K/2 \rfloor$ logical qubits. A register state $\ket{\phi_x}$ is initially at position $x$.

Suppose we have a set of control states labeled $\ket{x}$, $x = 0, \dots, K-1$. We can write these as $\lceil \log_2(K) \rceil$-qubit states where the state of each qubit is given by the corresponding digit of the binary expansion of $x$: 
$$\ket{x}= \ket{x_0} \ket{x_1} \dots \ket{x_{\lceil \log_2(K) \rceil-1}},$$
where $x_0$ is the smallest bit.
When we input state $\ket{x}$ to the $\SWAP$ network, it fetches the $x$th $b$-qubit register and places it in the $0$th register~\cite{Low_2024}. Explicitly, the operation that the $\SWAP$ network performs is as follows:
\begin{align}
    \SWAP_{\rm{net.}} \biggl[\ket{x} \bigotimes _{x=0}^{K-1}\ket{\phi_x}_x\biggr]= \ket{x}\otimes \ket{\phi_x}_0\otimes \cdots 
\end{align}
Here the $\cdots$ indicates that the output state of the registers $x=1,\cdots K-1$ is not important.
The number of $\CSWAP$ operations needed to perform the $\SWAP_{\rm{net.}}$ circuit is upper-bounded by $\mathcal{O}(bK)$, see Fig.~1(b) in Ref.~\cite{Low_2024}.
However, our $\CT$ gadget allows one to perform a $\SWAP$ network with just $\mathcal{O}(\log(K))$ Clifford measurements.

To achieve this, take two codeblocks that encode at least $k = b\lfloor\frac{K}{2}\rfloor$ logical qubits.
Following the $\SWAP$ network circuit in Ref.~\cite{Low_2024}, we perform $\CSWAP$ between all pairs of $b$-qubit registers $$(i_A, i_B) \in \{(i,i+2^j)~|~i\in 2^{j+1}\{0, 1, \dots, K-1\}\}$$ controlled by qubit $j\in \{0,\dots,\lceil \log_2(K)\rceil -1\}$ of the control state $\ket{x}$.
In particular, we use the $\CT$ gadget to parallelize all $\CSWAP$ operations that share a control qubit $j$.
In addition to the actual Clifford measurement, this requires that we perform Clifford $\SWAP$s to move the registers into the correct positions for each index $j$: we want each $(i,i+2^j)$ register pair to correspond to a logical qubit pair across the two codeblocks A and B. 

Since there are only $\lceil \log_2(K)\rceil$ control qubits in the state $\ket{x}$, and each $\CT$ gadget only needs one Clifford measurement, the whole $\SWAP$ network only requires $\mathcal{O}(\log_2(K))$ Clifford measurements, plus some additional Clifford gates. We have lost the dependence on the size $b$ of our stored registers; as long as we have a large enough codeblock to support our desired $b$, the gauging measurement deals with everything in parallel. 
Importantly, the number of Clifford measurements is \emph{exponentially} smaller compared to the number of $T$ states to be prepared via distillation or other means.

\subsubsection{ Parallel $\CSWAP^{\otimes k}$ for the hidden-cut problem}

The $\CSWAP^{\otimes k}$ circuit structure is useful in other applications in addition to building a quantum look-up table (QROM). In particular, we can use our $\CSWAP^{\otimes k}$ gadget in circuits that solve the state hidden-subgroup problem or hidden-cut problem.
We briefly summarize the problem here and refer the interested reader to Ref.~\cite{bouland2024statehiddensubgroupproblem} for details.

At a high level, the problem is as follows: given many copies of an $K$-qubit state $\ket{\psi}$, the goal is to find a bipartition of the qubits, each with $K/2$ qubits, such that the states are unentangled across this partition.
Specifically, $\ket{\psi}=\ket{\phi_1}_C\otimes \ket{\phi_2}_{\bar{C}}$ is a product across an unknown cut $$C\in {[K]\choose K/2}.$$ The factor states, $\ket{\phi_1}$ and $\ket{\phi_2}$ are $\epsilon$-far from being product states (see Def.~5 of Ref.~\cite{bouland2024statehiddensubgroupproblem}).
Since there are exponentially many ($2^{\Theta(K)}$) bipartitions of the state, a brute-force search for $C$ is not efficient. 

Ref.~\cite{bouland2024statehiddensubgroupproblem} introduces two algorithms that solve for $C$ in polynomial time.
Alg.~1 of Ref.~\cite{bouland2024statehiddensubgroupproblem} is a non-adaptive algorithm that consumes $\mathcal{O}(K^2/\epsilon^2)$ copies of $\ket{\psi}$, acting coherently on $\mathcal{O}(K/\epsilon^2)$ copies at a time.
In total, it uses $\mathcal{O}(K^3/\epsilon^2)$ $\CSWAP$ operations. 
Each $\CSWAP$ can be decomposed into a constant number of $T$ gates via the decomposition into $\CX$s and Toffoli gates~\cite{Jones_2013,nielsen2010quantum}.
Consequently, Alg.~1 requires $\mathcal{O}(K^3/\epsilon^2)$ T states.

We find that by applying our $\CT$ gadgets, Alg.~1 can be realized in $\mathcal{O}(K^2)$ Clifford measurements and in $\mathcal{O}(1)$ logical depth, where logical depth quantifies the depth of the compiled logical gadgets which may take $\mathcal{O}(d)$ syndrome extraction rounds each.
To achieve this, we note that the $K$ parallel $\mathcal{O}(K/\epsilon^2)$ $\CSWAP$s during the Fourier sampling circuit of Alg.~1 can be performed in parallel with $O(K)$ $\CSWAP^{\otimes K}$ gadgets.
Since this algorithm is non-adaptive in the sense that each circuit of the $O(K)$ shots does not depend on the output of another shot, we can run all $O(K)$ shots in parallel, each of which can be executed in a $\mathcal{O}(1)$ logical depth circuit.

Alg.~2 of Ref.~\cite{bouland2024statehiddensubgroupproblem} is an adaptive algorithm that consumes $\mathcal{O}(K/\epsilon^2)$ copies of $\ket{\psi}$, acting coherently on $\mathcal{O}(1/\epsilon^2)$ copies at a time.
In total, this algorithm uses $\mathcal{O}(K^2/\epsilon^2)$ $T$ states, and runs circuits of depth $\mathcal{O}(K^2)+\mathcal{O}(\log(1/\epsilon))$.
The Fourier sampling circuit in Alg.~2 uses $K$ parallel $\mathcal{O}(1/\epsilon^2)$ $\CSWAP$s.
Using Clifford measurements, we can perform these $\CSWAP$s with $\mathcal{O}(K)$ Clifford measurements.
However, because the shots are adaptive, even if each shot is constant depth, we must run all shots sequentially. 
As a result, the total logical depth required is $\mathcal{O}(K)$.

\ifdefined\ismainfile
\else
    \bibliography{bib}
    \end{document}
\fi

\section{Discussion}
In this work, we have described a general \emph{gauging measurement} protocol by which Clifford logicals of generic QLDPC codes may be measured fault-tolerantly. First, the Clifford operator is gauged, deforming the initial code into a non-Abelian Clifford stabilizer code. Next, the system is deformed back onto the initial code and the input code state is projected onto an eigenstate of the target Clifford logical. We have proven that the full measurement protocol is fault tolerant, given an input qubit Pauli stabilizer code with an order-$2$ transversal Clifford. This treatment of fault-tolerance for non-Abelian QLDPC codes goes beyond the heuristic discussions that have appeared previously in the literature. 

The logical Clifford magic states prepared by our protocol applied to general CSS QLDPC codes may be used to perform non-Clifford operations, either by conversion to a conventional $\ket{\mathsf{Tof}}$ or $\ket{CCZ}$ state, or through a gadget such as those we have introduced in Sec.~\ref{sec:applications}. In the applications of the gauging measurement protocol we have discussed, we were able to reduce the magic state cost for several useful algorithmic subroutines compared to prior methods. We have also discussed several examples of codes resulting from gauging transversal gates in Sec~\ref{sec:examples}.

Our work presents an approach to fault-tolerant magic state preparation, that is distinct from distillation, cultivation, and transversal non-Clifford strategies. Our method only requires the existence of a transversal Clifford operator, and as such the gauging measurement enables us to prepare magic states on a wide array of codes, including generic CSS codes. We avoid the need for costly post-selection by leveraging the distance and LDPC nature of the intermediate gauged code. 

The appearance of Clifford codes in our protocol introduces new complications as compared to Pauli measurement schemes. The stabilizer group of a Clifford stabilizer code may not commute outside of the codespace, detectors become randomized in the presence of certain faults. These randomized detectors no longer provide useful information about the fault history. We have developed new theoretical tools to analyze these complications; in particular, we have developed a framework through which the fault-tolerance properties of Clifford LDPC codes may be analyzed in the presence of such randomized detectors.

Our work on the fault-tolerance of procedures for logical Clifford measurement on QLDPC codes has uncovered a number of new directions which we discuss here, and leave to future exploration. 

\textbf{New useful magic states? }We have presented some applications of Clifford gauging measurements aside from magic state preparation; exploring the landscape of similar ``gauge-able'' gadgets is a promising direction. In particular, it would be interesting to better understand the kinds of structures in circuits that can be made more efficient through a gauging operation. The $\SWAP$ network example we discussed in this work indicates that algorithms containing many gates conditioned on one ancilla can be compressed in a straightforward manner. Are there other common circuit motifs one can produce useful resource states for via gauging?

\textbf{Going beyond qubits? }Currently, our fault distance proof is limited to the case of gauging $\mathbb{Z}_2$ Clifford gates possessed by qubit stabilizer codes. While this encompasses most relevant magic states and Clifford gates, there are interesting Clifford gates that do not fit these requirements. As an explicit example, take the order-3 facet gate $HS$ on the self-dual BB code we considered in Sec.~\ref{sec:examples}. Or we may go beyond the Clifford hierarchy and consider transversal gates  applied to \emph{qudit} codes, see the example of gauging charge conjugation symmetry on a qutrit BB code in Sec.~\ref{sec:examples}. In both cases, the resulting algebra of gauged code stabilizers has a more complex structure than the $\mathbb{Z}_2$ case. Specifically, there are some non-Abelian faults that drop their own detectors, making the spacetime cleaning procedure ambiguous. However, there may be hope to construct a decoder for such codes explicitly by modifying techniques leveraged for similar topological codes~\cite{lyons2026}.

\textbf{Exotic phases of matter? }One may ask, what is the general structure of the non-Pauli codes produced by gauging Clifford logical operators? Can they be organized into well-defined phases of matter, potentially through a topological defect network picture? Can we produce more exotic models by gauging nonabelian group symmetries, or even Hopf algebra symmetries? Can these techniques and constructions be extended to continuous variable systems?

\textbf{Decoding? }We have proved that many Clifford gauging protocols for stabilizer qLDPC codes have nontrivial spacetime distance. Putting these protocols into practice will require some sort of decoding while we remain in the intermediate Clifford stabilizer code. We leave a more detailed treatment of decoding strategies for the Clifford LDPC codes introduced here to future work. However, there are a few directions we believe seem promising to explore. In particular, decoders known to work well for expander codes, like small-set flip (SSF) \cite{Leverrier_2015,Fawzi_2018_2,Fawzi_2018} and windowed BP-OSD \cite{Panteleev_2021, roffe2020, huang2023improvednoisysyndromedecoding, gong2024lowlatencyiterativedecodingqldpc} could be generalized in a straightforward way to $\mathbb{Z}_2$ codes with $\mathbb{Z}_2$ gauged Clifford symmetries. We note that just-in-time decoders have been proven to work for non-Abelian \emph{topological} codes \cite{davydova2025universal, lyons2026}. These decoders possess a renormalization group-type structure which is tailored to the finite-dimensional nature of topological codes, but it remains to be seen whether they are suitable for expander geometry. However, one could envision combining the just-in-time structure with a better heuristic decoder like BP-OSD. We predict that it will be necessary to incorporate a window-decoding strategy alongside the BP-OSD, since non-Abelian faults that persist for any extensive period of time can hide an extensive number of Abelian faults, and thus reduce the spacetime distance of the procedure. 

\textbf{Numerical simulation? }After constructing a viable decoder, it remains to be seen what the overall resource overhead for the gauging measurement protocol looks like in practice. Recent work \cite{delafuente2026highthresholddecodingnonpaulicodes} has demonstrated that just-in-time decoding for a topological Clifford code supporting $D(D_4)$ anyons can perform comparably to the surface code under circuit-level noise. If we conjecture that the threshold for the Clifford measurement protocol is similar to the memory threshold of the underlying Clifford code, this suggests that the performance of gauging-based methods may be comparable to stabilizer ones. Recently introduced efficient Clifford code simulators may be leveraged to make the necessary numerics possible ~\cite{takada2026exactefficientsimulationnoisy, Surti_2026,delafuente2026efficientlyestimatingfailurerates}. 

\textbf{Fast, High-rate surgery? }We have not attempted to optimize any overhead or resource-costs for the protocol introduced here. It is natural to consider how to implement fast Clifford surgery in the presence of many logical representatives of the Clifford transversal in question, see Ref.~\cite{williamson2026fast}. Determining the properties that the input code and the gauging graph need to satisfy to allow for fault-tolerant fast surgery, as well as identifying promising code candidates, is a topic for future work. It is also interesting to consider how we can improve the rate of magic state preparation through parallel gauging methods: we presented a protocol to generate a set of $k$ disjoint logical magic states on a high-rate code with an unaddressable transversal gate in an $O(k)$ number of gauging measurements, performed serially. Given a code with addressable Clifford gates, it may be possible to create $O(k)$ magic states in parallel following a procedure similar to Ref.~\cite{cowtan2025parallel,Zhang_2025}, reducing time overhead. Incorporating Clifford gauging measurement into existing architectures, such as the fixed-connectivity extractor architecture \cite{he2025extractorsqldpcarchitecturesefficient} is also interesting. In this setting, we imagine that the base code has a transversal Clifford operator, such as a $\CNOT$ between codeblocks, that one can gauge to prepare magic states or perform a particular magic gate. All extra gauge qubits and qubit connectivity required to measure $\CNOT$ can be fixed in advance for each disjoint pair of codeblocks. Consequently, a pair of codeblocks would be able to dynamically change their role in the compilation from computational logical processing units to magic state factories. More generally, extractors can be generalized to support Clifford measurements on a code block by constructing an ancilla based on the graph desiderata presented above for the whole code (or a targeted subregion) rather than the Pauli measurement graph desiderata.

\bibliography{bib}

\appendix

\newpage
\ifdefined\ismainfile
\else
    \documentclass[11pt]{article}
    \input{preamble.tex}
    \begin{document}
    \title{Appendices}
    \date{\today}
    \maketitle
    \appendix
\fi

\makeatletter
\@ifundefined{appendixproofenv}{}{}
\@ifundefined{c@appendixproof}{\newcounter{appendixproof}}{}
\makeatother

\section{Alternate forms for $\mathbb{Z}_2$ stabilizers}

Here we derive an alternate form for the gauged stabilizers in the case $\T^2 = \id$. This alternate form replaces the generalized phase gates $\Sg$ which are supported on multiple qubits with an increased number of $CZ$ gates. The choice of which stabilizer generators to use may depend on which kind of gate is most convenient for the given purpose or platform.

First, we prove the following identity. Let $U^2 = (-1)^m$. Consider an edge of the gauging graph $e$ whose boundary consists of vertices $v, v'$. Then
\begin{equation}
    C_v U \xrightarrow[L_e = +1]{} C_e U \cdot C_{v'} U [C Z_{e,v'}]^m 
\end{equation}
\begin{proof}
    \begin{equation}
\begin{aligned}
    C_v U &= \frac{1 + Z_v}{2} \otimes \id + \frac{1 - Z_v}{2} \otimes U \\
    &\xrightarrow[L_e = +1]{} \frac{1 + Z_e Z_{v'}}{2} \otimes \id + \frac{1 - Z_e Z_{v'}}{2} \otimes (-1)^m U^2 U \\
    &= \left [ \left ( \frac{1 + Z_e}{2} \right ) \left ( \frac{1 + Z_{v'}}{2}\right ) + \left ( \frac{1 - Z_e}{2} \right ) \left ( \frac{1 - Z_{v'}}{2}\right ) \right ] \otimes \id \\
    & \quad \quad + \left [ \left ( \frac{1 + Z_e}{2} \right ) \left ( \frac{1 - Z_{v'}}{2}\right ) + \left ( \frac{1 - Z_e}{2} \right ) \left ( \frac{1 + Z_{v'}}{2}\right ) \right ] \otimes U  \\
    &= \left ( \frac{1 + Z_e}{2} \right ) \left ( \frac{1 + Z_{v'}}{2}\right ) \otimes \id  + \left ( \frac{1 - Z_e}{2} \right ) \left ( \frac{1 - Z_{v'}}{2}\right ) \otimes (-1)^{m n_{v'}} U^2 \\
    & \quad \quad + \left ( \frac{1 + Z_e}{2} \right ) \left ( \frac{1 - Z_{v'}}{2}\right ) \otimes U + \left ( \frac{1 - Z_e}{2} \right ) \left ( \frac{1 + Z_{v'}}{2}\right ) \otimes (-1)^{m n_{v'}} U  \\
    &= \left [\left ( \frac{1 + Z_e}{2} \right ) \otimes \id + \left ( \frac{1 - Z_{e}}{2}\right ) \otimes(-1)^{m n_{v'}} U \right ] \left [ \left ( \frac{1 + Z_{v'}}{2}\right ) \otimes \id + \left ( \frac{1 - Z_{v'}}{2}\right ) \otimes U\right ] \\
    &= \left [\left ( \frac{1 + Z_e}{2} \right ) \otimes \id + \left ( \frac{1 - Z_{e}}{2}\right ) \otimes U \right ] \left [ \left ( \frac{1 + Z_e}{2} \right ) \otimes \id + \left ( \frac{1 - Z_{e}}{2}\right ) \otimes Z_{v'}^m \right ] \\
    & \quad \quad \quad \cdot \left [ \left ( \frac{1 + Z_{v'}}{2}\right ) \otimes \id + \left ( \frac{1 - Z_{v'}}{2}\right ) \otimes U\right ] \\
    &= C_e U \cdot [CZ_{e,v'}]^m \cdot C_{v'} U
\end{aligned}
\end{equation}
\end{proof}
Similarly, if $v, v'$ are joined by a longer set of edges $p(v, v') = \{e_1, e_2, \cdots, e_n\}$, we can derive an equivalent identity. Let $\partial e_i = \{v_{i-1}, v_i\}$, and $v_0 \equiv v$, $v_n \equiv v'$. 
\begin{equation}
\begin{aligned}
    C_v U &= C_{v_1} U C_{e_1} U (CZ_{e_1, v_1})^m \\
    &= [C_{v_2} U C_{e_2} U (CZ_{e_2, v_2})^m] C_{e_1} U (CZ_{e_1, v_1})^m \\
    &= C_{v'} U \left ( \prod_{e \in p(v, v')} C_e U \right ) \prod_{i = 1}^n (CZ_{e_i, v_i})^m \\
\end{aligned}
\end{equation}
Note that, since $Z^2 = 1$, we can apply our identity to the last set of CZ terms, setting $m = 0$:
\begin{equation}
    \begin{aligned}
        CZ_{v_i, e_i} = CZ_{v', e_i} \prod_{j > i} CZ_{e_j, e_i} 
    \end{aligned}
\end{equation}
Plugging back into the general case: 
\begin{equation}
    C_v U = C_{v'} U \left ( \prod_{e \in p(v, v')} C_e U \right ) \left [\prod_{i = 1}^n (CZ_{v', e_i})^m \left  (\prod_{j > i} CZ_{e_j, e_i} \right )^m \right ]
\end{equation}
Thus, an equivalent form of the decorated stabilizers is given by: 
\begin{equation}
\begin{aligned}
    \Circo_2{[S]} &= S \cdot \prod_{v \in \invvertexset{S}} C_v (i^{\mu_v} (N_S)_v) \\
    &= S \cdot \prod_{v \in \invvertexset{S} \backslash v_0} \left ( \prod_{e \in p(v, v_0)} C_e (i^{\mu_v} N_S) \right ) \left (\prod_{j > i} CZ_{e_j, e_i} \right )^{\mu_v}
\end{aligned}
\end{equation}

\section{Deferred proofs for the gauging procedure}\label{app:gauging-procedure-proofs}

\refstepcounter{appendixproof}\label{proof:vertex-ancilla-dropout}
\begin{proof}[Proof~\theappendixproof\ of Eq.~\eqref{eq:vertex-ancilla-dropout}]
Recall the setting of Section~\ref{sec:gauging-procedure}: $\omega=e^{2\pi i/p}$, $\Zp$ is the generalized Pauli-$Z$ acting on the $p$-dimensional vertex and edge ancillas, and every dyad admits the expansion $\dyad{a}=\frac{1}{p}\sum_{b=0}^{p-1}\omega^{ab}\Zp^{b}$. After applying the cluster entangler, every edge $e=v \to v'$ of the gauging graph satisfies $L_e=\Zp_{v}^\dagger \Zp_e \Zp_{v'}=+1$, and $p(v,v_0)$ denotes a path of edges from the vertex $v$ to the root $v_0$ with orientation signs $\eta_p(e)=\pm1$.  Then:
\begin{equation}
    \begin{aligned}
        \sum_{a_0 = 0}^{p-1} \dyad{a_0}_{v_0} \otimes \dyad{\mathbf{a'} + a_0}_{v \neq v_0} &= \sum_{a_0=0}^{p-1} \left (\frac{1}{p} \sum_{n = 0}^{p-1} \omega^{a_0 n} \Zp^n_{v_0} \right ) \otimes \left (\frac{1}{p^{|v\neq v_0|} }\sum_{\mathbf{m} \in \mathbb{Z}_p^{|v\neq v_0|}} \omega^{(\mathbf{a'} + a_0)\cdot \mathbf{m}} \Zp^{\mathbf{m}}_{v\neq v_0} \right ) \\
        &= \frac{1}{p^{|v|}} \sum_{n, \mathbf{m}} \sum_{a_0} \omega^{(n + |m|) a_0 + \mathbf{a'}\cdot \mathbf{m}} \Zp_{v_0}^n \Zp_{v \neq v_0}^{\mathbf{m}} \\
        &\xrightarrow{L_e = +1} \frac{1}{p^{|v|}} \sum_{n, \mathbf{m}} \sum_{a_0} \omega^{(n + |m|) a_0 + \mathbf{a'}\cdot \mathbf{m}} \Zp_{v_0}^{n + |m|} \prod_{v \neq v_0} \prod_{e \in p(v, v_0)} \Zp_{e}^{\eta_p(e) m_v} \\
        &= \frac{1}{p^{|v \neq v_0|}} \sum_{a_0, \mathbf{m}} \left ( \frac{1}{p} \sum_{n'} \omega^{n' a_0} \Zp_{v_0}^{n'} \right ) \prod_{v \neq v_0} \prod_{e \in p(v, v_0)} \omega^{a'_v m_v} \Zp_{e}^{\eta_p(e) m_v} \\
        &=\frac{1}{p^{|v \neq v_0|}} \sum_{\mathbf{m}} \left ( \sum_{a_0} \dyad{a_0}_{v_0} \right )  \prod_{v \neq v_0} \prod_{e \in p(v, v_0)} \omega^{a'_v m_v} \Zp_{e}^{\eta_p(e) m_v} \\
        &=\frac{1}{p^{|v \neq v_0|}} \sum_{\mathbf{m}} \id_{v_0} \otimes \prod_{v \neq v_0} \prod_{e \in p(v, v_0)} \omega^{a'_v m_v} \Zp_{e}^{\eta_p(e) m_v} \\
        &=\id_{v_0} \otimes \prod_{v \neq v_0} \prod_{e \in p(v, v_0)} \dyad{\eta_p(e) a_v'}_e \\
    \end{aligned}
\end{equation}
\end{proof}

\refstepcounter{appendixproof}\label{proof:clifford-decorated-commutation-relations}
\begin{proof}[Proof~\theappendixproof\ of Corollary~\ref{cor:z2-commutation-relations}]
Recall from Eq.~\eqref{eq:z2-decorated-stabilizers} that $\gaugedS=\Circo_2[S]=S\cdot\bigl(\prod_{v\neq v_0}\mathsf{S}^{\mu_v}_{p(v,v_0)}\prod_{e\in p(v,v_0)}C_e(N_S)_v\bigr)$, with decorations $\mathsf{S}^{\mu_v}_{p(v,v_0)}$ and $C_e(N_S)_v$ along rooted paths, where $N_S=\ComGrp{S}{\T}$, $\ComGrp{S_v}{\T_v}=i^{\mu_v}(N_S)_v$, and $\sum_v\mu_v=0 \pmod 4$.

\[
\resizebox{\textwidth}{!}{$\displaystyle
\begin{aligned}
   & \ComGrp{\gaugedS}{\mathbb{A}_{v_0}} \nonumber\\
   =& \left ( \prod_{v \neq v_0} \mathsf{S}^{\mu_v}_{p(v, v_0)} \prod_{e \in p(v, v_0)} C_e (N_S)_v \right )  S  \T_{v_0} \prod_{e \in \delta v_0} X_e
      S  \left ( \prod_{v \neq v_0} \mathsf{S}^{-\mu_v}_{p(v, v_0)} \prod_{e \in p(v, v_0)} C_e (N_S)_v \right )  \T_{v_0} \prod_{e \in \delta v_0 } X_e \nonumber\\
    =&\left ( \prod_{v \neq v_0} \mathsf{S}^{\mu_v}_{p(v, v_0)} \prod_{e \in p(v, v_0)} C_e (N_S)_v \right )  ( S \T_{v_0} S \T_{v_0})
      \prod_{e \in \delta v_0 } X_e \left ( \prod_{v \neq v_0} \mathsf{S}^{-\mu_v}_{p(v, v_0)} \prod_{e \in p(v, v_0)} C_e (N_S)_v \right ) \prod_{e \in \delta v_0 } X_e \nonumber\\
    =& i^{\mu_{v_0}} (N_S)_{v_0} \left ( \prod_{v \neq v_0} \mathsf{S}^{\mu_v}_{p(v, v_0)} \prod_{e \in p(v, v_0)} C_e (N_S)_v \right )
     \prod_{e \in \delta v_0 } X_e \left ( \prod_{v \neq v_0} \mathsf{S}^{-\mu_v}_{p(v, v_0)} \prod_{e \in p(v, v_0)} C_e (N_S)_v \right ) \prod_{e \in \delta v_0 } X_e \nonumber\\
    =& i^{\mu_{v_0}}(N_S)_{v_0} \left ( \prod_{v \neq v_0} \mathsf{S}^{\mu_v}_{p(v, v_0)} \prod_{e \in p(v, v_0)} C_e (N_S)_v \right )
      \prod_{v \neq v_0} i^{\mu_v} \mathsf{S}^{-2\mu_v}_{p(v, v_0)} \mathsf{S}^{\mu_v}_{p(v, v_0)} \prod_{e \in \delta v_0 } X_e \prod_{e \in p(v, v_0)} C_e (N_S)_v \prod_{e \in \delta v_0 } X_e \nonumber\\
    =& i^{\sum_{v} \mu_{v}}(N_S)_{v_0}  \prod_{v \neq v_0} \mathsf{S}^{-2\mu_v}_{p(v, v_0)}
    \prod_{e \in p(v, v_0)} C_e (N_S)_v \prod_{e \in \delta v_0 } X_e \prod_{e \in p(v, v_0)} C_e (N_S)_v \prod_{e \in \delta v_0 } X_e \nonumber\\
    =& (N_S)_{v_0}  \prod_{v \neq v_0} \mathsf{S}^{-2\mu_v}_{p(v, v_0)} (N_S)_v
        = N_S \prod_{v \neq v_0} \prod_{e \in p(v, v_0)} Z^{\mu_v}_e
        = \mathbb{N}_S
\end{aligned}$}
\]
\[
\resizebox{\textwidth}{!}{$\displaystyle
\begin{aligned}
    &\ComGrp{\gaugedS}{\mathbb{A}_u} \nonumber\\
    =& \left ( \prod_{v \neq v_0} \mathsf{S}^{\mu_v}_{p(v, v_0)} \prod_{e \in p(v, v_0)} C_e (N_S)_v \right )  S  \T_{u} \prod_{e \in \delta u} X_e
      S  \left ( \prod_{v \neq v_0} \mathsf{S}^{-\mu_v}_{p(v, v_0)} \prod_{e \in p(v, v_0)} C_e (N_S)_v \right )  \T_{u} \prod_{e \in \delta u } X_e \nonumber\\
    =& \left ( \prod_{v \neq v_0} \mathsf{S}^{\mu_v}_{p(v, v_0)} \prod_{e \in p(v, v_0)} C_e (N_S)_v \right )  (i^{-\mu_u} (N_S)_u)
      \prod_{e \in \delta u } X_e \T_{u} \left ( \prod_{v \neq v_0} \mathsf{S}^{-\mu_v}_{p(v, v_0)} \prod_{e \in p(v, v_0)} C_e (N_S)_v \right ) \T_{u} \prod_{e \in \delta u } X_e \nonumber\\
    =& i^{-\mu_u} (N_S)_u \left ( \prod_{v \neq v_0} \mathsf{S}^{\mu_v}_{p(v, v_0)} \prod_{e \in p(v, v_0)} C_e (N_S)_v \right )
     \prod_{e \in \delta u } X_e \left ( \prod_{e \in p(u, v_0)} Z_e^{\mu_u}\right )\left (\prod_{v \neq v_0} \mathsf{S}^{-\mu_v}_{p(v, v_0)} \prod_{e \in p(v, v_0)} C_e (N_S)_v \right ) \prod_{e \in \delta u } X_e \nonumber\\
    =& i^{-\mu_u} (N_S)_u \left ( \prod_{v \neq v_0} \mathsf{S}^{\mu_v}_{p(v, v_0)} \prod_{e \in p(v, v_0)} C_e (N_S)_v \right )
       (-1)^{\mu_u} \left ( \prod_{e \in p(u, v_0)} Z_e^{\mu_u}\right ) \prod_{e \in \delta u } X_e \prod_{v \neq v_0} \mathsf{S}^{-\mu_v}_{p(v, v_0)}  \prod_{e \in p(v, v_0)} C_e (N_S)_v \prod_{e \in \delta u } X_e  \nonumber\\
    =& i^{\mu_u} (N_S)_u \left ( \prod_{v \neq v_0} \mathsf{S}^{\mu_v}_{p(v, v_0)} \prod_{e \in p(v, v_0)} C_e (N_S)_v \right ) \left ( \prod_{e \in p(u, v_0)} Z_e^{\mu_u}\right ) \nonumber\\
    &  \left (i^{-\mu_u} \mathsf{S}^{-2\mu_u}_{p(u, v_0)} \right ) \prod_{v \neq v_0} \mathsf{S}^{-\mu_v}_{p(v, v_0)} \prod_{e \in \delta u } X_e \prod_{e \in p(v, v_0)} C_e (N_S)_v \prod_{e \in \delta u } X_e \nonumber\\
    =& (N_S)^2_u \left ( \prod_{v \neq v_0} \mathsf{S}^{\mu_v}_{p(v, v_0)} \prod_{e \in p(v, v_0)} C_e (N_S)_v \right ) \left ( \prod_{e \in p(u, v_0)} Z_e^{\mu_u}\right )^2
      \prod_{v \neq v_0} \mathsf{S}^{-\mu_v}_{p(v, v_0)} \prod_{e \in \delta u } C_e (N_S)_v      = \id ,\qquad \text{for $u\neq v_0$.}
\end{aligned}$}
\]

Consider two decorated stabilizers that overlap on some degrees of freedom.
\[
\resizebox{\textwidth}{!}{$\displaystyle
\begin{aligned}
    \ComGrp{\gaugedS_i}{\gaugedS_j} &= \left ( \prod_{v \neq v_0} \mathsf{S}^{\mu_v}_{p(v, v_0)} \prod_{e \in p(v, v_0)} C_e(N_{S_i})_v \right )  S_i \left ( \prod_{u \neq u_0} \mathsf{S}^{\mu_u}_{p(u, u_0)} \prod_{e \in p(u, u_0)} C_e(N_{S_j})_u \right ) S_j S_i \nonumber\\
    &\qquad  \left ( \prod_{v \neq v_0} \mathsf{S}^{-\mu_v}_{p(v, v_0)} \prod_{e \in p(v, v_0)} C_e(N_{S_i})_v \right ) S_j \left ( \prod_{u \neq u_0} \mathsf{S}^{-\mu_u}_{p(u, u_0)} \prod_{e \in p(u, u_0)} C_e(N_{S_j})_u \right ) \nonumber\\
    &= \left ( \prod_{v \neq v_0} \mathsf{S}^{\mu_v}_{p(v, v_0)} \prod_{e \in p(v, v_0)} C_e(N_{S_i})_v \right )  S_i \left ( \prod_{u \neq u_0} \mathsf{S}^{\mu_u}_{p(u, u_0)} \prod_{e \in p(u, u_0)} C_e(N_{S_j})_u \right ) S_i \nonumber\\
    &\qquad  S_j \left ( \prod_{v \neq v_0} \mathsf{S}^{-\mu_v}_{p(v, v_0)} \prod_{e \in p(v, v_0)} C_e(N_{S_i})_v \right ) S_j \left ( \prod_{u \neq u_0} \mathsf{S}^{-\mu_u}_{p(u, u_0)} \prod_{e \in p(u, u_0)} C_e(N_{S_j})_u \right ) \\
    &= \left ( \prod_{v \neq v_0} \mathsf{S}^{\mu_v}_{p(v, v_0)} \prod_{e \in p(v, v_0)} C_e(N_{S_i})_v  \right )   \left ( \prod_{u \neq u_0} \mathsf{S}^{\mu_u}_{p(u, u_0)}  \prod_{e \in p(u, u_0)}  C_e((S_i)_u (N_{S_j})_u (S_i)_u)   \right )\nonumber\\
    &\qquad \left ( \prod_{v \neq v_0} \mathsf{S}^{-\mu_v}_{p(v, v_0)}  \prod_{e \in p(v, v_0)} C_e( (S_j)_v (N_{S_i})_v (S_j)_v)\right )  \left ( \prod_{u \neq u_0} \mathsf{S}^{-\mu_u}_{p(u, u_0)} \prod_{e \in p(u, u_0)} C_e(N_{S_j})_u \right ) \\
     &= \left ( \prod_{v \neq v_0} \mathsf{S}^{\mu_v}_{p(v, v_0)} \prod_{e \in p(v, v_0)} C_e(N_{S_i})_v  \right )   \left ( \prod_{u \neq u_0} \mathsf{S}^{\mu_u}_{p(u, u_0)}  \prod_{e \in p(u, u_0)} Z^{\gamma(i,j)}_e C_e(N_{S_j})_u   \right )\nonumber\\
    &\qquad \left ( \prod_{v \neq v_0} \mathsf{S}^{-\mu_v}_{p(v, v_0)}  \prod_{e \in p(v, v_0)} Z^{\gamma(j, i)}_e C_e(N_{S_i})_v\right )  \left ( \prod_{u \neq u_0} \mathsf{S}^{-\mu_u}_{p(u, u_0)} \prod_{e \in p(u, u_0)} C_e(N_{S_j})_u \right ) \\
    &= \prod_{v \neq u_0, v_0} \prod_{e \in p(v, u_0)} Z^{\gamma_v(i, j)}_e \prod_{f \in p(v, v_0)} Z_f^{\gamma_v(j, i)}, \quad \text{ as } \gamma \neq 0 \text{ in places where } S_i, S_j \text{ overlap.}
\end{aligned}$}
\]
In the above, we have defined $(N_{S_j})_u (S_i)_u = (-1)^{\gamma_u (i, j)} (S_i)_u (N_{S_j})_u$. Note that $\gamma_u(i, j) = \gamma_u(j, i)$.

The middle equality follows by conjugating the second commutator by $\T$ and using the symmetry of Pauli commutator signs.
The remaining uncancelled gauge-$Z$ support is precisely a cycle obtained by comparing the two paths $p(v, u_0)$ and $p(v, v_0)$ used for the decoration of either stabilizer, giving the flux check $\mathbb{B}_{p(i,j)}$.  

Finally, $\bp$ commutes with $\av$ by cycle-boundary parity, and $\bp$ commutes with $\gaugedS$ because the gauge support of $\gaugedS$ is $Z$-type or diagonal in the gauge computational basis.
\end{proof}

\ifdefined\ismainfile
\else
    \bibliographystyle{unsrt}
    \bibliography{bib}

@misc{freedman2021buildingmanifoldsquantumcodes,
      title={Building manifolds from quantum codes}, 
      author={Michael Freedman and Matthew B. Hastings},
      year={2021},
      eprint={2012.02249},
      archivePrefix={arXiv},
      primaryClass={math.DG},
      url={https://arxiv.org/abs/2012.02249}, 
}

@article{Zhang_2025,
   title={Time-Efficient Logical Operations on Quantum Low-Density Parity Check Codes},
   volume={134},
   ISSN={1079-7114},
   url={http://dx.doi.org/10.1103/PhysRevLett.134.070602},
   DOI={10.1103/physrevlett.134.070602},
   number={7},
   journal={Physical Review Letters},
   publisher={American Physical Society (APS)},
   author={Zhang, Guo and Li, Ying},
   year={2025},
   month=Feb }

@misc{swaroop2025universaladaptersquantumldpc,
      title={Universal adapters between quantum LDPC codes}, 
      author={Esha Swaroop and Tomas Jochym-O'Connor and Theodore J. Yoder},
      year={2025},
      eprint={2410.03628},
      archivePrefix={arXiv},
      primaryClass={quant-ph},
      url={https://arxiv.org/abs/2410.03628}, 
}

@article{chakraborty2026,
  title = {No-Go Theorem on Fault Tolerant Gadgets for Multiple Logical Qubits},
  author = {Chakraborty, Aranya and Gottesman, Daniel},
  journal = {PRX Quantum},
  volume = {7},
  issue = {3},
  pages = {033027},
  numpages = {12},
  year = {2026},
  month = {Aug},
  publisher = {American Physical Society},
  doi = {10.1103/y14y-7kp3},
  url = {https://link.aps.org/doi/10.1103/y14y-7kp3}
}

@article{Bulmash_2019,
  title = {Gauging fractons: Immobile non-Abelian quasiparticles, fractals, and position-dependent degeneracies},
  author = {Bulmash, Daniel and Barkeshli, Maissam},
  journal = {Phys. Rev. B},
  volume = {100},
  issue = {15},
  pages = {155146},
  numpages = {35},
  year = {2019},
  month = {Oct},
  publisher = {American Physical Society},
  doi = {10.1103/PhysRevB.100.155146},
  url = {https://link.aps.org/doi/10.1103/PhysRevB.100.155146}
}

@article{Zhu_2025,
  title = {Non-Clifford and Parallelizable Fault-Tolerant Logical Gates on Constant and Almost-Constant Rate Homological Quantum Low-Density Parity-Check Codes via Higher Symmetries},
  author = {Zhu, Guanyu and Sikander, Shehryar and Portnoy, Elia and Cross, Andrew W. and Brown, Benjamin J.},
  journal = {PRX Quantum},
  volume = {6},
  issue = {4},
  pages = {040361},
  numpages = {43},
  year = {2025},
  month = {Dec},
  publisher = {American Physical Society},
  doi = {10.1103/wcxs-w69t},
  url = {https://link.aps.org/doi/10.1103/wcxs-w69t}
}

@misc{zhu2026nonabelianqldpctqftformalism,
      title={Non-Abelian qLDPC: TQFT Formalism, Addressable Gauging Measurement and Application to Magic State Fountain on 2D Product Codes}, 
      author={Guanyu Zhu and Ryohei Kobayashi and Po-Shen Hsin},
      year={2026},
      eprint={2601.06736},
      archivePrefix={arXiv},
      primaryClass={quant-ph},
      url={https://arxiv.org/abs/2601.06736}, 
}

@inproceedings{Fawzi_2018_2, series={STOC ’18},
   title={Efficient decoding of random errors for quantum expander codes},
   url={http://dx.doi.org/10.1145/3188745.3188886},
   DOI={10.1145/3188745.3188886},
   booktitle={Proceedings of the 50th Annual ACM SIGACT Symposium on Theory of Computing},
   publisher={ACM},
   author={Fawzi, Omar and Grospellier, Antoine and Leverrier, Anthony},
   year={2018},
   month=June, pages={521–534},
   collection={STOC ’18} }

@inproceedings{Fawzi_2018,
   title={Constant Overhead Quantum Fault-Tolerance with Quantum Expander Codes},
   url={http://dx.doi.org/10.1109/FOCS.2018.00076},
   DOI={10.1109/focs.2018.00076},
   booktitle={2018 IEEE 59th Annual Symposium on Foundations of Computer Science (FOCS)},
   publisher={IEEE},
   author={Fawzi, Omar and Grospellier, Antoine and Leverrier, Anthony},
   year={2018},
   month=Oct, pages={743–754} }

@misc{cross2025improvedqldpcsurgerylogical,
      title={Improved QLDPC Surgery: Logical Measurements and Bridging Codes}, 
      author={Andrew W. Cross and Zhiyang He and Patrick J. Rall and Theodore J. Yoder},
      year={2025},
      eprint={2407.18393},
      archivePrefix={arXiv},
      primaryClass={quant-ph},
      url={https://arxiv.org/abs/2407.18393}, 
}

@article{Tantivasadakarn23,
  title = {Shortest Route to Non-Abelian Topological Order on a Quantum Processor},
  author = {Tantivasadakarn, Nathanan and Verresen, Ruben and Vishwanath, Ashvin},
  journal = {Phys. Rev. Lett.},
  volume = {131},
  issue = {6},
  pages = {060405},
  numpages = {5},
  year = {2023},
  month = {Aug},
  publisher = {American Physical Society},
  doi = {10.1103/PhysRevLett.131.060405},
  url = {https://link.aps.org/doi/10.1103/PhysRevLett.131.060405}
}

@misc{delafuente2026highthresholddecodingnonpaulicodes,
      title={High-threshold decoding of non-Pauli codes for 2D universality}, 
      author={Julio C. Magdalena de la Fuente and Noa Feldman and Jens Eisert and Andreas Bauer},
      year={2026},
      eprint={2604.02033},
      archivePrefix={arXiv},
      primaryClass={quant-ph},
      url={https://arxiv.org/abs/2604.02033}, 
}

@misc{lyons2026,
      title={Quantum computing with anyons is fault tolerant}, 
      author={Anasuya Lyons and Benjamin J. Brown},
      year={2026},
      eprint={2602.11258},
      archivePrefix={arXiv},
      primaryClass={quant-ph},
      url={https://arxiv.org/abs/2602.11258}, 
}

@INPROCEEDINGS{Leverrier_2015,
  author={Leverrier, Anthony and Tillich, Jean-Pierre and Zémor, Gilles},
  booktitle={2015 IEEE 56th Annual Symposium on Foundations of Computer Science}, 
  title={Quantum Expander Codes}, 
  year={2015},
  volume={},
  number={},
  pages={810-824},
  doi={10.1109/FOCS.2015.55}}

@misc{verresen2022efficientlypreparingschrodingerscat,
      title={Efficiently preparing Schr\"odinger's cat, fractons and non-Abelian topological order in quantum devices}, 
      author={Ruben Verresen and Nathanan Tantivasadakarn and Ashvin Vishwanath},
      year={2022},
      eprint={2112.03061},
      archivePrefix={arXiv},
      primaryClass={quant-ph},
      url={https://arxiv.org/abs/2112.03061}, 
}

@article{Ren_2025,
  title = {Efficient Preparation of Solvable Anyons with Adaptive Quantum Circuits},
  author = {Ren, Yuanjie and Tantivasadakarn, Nathanan and Williamson, Dominic J.},
  journal = {Phys. Rev. X},
  volume = {15},
  issue = {3},
  pages = {031060},
  numpages = {26},
  year = {2025},
  month = {Aug},
  publisher = {American Physical Society},
  doi = {10.1103/b9hf-gx4f},
  url = {https://link.aps.org/doi/10.1103/b9hf-gx4f}
}

@article{Tantivasadakarn_2023,
  title = {Hierarchy of Topological Order From Finite-Depth Unitaries, Measurement, and Feedforward},
  author = {Tantivasadakarn, Nathanan and Vishwanath, Ashvin and Verresen, Ruben},
  journal = {PRX Quantum},
  volume = {4},
  issue = {2},
  pages = {020339},
  numpages = {25},
  year = {2023},
  month = {Jun},
  publisher = {American Physical Society},
  doi = {10.1103/PRXQuantum.4.020339},
  url = {https://link.aps.org/doi/10.1103/PRXQuantum.4.020339}
}

@article{cowtan2025parallel,
   title={Parallel Logical Measurements via Quantum Code Surgery},
   volume={7},
   ISSN={2691-3399},
   url={http://dx.doi.org/10.1103/gj8x-n5gg},
   DOI={10.1103/gj8x-n5gg},
   number={2},
   journal={PRX Quantum},
   publisher={American Physical Society (APS)},
   author={Cowtan, Alexander and He, Zhiyang and Williamson, Dominic J. and Yoder, Theodore J.},
   year={2026},
   month=May }

@article{Fowler_2012,
   title={Surface codes: Towards practical large-scale quantum computation},
   volume={86},
   ISSN={1094-1622},
   url={http://dx.doi.org/10.1103/PhysRevA.86.032324},
   DOI={10.1103/physreva.86.032324},
   number={3},
   journal={Physical Review A},
   publisher={American Physical Society (APS)},
   author={Fowler, Austin G. and Mariantoni, Matteo and Martinis, John M. and Cleland, Andrew N.},
   year={2012},
   month=Sept }

@article{williamson2024lowoverheadfaulttolerantquantumcomputation,
   title={Low-overhead fault-tolerant quantum computation by gauging logical operators},
   volume={22},
   ISSN={1745-2481},
   url={http://dx.doi.org/10.1038/s41567-026-03220-8},
   DOI={10.1038/s41567-026-03220-8},
   number={4},
   journal={Nature Physics},
   publisher={Springer Science and Business Media LLC},
   author={Williamson, Dominic J. and Yoder, Theodore J.},
   year={2026},
   month=Apr, pages={598–603} }

@article{davydova2025universal,
   title={Universal Fault-Tolerant Quantum Computation in 2D without Getting Tied in Knots},
   volume={16},
   ISSN={2160-3308},
   url={http://dx.doi.org/10.1103/vrty-qs5h},
   DOI={10.1103/vrty-qs5h},
   number={3},
   journal={Physical Review X},
   publisher={American Physical Society (APS)},
   author={Davydova, Margarita and Bauer, Andreas and Magdalena de la Fuente, Julio C. and Webster, Mark and Williamson, Dominic J. and Brown, Benjamin J.},
   year={2026},
   month=July }

@misc{he2025extractorsqldpcarchitecturesefficient,
      title={Extractors: QLDPC Architectures for Efficient Pauli-Based Computation}, 
      author={Zhiyang He and Alexander Cowtan and Dominic J. Williamson and Theodore J. Yoder},
      year={2025},
      eprint={2503.10390},
      archivePrefix={arXiv},
      primaryClass={quant-ph},
      url={https://arxiv.org/abs/2503.10390}, 
}

@article{Ide_2025,
  title = {Fault-Tolerant Logical Measurements via Homological Measurement},
  author = {Ide, Benjamin and Gowda, Manoj G. and Nadkarni, Priya J. and Dauphinais, Guillaume},
  journal = {Phys. Rev. X},
  volume = {15},
  issue = {2},
  pages = {021088},
  numpages = {23},
  year = {2025},
  month = {Jun},
  publisher = {American Physical Society},
  doi = {10.1103/PhysRevX.15.021088},
  url = {https://link.aps.org/doi/10.1103/PhysRevX.15.021088}
}

@article{dennis2001toward,
   title={Toward fault-tolerant quantum computation without concatenation},
   volume={63},
   ISSN={1094-1622},
   url={http://dx.doi.org/10.1103/PhysRevA.63.052314},
   DOI={10.1103/physreva.63.052314},
   number={5},
   journal={Physical Review A},
   publisher={American Physical Society (APS)},
   author={Dennis, Eric},
   year={2001},
   month=Apr }

@misc{LiangChen2025Bivariate,
      title={Self-dual bivariate bicycle codes with transversal Clifford gates}, 
      author={Zijian Liang and Yu-An Chen},
      year={2026},
      eprint={2510.05211},
      archivePrefix={arXiv},
      primaryClass={quant-ph},
      url={https://arxiv.org/abs/2510.05211}, 
}

@misc{Xu2025batched,
      title={Batched high-rate logical operations for quantum LDPC codes}, 
      author={Qian Xu and Hengyun Zhou and Dolev Bluvstein and Madelyn Cain and Marcin Kalinowski and John Preskill and Mikhail D. Lukin and Nishad Maskara},
      year={2025},
      eprint={2510.06159},
      archivePrefix={arXiv},
      primaryClass={quant-ph},
      url={https://arxiv.org/abs/2510.06159}, 
}

@article{Bravyi2024faulttolerant,
   title={High-threshold and low-overhead fault-tolerant quantum memory},
   volume={627},
   ISSN={1476-4687},
   url={http://dx.doi.org/10.1038/s41586-024-07107-7},
   DOI={10.1038/s41586-024-07107-7},
   number={8005},
   journal={Nature},
   publisher={Springer Science and Business Media LLC},
   author={Bravyi, Sergey and Cross, Andrew W. and Gambetta, Jay M. and Maslov, Dmitri and Rall, Patrick and Yoder, Theodore J.},
   year={2024},
   month=Mar, pages={778–782} }

@article{Low_2024,
   title={Trading T gates for dirty qubits in state preparation and unitary synthesis},
   volume={8},
   ISSN={2521-327X},
   url={http://dx.doi.org/10.22331/q-2024-06-17-1375},
   DOI={10.22331/q-2024-06-17-1375},
   journal={Quantum},
   publisher={Verein zur Forderung des Open Access Publizierens in den Quantenwissenschaften},
   author={Low, Guang Hao and Kliuchnikov, Vadym and Schaeffer, Luke},
   year={2024},
   month=jun, pages={1375} }

@article{Zhang_2022,
   title={Quantum State Preparation with Optimal Circuit Depth: Implementations and Applications},
   volume={129},
   ISSN={1079-7114},
   url={http://dx.doi.org/10.1103/PhysRevLett.129.230504},
   DOI={10.1103/physrevlett.129.230504},
   number={23},
   journal={Physical Review Letters},
   publisher={American Physical Society (APS)},
   author={Zhang, Xiao-Ming and Li, Tongyang and Yuan, Xiao},
   year={2022},
   month=nov }

@misc{liu2025blockencodinglowgate,
      title={Low-gate-count block encodings for second-quantized fermionic Hamiltonians}, 
      author={Diyi Liu and Shuchen Zhu and Lin Lin and Guang Hao Low and Chao Yang},
      year={2026},
      eprint={2510.08644},
      archivePrefix={arXiv},
      primaryClass={quant-ph},
      url={https://arxiv.org/abs/2510.08644}, 
}

@misc{bouland2024statehiddensubgroupproblem,
      title={The state hidden subgroup problem and an efficient algorithm for locating unentanglement}, 
      author={Adam Bouland and Tudor Giurgica-Tiron and John Wright},
      year={2024},
      eprint={2410.12706},
      archivePrefix={arXiv},
      primaryClass={quant-ph},
      url={https://arxiv.org/abs/2410.12706}, 
}

@article{Jones_2013,
   title={Low-overhead constructions for the fault-tolerant Toffoli gate},
   volume={87},
   ISSN={1094-1622},
   url={http://dx.doi.org/10.1103/PhysRevA.87.022328},
   DOI={10.1103/physreva.87.022328},
   number={2},
   journal={Physical Review A},
   publisher={American Physical Society (APS)},
   author={Jones, Cody},
   year={2013},
   month=Feb }

@book{nielsen2010quantum,
  title={Quantum computation and quantum information},
  author={Nielsen, Michael A and Chuang, Isaac L},
  year={2010},
  publisher={Cambridge University Press}
}

@article{Criger2016NoiseThresholds422ToricCode,
  title = {Noise thresholds for the {[[4,2,2]]}-concatenated toric code},
  author = {Criger, Ben and Terhal, Barbara M.},
  journal = {Quantum Information and Computation},
  volume = {16},
  number = {15--16},
  pages = {1261--1281},
  year = {2016},
  doi = {10.26421/QIC16.15-16},
  eprint = {1604.04062},
  archivePrefix = {arXiv},
  primaryClass = {quant-ph}
}

@article{Bombin2006TopologicalQuantumDistillation,
   title={Topological Quantum Distillation},
   volume={97},
   ISSN={1079-7114},
   url={http://dx.doi.org/10.1103/PhysRevLett.97.180501},
   DOI={10.1103/physrevlett.97.180501},
   number={18},
   journal={Physical Review Letters},
   publisher={American Physical Society (APS)},
   author={Bombin, H. and Martin-Delgado, M. A.},
   year={2006},
   month=Oct }

@article{Bravyi2010MajoranaFermionCodes,
   title={Majorana fermion codes},
   volume={12},
   ISSN={1367-2630},
   url={http://dx.doi.org/10.1088/1367-2630/12/8/083039},
   DOI={10.1088/1367-2630/12/8/083039},
   number={8},
   journal={New Journal of Physics},
   publisher={IOP Publishing},
   author={Bravyi, Sergey and Terhal, Barbara M and Leemhuis, Bernhard},
   year={2010},
   month=Aug, pages={083039} }

@article{Breuckmann_2026,
   title={Cups and Gates I: Cohomology Invariants and Logical Quantum Operations},
   volume={407},
   ISSN={1432-0916},
   url={http://dx.doi.org/10.1007/s00220-026-05570-z},
   DOI={10.1007/s00220-026-05570-z},
   number={5},
   journal={Communications in Mathematical Physics},
   publisher={Springer Science and Business Media LLC},
   author={Breuckmann, Nikolas P. and Davydova, Margarita and Eberhardt, Jens N. and Tantivasadakarn, Nathanan},
   year={2026},
   month=Apr }

@misc{sahay2026computingqldpccodesclimbing,
      title={Computing with qLDPC Codes by Climbing the Chain Map Hierarchy}, 
      author={Rahul Sahay and David M. Long and Vedika Khemani},
      year={2026},
      eprint={2609.02999},
      archivePrefix={arXiv},
      primaryClass={quant-ph},
      url={https://arxiv.org/abs/2609.02999}, 
}

@misc{manjunath2026universalquantumcomputationgroup,
      title={Universal quantum computation with group surface codes}, 
      author={Naren Manjunath and Vieri Mattei and Apoorv Tiwari and Tyler D. Ellison},
      year={2026},
      eprint={2603.05502},
      archivePrefix={arXiv},
      primaryClass={quant-ph},
      url={https://arxiv.org/abs/2603.05502}, 
}

@article{warman2026constantdepthcliffordhierarchygatesnonabelian,
   title={Constant-Depth Clifford-Hierarchy Gates via Non-Abelian Surface Codes},
   volume={7},
   ISSN={2691-3399},
   url={http://dx.doi.org/10.1103/kfxl-rv7c},
   DOI={10.1103/kfxl-rv7c},
   number={3},
   journal={PRX Quantum},
   publisher={American Physical Society (APS)},
   author={Warman, Alison and Schäfer-Nameki, Sakura},
   year={2026},
   month=Aug }

@misc{jing2026integerlinearprogrammingdecoder,
      title={Integer Linear Programming Decoder for Abelian and Non-Abelian Topological Codes}, 
      author={Dian Jing and Aubrey Zhang and Liang Jiang and Ruben Verresen},
      year={2026},
      eprint={2608.18512},
      archivePrefix={arXiv},
      primaryClass={quant-ph},
      url={https://arxiv.org/abs/2608.18512}, 
}

@article{Bauer_2026,
  title = {Planar Fault-Tolerant Circuits for Non-Clifford Gates on the 2D Color Code},
  author = {Bauer, Andreas and Magdalena de la Fuente, Julio C.},
  journal = {PRX Quantum},
  volume = {7},
  issue = {2},
  pages = {020367},
  numpages = {57},
  year = {2026},
  month = {Jun},
  publisher = {American Physical Society},
  doi = {10.1103/p7c9-x1m9},
  url = {https://link.aps.org/doi/10.1103/p7c9-x1m9}
}

@article{jing-intrinsic-heralding,
  title = {Intrinsic Heralding and Optimal Decoders for Non-Abelian Topological Order},
  author = {Jing, Dian and Sala, Pablo and Jiang, Liang and Verresen, Ruben},
  journal = {Phys. Rev. Lett.},
  volume = {136},
  issue = {12},
  pages = {120405},
  numpages = {9},
  year = {2026},
  month = {Mar},
  publisher = {American Physical Society},
  doi = {10.1103/ccj7-ctd8},
  url = {https://link.aps.org/doi/10.1103/ccj7-ctd8}
}

@article{Kobayashi_2026,
  title = {Clifford Hierarchy Stabilizer Codes: Transversal Non-Clifford Gates and Magic States},
  author = {Kobayashi, Ryohei and Zhu, Guanyu and Hsin, Po-Shen},
  journal = {Phys. Rev. Lett.},
  volume = {136},
  issue = {25},
  pages = {250802},
  numpages = {7},
  year = {2026},
  month = {Jun},
  publisher = {American Physical Society},
  doi = {10.1103/j2q4-bnc6},
  url = {https://link.aps.org/doi/10.1103/j2q4-bnc6}
}

@misc{huang2026hybridlatticesurgerynonclifford,
      title={Hybrid Lattice Surgery: Non-Clifford Gates via Non-Abelian Surface Codes}, 
      author={Sheng-Jie Huang and Alison Warman and Sakura Schafer-Nameki and Yanzhu Chen},
      year={2026},
      eprint={2510.20890},
      archivePrefix={arXiv},
      primaryClass={quant-ph},
      url={https://arxiv.org/abs/2510.20890}, 
}

@article{Chamberland_2019,
   title={Fault-tolerant magic state preparation with flag qubits},
   volume={3},
   ISSN={2521-327X},
   url={http://dx.doi.org/10.22331/q-2019-05-20-143},
   DOI={10.22331/q-2019-05-20-143},
   journal={Quantum},
   publisher={Verein zur Forderung des Open Access Publizierens in den Quantenwissenschaften},
   author={Chamberland, Christopher and Cross, Andrew W.},
   year={2019},
   month=May, pages={143} }

@article{chamberland2020very,
   title={Very low overhead fault-tolerant magic state preparation using redundant ancilla encoding and flag qubits},
   volume={6},
   ISSN={2056-6387},
   url={http://dx.doi.org/10.1038/s41534-020-00319-5},
   DOI={10.1038/s41534-020-00319-5},
   number={1},
   journal={npj Quantum Information},
   publisher={Springer Science and Business Media LLC},
   author={Chamberland, Christopher and Noh, Kyungjoo},
   year={2020},
   month=Oct }

@article{Scruby_2022,
   title={Numerical Implementation of Just-In-Time Decoding in Novel Lattice Slices Through the Three-Dimensional Surface Code},
   volume={6},
   ISSN={2521-327X},
   url={http://dx.doi.org/10.22331/q-2022-05-24-721},
   DOI={10.22331/q-2022-05-24-721},
   journal={Quantum},
   publisher={Verein zur Forderung des Open Access Publizierens in den Quantenwissenschaften},
   author={Scruby, T. R. and Browne, D. E. and Webster, P. and Vasmer, M.},
   year={2022},
   month=May, pages={721} }

@misc{bauer2026findingdiagonallogicalgates,
      title={Finding diagonal logical gates in CSS codes and circuits}, 
      author={Andreas Bauer},
      year={2026},
      eprint={2607.26477},
      archivePrefix={arXiv},
      primaryClass={quant-ph},
      url={https://arxiv.org/abs/2607.26477}, 
}

@article{Menon_2026,
  title = {Magic Tricycles: Efficient Magic-State Generation with Finite Block-Length Quantum LDPC Codes},
  author = {Menon, Varun and Bonilla Ataides, J. Pablo and Mehta, Rohan and Gu, Andi and Tan, Daniel Bochen and Lukin, Mikhail D.},
  journal = {Phys. Rev. X},
  volume = {16},
  issue = {2},
  pages = {021014},
  numpages = {37},
  year = {2026},
  month = {Apr},
  publisher = {American Physical Society},
  doi = {10.1103/ghhp-cytl},
  url = {https://link.aps.org/doi/10.1103/ghhp-cytl}
}

@misc{jacob2026singleshotdecodingfaulttolerantgates,
      title={Single-Shot Decoding and Fault-tolerant Gates with Trivariate Tricycle Codes}, 
      author={Abraham Jacob and Campbell McLauchlan and Dan E. Browne},
      year={2026},
      eprint={2508.08191},
      archivePrefix={arXiv},
      primaryClass={quant-ph},
      url={https://arxiv.org/abs/2508.08191}, 
}

@misc{gidney2024magic,
      title={Magic state cultivation: growing T states as cheap as CNOT gates}, 
      author={Craig Gidney and Noah Shutty and Cody Jones},
      year={2024},
      eprint={2409.17595},
      archivePrefix={arXiv},
      primaryClass={quant-ph},
      url={https://arxiv.org/abs/2409.17595}, 
}

@article{gupta2024encoding,
   title={Encoding a magic state with beyond break-even fidelity},
   volume={625},
   ISSN={1476-4687},
   url={http://dx.doi.org/10.1038/s41586-023-06846-3},
   DOI={10.1038/s41586-023-06846-3},
   number={7994},
   journal={Nature},
   publisher={Springer Science and Business Media LLC},
   author={Gupta, Riddhi S. and Sundaresan, Neereja and Alexander, Thomas and Wood, Christopher J. and Merkel, Seth T. and Healy, Michael B. and Hillenbrand, Marius and Jochym-O’Connor, Tomas and Wootton, James R. and Yoder, Theodore J. and Cross, Andrew W. and Takita, Maika and Brown, Benjamin J.},
   year={2024},
   month=Jan, pages={259–263} }

@article{hetenyi2026constant,
   title={Constant-Depth Magic State Cultivation with Clifford Measurements by Gauging},
   volume={137},
   ISSN={1079-7114},
   url={http://dx.doi.org/10.1103/f7fz-6zsy},
   DOI={10.1103/f7fz-6zsy},
   number={14},
   journal={Physical Review Letters},
   publisher={American Physical Society (APS)},
   author={Hetényi, Bence and Brown, Benjamin J. and Williamson, Dominic J.},
   year={2026},
   month=Sept }

@article{brown2020fault,
   title={A fault-tolerant non-Clifford gate for the surface code in two dimensions},
   volume={6},
   ISSN={2375-2548},
   url={http://dx.doi.org/10.1126/sciadv.aay4929},
   DOI={10.1126/sciadv.aay4929},
   number={21},
   journal={Science Advances},
   publisher={American Association for the Advancement of Science (AAAS)},
   author={Brown, Benjamin J.},
   year={2020},
   month=May }

@misc{williamson2026fast,
      title={Fast magic state preparation by gauging higher-form transversal gates in parallel}, 
      author={Dominic J. Williamson},
      year={2026},
      eprint={2601.22939},
      archivePrefix={arXiv},
      primaryClass={quant-ph},
      url={https://arxiv.org/abs/2601.22939}, 
}

@misc{christos2026nonabelianquantumlowdensityparity,
      title={Non-Abelian Quantum Low-Density Parity Check Codes and Non-Clifford Operations from Gauging Logical Gates via Measurements}, 
      author={Maine Christos and Chiu Fan Bowen Lo and Vedika Khemani and Rahul Sahay},
      year={2026},
      eprint={2602.12228},
      archivePrefix={arXiv},
      primaryClass={quant-ph},
      url={https://arxiv.org/abs/2602.12228}, 
}

@misc{panteleev2022asymptotically,
      title={Asymptotically Good Quantum and Locally Testable Classical LDPC Codes}, 
      author={Pavel Panteleev and Gleb Kalachev},
      year={2022},
      eprint={2111.03654},
      archivePrefix={arXiv},
      primaryClass={cs.IT},
      url={https://arxiv.org/abs/2111.03654}, 
}

@article{Moussa_2016,
   title={Transversal Clifford gates on folded surface codes},
   volume={94},
   ISSN={2469-9934},
   url={http://dx.doi.org/10.1103/PhysRevA.94.042316},
   DOI={10.1103/physreva.94.042316},
   number={4},
   journal={Physical Review A},
   publisher={American Physical Society (APS)},
   author={Moussa, Jonathan E.},
   year={2016},
   month=Oct }

@article{roffe2020,
  title = {Decoding across the quantum low-density parity-check code landscape},
  author = {Roffe, Joschka and White, David R. and Burton, Simon and Campbell, Earl},
  journal = {Phys. Rev. Res.},
  volume = {2},
  issue = {4},
  pages = {043423},
  numpages = {13},
  year = {2020},
  month = {Dec},
  publisher = {American Physical Society},
  doi = {10.1103/PhysRevResearch.2.043423},
  url = {https://link.aps.org/doi/10.1103/PhysRevResearch.2.043423}
}

@misc{huang2023improvednoisysyndromedecoding,
      title={Improved Noisy Syndrome Decoding of Quantum LDPC Codes with Sliding Window}, 
      author={Shilin Huang and Shruti Puri},
      year={2023},
      eprint={2311.03307},
      archivePrefix={arXiv},
      primaryClass={quant-ph},
      url={https://arxiv.org/abs/2311.03307}, 
}

@misc{gong2024lowlatencyiterativedecodingqldpc,
      title={Toward Low-latency Iterative Decoding of QLDPC Codes Under Circuit-Level Noise}, 
      author={Anqi Gong and Sebastian Cammerer and Joseph M. Renes},
      year={2024},
      eprint={2403.18901},
      archivePrefix={arXiv},
      primaryClass={quant-ph},
      url={https://arxiv.org/abs/2403.18901}, 
}

@misc{knill2004faulttolerantpostselectedquantumcomputation,
      title={Fault-Tolerant Postselected Quantum Computation: Schemes}, 
      author={E. Knill},
      year={2004},
      eprint={quant-ph/0402171},
      archivePrefix={arXiv},
      primaryClass={quant-ph},
      url={https://arxiv.org/abs/quant-ph/0402171}, 
}

@misc{wills2024constantoverheadmagicstatedistillation,
      title={Constant-Overhead Magic State Distillation}, 
      author={Adam Wills and Min-Hsiu Hsieh and Hayata Yamasaki},
      year={2024},
      eprint={2408.07764},
      archivePrefix={arXiv},
      primaryClass={quant-ph},
      url={https://arxiv.org/abs/2408.07764}, 
}

@misc{meier2012magicstatedistillationfourqubitcode,
      title={Magic-state distillation with the four-qubit code}, 
      author={Adam M. Meier and Bryan Eastin and Emanuel Knill},
      year={2012},
      eprint={1204.4221},
      archivePrefix={arXiv},
      primaryClass={quant-ph},
      url={https://arxiv.org/abs/1204.4221}, 
}

@article{Campbell_2017,
   title={Roads towards fault-tolerant universal quantum computation},
   volume={549},
   ISSN={1476-4687},
   url={http://dx.doi.org/10.1038/nature23460},
   DOI={10.1038/nature23460},
   number={7671},
   journal={Nature},
   publisher={Springer Science and Business Media LLC},
   author={Campbell, Earl T. and Terhal, Barbara M. and Vuillot, Christophe},
   year={2017},
   month=Sept, pages={172–179} }

@article{Panteleev_2021,
   title={Degenerate Quantum LDPC Codes With Good Finite Length Performance},
   volume={5},
   ISSN={2521-327X},
   url={http://dx.doi.org/10.22331/q-2021-11-22-585},
   DOI={10.22331/q-2021-11-22-585},
   journal={Quantum},
   publisher={Verein zur Forderung des Open Access Publizierens in den Quantenwissenschaften},
   author={Panteleev, Pavel and Kalachev, Gleb},
   year={2021},
   month=Nov, pages={585} }

@misc{takada2026exactefficientsimulationnoisy,
      title={Exact efficient simulation of noisy logical magic states using Clifford stabilizers}, 
      author={Yugo Takada and Stephen D. Bartlett and Dominic J. Williamson},
      year={2026},
      eprint={2609.16929},
      archivePrefix={arXiv},
      primaryClass={quant-ph},
      url={https://arxiv.org/abs/2609.16929}, 
}

@misc{delafuente2026efficientlyestimatingfailurerates,
      title={Efficiently estimating failure rates of fault-tolerant logical non-Clifford blocks}, 
      author={Julio C. Magdalena de la Fuente and Thomas R. Scruby},
      year={2026},
      eprint={2609.19485},
      archivePrefix={arXiv},
      primaryClass={quant-ph},
      url={https://arxiv.org/abs/2609.19485}, 
}

@article{Surti_2026,
  title = {Efficient Simulation of Logical Magic State Preparation Protocols},
  author = {Surti, Samyak and Daguerre, Lucas and Kim, Isaac H.},
  journal = {PRX Quantum},
  volume = {7},
  issue = {2},
  pages = {020329},
  numpages = {32},
  year = {2026},
  month = {May},
  publisher = {American Physical Society},
  doi = {10.1103/fby6-xjbm},
  url = {https://link.aps.org/doi/10.1103/fby6-xjbm}
}

@article{Quintavalle_2023,
   title={Partitioning qubits in hypergraph product codes to implement logical gates},
   volume={7},
   ISSN={2521-327X},
   url={http://dx.doi.org/10.22331/q-2023-10-24-1153},
   DOI={10.22331/q-2023-10-24-1153},
   journal={Quantum},
   publisher={Verein zur Forderung des Open Access Publizierens in den Quantenwissenschaften},
   author={Quintavalle, Armanda O. and Webster, Paul and Vasmer, Michael},
   year={2023},
   month=Oct, pages={1153} }

@article{Breuckmann2024,
   title={Fold-Transversal Clifford Gates for Quantum Codes},
   volume={8},
   ISSN={2521-327X},
   url={http://dx.doi.org/10.22331/q-2024-06-13-1372},
   DOI={10.22331/q-2024-06-13-1372},
   journal={Quantum},
   publisher={Verein zur Forderung des Open Access Publizierens in den Quantenwissenschaften},
   author={Breuckmann, Nikolas P. and Burton, Simon},
   year={2024},
   month=June, pages={1372} }

@article{bravyi2005universal,
  title={Universal quantum computation with ideal Clifford gates and noisy ancillas},
  author={Bravyi, Sergey and Kitaev, Alexei},
  journal={Physical Review A—Atomic, Molecular, and Optical Physics},
  volume={71},
  number={2},
  pages={022316},
  year={2005},
  publisher={APS}
}

@article{cohen2022low,
   title={Low-overhead fault-tolerant quantum computing using long-range connectivity},
   volume={8},
   ISSN={2375-2548},
   url={http://dx.doi.org/10.1126/sciadv.abn1717},
   DOI={10.1126/sciadv.abn1717},
   number={20},
   journal={Science Advances},
   publisher={American Association for the Advancement of Science (AAAS)},
   author={Cohen, Lawrence Z. and Kim, Isaac H. and Bartlett, Stephen D. and Brown, Benjamin J.},
   year={2022},
   month=May }

@misc{chang2026constant,
      title={Constant-Time Surgery on 2D Hypergraph Product Codes with Near-Constant Space Overhead}, 
      author={Kathleen Chang and Zhiyang He and Theodore J. Yoder and Guanyu Zhu and Tomas Jochym-O'Connor},
      year={2026},
      eprint={2603.02157},
      archivePrefix={arXiv},
      primaryClass={quant-ph},
      url={https://arxiv.org/abs/2603.02157}, 
}

@misc{cowtan2025fast,
      title={Fast and fault-tolerant logical measurements: Auxiliary hypergraphs and transversal surgery}, 
      author={Alexander Cowtan and Zhiyang He and Dominic J. Williamson and Theodore J. Yoder},
      year={2025},
      eprint={2510.14895},
      archivePrefix={arXiv},
      primaryClass={quant-ph},
      url={https://arxiv.org/abs/2510.14895}, 
}

@misc{zheng2025high,
      title={High-Rate Surgery: towards constant-overhead logical operations}, 
      author={Guo Zheng and Liang Jiang and Qian Xu},
      year={2025},
      eprint={2510.08523},
      archivePrefix={arXiv},
      primaryClass={quant-ph},
      url={https://arxiv.org/abs/2510.08523}, 
}

@misc{li2026transversal,
      title={Transversal dimension jump for product qLDPC codes}, 
      author={Christine Li and John Preskill and Qian Xu},
      year={2026},
      eprint={2510.07269},
      archivePrefix={arXiv},
      primaryClass={quant-ph},
      url={https://arxiv.org/abs/2510.07269}, 
}

@misc{baspin2025fast,
      title={Fast surgery for quantum LDPC codes}, 
      author={Nouédyn Baspin and Lucas Berent and Lawrence Z. Cohen},
      year={2025},
      eprint={2510.04521},
      archivePrefix={arXiv},
      primaryClass={quant-ph},
      url={https://arxiv.org/abs/2510.04521}, 
}

@misc{tan2025single,
      title={Single-Shot Universality in Quantum LDPC Codes via Code-Switching}, 
      author={Shi Jie Samuel Tan and Yifan Hong and Ting-Chun Lin and Michael J. Gullans and Min-Hsiu Hsieh},
      year={2025},
      eprint={2510.08552},
      archivePrefix={arXiv},
      primaryClass={quant-ph},
      url={https://arxiv.org/abs/2510.08552}, 
}

@article{bravyi2010tradeoffs,
   title={Tradeoffs for Reliable Quantum Information Storage in 2D Systems},
   volume={104},
   ISSN={1079-7114},
   url={http://dx.doi.org/10.1103/PhysRevLett.104.050503},
   DOI={10.1103/physrevlett.104.050503},
   number={5},
   journal={Physical Review Letters},
   publisher={American Physical Society (APS)},
   author={Bravyi, Sergey and Poulin, David and Terhal, Barbara},
   year={2010},
   month=Feb }

@misc{leverrier2022quantum,
      title={Quantum Tanner codes}, 
      author={Anthony Leverrier and Gilles Zémor},
      year={2022},
      eprint={2202.13641},
      archivePrefix={arXiv},
      primaryClass={quant-ph},
      url={https://arxiv.org/abs/2202.13641}, 
}

@misc{dinur2022goodquantumldpccodes,
      title={Good Quantum LDPC Codes with Linear Time Decoders}, 
      author={Irit Dinur and Min-Hsiu Hsieh and Ting-Chun Lin and Thomas Vidick},
      year={2022},
      eprint={2206.07750},
      archivePrefix={arXiv},
      primaryClass={quant-ph},
      url={https://arxiv.org/abs/2206.07750}, 
}

@article{sajith2025noncliffordgatesstabilizercodes,
  title = {Non-Clifford Gates between Stabilizer Codes via Non-Abelian Topological Order},
  author = {Sajith, Rohith and Song, Zijian and Roberts, Brenden and Menon, Varun and Li, Yabo},
  journal = {PRX Quantum},
  volume = {7},
  issue = {1},
  pages = {010361},
  numpages = {24},
  year = {2026},
  month = {Mar},
  publisher = {American Physical Society},
  doi = {10.1103/qttm-y8t1},
  url = {https://link.aps.org/doi/10.1103/qttm-y8t1}
}

@misc{huang2025generatinglogicalmagicstates,
      title={Generating logical magic states with the aid of non-Abelian topological order}, 
      author={Sheng-Jie Huang and Yanzhu Chen},
      year={2025},
      eprint={2502.00998},
      archivePrefix={arXiv},
      primaryClass={quant-ph},
      url={https://arxiv.org/abs/2502.00998}, 
}

@misc{vaknin2026efficientmagicstatecultivation,
      title={Efficient Magic State Cultivation on the Surface Code}, 
      author={Yotam Vaknin and Shoham Jacoby and Arne Grimsmo and Alex Retzker},
      year={2026},
      eprint={2502.01743},
      archivePrefix={arXiv},
      primaryClass={quant-ph},
      url={https://arxiv.org/abs/2502.01743}, 
}

@article{Tantivasadakarn_2024,
  title = {Long-Range Entanglement from Measuring Symmetry-Protected Topological Phases},
  author = {Tantivasadakarn, Nathanan and Thorngren, Ryan and Vishwanath, Ashvin and Verresen, Ruben},
  journal = {Phys. Rev. X},
  volume = {14},
  issue = {2},
  pages = {021040},
  numpages = {22},
  year = {2024},
  month = {Jun},
  publisher = {American Physical Society},
  doi = {10.1103/PhysRevX.14.021040},
  url = {https://link.aps.org/doi/10.1103/PhysRevX.14.021040}
}

@article{Williamson_2021,
  title = {Type-II fractons from coupled spin chains and layers},
  author = {Williamson, Dominic J. and Devakul, Trithep},
  journal = {Phys. Rev. B},
  volume = {103},
  issue = {15},
  pages = {155140},
  numpages = {14},
  year = {2021},
  month = {Apr},
  publisher = {American Physical Society},
  doi = {10.1103/PhysRevB.103.155140},
  url = {https://link.aps.org/doi/10.1103/PhysRevB.103.155140}
}

@misc{kitaev1995quantummeasurementsabelianstabilizer,
      title={Quantum measurements and the Abelian Stabilizer Problem}, 
      author={A. Yu. Kitaev},
      year={1995},
      eprint={quant-ph/9511026},
      archivePrefix={arXiv},
      primaryClass={quant-ph},
      url={https://arxiv.org/abs/quant-ph/9511026}, 
}

@misc{bombin2018transversal,
      title={Transversal gates and error propagation in 3D topological codes}, 
      author={Hector Bombin},
      year={2018},
      eprint={1810.09575},
      archivePrefix={arXiv},
      primaryClass={quant-ph},
      url={https://arxiv.org/abs/1810.09575}, 
}

@misc{yuan2026parsimonious,
      title={Parsimonious Quantum Low-Density Parity-Check Code Surgery}, 
      author={Andrew C. Yuan and Alexander Cowtan and Zhiyang He and Ting-Chun Lin and Dominic J. Williamson},
      year={2026},
      eprint={2603.05082},
      archivePrefix={arXiv},
      primaryClass={quant-ph},
      url={https://arxiv.org/abs/2603.05082}, 
}

@article{Kitaev_2003,
   title={Fault-tolerant quantum computation by anyons},
   volume={303},
   ISSN={0003-4916},
   url={http://dx.doi.org/10.1016/S0003-4916(02)00018-0},
   DOI={10.1016/s0003-4916(02)00018-0},
   number={1},
   journal={Annals of Physics},
   publisher={Elsevier BV},
   author={Kitaev, A.Yu.},
   year={2003},
   month=Jan, pages={2–30} }

@article{Prem_2019,
   title={Gauging permutation symmetries as a route to non-Abelian fractons},
   volume={7},
   ISSN={2542-4653},
   url={http://dx.doi.org/10.21468/SciPostPhys.7.5.068},
   DOI={10.21468/scipostphys.7.5.068},
   number={5},
   journal={SciPost Physics},
   publisher={Stichting SciPost},
   author={Prem, Abhinav and Williamson, Dominic},
   year={2019},
   month=Nov }
    \end{document}
\fi

\clearpage 

\section{Notation}

\begingroup
\renewcommand{\arraystretch}{1.25}
\setlength{\LTleft}{0pt}
\setlength{\LTright}{0pt}

{\small
\begin{longtable}{@{}>{\raggedright\arraybackslash}p{0.185\textwidth}@{\hspace{6pt}}>{\raggedright\arraybackslash}p{0.175\textwidth}@{\hspace{6pt}}p{0.075\textwidth}@{\hspace{6pt}}>{\raggedright\arraybackslash}p{0.475\textwidth}@{}}
\toprule
\textbf{Notation} & \textbf{Name} & \textbf{Sec.} & \textbf{Meaning} \\
\midrule
\endfirsthead
\multicolumn{4}{@{}l}{\textit{Table of notations (continued)}}\\
\toprule
\textbf{Notation} & \textbf{Name} & \textbf{Sec.} & \textbf{Meaning} \\
\midrule
\endhead
\midrule
\multicolumn{4}{r@{}}{\textit{continued on next page}}\\
\endfoot
\bottomrule
\endlastfoot

$\T$ &
transversal gate &
all &
Transversal logical gate being gauged, with $\T^p = \id$. \\

$q, p$ & qudit dimensions & all & the original code has qudits with dimension $q$, and the added gauge qudits have dimension $p$. \\

$\physq$ &
matter-qudit set &
all &
The set of matter qudits of the input/original code. \\

$v\in \vertexV$ &
vertex qudits &
\ref{sec:gauging-procedure} &
Vertex qudits are ancillas that measure vertex stabilizers. They form vertices of the gauging graph $\graphG$. Additionally, a vertex $v$ indexes a set of matter qudits $\vertexset{v}$ and its vertex stabilizer $\av$\\

$\Sigma:\physq\to\vertexV$ &
vertex-set map &
\ref{sec:gauging-procedure} &
Assignment of each matter qudit to its vertex set. Vertex sets are constant size and respect the structure of the transversal gate $\T$. \\

$\vertexset{v}$ &
vertex set &
\ref{sec:gauging-procedure} &
The block of matter qudits assigned to vertex $v$. \\

$\T_v$ &
on-site symmetry &
\ref{sec:gauging-procedure} &
Action of the transversal gate $\T$ on vertex set $\vertexset{v}$. \\

$\Agrp$ &
symmetry group &
\ref{sec:space-distance} &
A finite Abelian group $\Agrp$. When $\Agrp$ is the cyclic group of order $p$, $\T^p=\id$. \\

$\widehat{\Agrp}$ &
representation group &
\ref{sec:space-distance} &
Additively written group of irreducible representations (characters) of $\Agrp$; the coefficient group of all representation-label chains. \\

$\T^{\bm a}$,\ $\T^g=\T^{g\id}$ &
symmetry operators &
\ref{sec:space-distance} &
$\T^{\bm a}=\prod_{v\in \vertexV}\T_v^{a_v}$ for a vertex labeling $\bm a\in \Agrp[\vertexV]$; the constant labeling where $(g\id)_v = g\ \forall v\in \vertexV$ gives the global symmetry $\T^g$. \\

$\bm\chi\in \widehat{\Agrp}[\vertexV], \omega^{\bm\chi}(\bm a)$  &
eigenoperator label on each vertex &
\ref{sec:space-distance} &
$\bm\chi$ is a labeling of vertices by $\widehat{\Agrp}$ coefficients.  $\omega^{\bm\chi}(\bm a) := \prod_v \omega^{\chi_v}(a_v)$. \\

$\Xp^a,\Zp^\chi$ &
gauge Pauli operators &
\ref{sec:gauging-procedure} &
Generalized Pauli (shift/clock) operators on $|\Agrp|$-dimensional gauge qudits; $\Xp^a\ket{x}=\ket{x+a}$, $\Zp^\chi\ket{x}=\omega^\chi(x)\ket{x}$, with strings $\Xp^{\bm a}$ and $\Zp^{\bm\alpha}$. \\

$\av^g$,\ $A(\bm a)$ &
vertex stabilizer, product of vertex stabilizers according to $\bm a\in \Agrp[\vertexV]$&
\ref{sec:gauging-procedure} & $\av^g=\T_v^g\otimes\Xp^{\delta_1 g\bm e_v}$; general finite-Abelian form $A(\bm a):=\prod_{v\in\vertexV}\av^{a_v}=\T^{\bm a}\otimes\Xp^{\delta_1\bm a}$.  For fixed $g$, the product over all vertices recovers the symmetry $\T^g$. \\

$\bp$,\ $\gaugedB_f(\chi)$ &
cycle stabilizer &
\ref{sec:gauging-procedure} &
Cycle stabilizer $\bp=\prod_{e\in\partial_2 f}\cZ_e^{\eta_f(e)}$ for a face $f\in\faceF$, whose boundary is a cycle: $\partial_1\partial_2 f=0$; we also write $\gaugedB_p(\chi)=\Zp^{\chi\,\partial_2 f}$. \\

$\graphG_\bullet, \graphG^\bullet$ &
Gauging graph chain and cochain complex. &
\ref{sec:space-distance} &
Chains are labeled by $\widehat{\Agrp}$ coefficients, cochains are labeled by $\Agrp$ coefficients. $\graphG^\bullet = \Agrp[\vertexV] \xrightarrow{\delta_{1}} 
    \Agrp[\edgeE] \xrightarrow{\delta_2} 
    \Agrp[\faceF]$, $\graphG_\bullet = \widehat{\Agrp}[\vertexV] \xleftarrow{\partial_{1}}
    \widehat{\Agrp}[\edgeE] \xleftarrow{\partial_2} 
    \widehat{\Agrp}[\faceF] $\\

$\mathcal H_{\mathrm m}$, $\mathcal H_{\mathrm g}$ &
matter and gauge Hilbert spaces &
\ref{sec:space-distance} &
$\mathcal H_{\mathrm m}$ is the Hilbert space of the original matter qudits, and
$\mathcal H_{\mathrm g}:=\bigotimes_{e\in\edgeE}\mathbb C[\Agrp]$
is the Hilbert space of the gauge qudits. \\

$\cC$,\ $\cC'$ &
base code or original code, gauged code. &
\ref{sec:gauging-procedure} &
The input stabilizer code $\cC$ with stabilizer group $\mathcal S=\langle S_i\rangle$, and the deformed code $\cC'=\langle\gaugedS_i,\av,\bp\rangle$ produced by gauging.  \\
$\cC_\mathrm{sym.}$ &
Symmetric codespace &
\ref{sec:space-distance} &
symmetric codespace is defined $\cC_\mathrm{sym.}:=\cC\cap \mathcal{H}_\mathrm{sym.}$.  \\

$\Gamma$ &
gauging map &
\ref{sec:space-distance} &
$\Gamma:\mathcal H_{\rm sym.}\xrightarrow{}\mathcal H_{\rm m}\otimes \mathcal{H}_{\rm g}$.\\

$\Gamma[M]$ & gauging map on matter operator $M$ & \ref{sec:space-distance} &
Sends a matter operator to its decorated counterpart. When $M$ is a symmetrized stabilizer, it maps it to its decorated version $\overline{\gaugedS}=\Gamma[\bar{S}]$. \\

$\aS{i}$, $\nS{i}$ &
Abelian and non-Abelian stabilizers &
\ref{sec:spacetime-distance} &
$\mathbb{Z}_2$ gauged code stabilizers can be partitioned into non-Abelian stabilizers $\nS{}$ that always commute, while Abelian stabilizers satisfy $\ComGrp{\aS{i}}{\aS{j}} \in \langle \nS{} \rangle$.\\

$\mathcal H_{\mathrm{sym.}}$ &
symmetric subspace &
\ref{sec:space-distance} &
$\mathcal H_{\mathrm{sym.}}
:=\{\ket{\psi}\in\mathcal H_{\mathrm m}:
\T^g\ket{\psi}=\ket{\psi}\ \forall g\in\Agrp\}$. \\

$\ComGrp{A}{B}$ &
group commutator &
\ref{sec:gauging-procedure} &
Group/multiplicative commutator $\ComGrp{A}{B}=ABA^\dagger B^\dagger$. \\

$\supp{O}$, $\operatorname{Wt}(O)$ &
support and weight &
\ref{sec:space-distance} &
$\supp{O}$ is the set of matter and gauge qudits on which $O$ acts nontrivially, and
$\operatorname{Wt}(O):=|\supp{O}|$. 
\\

$d$;\ $d_{\rm g}$ &
original and gauged code distance &
\ref{sec:spacetime-distance} &
The minimum weights of operators with nonscalar projected action on $\cC$ and $\cC'$ respectively. \\

$\Pi_{\cC}$, $\Pi_{\cC_\mathrm{sym.}}$, $\Pi_{\cC'}$ & code projectors & \ref{sec:space-distance} & Projectors onto the original, symmetric, and gauged codespace.\\

$M_{\bm\chi}$ & matter eigenoperator & \ref{sec:space-distance} & The component of a matter operator $M$ that satisfies $\T^{\bm a} M_{\bm\chi} (\T^{\bm a})^\dagger= \omega^{\bm\chi}(\bm a) M_{\bm\chi}.$\\

$\sum_{\bm\chi}M_{\bm\chi}$ & Eigenoperator decomposition & \ref{sec:space-distance} & Every matter operator $M$ can be written as $M=\sum_{\bm\chi\in \widehat{\Agrp}[\vertexV]} M_{\bm\chi}$\\

$\E{i}{j}\in E$ &
gauge-qubit edge &
\ref{sec:gauging-procedure} &
Gauge qubit on the edge adjacent to vertices $i, j$. \\

$\fa$, $\fn$ &
Abelian and non-Abelian faults &
\ref{sec:spacetime-distance} &
Elementary space faults on a single vertex set: $\fa$ commutes with all non-Abelian stabilizers \\

$\graphG=(\vertexV, \edgeE)= (\vertexV,\edgeE,\faceF)$ &
gauging graph &
\ref{sec:gauging-procedure} &
Graph whose vertices host the vertex ancillas (one per vertex set) and whose edges host the gauge qudits; required to have $O(1)$ degree and a sparse cycle basis. \\

$S$, $\gaugedS$ &
ungauged, gauged stabilizer &
\ref{sec:gauging-procedure} &
Stabilizer of the original code and its gauged counterpart. \\

$\sym$, $\overline{\gaugedS}$ &
ungauged, gauged symmetrized stabilizer &
\ref{sec:gauging-procedure} &
$\sym = \frac{1}{p}\sum_{a=0}^{p-1}\T^{a}S(\T^{a})^\dagger$, the normalized average over the symmetry group. $\overline{\gaugedS}$ denotes its gauged counterpart. \\

$h_{\Agrp}(\graphG)$ &
generalized Cheeger constant &
\ref{sec:space-distance} &
(Cheeger) expansion constant of the gauging graph.\\

$\nu$ & constant denoting the maximum number of qudits in any vertex set. & \ref{sec:space-distance} & $\nu :=\max_{v\in \vertexV}|\vertexset{v}|$\\

$\mu_v$ &
 &
\ref{sec:gauging-procedure} &
Integer defined by $\ComGrp{S_v}{\T_v}=i^{\mu_v}(N_S)_v$, constrained by $\sum_v\mu_v=0\pmod 4$. \\

$N_S$, $(N_S)_v$, $\mathbb{N}_S$ &
 &
\ref{sec:gauging-procedure} &
$N_S=\ComGrp{S}{\T}$, its restriction to vertex set $v$, and its gauged form $\mathbb{N}_S=\Circo_2[N_S]$. \\

$\Pi_{\rm sym.}$ &
symmetry projector &
\ref{sec:space-distance} &
$\Pi_{\rm sym.}=\frac{1}{|\Agrp|}\sum_{g\in \Agrp}\T^g$ \\

$\mathsf{S}$ &
generalized phase gate &
\ref{sec:gauging-procedure} &
$\mathsf{S}_P = \frac{1 + \prod_{e \in P} Z_e}{2} + i \frac{1 - \prod_{e \in P} Z_e}{2}$. \\

\end{longtable}
}
\endgroup

\end{document}